**Master's thesis**

# Minimal dark matter models with radiative neutrino masses

## From Lagrangians to observables

Simon May

1st June 2018

Advisors: Prof. Dr. Michael Klasen, Dr. Karol Kovařík
Institut für Theoretische Physik
Westfälische Wilhelms-Universität Münster

# Contents

# 1 Introduction

> Theoretical [particle] physics has been starved of new data for more than an entire generation. How can a theoretician choose a good model in the absence of data?
>
> *Chris Lee* [Lee18]

Currently, theoretical particle physics is in a somewhat strange situation. The *Standard Model of particle physics* (SM) has been extremely successful both in explaining observed data from particle physics experiments as well as predicting new phenomena (sometimes decades in advance), with the discovery of the Higgs boson [ATL12; CMS12] (Nobel Prize 2013) as its latest and final great victory. However, from a theoretical perspective, it has long been known to be incomplete. For a long time, physicists have been waiting for data that *contradicts* the Standard Model in search of hints towards a new direction, yet measurements only seem to confirm its predictions with ever-increasing precision. Thus, even though the Standard Model is particle physics' greatest achievement, everyone wants it to fail. In the face of such a lack of radical new evidence, many potential solutions to fill the Standard Model's holes have appeared, but without any experimental guidance, choosing between them becomes speculation at some point or another. The times of a "particle zoo", where new particles were discovered in collider experiments every few years and all that was left to do was to figure out a theory that fits it all together, are long gone. Confronted with such a challenge, a diverse set of strategies for finding new theories seems all the more important.

Even brushing aside the Standard Model's complete and glaring lack of any description of gravity and associated issues like the hierarchy problem, which suggest that it really ought to be part of a larger, unified theory, two other important empirical facts are completely missing: dark matter and neutrino masses. While alternative attempts at explanations for the phenomena ascribed to dark matter still remain, the available evidence clearly favors it (e. g. in the form of *weakly interacting massive* *particles*: *WIMPs*). Further, the discovery of neutrino oscillations [SKK98; SNO02] and thus the confirmation that neutrinos do, in fact, possess mass, has been the subject of a recent Nobel Prize (2015). Taken together, these unexplained phenomena represent quite a sizable hole in the Standard Model's depiction of reality.

This thesis will study minimal extensions of the Standard Model which are capable of both delivering a candidate for dark matter (assuming that it is indeed composed of particles as we know them) and a mechanism which yields massive neutrinos. Approaches like *grand unified theories* (GUT) or *supersymmetry* (SUSY) could be characterized as "top-down", first establishing a new, larger symmetry or general principle and then deriving the phenomenological implications, whereas the "minimal models" under consideration here are "bottom-up", adding only a few fields to the Standard Model to provide dark matter and massive neutrinos, but without any profound prior justifications.

The primary goal is to obtain viable descriptions of both phenomena within the observed experimental constraints from direct detection, indirect detection and collider experiments. Different models provide different testable predictions (such as characteristic signatures in collider experiments) which can be analyzed as new experimental data become available. From a theoretical point of view, such minimal models can be effective theories in some limit. An extensive overview and classification is available in [RZY13]. As an added bonus, some of these models can accommodate even more desirable features, such as gauge coupling unification [Hag+16].

This work will approach these minimal models from two angles: First, general characteristics of models with $\leq 4$ additional fields (which are singlets, doublets or triplets under $SU(2)$), stabilized by a discrete $\mathbb{Z}_2$ symmetry, will be developed further (such as constructing their most general Lagrangian). Second, a promising representative model (T1-3-B with $\alpha = 0$) is analyzed in more detail. The model's phenomenology is studied using established tools like `SARAH` [Sta08; Sta14; Sta15] and `SPheno` [Por03; PS12] and observables like the dark matter relic density are computed using `micrOMEGAs` [Bel+06; Bel+07; Bel+09; Bel+11; Bel+14; Bar+18].

# Experimental and observational evidence

# 2

Dark matter and the issue of the neutrinos' masses are two prime examples of questions in the field of astroparticle physics. While dark matter originally arose as a problem in astrophysics, it is now a big focus in particle physics as well. Any extension of the Standard Model generally tries to accommodate dark matter in some way, based on the idea that the matter is made of elementary particles whose interactions, other than gravity, are unknown. Likewise, the questions of whether neutrinos have mass, and, more recently, why their masses are so small (but non-zero) is an open problem in particle physics. However, neutrinos are also important messengers in astrophysics, where they are produced, for example, in supernovae.

## 2.1. Dark matter

The history of dark matter already goes back quite a while. It first arose when issues with the velocity of luminous matter in galaxies and galaxy clusters were noticed, first by Fritz Zwicky, who measured the velocities of galaxies in the Coma cluster [Sch15a]. Routinely, galaxies or stars were found to have much larger velocities than expected from the mass of the matter surrounding them.

For example, the rotation curve of a galaxy may look as shown in fig. 2.1. The force of gravity keeps objects at a distance $r$ from the center in orbit with a velocity of

$$v = \sqrt{\frac{GM(r)}{r}} \tag{2.1}$$

where $M(r)$ is the mass contained within a sphere of radius $r$. Since most of the mass is concentrated within a galaxy's center, $M(r)$ is roughly constant for large distances

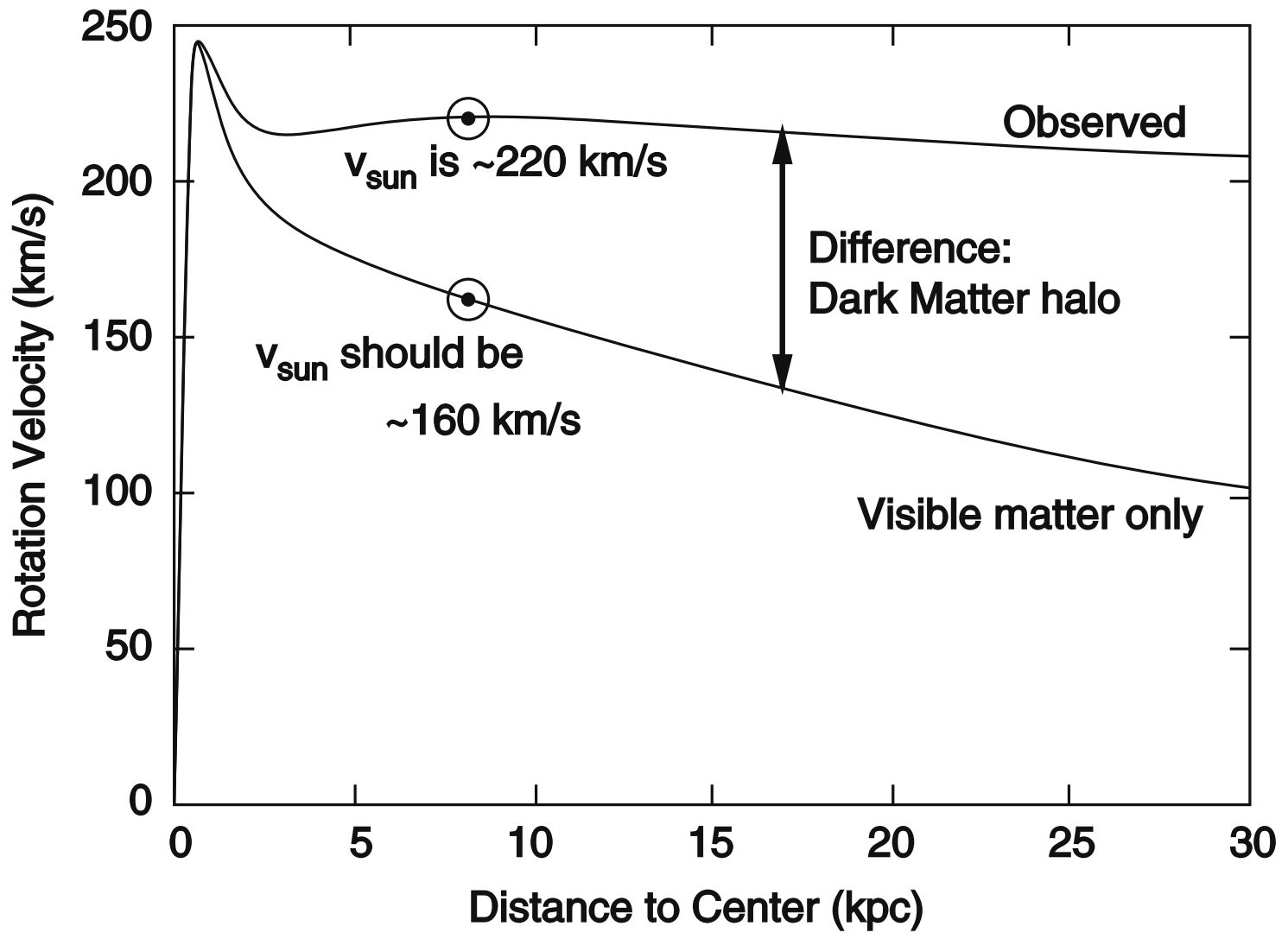

**Figure 2.1.:** Expected and observed rotation curve of the Milky Way galaxy. Taken from [Sch15a].

and one would expect the velocity to behave like

$$v \sim \frac{1}{\sqrt{r}} \quad \text{for large } r \tag{2.2}$$

However, as shown in fig. 2.1, the actual velocity distribution is rather different: Away from the center, it flattens out, remaining at a constant value all the way out to the edge of the galaxy.

This effect cannot be explained by the luminous matter that can be observed, such as stars, which are concentrated in the center. Rather, large amounts of mass seem to be distributed in a *halo* throughout the galaxy, but in a form which does not interact with light and thus remains undetected. This hypothetical unknown matter is called *dark matter*.

Over time, more and more evidence has accumulated in support of the idea that the universe contains a large amount of this unidentified non-luminous matter. Detection has proven difficult since the only known interaction this matter participates in is gravitational.

However, this does not mean that the presence of dark matter is not showing some effects. One of the clearest demonstrations is data obtained from gravitational lensing, with the prime example being the so-called *Bullet Cluster* (1E0657-558) [CGM04].

An image of the Bullet Cluster is shown in fig. 2.2. It actually consists of two galaxy clusters which are in the process of collision. Highlighted in red is the luminous matter (hot gas), which can be observed using X-ray telescopes. Yet, the blue regions are where most of the cluster's mass is concentrated according to measurements of gravitational lensing. The presence of dark matter provides an explanation for this seemingly strange discrepancy: During the collision of the clusters, the gas contained within them interacted electromagnetically and was thus slowed down. The dark matter, which does not interact (or at least only weakly), instead just passed through without being affected. This has caused the cluster's gas and dark matter to separate, resulting in the observed scenario.

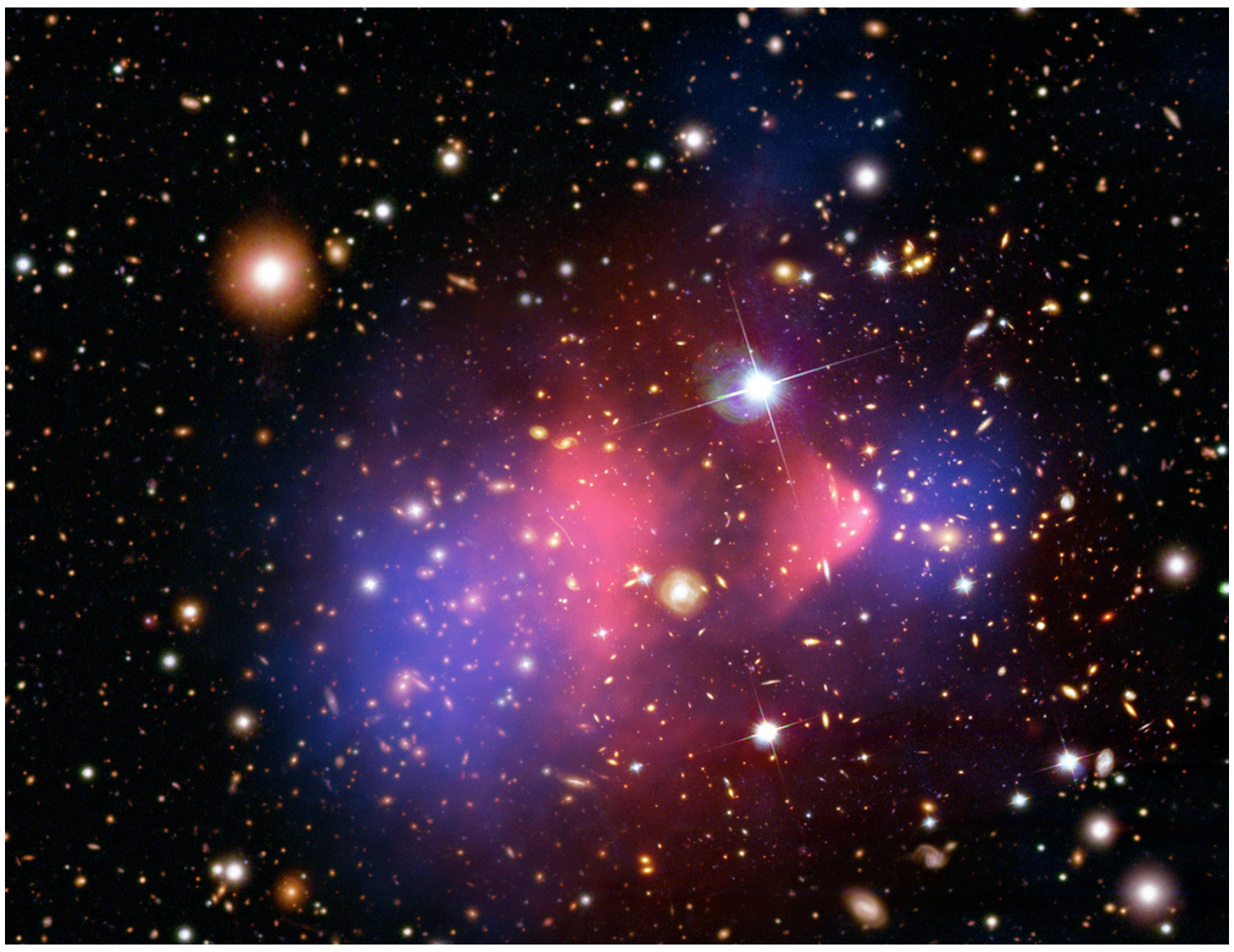

**Figure 2.2.:** Gravitational lensing map of the Bullet Cluster. Source: [Opt16].

Another very important point is the role of dark matter in forming the structure that can be observed in the universe today, even in the *cosmic microwave background* (CMB). Without cold (i. e. moving at non-relativistic speeds, which puts a lower bound on the mass of individual dark matter particles) dark matter, the structure of in the universe could not have formed. Observations like this have come together over time to form the Standard Model of cosmology, the ΛCDM model, where Λ stands for the cosmological constant (also called *dark energy*) and CDM means *cold dark matter*. The surprising result is that of the total energy content of the universe, only around 4 % is contributed by the known baryonic matter. Dark matter, on the other hand, is five times more abundant at around 25 % – yet its identity is still unknown thus far. Finally, dark energy makes up the overwhelming majority at 70 %, and it is understood even less than dark matter [Sch15a, p. 5]. From this perspective,

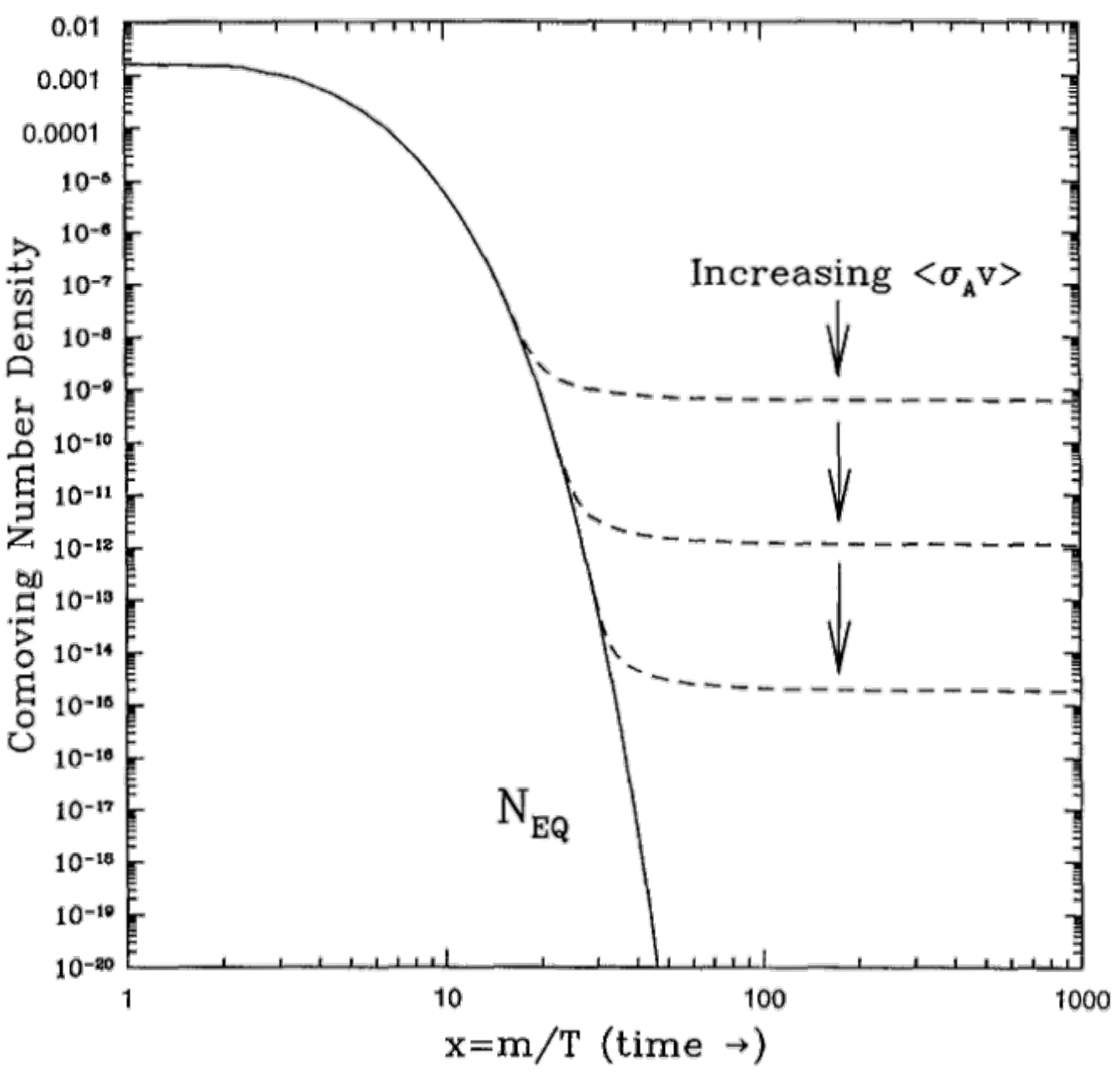

**Figure 2.3.:** Freeze-out mechanism for weakly interacting massive particles. Taken from [JKG96].

it seems as though we still do not know much about the universe at all.

The most favored paradigm for the nature of dark matter is the idea that it could be made of weakly interacting massive particles (WIMPs). Assuming that such particles were produced in thermal equilibrium after the Big Bang, they would be subject to annihilation once the universe cooled down beyond the point where these particles can be produced, steadily reducing their density. However, as the universe expands, annihilation between these particles eventually ceases, leaving behind a *thermal relic*. This mechanism, called the *freeze-out mechanism*, is illustrated in fig. 2.3. The number density $n$ of dark matter particles produced in this way is described by the Boltzmann equation [Ple17]

$$\dot{n}(t) + 3H(t)n(t) = -\langle\sigma v\rangle\Big(n(t)^2 - n_{\text{eq}}(t)^2\Big) \tag{2.3}$$

where $n(t)$ is the number density at the time $t$, $n_{\text{eq}}(t)$ is the number density if there were thermal equilibrium, $H$ is the Hubble parameter and $\langle\sigma v\rangle$ is the thermal average of the product of annihilation cross section $\sigma$ and velocity $v$. Programs like `micrOMEGAs` solve the Boltzmann equation numerically to compute what dark matter density a given model predicts.

The current experimentally determined value for the *dark matter relic density*, that

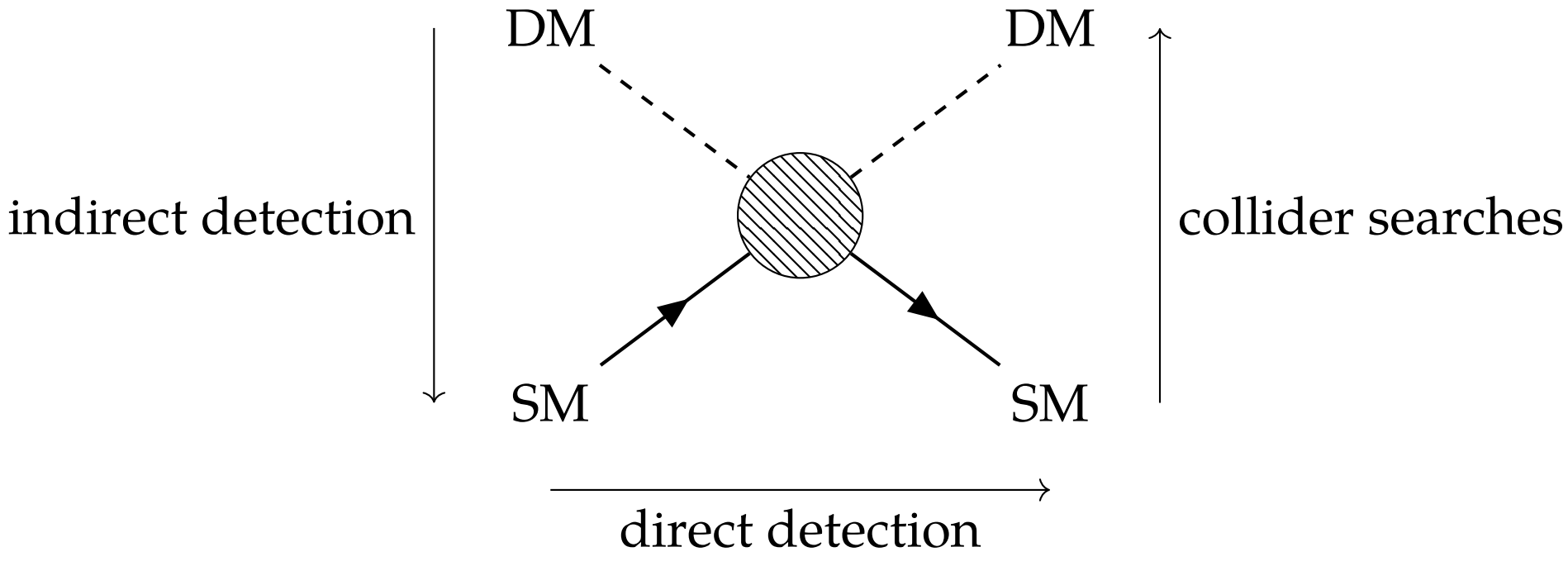

**Figure 2.4.:** A Feynman diagram illustrating the three different ways dark matter (consisting of elementary particles) can be searched for.

is, the density of dark matter in the current universe as a whole, is

$$\Omega_{\mathrm{DM}} h^2 = 0.1186 \pm 0.0020 \tag{2.4}$$

[Planck16; PDG16] where $\Omega_{\mathrm{DM}}$ is the density of dark matter in units of the critical density

$$\rho_{\mathrm{c}} = \frac{3H^2}{8\pi G} \tag{2.5}$$

and $h$ is the Hubble constant (the Hubble parameter at the current time) in units of $100\,\mathrm{km\,s^{-1}\,Mpc^{-1}}$:

$$H_0 = 100h\,\frac{\mathrm{km/s}}{\mathrm{Mpc}} \tag{2.6}$$

It turns out that the correct dark matter density can be obtained via particles whose annihilations are mediated by the weak interaction and have masses of around the electroweak scale (a few hundred GeV to some TeV). This is sometimes called the *WIMP miracle* [Ple17].

After a new particle, such as a WIMP, has been postulated to explain dark matter, the next question is how it could be detected. The three main options are shown in fig. 2.4. Depending on the direction in which this Feynman diagram is read, different kinds of processes result:

**Indirect detection:** Going from top to bottom, the process is the annihilation of two dark matter particles to Standard Model particles. Searches operating from this perspective are called indirect detection searches because dark matter is not observed directly, but through the products of its annihilation. Even though two dark matter particles would only rarely annihilate, astrophysical sources of, for example, gamma rays or neutrinos which cannot be explained by other means can be hints towards the nature of dark matter.

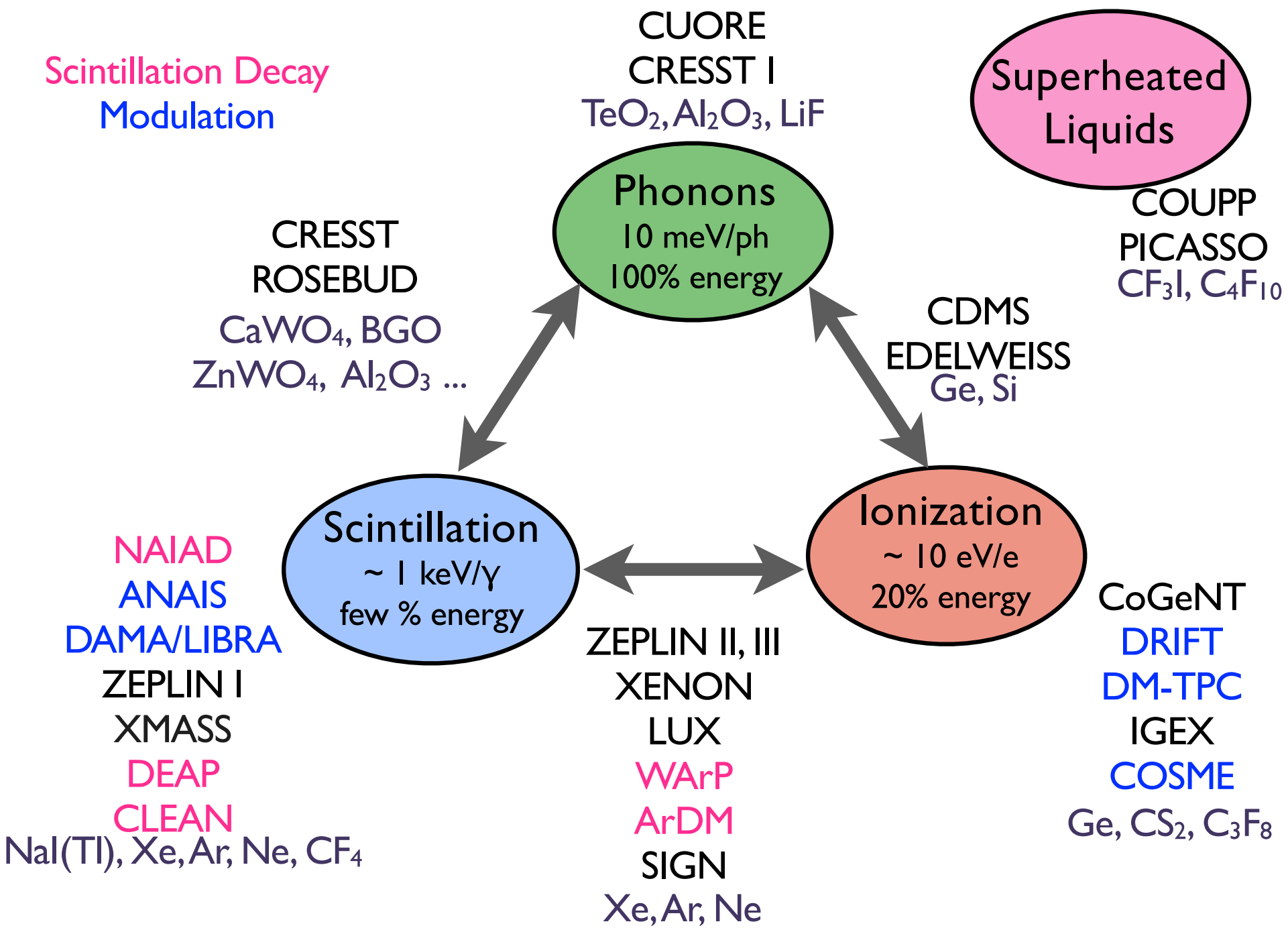

**Figure 2.5.:** An overview of the channels of dark matter direct detection, experiments making use of these channels and detector materials used in each case. Taken from [Saa13].

**Direct detection:** From left to right, the diagram shows the elastic scattering of a dark matter particle on a Standard Model particle, such as an atomic nucleus. In the case of a WIMP, this scattering process can be mediated by the Higgs boson (spin-independent) and the $Z$ boson (spin-dependent).

**Collider searches:** Reading from bottom to top results in the annihilation of two Standard Model particles into dark matter particles. Such a process could happen at particle accelerators: Given enough energy, the collision of two particles can result in the creation of a heavy particle, such as a WIMP. This could be noticed by the fact that the dark matter particle will escape from the detector, which will then report that some energy has gone “missing” (not been deposited in the detector).

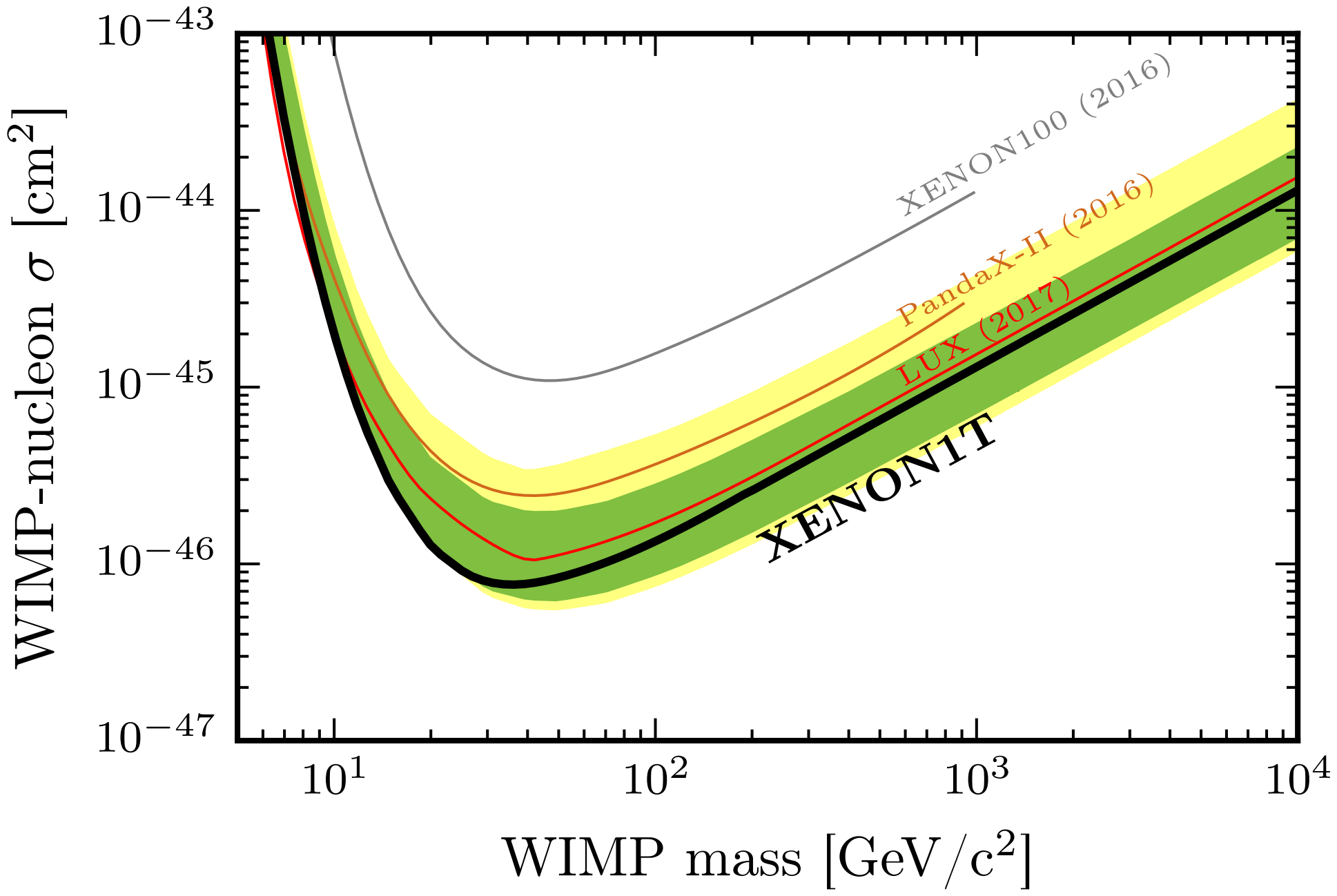

**Figure 2.6.:** Current results of dark matter direct detection experiments. The curve shows the upper bound on the spin-independent cross section of a hypothetical WIMP dark matter particle on a nucleon, as set by different experiments. Adapted from [XENON17].

Experimental dark matter searches are working in all three directions. The most stringent and tangible bounds are arguably set by direct detection searches at the moment, because they can provide a concrete value (the limit on the scattering cross section) which a model of dark matter must match. An overview of current efforts in direct detection of dark matter is shown in fig. 2.5. When a dark matter particle scatters off an atomic nucleus, there are different ways in which the recoil could be detected. Correspondingly, different experimental methods have developed making use of one or more of these detection channels.

As an example these experiments' results and in order to show some recent progress, fig. 2.6 displays the current limits on the scattering cross section between a WIMP and a nucleon. When a theoretical model is developed, the value which it predicts for this cross section can be calculated and compared to the experimental bounds. If it is too large, the proposed model is falsified by experiment.

## 2.2. Neutrino oscillations

In 2015, the Nobel prize was awarded "for the discovery of neutrino oscillations, which shows that neutrinos have mass". These oscillations arise from a mismatch of the quantum states which are created in electroweak interactions ("interaction eigenstates") and those which have a well-defined mass and propagate through space ("mass eigenstates"). This means that a state produced in an interaction is actually a superposition of those states with definite masses. Since the time evolution of a free particle state depends on its energy/mass, this mixture of states changes as the neutrino propagates – it oscillates between different states. For example, the probability of oscillation between two "neutrino species" (interaction eigenstates) $\nu$, $\nu'$ is given by

$$P(\nu \to \nu') = \sin(2\theta)^2 \sin\left(\frac{\Delta m^2 L}{4E}\right)^2 \tag{2.7}$$

with the mixing angle $\theta$, the traveled distance $L$ and the energy $E$. As is clear from the appearance of the difference of squared neutrino masses $\Delta m^2$, this would simply be zero if neutrinos were massless. For three generations of neutrinos, the oscillation probability becomes more complicated and depends on the *Pontecorvo–Maki–Nakagawa–Sakata matrix* (PMNS matrix), which describes how the neutrino mass eigenstates are related to their interaction eigenstates.

It should be noted that fundamentally, this exact same phenomenon has been well-known for the quarks for a long time, albeit under a different name and from a different perspective. There, the relation between the interaction and mass eigenstates is described by the *Cabibbo–Kobayashi–Maskawa matrix* (CKM matrix). For the neutrinos, there is one key difference, though: Their masses are many orders of magnitude smaller than those of the other fermions. This means that they oscillate on completely different length scales. Coupled with the fact that neutrinos were historically taken to be massless, this has lead to the point that what are usually called "the neutrinos" are their interaction eigenstates, while "the quarks" are identified with their mass eigenstates instead. These opposite points of view are responsible for the fact that neutrinos are said to oscillate, while weak interactions are said to "violate quark flavor". The difference merely comes down to whether one focuses on the states that change when a particle propagates (interaction eigenstates) or those that change in interactions (mass eigenstates). In the same way, the CKM and PMNS matrices both have exactly the same role: They are simply the fermion mixing

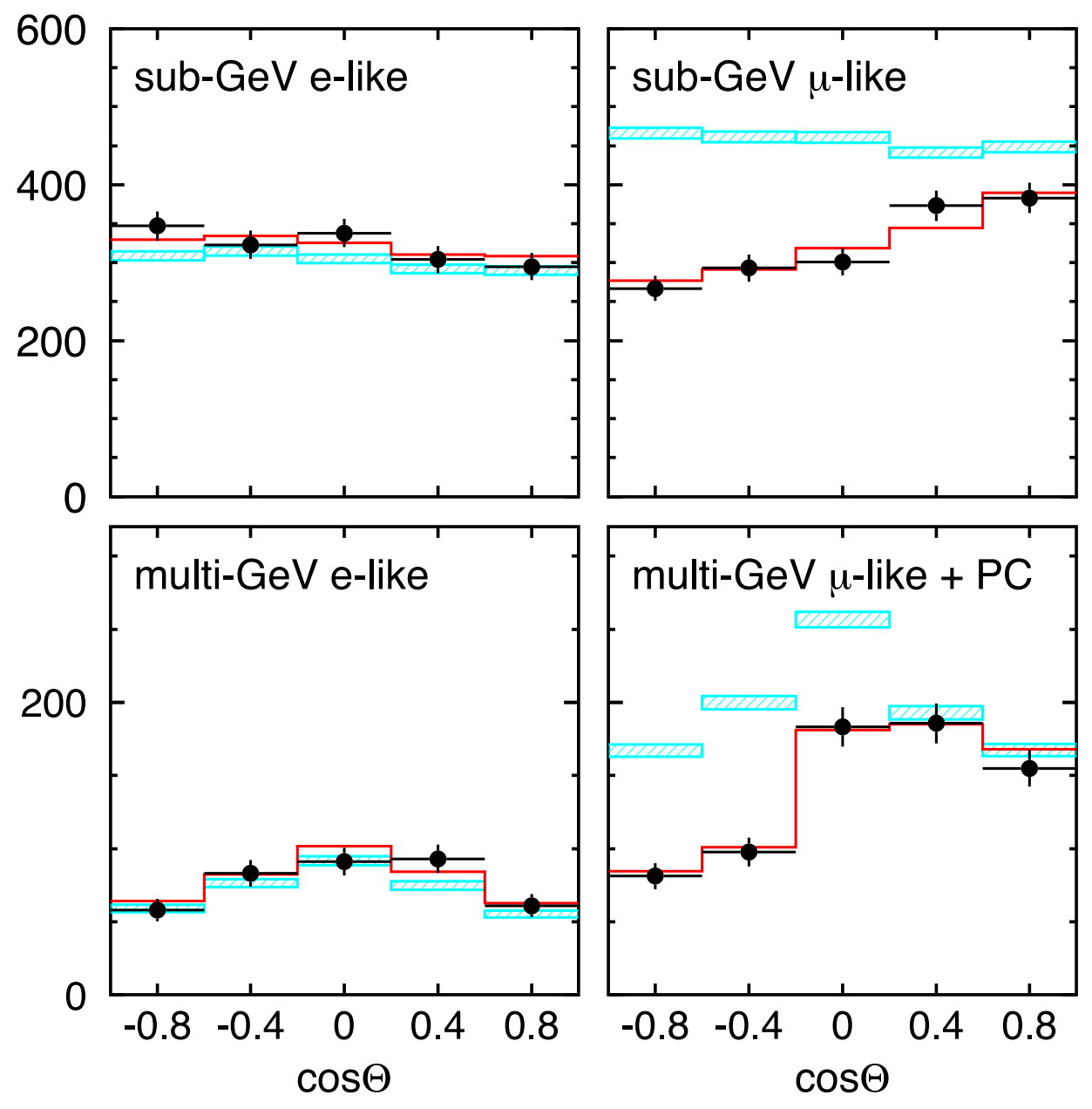

**Figure 2.7.:** Measurement of neutrino oscillations by the Super-Kamiokande experiment. Taken from [nob15].

matrices of the Standard Model (one for the quarks, one for the leptons).[1]

Oscillations of neutrinos were detected by measuring the neutrino flux and observing that, depending on the distance traveled, fewer neutrinos of a given species arrive. For example, fig. 2.7 shows measurements by Super-Kamiokande, which co-received the Nobel Prize 2015. It measures interactions from atmospheric neutrinos (produced by cosmic rays) using a large water tank with photomultipliers which has been placed in a mine. The amount of muon neutrinos coming from below, through the earth, was much lower than that coming from above. Since neutrinos barely interact at all with other particles, this suggests that the missing neutrinos have been converted to another species of neutrino which cannot be detected by this experiment.

The other half of the Nobel Prize went to the Sudbury Neutrino Observatory (SNO).

[1]The attentive reader may wonder what happens to the charged leptons, since the CKM matrix relates to the quarks and the PMNS matrix to the neutrinos. They are often not mentioned because the mismatch between mass and interaction eigenstates does not depend on the change-of-basis matrix for the different kinds of quarks or leptons individually, but only on a combination of both, which makes it possible to identify both kinds of states for one type of quark and lepton. Conventionally, this is done for the up-type quarks (placing all the mixing in the down-type quarks) and the charged leptons (with massless neutrinos, it was even possible to identify the two kinds of states for all leptons).

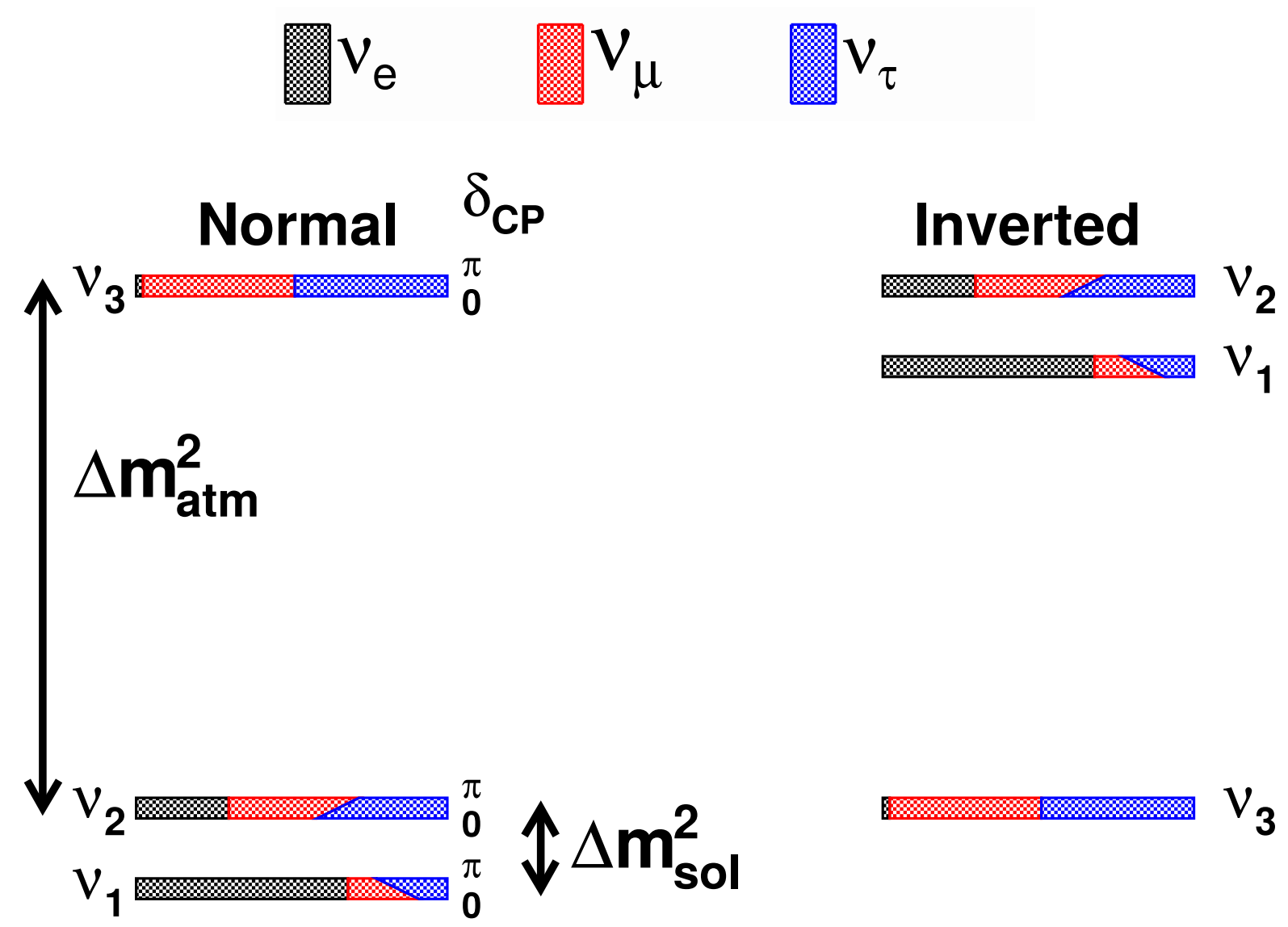

**Figure 2.8.:** Illustration of the measured neutrino mass parameters and the two possible neutrino mass hierarchies (normal hierarchy and inverted hierarchy). Taken from [QV15].

This experiment focused on solar neutrinos, produced in nuclear reactions within the sun. For a long time, the neutrino flux coming from the sun was much lower than predicted by calculations of these reactions, although it was not known whether something happened to the neutrinos or whether the solar models were incorrect. This was known as the *solar neutrino problem*. SNO also used a water Cherenkov detector, but the water in this case was heavy water (containing deuterons). This allowed for reactions with different detection channels which were either sensitive to only electron neutrinos (which are produced in the sun) or all neutrino flavors. The "missing" solar neutrinos were thus shown to have oscillated to $\mu$ and $\tau$ neutrinos.

The current knowledge about the layout of neutrino masses and states is illustrated in fig. 2.8. The absolute values of the differences between the squared masses have been measured, indicating that two neutrino masses are relatively similar, while the third one is further away. However, the *sign* of $\Delta m^2_{32} = \Delta m^2_{\text{atm}}$ is not known. This means that the third neutrino $\nu_3$ could either be heavier ("normal hierarchy") or lighter ("inverted hierarchy") than the others. The experimentally determined values of these mass differences are [PDG16]

$$\Delta m^2_{21} = \Delta m^2_2 - \Delta m^2_1 = (7.53 \pm 0.18) \times 10^{-5}\,\text{eV}^2 \tag{2.8}$$

$$|\Delta m^2_{32}| = |\Delta m^2_3 - \Delta m^2_2| = (2.44 \pm 0.06) \times 10^{-3}\,\text{eV}^2 \qquad \text{(normal hierarchy)} \tag{2.9}$$

$$|\Delta m^2_{32}| = |\Delta m^2_3 - \Delta m^2_2| = (2.51 \pm 0.06) \times 10^{-3}\,\mathrm{eV}^2 \qquad \text{(inverted hierarchy)} \tag{2.10}$$

They were determined using data from atmospheric neutrinos (e. g. Super-Kamiokande) and solar neutrinos (e. g. SNO). In addition, cosmological observations, such as of the CMB, galaxy clusters and supernovae, impose an upper bound on the sum of all neutrino masses [PDG16]

$$\sum_i m_{\nu_i} < 0.23\,\mathrm{eV} \tag{2.11}$$

Experiments on the $\beta$ decay of tritium also set an upper bound on the neutrino masses, but this is currently not as constraining with $m_{\nu_i} < 2\,\mathrm{eV}$.

Once it was established that neutrinos have mass, however, the following question arose: Why are these masses so tiny? Compared to the mass of the top quark, which is larger than 100 GeV, the masses of the fermions in the Standard Model span almost twelve orders of magnitude. It seems at least a bit suspicious that certain parameters of the Standard Model would just be suppressed to this degree arbitrarily or "by chance", inspiring many theories for "naturally small" neutrino masses.

# Gauge theories and the Standard Model of particle physics

# 3

## 3.1. Mathematical background

A common point of confusion when working and communicating in physics is that many concepts remain vaguely-defined, often leaving people unfamiliar with all the unwritten conventions and tacit assumptions of a given sub-field or community uncertain and wondering what, precisely, is the meaning of even important statements and equations. As a main endeavor of this work is to illuminate "how things are done and why" in model building for particle physics (at least pertaining to a certain class of models), it seems fitting and necessary to lay down a clear foundation on the mathematical level as well. Of course, not every basic concept will be defined here – elementary analysis and linear algebra are presupposed. Instead, the focus is on ideas which are not clearly defined even in advanced physics courses or which have several meanings in mathematics. Similar definitions can be found in the many textbooks on algebra and group theory, such as [Rom07; Hal15; Jon98; FH04]. A more elementary overview from a physical point of view is available in [Sch15b].

### 3.1.1. Group and representation theory

**Definition 1** (Group)**.** A *group* is a set $G$ together with an operation $\cdot : G \times G \to G$, denoted $a \cdot b$ or $ab$. Any group $(G, \cdot)$ must satisfy the following axioms:

**Associativity** $(a \cdot b) \cdot c = a \cdot (b \cdot c) \quad \forall a, b, c \in G.$

**Identity element** $\exists e \in G : \; e \cdot a = a \cdot e = a \quad \forall a \in G.$

**Inverse element** $\forall a \in G \; \exists b \in G : \; a \cdot b = b \cdot a = e$, where $e$ is the identity element.

The group axioms automatically imply that the *identity element* $e$ and the *inverse element* $a^{-1}$ corresponding to a group element $a$ are unique.

Most of the time, the set $G$ is used synonymously for the group $(G, \cdot)$ as a whole, with the group operation implicit. For matrix groups, the group operation is usually the ordinary matrix multiplication. Crucially, the group operation is not required to be commutative $(ab \neq ba)$. The special category of groups where this is the case are called *abelian groups*.

**Definition 2** (Direct product of groups)**.** Given two groups $(G, *)$ and $(H, \bullet)$, the *(direct) product of the groups* $(G \times H, \cdot)$ is a group defined as follows:

- The resulting group's underlying set is the Cartesian product of $G$ and $H$, $G \times H = \{(g, h) \mid g \in G, h \in H\}$.
- The binary operation on $G \times H$ is defined component-wise: $(g_1, h_1) \cdot (g_2, h_2) = (g_1 * g_2, h_1 \bullet h_2) \quad \forall g_1, g_2 \in G, h_1, h_2 \in H$.

Again, $G \times H$ is used as a substitute for the group as a whole because the group operation is clear from definition 2. Essentially, the direct product generalizes the Cartesian product of sets to groups so that the result is not only a set, but again a group.

An extremely important class of groups in physics are the *Lie groups*, which are continuous groups whose elements depend analytically on a set of parameters. This allows every group element to be written using the exponential map. Therefore, there is an infinitesimal neighborhood that can be associated with every element of a Lie group.

**Definition 3** (Lie group)**.** A *Lie group* $(G, \cdot)$ is a group where $G$ is also a finite-dimensional smooth manifold such that the group operations of multiplication and inversion are smooth maps, i. e. $G \times G \to G, (a, b) \mapsto a^{-1}b$ is a smooth map.

Due to their properties, Lie groups are strongly connected with another concept called *Lie algebras*. A group's Lie algebra is exactly the "infinitesimal approximation" around the group's identity element.

**Definition 4** (Lie algebra)**.** A *Lie algebra* is a vector space $\mathfrak{g}$ over some field $F$ together with a binary operation $[\cdot, \cdot] : \mathfrak{g} \times \mathfrak{g} \to \mathfrak{g}$ called the *Lie bracket* that satisfies the following axioms:

**Bilinearity** $[aX + bY, Z] = a[X, Z] + b[Y, Z], [Z, aX + bY] = a[Z, X] + b[Z, Y]$
$\forall a, b \in F$ and $\forall X, Y, Z \in \mathfrak{g}$.

**Antisymmetry** $[X, Y] = -[Y, X] \quad \forall X, Y \in \mathfrak{g}$.

**Jacobi identity** $[X, [Y, Z]] + [Z, [X, Y]] + [Y, [Z, X]] = 0 \quad \forall X, Y, Z \in \mathfrak{g}$.

Antisymmetry of the Lie bracket directly implies that $[X, X] = 0 \; \forall X \in \mathfrak{g}$. Similar to groups, Lie algebras are usually referred to only by their vector space $\mathfrak{g}$ – since we will deal exclusively with matrix Lie groups, the Lie bracket will always correspond to the familiar commutator of matrices.

It turns out that many groups are "the same" or "equal" to another in the sense that their elements have simply been relabeled. However, it can be surprisingly difficult to find out if this is the case.

**Definition 5** ((Lie) group and Lie algebra isomorphism)**.** Given two groups $(G, *)$ and $(H, \bullet)$, a *group isomorphism* from $(G, *)$ to $(H, \bullet)$ is a bijective group homomorphism $\Phi : G \to H$.[1] That is, $\Phi$ must fulfill $\Phi(g_1 * g_2) = \Phi(g_1) \bullet \Phi(g_2) \; \forall g_1, g_2 \in G$.

Similarly, given two Lie algebras $\mathfrak{g}$ and $\mathfrak{h}$, a *Lie algebra isomorphism* from $\mathfrak{g}$ to $\mathfrak{h}$ is a bijective Lie algebra homomorphism $\varphi : \mathfrak{g} \to \mathfrak{h}$. That is, $\varphi$ must fulfill $\varphi([X, Y]) = [\varphi(X), \varphi(Y)] \; \forall X, Y \in G$.

Two groups or Lie algebras are *isomorphic* if there exists an isomorphism between them. This is written as $(G, *) \cong (H, \bullet)$ (or $G \cong H$) or $\mathfrak{g} \cong \mathfrak{h}$, respectively.

Of course, there are analogous concepts of isomorphisms for other algebraic structures. In general, an isomorphism is a bijective homomorphism, i. e. a bijective map that preserves the structure of a given object.

For matrix Lie groups, the aforementioned connection between Lie groups and algebras manifests as follows:

- Given a matrix Lie algebra $\mathfrak{g}$, the corresponding Lie group is

$$G = \exp(\mathfrak{g}) = \{\exp(X) \mid X \in \mathfrak{g}\} \tag{3.1}$$

- Given a matrix Lie group $G$, the corresponding Lie algebra is

$$\mathfrak{g} = \{X \in M(n, \mathbb{C}) \mid \exp(tX) \in G \; \forall t \in \mathbb{R}\} \tag{3.2}$$

  Put differently, if $\exp(tX)$ is a one-parameter subgroup of $G$, then $X$ is an element of the Lie algebra.

[1] For Lie groups, $\Phi$ and its inverse $\Phi^{-1}$ are additionally required to be continuous in order for $\Phi$ to be called a *Lie group isomorphism*.

where $\exp(X) = e^X = \sum_{n=0}^{\infty} \frac{X^n}{n!}$ is the matrix exponential. From the perspective of differential geometry, the Lie algebra is the tangent space at the identity of the manifold corresponding to the group. This can be illustrated in a way similar to (3.2):

$$X = \frac{\mathrm{d}}{\mathrm{d}t} e^{tX} \bigg|_{t=0} \tag{3.3}$$

However, the correspondence between a Lie group and its algebra is not one-to-one. Several (non-isomorphic) groups can have "the same" (isomorphic) Lie algebras, but applying the exponential map to this algebra will always result in the same group. This distinguished group sets itself apart from the others by possessing the topological property of being *simply connected*.

**Definition 6** (Universal covering group)**.** Given a connected matrix Lie group $G$, the *universal covering group* of $G$ is a simply connected matrix Lie group $H$ given by a Lie group homomorphism $\varphi : H \to G$ ($\varphi(gh) = \varphi(g)\varphi(h)\ \forall g \in G, h \in H$) such that the associated Lie algebras $\mathfrak{g}$ and $\mathfrak{h}$ are isomorphic.

Two very important examples of this are $\mathrm{SU}(2)$, which is the universal cover of $\mathrm{SO}(3)$ ($\mathfrak{so}(3) \cong \mathfrak{su}(2)$ with $\exp(\mathfrak{su}(2)) = \mathrm{SU}(2)$) and $\mathrm{SL}(2,\mathbb{C})$, which is the universal cover of the proper orthochronous Lorentz group $\mathrm{SO}(1,3)^+$ ($\mathfrak{so}(1,3) \cong \mathfrak{sl}(2,\mathbb{C})$ with $\exp(\mathfrak{sl}(2,\mathbb{C})) = \mathrm{SL}(2,\mathbb{C})$). Both of these are double covers, i. e. two elements of the universal cover correspond to one element of the non-universal group. That means that the universal covering groups of $\mathrm{SO}(3)$ and $\mathrm{SO}(1,3)^+$ are also their spin groups: $\mathrm{Spin}(3) \cong \mathrm{SU}(2)$, $\mathrm{Spin}(1,3) \cong \mathrm{SL}(2,\mathbb{C})$. This insight explains why studying the representations (see definition 7) of $\mathrm{SU}(2)$ yields the spinors in 3 spatial dimensions (Pauli spinors) and studying the representations of $\mathrm{SL}(2,\mathbb{C})$ yields the spinors in $1+3$ spacetime dimensions (Weyl spinors, Dirac spinors, Majorana spinors).

From the perspective of physics, an extremely important branch of group theory is representation theory, which deals with the question of how the group elements can be represented by linear transformations (matrices) on vector spaces.

**Definition 7** (Representation of a group)**.** A *representation of a group* $G$ on a vector space $V$ (called the *representation space*) is a group homomorphism $\rho : G \to \mathrm{GL}(V)$. That is, $\rho$ associates a matrix to each group element and must fulfill $\rho(g_1 g_2) = \rho(g_1)\rho(g_2)\ \forall g_1, g_2 \in G$.

**Definition 8** (Representation of a Lie algebra)**.** A *representation of a Lie algebra* $\mathfrak{g}$ on a vector space $V$ is a Lie algebra homomorphism $r : \mathfrak{g} \to \mathfrak{gl}(V)$, where $\mathfrak{gl}(V)$ is the vector space of linear maps $V \to V$ (matrices). That is, $\rho$ must fulfill $r([X,Y]) = r(X)\rho(Y) - r(Y)\rho(X)\ \forall X, Y \in \mathfrak{g}$, i. e. the Lie bracket on $\mathfrak{gl}(V)$ is given by the commutator of matrices.

Once again, common usage tends to omit details and both the representative matrices $\rho(g)$ and the (elements of the) representation space $V$ are themselves often called the "representation", with the map $\rho$ (hopefully) known implicitly. In physics, it is also often said that an object is "in the $X$ representation", which this means that the object is an element of the representation space which belongs to the $X$ representation.

For finite-dimensional representations, there is a direct correspondence between a representation of a Lie group and a representation of the corresponding Lie algebra. Given a representation $\rho$ of a Lie group, the representation $r$ of the Lie algebra is simply the derivative of the group representation at the identity, just as for matrix Lie groups themselves:

$$\rho(e^X) = e^{r(X)} \Rightarrow r(X) = \frac{\mathrm{d}}{\mathrm{d}t}\rho(e^{tX})\bigg|_{t=0} \tag{3.4}$$

Thus, the word "representation" can be used unqualified in this context, referring both to the representation of the Lie group and to the corresponding induced representation of the Lie algebra. In fact, one usually obtains the representations of a Lie group by looking at the Lie algebra representations and generating the group elements using the exponential map.[2]

**Definition 9** (Isomorphic representations)**.** Given a group $G$, two representations $\rho_1 : G \to \mathrm{GL}(V)$ and $\rho_2 : G \to \mathrm{GL}(W)$ are said to be *isomorphic* or *equivalent*, $\rho_1 \cong \rho_2$, if there exists a vector space isomorphism $S : V \to W$ such that

$$S^{-1}\rho_2(g)S = \rho_1(g) \quad \forall g \in G$$

If $G$ is a Lie group, the associated Lie algebra representations of $\rho_1$ and $\rho_2$ are isomorphic exactly if $\rho_1 \cong \rho_2$.

In other words, two representations related simply by a change of basis are isomorphic or "equivalent"; any representation $\rho$ is isomorphic to a similarity transform of itself.

[2]However, it should be noted that this procedure does not only yield the representations of the original group, but of its universal covering group.

**Definition 10** (Adjoint[3] representation)**.** Given a matrix Lie group $G$ with Lie algebra $\mathfrak{g}$, define the linear maps $\mathrm{Ad}_g : \mathfrak{g} \to \mathfrak{g}, X \mapsto gXg^{-1}$ and $\mathrm{ad}_X : \mathfrak{g} \to \mathfrak{g}, Y \mapsto [X, Y]$ with $g \in G, X \in \mathfrak{g}$ (*adjoint map*). The *adjoint representation of the Lie group $G$* is the map $\mathrm{Ad} : G \to \mathrm{GL}(\mathfrak{g}), g \mapsto \mathrm{Ad}_g$. Correspondingly, the *adjoint representation of the Lie algebra* $\mathfrak{g}$ is the map $\mathrm{ad} : \mathfrak{g} \to \mathfrak{gl}(\mathfrak{g}), X \mapsto \mathrm{ad}_X$.

In other words, the representation space of the adjoint group representation is the Lie algebra itself and the group elements $g$ act on the algebra elements $X$ through the adjoint action $gXg^{-1}$. Then, as always, the algebra representation is induced by the group representation through the derivative at the identity:

$$\mathrm{ad}_X = \frac{\mathrm{d}}{\mathrm{d}t} \mathrm{Ad}_{\exp(tX)} \bigg|_{t=0}$$

That the group elements $g$ and algebra elements $X$ are represented by $\mathrm{Ad}_g$ and $\mathrm{ad}_X$ can be imagined more easily by taking this to mean that they are represented by the matrices corresponding to these linear maps.

In physics, it is often said that a Lie algebra's basis vectors are the (*infinitesimal*) *generators* of the group because all group elements can be "generated" by their linear combinations using the exponential map.

**Definition 11** (Generators and structure constants)**.** Given a Lie algebra $\mathfrak{g}$, the elements of a basis $\{X_i\}$ of $\mathfrak{g}$ are also called the *generators* of the corresponding group $G = \exp(\mathfrak{g})$ and the constants $f_{ij}{}^k$ specified through $[X_i, X_j] = f_{ij}{}^k X_k$ are called the *structure constants*.

If the $X_i$ are matrices, they are also said to fulfill the *commutation relations* given in definition 11.

A Lie algebra is uniquely determined[4] through its structure constants. A useful fact to keep in mind is that given a basis ("generators") of $\mathfrak{g}$ specified through $[X_i, X_j] = f_{ij}{}^k X_k$, the corresponding basis matrices $X_i^{(\mathrm{A})}$ in the adjoint representation of the Lie algebra will be given by the structure constants through $(X_i^{(\mathrm{A})})^j{}_k \sim f_i{}^j{}_k$.

**Definition 12** (Direct sum[5])**.** Given two vector spaces $V$ and $W$ over a field $F$, the *direct sum $V \oplus W$* is a vector space defined as follows:

[3] It should be noted that "adjoint" in this case has nothing to do with the Hermitian adjoint $A^\dagger$ of a linear operator $A$.

[4] up to isomorphism

[5] There is a related notion called the *direct product* of vector spaces (cf. 2 for the concept on groups). However, these concepts only differ when infinite sums/products of vector spaces are considered.

- The resulting vector space's underlying set is the Cartesian product of $V$ and $W$, $V \times W = \{(v, w) \mid v \in V, w \in W\}$.
- Addition on $V \oplus W$ is defined component-wise:
  $(v_1, w_1) + (v_2, w_2) = (v_1 + v_2, w_1 + w_2) \quad \forall v_1, v_2 \in V, w_1, w_2 \in W$.
- Scalar multiplication on $V \oplus W$ is defined component-wise:
  $c(v, w) = (cv, cw) \quad \forall v \in V, w \in W, c \in F$.

For two matrices $A$ and $B$, a related operation called the *direct sum of matrices* $A \oplus B$ is also defined: It is the block matrix

$$A \oplus B = \begin{pmatrix} A & 0 \\ 0 & B \end{pmatrix}$$

The *direct sum of two Lie algebras* $\mathfrak{g}_1$ and $\mathfrak{g}_2$ is the vector space $\mathfrak{g}_1 \oplus \mathfrak{g}_2$ together with the Lie bracket given by

$$[(X_1, X_2), (Y_1, Y_2)] = ([X_1, Y_1], [X_2, Y_2]) \quad \forall X_1, Y_1 \in \mathfrak{g}_1, X_2, Y_2 \in \mathfrak{g}_2$$

The direct sum generalizes the Cartesian product of sets to vector spaces just as the direct product does to groups. The resulting vectors can not only be thought of as tuples $(v, w)$, but, given a basis on the vector space, as one large (column) vector with the entries of both $v$ and $w$ ($m = \dim(V)$, $n = \dim(W)$):

$$(v, w) \mapsto \begin{pmatrix} v_1 \\ \vdots \\ v_m \\ w_1 \\ \vdots \\ w_n \end{pmatrix}$$

As a converse to the definition of the direct sum of Lie algebras, one can recognize whether an algebra decomposes into a direct sum if there are subspaces where the Lie bracket of elements from different subspaces is always zero. An important fact concerning the connection of direct sums of Lie algebras and their corresponding group is that [Hal15, theorem 5.11]

$$\exp(\mathfrak{g}_1 \oplus \mathfrak{g}_2) \cong \exp(\mathfrak{g}_1) \times \exp(\mathfrak{g}_2) \tag{3.5}$$

An important distinguishing factor of representations is whether they have any non-trivial sub-representations.

**Definition 13** (Reducible and irreducible representations)**.** Given a group $G$ and a vector space $V$, let $\rho : G \to \mathrm{GL}(V)$ be a representation of $G$. A subspace $W$ of $V$ is called *invariant* if $\rho(g)w \in W \; \forall g \in G, w \in W$. Excluding the trivial cases $W = \{0\}$ and $W = V$, a representation which has invariant subspaces is called *reducible*, while a representation without invariant subspaces is called *irreducible*.

A representation is called *completely reducible* if it is (isomorphic to) a direct sum of irreducible representations.

A reducible representation can always be written in block-matrix form

$$\rho(g) = \begin{pmatrix} \rho^{(1)}(g) & N(g) \\ 0 & \rho^{(2)}(g) \end{pmatrix} \quad \forall g \in G$$

with $V = V_1 \oplus V_2$ and $\rho^{(i)} : G \to V_i$. In this case, the invariant subspace is $W = V_1$ since $\rho(g)(v^{(1)}, 0) \in V_1$. If, additionally, $N(g) = 0$, both $V_1$ and $V_2$ are invariant subspaces and $\rho(g)$ decomposes into a direct sum $\rho(g) = \rho^{(1)}(g) \oplus \rho^{(2)}(g)$.

An important result obtained through a method called "Weyl's unitarian trick" or "unitary trick" states that every finite-dimensional representation of a compact matrix Lie group is completely reducible, i. e. a direct sum of irreducible representations. This means that for such groups, all representations can be constructed from just the irreducible ones using direct sums. Examples of compact groups are $\mathrm{O}(N)$, $\mathrm{SO}(N)$, $\mathrm{U}(N)$ and $\mathrm{SU}(N)$.

**Definition 14** (Conjugate, unitary, real, pseudo-real and complex representations)**.** Given a group $G$ and a representation $\rho : G \to \mathrm{GL}(V)$, the following definitions are made:

- The *complex conjugate* representation corresponding to $\rho$ is the representation $\rho^* : G \to \mathrm{GL}(V), g \mapsto \rho(g)^*$.
- $\rho$ is a *unitary representation* if $\rho(g)^\dagger = \rho(g)^{-1} \; \forall g \in G$.
- $\rho$ is exactly one of the following:
  - A *real representation* if it is isomorphic to a representation $\rho' : G \to \mathrm{GL}(\dim(V), \mathbb{R})$, that is, if there is a similarity transformation $\rho'(g) = S^{-1}\rho(g)S$ such that $\rho'$ is real ($\rho'^* = \rho'$) $\forall g \in G$;

– a *pseudo-real representation* if it is isomorphic to its complex conjugate representation $\rho^*$, but is not real, i. e. if there is a similarity transformation such that $\rho(g)^* = S^{-1}\rho(g)S \ \forall g \in G$;

– a *complex representation* otherwise.

Both real and pseudo-real representations are isomorphic to their complex conjugate representations, and conversely, all representations which are isomorphic to their conjugates are either real or pseudo-real. According to the definition, the matrices of a real representation can always be made real through a basis transformation. A group's adjoint representation is always real. For symmetry groups in quantum theories, unitary representations are usually needed so that the probability amplitude $\langle\varphi|\psi\rangle$ remains invariant under transformations.

It should be stressed again that all definitions concerning representations of Lie groups (such as reducible, irreducible, conjugate, unitary, real, pseudo-real, complex) can be made analogously for representations of Lie algebras.

### 3.1.2. Tensors

**Definition 15** (Tensor product)**.** Given two vector spaces $V$ and $W$ over a field $F$, the *tensor product* $V \otimes W$ is the vector space whose elements and operations are constructed as follows:

- From the Cartesian product $V \times W$, the vector space $F(V \times W)$ over $F$ with basis $U \times V$ (also called the free vector space on $V \times W$) is formed. This is the vector space of tuples $(v, w)$ $(v \in V, w \in W)$ and their linear combinations, where no further properties of addition or multiplication are assumed (in particular, they are *not* defined component-wise).

- The vectors of $V \otimes W$ are the equivalence classes generated by the following relations on $F(V \times W)$:

$$\begin{aligned}(v_1, w) + (v_2, w) &\sim (v_1 + v_2, w)\\ (v, w_1) + (v, w_2) &\sim (v, w_1 + w_2)\\ c(v, w) &\sim (cv, w)\\ c(v, w) &\sim (v, cw)\end{aligned}$$

$\forall v, v_1, v_2 \in V, w, w_1, w_2 \in W, c \in F$, i. e. the tensor product is the quotient space $V \otimes W = F(V \times W)/Z$, where $Z$ is the subspace of $F(V \times W)$ whose elements are equivalent to 0. In other words, each vector on one side of one of these relations is treated as equivalent to the vector on the other side of the relation and the two are identified with each other.

The *tensor product of vectors* $v \otimes w \in V \otimes W$ ($v \in V, w \in W$) is used to denote individual elements of a tensor product space. It is bilinear, which means that it has the following properties directly reflecting the equivalence relations above:

- Distributivity:

$$(v_1 + v_2) \otimes w = v_1 \otimes w + v_2 \otimes w$$
$$v \otimes (w_1 + w_2) = v \otimes w_1 + v \otimes w_2$$

  $\forall v, v_1, v_2 \in V, w, w_1, w_2 \in W.$

- $c(v \otimes w) = (cv) \otimes w = v \otimes (cw) \quad \forall v \in V, w \in W, c \in F.$

The abstract definition is a bit technical, but the idea is as follows: The space $F(V \times W)$ consists of all formal sums of the form $\sum_i c_i(v_i, w_i)$ where a sum like $(v_1, w_1) + (v_2, w_2)$ does *not* have any further meaning and cannot be "simplified" (e. g. through component-wise addition). This is what is needed for tensors, though, since a sum of tensor products $v_1 \otimes w_1 + v_2 \otimes w_2$ can also not, in general, be simplified. In this sense, it is really only a means to an end to obtain a vector space of generic tuples $(v, w)$. The essential properties of the tensor product are then enforced through the equivalence relations, giving meaning to addition and scalar multiplication on the heretofore meaningless generic tuples. Given bases of $V$ and $W$, the components of a tensor product $v \otimes w$ are simply given by the components of the individual vectors: $(v \otimes w)_{ij} = v_i w_j$.

It is useful to keep in mind the dimensions of the direct sum and tensor product of two vector spaces $V$ and $W$:

$$\dim(V \oplus W) = \dim(V) + \dim(W) \tag{3.6}$$
$$\dim(V \otimes W) = \dim(V) \dim(W) \tag{3.7}$$

**Definition 16** (Tensor)**.** A *tensor of type* $(m, n)$ on a vector space $V$ over a field $F$ is an element of the vector space

$$T^m_n(V) = \underbrace{V \otimes \ldots \otimes V}_{m} \otimes \underbrace{V^* \otimes \ldots \otimes V^*}_{n} = V^{\otimes m} \otimes V^{*\otimes n}$$

where $V^*$ is the dual space of $V$ (i. e. the vector space of linear functionals – linear maps $V \to F$). Such a tensor is said to be of *order* $m + n$.

More concretely, tensors are objects of the form

$$\sum_{i=1}^{n} v_{i1} \otimes \ldots \otimes v_{im} \otimes w_{i1} \otimes \ldots \otimes w_{in} \quad (v_{ij} \in V, w_{ij} \in V^*) \tag{3.8}$$

The tensor product is also called *outer product* for vectors and *Kronecker product* for matrices. If one or both of the operands of a tensor product are already tensors, the result is a higher-order tensor.

Tensors can be viewed in a number of different ways. Given a basis $\{\vec{e}_i\}$ of $V$ and the corresponding dual basis $\{\vec{e}^i\}$, a tensor $T$ can be written in terms of components:

$$T = T^{i_1 \cdots i_m}_{j_1 \cdots j_n} \vec{e}_{i_1} \otimes \ldots \otimes \vec{e}_{i_m} \otimes \vec{e}^{j_1} \otimes \ldots \otimes \vec{e}^{j_n} \tag{3.9}$$

In this form, the familiar covariant (lowered indices) and contravariant (raised indices) transformation behavior of the tensor components under a change of basis becomes clear. Even though tensors are defined here to have all their contravariant indices before their covariant ones, tensors are sometimes defined analogously with mixed co- and contravariant indices.

Further, a tensor $T$ directly corresponds to a multilinear map

$$T : \underbrace{V^* \times \ldots \times V^*}_{m} \times \underbrace{V \times \ldots \times V}_{n} \to F = T : V^m \times V^{*n} \to F \tag{3.10}$$

where the components can now be obtained by applying the map to the basis:

$$T^{i_1 \cdots i_m}_{j_1 \cdots j_n} = T(\vec{e}^{i_1}, \ldots, \vec{e}^{i_m}, \vec{e}_{j_1}, \ldots, \vec{e}_{j_n}) \tag{3.11}$$

The order of $V$ and $V^*$ is reversed here because the domain of the map does not specify the tensor itself, but the kinds of objects that it accepts in order to form a scalar. From this point of view, it becomes clear that tensors are a generalization of matrices (linear maps). In fact, one can make the following identifications:

- $T^0_0(V) = F$ (scalars)
- $T^1_0(V) = V$ (vectors)
- $T^0_1(V) = V^*$ (linear functionals/"row vectors")

- $T^1_1(V) = M(\dim(V), F)$ (matrices)

As mentioned before, all representations of a compact matrix Lie group are just direct sums of irreducible representations. However, for groups such as $\mathrm{SU}(N)$, one can also form tensor products of certain *fundamental representations* so that every irreducible representation appears in the resulting direct sum decomposition. This procedure of obtaining all the irreducible representations from just a small number of fundamental ones is also known as *Clebsch–Gordan decomposition* and of course very familiar from the addition of angular momenta in quantum mechanics.

**Definition 17** (Direct sum representation)**.** Given a group $G$ and two representations $\rho_1 : G \to \mathrm{GL}(V_1)$ and $\rho_2 : G \to \mathrm{GL}(V_2)$, the *direct sum of the representations* is a representation

$$\rho_1 \oplus \rho_2 : G \to V_1 \oplus V_2, g \mapsto \rho_1(g) \oplus \rho_2(g)$$

In other words, $g \in G$ acts component-wise on each of the subspaces with

$$(\rho_1 \oplus \rho_2)(g)\,(v_1, v_2) = (\rho_1(g)v_1, \rho_2(g)v_2) \quad \forall v_1 \in V_1, v_2 \in V_2$$

Correspondingly, if $\mathfrak{g}$ is the Lie algebra of $G$ and $r_1 : \mathfrak{g} \to \mathfrak{gl}(V_1)$ and $r_2 : \mathfrak{g} \to \mathfrak{gl}(V_2)$ are the associated representations, their corresponding direct sum is a representation

$$r_1 \oplus r_2 : \mathfrak{g} \to V_1 \oplus V_2, X \mapsto r_1(X) \oplus r_2(X)$$

**Definition 18** (Tensor product representation)**.** Given a group $G$ and two representations $\rho_1 : G \to \mathrm{GL}(V_1)$ and $\rho_2 : G \to \mathrm{GL}(V_2)$, the *tensor product of the representations* is a representation

$$\rho_1 \otimes \rho_2 : G \to V_1 \otimes V_2, g \mapsto \rho_1(g) \otimes \rho_2(g)$$

In other words, $g \in G$ acts "factor-wise" on the tensor space with

$$(\rho_1 \otimes \rho_2)(g)\, v_1 \otimes v_2 = \rho_1(g)v_1 \otimes \rho_2(g)v_2 \quad \forall v_1 \in V_1, v_2 \in V_2$$

Correspondingly, if $\mathfrak{g}$ is the Lie algebra of $G$ and $r_1 : \mathfrak{g} \to \mathfrak{gl}(V_1)$ and $r_2 : \mathfrak{g} \to \mathfrak{gl}(V_2)$ are the associated representations, their corresponding tensor product is a representation

$$r_1 \otimes r_2 : \mathfrak{g} \to V_1 \otimes V_2, X \mapsto r_1(X) \otimes \mathbb{1} + \mathbb{1} \otimes r_2(X)$$

Tensor products of representations of not one, but two different groups and algebras can be defined analogously. These are then representations of $G_1 \times G_2$ and $\mathfrak{g}_1 \oplus \mathfrak{g}_2$, respectively, instead of $G$ and $\mathfrak{g}$. Unfortunately, this is ambiguous when $G_1 = G_2, \mathfrak{g}_1 = \mathfrak{g}_2$, in which case it must explicitly be stated which version is intended.

In other words, definition 18 demonstrates that

$$\rho_1(e^X) \oplus \rho_2(e^X) = e^{r_1(X) \oplus r_2(X)} \tag{3.12}$$

$$\rho_1(e^X) \otimes \rho_2(e^X) = e^{r_1(X) \otimes \mathbb{1} + \mathbb{1} \otimes r_2(X)} \tag{3.13}$$

A related identity following from this notation is, for any square matrices $A$ and $B$,

$$e^{A \otimes \mathbb{1} + \mathbb{1} \otimes B} = e^A \otimes e^B \tag{3.14}$$

## 3.2. Representations of the Lorentz group

An overview and detailed discussion of the Lorentz group, its universal covering group $\mathrm{SL}(2, \mathbb{C})$ and their representations are given in [Sch15b]. A lot of information on Weyl and Dirac spinors is also available in [DHM10].

The full Lorentz group

$$\mathrm{O}(1,3) = \{\Lambda \in \mathrm{GL}(4, \mathbb{R}) \mid \Lambda^\mathsf{T} \eta \Lambda = \eta\} \tag{3.15}$$

consists of four disjoint connected components, only one of which contains the identity element and is thus the group whose representations are obtained from representations of the Lie algebra. This identity component is the proper orthochronous Lorentz group $\mathrm{SO}(1,3)^+$, which puts the additional constraints on the Lorentz transformations $\Lambda$ that $\det(\Lambda) = 1$ (no spatial reflections) and $\Lambda^0{}_0 > 0$ (no time reversal). The other components can be obtained from this identity component using the discrete transformations $P$ (*parity* or space inversion) and $T$ (*time reversal*):

$$P = \mathrm{diag}(1, -1, -1, -1) \qquad T = \mathrm{diag}(-1, 1, 1, 1) \tag{3.16}$$

so that the full Lorentz group has the components

$$\mathrm{O}(1,3) = \mathrm{SO}(1,3)^+ \cup P\,\mathrm{SO}(1,3)^+ \cup T\,\mathrm{SO}(1,3)^+ \cup PT\,\mathrm{SO}(1,3)^+ \tag{3.17}$$

Clearly, this allows one to focus on $SO(1,3)^+$, while the other components are simply the result of finding the elements $\rho(P)$ and $\rho(T)$ (for a representation $\rho$) and applying them to a representation of $SO(1,3)^+$.

Since the easiest way to find the representations of a Lie group is through the representations of its algebra, one works extensively with the algebras of these two groups. The relationship between the Lie algebras of the Lorentz group and $SL(2,\mathbb{C})$ is the following:

$$\mathfrak{o}(1,3) = \mathfrak{so}(1,3) = \left\{ \lambda \in \mathfrak{gl}(4,\mathbb{R}) \mid \eta\lambda^{\mathsf{T}}\eta = -\lambda \right\} \tag{3.18}$$

$$\mathfrak{sl}(2,\mathbb{C}) = \{ X \in \mathfrak{gl}(2,\mathbb{C}) \mid \mathrm{Tr}(X) = 0 \} \tag{3.19}$$

$$\mathfrak{sl}(2,\mathbb{C}) \cong \mathfrak{so}(1,3) \tag{3.20}$$

However, as explained with definition 6, exponentiating representations of a group's Lie algebra will result in the representations of its universal covering group, not only those of the original group. Thus, the term "representation of the Lorentz group" is usually an abbreviation for "representation of the universal covering group of the (proper orthochronous) Lorentz group", including spinor representations.

The Lie algebra of the Lorentz group $\mathfrak{so}(1,3)$ can be defined through the structure constants[6,7]

$$[M^{\mu\nu}, M^{\rho\sigma}] = \eta^{\mu\sigma}M^{\nu\rho} + \eta^{\nu\rho}M^{\mu\sigma} - \eta^{\mu\rho}M^{\nu\sigma} - \eta^{\nu\sigma}M^{\mu\rho} \tag{3.21}$$

A basis for this Lie algebra is given by the matrices $M^{\mu\nu}$ defined as

$$(M^{\mu\nu})^{\alpha}{}_{\beta} = \eta^{\mu\alpha}\delta^{\nu}_{\beta} - \eta^{\nu\alpha}\delta^{\mu}_{\beta} \tag{3.22}$$

which fulfill $M^{\mu\nu} = -M^{\nu\mu}$. This allows writing a general Lorentz transformation as

$$\Lambda = \exp\left(\frac{1}{2}\omega_{\mu\nu}M^{\mu\nu}\right) \tag{3.23}$$

[6] The Lie algebra $\mathfrak{so}(1,3)$ is six-dimensional, so the basis can be written with one index which runs from 1 to 6 or with two antisymmetric indices which run from 1 to 4 (or 0 to 3). For the Lorentz group, the latter option is more convenient because it allows the use of the familiar four-vector index formalism.

[7] As before, the "mathematician's convention" (without inserting any additional factors of $i$) will be used. Accordingly, (3.21) and the following expressions differ from the usual definition in the "physicist's convention" by a factor of $i$. For more information about the two different conventions, see appendix A.

with $\omega_{\mu\nu} = -\omega_{\nu\mu} \in \mathbb{R}$. The matrices $\frac{1}{2}\omega_{\mu\nu}M^{\mu\nu}$ defined in this way have the same general form as the matrices $\lambda$ of $\mathfrak{so}(1,3)$ in (3.18), confirming that they indeed span the same vector space.

For a physical interpretation, it is useful to look at the purely spatial basis elements and those with a time component independently. Defining

$$J^i = \frac{1}{2}\varepsilon^{ijk}M_{jk} \qquad K^i = M^{0i} \tag{3.24}$$

which fulfill the commutation relations

$$[J^i, J^j] = \varepsilon^{ijk}J^k \quad [J^i, K^j] = \varepsilon^{ijk}K^k \quad [K^i, K^j] = -\varepsilon^{ijk}J^k \tag{3.25}$$

one recognizes the structure constants of $\mathfrak{so}(3)$ in the commutator $[J^i, J^j]$, indicating that Lorentz transformations involving only the generators $J^i$ are simply the familiar spatial rotations in three dimensions, with the rotation axis given by the index $i$. On the other hand, transformations using only the $K^i$ turn out to have the form of Lorentz boosts in the $i$ direction. Thus, the $J^i$ can be identified as the *generators of rotations* and the $K^i$ as the *generators of boosts*. Using $\omega_{ij} = \varepsilon_{ijk}\alpha_k$, $\omega_{0i} = \eta_i$, the Lorentz transformations take the form

$$\Lambda = \exp\left(\vec{\alpha}\cdot\vec{J} + \vec{\eta}\cdot\vec{K}\right) \tag{3.26}$$

where $\alpha^i$ are the rotation angles and $\eta^i$ are the rapidities with $v^i = \tanh(\eta^i)$.

While (3.26) is helpful in highlighting the physical interpretation of the Lorentz group, it limits the analysis to the defining representation where the original Lorentz transformation matrices act on four-vectors. More insight is gained by switching to a basis

$$\left.\begin{matrix} T^i_{\mathrm{L}} \\ T^i_{\mathrm{R}} \end{matrix}\right\} = \frac{1}{2}(J^i \pm iK^i) \qquad \text{or, inversely,} \qquad \left.\begin{matrix} J^i \\ iK^i \end{matrix}\right\} = T^i_{\mathrm{L}} \pm T^i_{\mathrm{R}} \tag{3.27}$$

which fulfills

$$[T^i_{\mathrm{L}}, T^j_{\mathrm{L}}] = \varepsilon^{ijk}T^k_{\mathrm{L}} \quad [T^i_{\mathrm{R}}, T^j_{\mathrm{R}}] = \varepsilon^{ijk}T^k_{\mathrm{R}} \quad [T^i_{\mathrm{L}}, T^j_{\mathrm{R}}] = 0 \tag{3.28}$$

In this basis, (3.26) takes the form

$$\Lambda = \exp\left((\vec{\alpha} - i\vec{\eta})\cdot\vec{T}_{\mathrm{L}} + (\vec{\alpha} + i\vec{\eta})\cdot\vec{T}_{\mathrm{R}}\right) = \exp\left((\vec{\alpha} - i\vec{\eta})\cdot\vec{T}_{\mathrm{L}}\right)\exp\left((\vec{\alpha} + i\vec{\eta})\cdot\vec{T}_{\mathrm{R}}\right) \tag{3.29}$$

where the last step can be shown using the Baker–Campbell–Hausdorff formula.

This definition hides a detail which might seem inconspicuous at first: In defining (3.27), complex linear combinations of the basis vectors have been introduced, and hence this is no longer a real, but a complex vector space. In fact, with this basis, the passage to the *complexification* $\mathfrak{so}(1,3)_{\mathbb{C}} = \mathfrak{so}(1,3) \oplus i\mathfrak{so}(1,3)$ of $\mathfrak{so}(1,3)$ has been performed. Fortunately, there is no need to worry since the representations of this complexified algebra directly correspond to representations of the original one. For example, given a representation in this new basis, one can go back to a representation on a real vector space by returning to the original basis of $J^i$ and $K^i$ (using (3.27)) and restricting to real coefficients in linear combinations again. In fact, the representations of SU(2) are typically obtained through a very similar procedure by going to the complexified Lie algebra $\mathfrak{su}(2)_{\mathbb{C}}$ ("ladder operators"). A lot more information on the technical details of this procedure can be found in [Kna86; Hal15].

To summarize, it is evident from (3.28) that the (complexified) Lorentz algebra decomposes into the direct sum of two copies of the (complexified) Lie algebra $\mathfrak{su}(2)_{\mathbb{C}}$, whose representations are already known:[8]

$$\mathfrak{so}(1,3)_{\mathbb{C}} \cong \mathfrak{sl}(2,\mathbb{C})_{\mathbb{C}} \cong \mathfrak{su}(2)_{\mathbb{C}} \oplus \mathfrak{su}(2)_{\mathbb{C}} \tag{3.30}$$

In this way, it becomes clear that the Lie algebra of the Lorentz group really only consists of two independent copies of $\mathfrak{su}(2)$, the familiar Lie algebra of SU(2). Hence, all irreducible representations of $\mathfrak{sl}(2,\mathbb{C}) \cong \mathfrak{so}(1,3)$ and thus, of $\mathrm{SL}(2,\mathbb{C})$ and $\mathrm{SO}(1,3)^+$, which it covers, are found simply through all (tensor product) pairs[9] of irreducible representations of $\mathfrak{su}(2)$, which correspond to the irreducible representations of $\exp(\mathfrak{su}(2) \oplus \mathfrak{su}(2)) = \mathrm{SU}(2) \times \mathrm{SU}(2)$. Accordingly, they are labeled

$$(j_{\mathrm{L}}, j_{\mathrm{R}}) \quad \text{with } j_{\mathrm{L}}, j_{\mathrm{R}} \in \frac{\mathbb{N}}{2} \tag{3.31}$$

where $j_{\mathrm{L}}$ and $j_{\mathrm{R}}$ label the well-known irreducible representations of SU(2).[10] Their

[8] The complexification of $\mathfrak{su}(2)$ is, in fact, just the Lie algebra $\mathfrak{sl}(2,\mathbb{C}) \cong \mathfrak{so}(1,3)$ again: $\mathfrak{sl}(2,\mathbb{C}) = \mathfrak{su}(2)_{\mathbb{C}} = \mathfrak{su}(2) \oplus i\mathfrak{su}(2)$.

[9] See, for example, [Ros02, section 6.3, proposition 10]: "The irreducible representations of $G \times H$ are precisely the representations $\pi \otimes \rho$ where $\pi$ is an irreducible representation of $G$ and $\rho$ is an irreducible representation of $H$."

[10] A basis for $\mathfrak{su}(2)$ is given by $[J^i, J^j] = \varepsilon^{ij}{}_k J^k$, with the Casimir element $\vec{J}^2 = -j(j+1)\mathbb{1}$, $j \in \frac{\mathbb{N}}{2}$ ("angular momentum quantum number"). The irreducible representations have dimension $2j+1$.

dimensions are given by

$$\dim((j_\mathrm{L}, j_\mathrm{R})) = (2j_\mathrm{L} + 1)(2j_\mathrm{R} + 1) \tag{3.32}$$

The explicit representations $r_{(j_\mathrm{L}, j_\mathrm{R})}$ of the algebra and $\rho_{(j_\mathrm{L}, j_\mathrm{R})}$ of the group are (cf. (3.27))

$$\begin{aligned}
r_{(j_\mathrm{L},j_\mathrm{R})}(T^i_\mathrm{L}) &= J^i_{(j_\mathrm{L})} \otimes \mathbb{1}_{2j_\mathrm{R}+1} && (3.33)\\
r_{(j_\mathrm{L},j_\mathrm{R})}(T^i_\mathrm{R}) &= \mathbb{1}_{2j_\mathrm{L}+1} \otimes J^i_{(j_\mathrm{R})} && (3.34)\\
r_{(j_\mathrm{L},j_\mathrm{R})}(J^i) &= r_{(j_\mathrm{L},j_\mathrm{R})}(T^i_\mathrm{L}) + r_{(j_\mathrm{L},j_\mathrm{R})}(T^i_\mathrm{R}) \\
&= J^i_{(j_\mathrm{L})} \otimes \mathbb{1}_{2j_\mathrm{R}+1} + \mathbb{1}_{2j_\mathrm{L}+1} \otimes J^i_{(j_\mathrm{R})} && (3.35)\\
r_{(j_\mathrm{L},j_\mathrm{R})}(K^i) &= -i\big(r_{(j_\mathrm{L},j_\mathrm{R})}(T^i_\mathrm{L}) - r_{(j_\mathrm{L},j_\mathrm{R})}(T^i_\mathrm{R})\big) \\
&= -i\big(J^i_{(j_\mathrm{L})} \otimes \mathbb{1}_{2j_\mathrm{R}+1} - \mathbb{1}_{2j_\mathrm{L}+1} \otimes J^i_{(j_\mathrm{R})}\big) && (3.36)\\
r_{(j_\mathrm{L},j_\mathrm{R})}(X) &= \alpha^i\, r_{(j_\mathrm{L},j_\mathrm{R})}(J^i) + \eta^i\, r_{(j_\mathrm{L},j_\mathrm{R})}(K^i) \qquad (X = \vec{\alpha}\cdot\vec{J} + \vec{\eta}\cdot\vec{K} \in \mathfrak{so}(1,3)) && (3.37)\\
\rho_{(j_\mathrm{L},j_\mathrm{R})}(\Lambda) &= \exp\big(r_{(j_\mathrm{L},j_\mathrm{R})}(\vec{\alpha}\cdot\vec{J} + \vec{\eta}\cdot\vec{K})\big) \\
&= \exp\big(\alpha^i r_{(j_\mathrm{L},j_\mathrm{R})}(J^i) + \eta^i r_{(j_\mathrm{L},j_\mathrm{R})}(K^i)\big) \qquad (\Lambda = \exp(X) \in \mathrm{SL}(2,\mathbb{C})) && (3.38)
\end{aligned}$$

where $J^i_{(j)}$ are the usual basis elements of the irreducible representation of $\mathfrak{su}(2)$ with dimension $2j + 1$ (see footnote 10), also denoted as $\mathbf{n}$ with $n = 2j + 1$. In other words, they are the angular momentum operators for an angular momentum of $j$. In analogy with the familiar Clebsch–Gordan decomposition of SU(2), the representations $(j_\mathrm{L}, j_\mathrm{R})$ can be related to objects with spin $j_\mathrm{L} + j_\mathrm{R}$ (cf. sections 3.2.1, 3.2.2 and 3.2.5).

### 3.2.1. Scalars: The $(0,0)$ representation

The simplest representation of the Lorentz group is the trivial one, $(0,0)$, where the transformation matrices are simply identity matrices. The singlet representation is used for both copies of SU(2):[11]

$$\rho_\mathrm{S}(\Lambda) = \exp\big(r(T^i_\mathrm{L})\big) = \exp\big(r(T^i_\mathrm{R})\big) = \mathbb{1} \Rightarrow r_\mathrm{S}(T^i_\mathrm{L}) = r_\mathrm{S}(T^i_\mathrm{R}) = 0 \tag{3.39}$$

[11] In more detail, (3.39) would be written as

$$\begin{aligned}
r_\mathrm{S}(T^i_\mathrm{L}) &= r_\mathrm{S}(T^i_\mathrm{R}) = r_\mathrm{S}(J^i) = r_\mathrm{S}(K^i) = 0\\
r_\mathrm{S}(X) &= r_\mathrm{S}\big(\vec{\alpha}\cdot\vec{J} + \vec{\eta}\cdot\vec{K}\big) = \alpha^i r_\mathrm{S}(J^i) + \eta^i r_\mathrm{S}(K^i) = 0 + 0 = 0\\
\rho_\mathrm{S}(\Lambda) &= \rho_\mathrm{S}(e^X) = \exp(r_\mathrm{S}(X)) = \exp\big(r_\mathrm{S}\big(\vec{\alpha}\cdot\vec{J} + \vec{\eta}\cdot\vec{K}\big)\big) = \exp(0) = \mathbb{1}
\end{aligned}$$

(denoting $r_S = r_{(0,0)}$, $\rho_S = \rho_{(0,0)}$). This representation is one-dimensional (3.32), which also directly implies (3.39) because this is the only way the commutation relations can be fulfilled in one dimension. Objects from this representation space are *Lorentz scalars* – they do not transform at all:[12]

$$\phi \in (0,0) \Rightarrow \phi \mapsto \mathbb{1}\phi = \phi \tag{3.40}$$

Constructing a Lorentz-invariant Lagrangian can thus be viewed as the task of finding out how the different representations of a theory can be combined into a new one which is isomorphic to the scalar representation.

### 3.2.2. Weyl spinors: The $(\frac{1}{2},0)$ and $(0,\frac{1}{2})$ representations

The smallest non-trivial representations are two-dimensional. They arise by choosing the doublet representation for one copy of SU(2) and the singlet representation for the other. Although both representations are similar, they are not identical! In fact, the representations $\left(\frac{1}{2},0\right)$ and $\left(0,\frac{1}{2}\right)$ are conjugate to each other.[13]

For $\left(\frac{1}{2},0\right)$, the representation (denoted $r_L = r_{\left(\frac{1}{2},0\right)}$, $\rho_L = \rho_{\left(\frac{1}{2},0\right)}$) is[14]

$$r_L(T_L^i) = -\frac{i}{2}\sigma^i \qquad r_L(T_R^i) = 0 \tag{3.41}$$

$$\Rightarrow r_L(J^i) = -\frac{i}{2}\sigma^i \qquad r_L(K^i) = -\frac{1}{2}\sigma^i \tag{3.42}$$

$$\rho_L(\Lambda) = \exp\bigl(\alpha^i r(J^i) + \eta^i r(K^i)\bigr) = \exp\left(\frac{1}{2}(-\vec{\eta} - i\vec{\alpha}) \cdot \vec{\sigma}\right) \tag{3.43}$$

In other words, rotations on objects of this representation are given by the transformation $\rho_L\left(e^{\vec{\alpha}\cdot\vec{J}}\right) = e^{-\frac{i}{2}\vec{\alpha}\cdot\vec{\sigma}}$, which are just the matrices SU(2), while boosts take the form $\rho_L\left(e^{\vec{\eta}\cdot\vec{K}}\right) = e^{-\frac{1}{2}\vec{\eta}\cdot\vec{\sigma}}$. Objects from this representation space are called *left-handed*

[12] The notation $\phi \in (j_1, j_2)$ is supposed to mean that $\phi$ is an element of the representation space of the $(j_1, j_2)$ representation.

[13] In general, $(j_1, j_2)$ and $(j_2, j_1)$ are conjugates. However, it should be noted that this means that the conjugate of $(j_1, j_2)$ is *isomorphic* to $(j_2, j_1)$. Depending on the bases chosen for the different representations, the two may only be equal after a basis transformation.

[14] The tensor products appearing in (3.33) to (3.36) simply disappear because $V \otimes \mathbb{R} \cong V$ for any vector space $V$.

*Weyl spinors*[15] and they transform as follows:

$$\psi_{\mathrm{L}} \in \left(\tfrac{1}{2}, 0\right) \Rightarrow \psi_{\mathrm{L}} \mapsto \exp\left(\frac{1}{2}(-\vec{\eta} - i\vec{\alpha}) \cdot \vec{\sigma}\right)\psi_{\mathrm{L}} \tag{3.44}$$

Using the definition $\sigma^{\mu\nu} = \frac{1}{4}(\sigma^{\mu}\bar{\sigma}^{\nu} - \sigma^{\nu}\bar{\sigma}^{\mu})$, this transformation can also be rewritten as[16]

$$\psi_{\mathrm{L}} \mapsto \exp\left(\frac{1}{2}\omega_{\mu\nu}\sigma^{\mu\nu}\right)\psi_{\mathrm{L}} \tag{3.45}$$

On the other hand, for $\left(0, \frac{1}{2}\right)$, the roles of $\vec{T}_{\mathrm{L}}$ and $\vec{T}_{\mathrm{R}}$ are reversed. The representation (denoted $r_{\mathrm{R}} = r_{\left(0,\frac{1}{2}\right)}$, $\rho_{\mathrm{R}} = \rho_{\left(0,\frac{1}{2}\right)}$) is

$$r_{\mathrm{R}}(T^i_{\mathrm{L}}) = 0 \qquad r_{\mathrm{R}}(T^i_{\mathrm{R}}) = -\frac{i}{2}\sigma^i \tag{3.46}$$

$$\Rightarrow r_{\mathrm{R}}(J^i) = -\frac{i}{2}\sigma^i \qquad r_{\mathrm{R}}(K^i) = \frac{1}{2}\sigma^i \tag{3.47}$$

$$\rho_{\mathrm{R}}(\Lambda) = \exp\left(\alpha^i r(J^i) + \eta^i r(K^i)\right) = \exp\left(\frac{1}{2}(\vec{\eta} - i\vec{\alpha}) \cdot \vec{\sigma}\right) \tag{3.48}$$

Therefore, the rotations are given by $\rho_{\mathrm{R}}\left(e^{\vec{\alpha}\cdot\vec{J}}\right) = e^{-\frac{i}{2}\vec{\alpha}\cdot\vec{\sigma}}$, just as for $\left(0, \frac{1}{2}\right)$. However, a sign has changed in the boost transformations: $\rho_{\mathrm{R}}\left(e^{\vec{\eta}\cdot\vec{K}}\right) = e^{\frac{1}{2}\vec{\eta}\cdot\vec{\sigma}}$. Objects from this representation space are called *right-handed Weyl spinors* and they transform as follows:

$$\psi_{\mathrm{R}} \in \left(0, \tfrac{1}{2}\right) \Rightarrow \psi_{\mathrm{R}} \mapsto \exp\left(\frac{1}{2}(\vec{\eta} - i\vec{\alpha}) \cdot \vec{\sigma}\right)\psi_{\mathrm{R}} \tag{3.49}$$

As in (3.45), using the definition $\bar{\sigma}^{\mu\nu} = \frac{1}{4}(\bar{\sigma}^{\mu}\sigma^{\nu} - \bar{\sigma}^{\nu}\sigma^{\mu})$, this transformation can be written as

$$\psi_{\mathrm{R}} \mapsto \exp\left(\frac{1}{2}\omega_{\mu\nu}\bar{\sigma}^{\mu\nu}\right)\psi_{\mathrm{R}} \tag{3.50}$$

Under parity transformations (see (3.16)), $\vec{J}$ and $\vec{K}$ behave as an axial vector and a vector, respectively:

$$P\vec{J}P^{-1} = \vec{J} \qquad P\vec{K}P^{-1} = -\vec{K} \qquad \Rightarrow P\vec{T}_{\mathrm{L}}P^{-1} = \vec{T}_{\mathrm{R}} \tag{3.51}$$

[15] The names "left-handed" and "right-handed" come from the behavior of these spinors under parity transformations.

[16] While $J^i$ and $K^i$ are represented in (3.42) and (3.47) by the Pauli matrices (with an additional factor), the matrices $\sigma^{\mu\nu}$ and $\bar{\sigma}^{\mu\nu}$ (see (3.50)) are the corresponding representations of $M^{\mu\nu}$ from (3.23).

This is exactly the difference between $\left(\frac{1}{2},0\right)$ and $\left(0,\frac{1}{2}\right)$: $\vec{J}$ stays the same while $\vec{K}$ switches sign, see (3.42) and (3.47). Hence, $Pr_{\mathrm{L}}P^{-1} = r_{\mathrm{R}}$ and $P\rho_{\mathrm{L}}P^{-1} = \rho_{\mathrm{R}}$, i. e. a parity transformation swaps the left- and right-handed representations.

### 3.2.3. Dirac spinors: The $(\frac{1}{2},0) \oplus (0,\frac{1}{2})$ representation

The solutions of the (free) *Dirac equation*[17]

$$(i\partial_\mu \gamma^\mu - m)\psi_{\mathrm{D}} = 0 \tag{3.52}$$

where $\gamma^\mu$ must satisfy the Clifford algebra

$$\{\gamma^\mu, \gamma^\nu\} = 2\eta^{\mu\nu}\mathbb{1} \tag{3.53}$$

are (at least) four-dimensional *Dirac spinors* or *bispinors*.[18] These spinors correspond to the $(\frac{1}{2},0) \oplus (0,\frac{1}{2})$ representation of the Lorentz group. However, as clearly evident from the appearance of the direct sum, this is a reducible representation which can be "broken down" further into the two Weyl spinor representations. From this point of view, a Dirac spinor is nothing more than an object which consists of one left-handed and one right-handed Weyl spinor as its components.

This decomposition of Dirac spinors becomes most evident in the *Weyl basis* (also called the "chiral basis") where the $\gamma$ matrices take the following form:[19]

$$\gamma^\mu = \begin{pmatrix} 0 & \sigma^\mu \\ \bar{\sigma}^\mu & 0 \end{pmatrix} \qquad \gamma_5 = i\gamma^0\gamma^1\gamma^2\gamma^3 = \begin{pmatrix} -\mathbb{1}_2 & 0 \\ 0 & \mathbb{1}_2 \end{pmatrix} \tag{3.54}$$

Since $\gamma_5$ (which is related to the projectors to left-handed and right-handed components) is diagonal in this basis, a Dirac spinor $\psi_{\mathrm{D}}$ is of the form

$$\psi_{\mathrm{D}} = \begin{pmatrix} \psi_{\mathrm{L}} \\ \psi_{\mathrm{R}} \end{pmatrix} \tag{3.55}$$

where $\psi_{\mathrm{L}}$ is a left-handed Weyl spinor (representation $(\frac{1}{2},0)$) and $\psi_{\mathrm{R}}$ is a right-handed Weyl spinor (representation $(0,\frac{1}{2})$). This demonstrates that a Dirac spinor has twice as many independent variables ("degrees of freedom") as a Weyl spinor – it quite literally just consists of two Weyl spinors. The projection operators simply project to the upper or lower two-component spinor in this basis:

$$\psi_{\mathrm{D,L}} = P_{\mathrm{L}}\psi_{\mathrm{D}} = \frac{1}{2}(1-\gamma_5)\psi_{\mathrm{D}} = \begin{pmatrix}\psi_{\mathrm{L}}\\ 0\end{pmatrix} \quad \psi_{\mathrm{D,R}} = P_{\mathrm{R}}\psi_{\mathrm{D}} = \frac{1}{2}(1+\gamma_5)\psi_{\mathrm{D}} = \begin{pmatrix}0\\ \psi_{\mathrm{R}}\end{pmatrix} \tag{3.56}$$

[17] The object $\psi_{\mathrm{D}}$ in the Dirac equation is of course not a strictly finite-dimensional object, but a field $\psi_{\mathrm{D}}(x)$. Field representations are discussed further in section 3.2.6.

[18] The name "bispinors" is used because they have twice as many components as Pauli spinors with their two states of "spin up" and "spin down".

[19] With $(\sigma^\mu) = (\mathbb{1}_2, \vec{\sigma})$, $(\bar{\sigma}^\mu) = (\mathbb{1}_2, -\vec{\sigma})$ (cf. appendix B).

Thus, the group transformation acts individually on the two Weyl spinors, illustrating the decomposition into the direct sum:[20]

$$\begin{aligned}\psi_{\mathrm{D}} \mapsto (\rho_{\mathrm{L}} \oplus \rho_{\mathrm{R}})\psi_{\mathrm{D}} &= \begin{pmatrix}\rho_{\mathrm{L}} & 0\\ 0 & \rho_{\mathrm{R}}\end{pmatrix}\psi_{\mathrm{D}} = \begin{pmatrix}\rho_{\mathrm{L}}\psi_{\mathrm{L}}\\ \rho_{\mathrm{R}}\psi_{\mathrm{R}}\end{pmatrix}\\ &= \begin{pmatrix}\exp\left(\frac{1}{2}\omega_{\mu\nu}\sigma^{\mu\nu}\right)\psi_{\mathrm{L}}\\ \exp\left(\frac{1}{2}\omega_{\mu\nu}\bar{\sigma}^{\mu\nu}\right)\psi_{\mathrm{R}}\end{pmatrix} = \exp\left(\frac{1}{2}\omega_{\mu\nu}\gamma^{\mu\nu}\right)\psi_{\mathrm{D}}\end{aligned} \tag{3.57}$$

with $\gamma^{\mu\nu} = \sigma^{\mu\nu} \oplus \bar{\sigma}^{\mu\nu} = \frac{1}{4}[\gamma^\mu, \gamma^\nu]$. A parity transformation of a Dirac spinor yields another Dirac spinor, as opposed to Weyl spinors, where it exchanges left- and right-handed spinors.

It should already be noted at this point that for massless fermions, any distinction between two-component (Weyl) and four-component (Dirac/Majorana) spinors is irrelevant. In the massless case $m = 0$, the Dirac equation (3.52) decouples into two independent equations for $\psi_{\mathrm{L}}$ and $\psi_{\mathrm{R}}$, called the *Weyl equations*:

$$i\partial_\mu\gamma^\mu\psi = \begin{pmatrix}0 & \sigma^\mu\\ \bar{\sigma}^\mu & 0\end{pmatrix}\begin{pmatrix}\psi_{\mathrm{L}}\\ \psi_{\mathrm{R}}\end{pmatrix} = \begin{cases} i\partial_\mu\sigma^\mu\psi_{\mathrm{R}} = 0\\ i\partial_\mu\bar{\sigma}^\mu\psi_{\mathrm{L}} = 0\end{cases} \tag{3.58}$$

Hence, $\psi_{\mathrm{L}}$ and $\psi_{\mathrm{R}}$ can be treated completely independently.

In order to form Lorentz-invariant terms in the Lagrangian, two operations, "Dirac conjugation" (denoted as $\bar{\psi}_{\mathrm{D}}$) and charge conjugation ($\psi^{\mathrm{c}}_{\mathrm{D}}$) are defined for Dirac spinors [DHM10, app. G]:

$$\bar{\psi}_{\mathrm{D}} = \psi^\dagger_{\mathrm{D}}\beta \quad \text{with } \beta\gamma^\mu\beta^{-1} = (\gamma^\mu)^\dagger \text{ (and } \beta^\dagger = \beta) \tag{3.59}$$

$$\psi^{\mathrm{c}}_{\mathrm{D}} = C\bar{\psi}^{\mathrm{T}}_{\mathrm{D}} = C\beta^{\mathrm{T}}\psi^*_{\mathrm{D}} \quad \text{with } C^{-1}\gamma^\mu C = -(\gamma^\mu)^{\mathrm{T}} \text{ (and } (\psi^{\mathrm{c}}_{\mathrm{D}})^{\mathrm{c}} = \psi_{\mathrm{D}}) \tag{3.60}$$

The conditions in parentheses are required so that $(\bar{\psi}_{\mathrm{D}}\psi_{\mathrm{D}})^\dagger = \bar{\psi}_{\mathrm{D}}\psi_{\mathrm{D}}$ and charge conjugation behaves like a conjugation operation (double application is the identity operation). With these conditions, it can be shown that

$$C^{\mathsf{T}} = -C \tag{3.61}$$

[20] The argument $\Lambda$ to $\rho_{\mathrm{L}}$ and $\rho_{\mathrm{R}}$ is omitted here to reduce redundancy, i. e. $\rho_{\mathrm{L}} = \rho_{\mathrm{L}}(\Lambda)$, $\rho_{\mathrm{R}} = \rho_{\mathrm{R}}(\Lambda)$.

Terms written like this use the prevalent four-component spinor formalism. As long as a "barred" Dirac spinor in a product is always matched with an "unbarred" one, the result will be a Lorentz scalar. In the commonly-used bases for the $\gamma$ matrices, the condition for $\beta$ in (3.59) is usually fulfilled by $\beta = \gamma^0$ because it implies the hermicity conditions $(\gamma^0)^\dagger = \gamma^0$, $(\gamma^i)^\dagger = -\gamma^i$. In the Weyl basis, these two matrices take the form

$$\beta = \gamma^0 = \begin{pmatrix} 0 & \mathbb{1}_2 \\ \mathbb{1}_2 & 0 \end{pmatrix} \qquad C = -i\gamma^2\gamma^0 = \begin{pmatrix} -i\sigma^2 & 0 \\ 0 & i\sigma^2 \end{pmatrix} \tag{3.62}$$

Although the Dirac spinor notation may be more familiar to many and has historically and commonly been used because of its connection to the Dirac equation, Weyl spinors are the more fundamental objects and their use can often simplify the treatment of a theory, especially for chiral theories (like the Standard Model) or theories with Majorana fermions. The Weyl spinor formalism and a translation between both formalisms are presented in section 3.3.

### 3.2.4. Majorana spinors

Majorana fermions are the fermionic (spin 1/2) equivalent of real Lorentz scalars compared to Dirac fermions and complex Lorentz scalars. Just like real scalars, they are neutral with respect to all global symmetries of a theory. In a basis (such as the *Majorana basis*) where the $\gamma^\mu$ are purely imaginary, $(\gamma^\mu)^* = -\gamma^\mu$, there are purely real solutions to the Dirac equation (3.52), which represent Majorana fermions [Pal11].

Majorana fermions can be described using Majorana *spinors*, which are defined as Dirac spinors with the additional constraint

$$\psi_{\mathrm{M}}^{\mathrm{c}} = \psi_{\mathrm{M}} \Rightarrow \psi_{\mathrm{R}} = i\sigma^2\psi_{\mathrm{L}}^* \tag{3.63}$$

Such a spinor thus has the form

$$\psi_{\mathrm{M}} = \begin{pmatrix} \psi_{\mathrm{L}} \\ i\sigma^2\psi_{\mathrm{L}}^* \end{pmatrix} \quad \text{or} \quad \psi_{\mathrm{M}} = \begin{pmatrix} -i\sigma^2\psi_{\mathrm{R}}^* \\ \psi_{\mathrm{R}} \end{pmatrix} \tag{3.64}$$

However, as becomes apparent from these definitions, the condition (3.63) cuts the degrees of freedom of a Dirac spinor in half. A Majorana fermion can thus equivalently be described by either a four-component Majorana spinor or by a single two-component Weyl spinor with its conjugate. The choice depends on whether one is working in the four-component spinor formalism or in the two-component one presented in section 3.3.

### 3.2.5. Lorentz vectors: The $(\frac{1}{2},\frac{1}{2})$ representation

The next-largest irreducible representation of the Lorentz group after the Weyl spinors is $(\frac{1}{2},\frac{1}{2})$. Since the irreducible representations are tensor products, this representation is really

$$(\tfrac{1}{2},\tfrac{1}{2}) = (\tfrac{1}{2},0) \otimes (0,\tfrac{1}{2}) \tag{3.65}$$

so the representation space consists of tensor products with one left-handed and one right-handed Weyl spinor each. Since these are order-2 tensors, they can be written as matrices. Explicitly, the representation (denoted $r_\mathrm{V} = r_{(\frac{1}{2},\frac{1}{2})}, \rho_\mathrm{V} = \rho_{(\frac{1}{2},\frac{1}{2})}$)

$$r_\mathrm{V}(T^i_\mathrm{L}) = -\frac{i}{2}\sigma^i \otimes \mathbb{1}_2 \qquad r_\mathrm{V}(T^i_\mathrm{R}) = \mathbb{1}_2 \otimes -\frac{i}{2}\sigma^i \tag{3.66}$$

$$\Rightarrow r_\mathrm{V}(J^i) = -\frac{i}{2}\big(\sigma^i \otimes \mathbb{1}_2 + \mathbb{1}_2 \otimes \sigma^i\big) \qquad r_\mathrm{V}(K^i) = -\frac{1}{2}\big(\sigma^i \otimes \mathbb{1}_2 - \mathbb{1}_2 \otimes \sigma^i\big) \tag{3.67}$$

$$\rho_\mathrm{V}(\Lambda) = \exp\left(\frac{1}{2}(-\vec{\eta} - i\vec{\alpha}) \cdot \vec{\sigma}\right) \otimes \exp\left(\frac{1}{2}(\vec{\eta} - i\vec{\alpha}) \cdot \vec{\sigma}\right) \tag{3.68}$$

$$= \exp\left(\frac{1}{2}\omega_{\mu\nu}\sigma^{\mu\nu}\right) \otimes \exp\left(\frac{1}{2}\omega_{\mu\nu}\bar{\sigma}^{\mu\nu}\right) \tag{3.69}$$

$(\frac{1}{2},\frac{1}{2})$ is the only four-dimensional irreducible representation, which already provides a first hint that it must be connected to Lorentz vectors, the defining representation of the Lorentz group. That they are really equivalent can be seen by looking at a representation on the (real) vector space of Hermitian $2 \times 2$ matrices ($M^\dagger = M$), a basis of which is given by the matrices $\sigma^\mu$. A general Hermitian matrix $V$ is then given by[21]

$$V_{a\dot{b}} = V_\mu(\sigma^\mu)_{a\dot{b}} = \begin{pmatrix} V_0 + V_3 & V_1 - iV_2 \\ V_1 + iV_2 & V_0 - V_3 \end{pmatrix}_{a\dot{b}} \tag{3.70}$$

[21]The right-handed spinor indices are already denoted using dots as in accord with section 3.3.

It can be shown that in this basis, the tensor transformation[22]

$$V \in (\tfrac{1}{2}, \tfrac{1}{2}) \Rightarrow V_{a\dot{b}} \mapsto \exp\left(\frac{1}{2}\omega_{\mu\nu}\sigma^{\mu\nu}\right)_a^{\ c} \exp\left(\frac{1}{2}\omega_{\mu\nu}\bar{\sigma}^{\mu\nu}\right)_{\dot{b}}^{\ \dot{d}} V_{c\dot{d}} \tag{3.71}$$

has the same effect as the vector transformation

$$V^\mu \mapsto \Lambda^\mu{}_\nu V^\nu \tag{3.72}$$

when translated to the components $V^\mu$, where $\Lambda$ are indeed the original Lorentz transformation matrices from (3.23). Thus, the suggestive notation $V^\mu$ used in (3.70) for the components of $V$ with respect to the basis $\sigma_\mu$ is justified and the $(\frac{1}{2}, \frac{1}{2})$ representation can indeed equivalently be viewed either as a $2 \times 2$ matrix formed from a left- and a right-handed Weyl spinor or as 4-component Lorentz vector.

### 3.2.6. Field representations

So far, only finite-dimensional representations of the Lorentz group have been considered. These consist of matrices $\rho(\Lambda)$ acting on constant vectors $\psi$ from some vector space:

$$\psi \mapsto \rho(\Lambda)\psi \tag{3.73}$$

However, in field theories like the Standard Model (see section 3.4), the objects of the representation space are not constant, but fields, i. e. functions $\psi(x)$ of spacetime. Representations on a function space are generally infinite-dimensional because an infinite number of basis functions is usually necessary to construct a function (as, for example, with the Fourier transform, which uses the basis functions $e^{ik\cdot x}$).

Fortunately, the finite-dimensional representations and the infinite-dimensional field behavior are independent (every component of a finite-dimensional representation is simply a field by itself) and can thus be treated independently. The most basic field is the scalar field $\phi(x)$, whose values do not have additional structure (and thus behavior under transformations). If the field value $\phi(x)$ is observed in one reference frame, the coordinates in a transformed frame are $x' = \Lambda x$ (where $\Lambda$ are the original "vector" Lorentz transformations from (3.23)) and the observed value is $\phi'(x')$.[23] However, $x$ and $x'$ correspond to the same point in spacetime (just in different reference frames), so it must be true that $\phi'(x') = \phi(x)$. Put differently, the transformation of the field $\phi$ is

$$\phi(x) \mapsto \phi(\Lambda^{-1}x) \tag{3.74}$$

[22] It should be noted that the transformation $\exp\left(\frac{1}{2}\omega_{\mu\nu}\bar{\sigma}^{\mu\nu}\right)_{\dot{b}}^{\ \dot{d}}$ with the indices given as specified is not the same as the one normally obtained from $\rho_R = \exp\left(\frac{1}{2}\omega_{\mu\nu}\bar{\sigma}^{\mu\nu}\right)$, but really $\sigma^2\rho_R\sigma^2$. This change is necessary because the right-handed part ("dotted index") of the Pauli matrices (when interpreted as a tensor of Weyl spinors) is not exactly identical to a right-handed Weyl spinor, but corresponds to a spinor of the form $-i\sigma^2\psi_R^*$ (the index has been lowered). All of this becomes clear in section 3.3.

Combining arbitrary finite-dimensional representations and fields, a general (multi-component) Lorentz field $\psi(x)$ should behave as

$$\psi(x) \mapsto \rho(\Lambda)\psi(\Lambda^{-1}x) \tag{3.75}$$

One can think about the infinite-dimensional part as coming from the representation

$$r_\infty(M^{\mu\nu}) = x^\mu \partial^\nu - x^\nu \partial^\mu \tag{3.76}$$

which fulfills (3.21) as necessary. (3.74) then takes the form[24]

$$\phi(x) \mapsto \exp\left(\frac{1}{2}\omega_{\mu\nu} r_\infty(M^{\mu\nu})\right)\phi(x) = \phi(\Lambda^{-1}x) \tag{3.77}$$

Together with a finite-dimensional representation $r(M^{\mu\nu})$, a general field transforms as

$$\psi(x) \mapsto \exp\left(\frac{1}{2}\omega_{\mu\nu} r(M^{\mu\nu})\right)\exp\left(\frac{1}{2}\omega_{\mu\nu} r_\infty(M^{\mu\nu})\right)\psi(x) = \exp\left(\frac{1}{2}\omega_{\mu\nu} r_{\mathrm{f}}(M^{\mu\nu})\right) \tag{3.78}$$

where[25] $r_{\mathrm{f}}(M^{\mu\nu}) = r(M^{\mu\nu}) + r_\infty(M^{\mu\nu})$ is called a field representation of the Lorentz group.

[23] When discussing transformations, there is always the question whether active or passive transformations are intended. With active transformations, the transformed object is actually changed – it is different from the original object. With passive transformations, only the coordinate system (basis) is changed, but the object stays the same; it seems "different", though, because the coordinates used to describe it change. Given an active transformation, its *inverse* must be used as a passive transformation to achieve the same effect. The confusion between the two is unfortunately exacerbated by the fact that for functions (fields) there is a difference between defining a new function (active) and simply using different arguments to the same function (passive). In this work, only active transformations are used.

[24] For example, one can verify that $r_\infty(M^{\mu\nu})x^\alpha = (x^\mu\partial^\nu - x^\nu\partial^\mu)x^\alpha = (\eta^{\nu\alpha}\delta^\mu_\beta - \eta^{\mu\alpha}\delta^\nu_\beta)x^\beta = -(M^{\mu\nu})^\alpha{}_\beta x^\beta$ with $M^{\mu\nu}$ as in (3.22).

[25] By the Baker–Campbell–Hausdorff formula, this is possible because $r(M^{\mu\nu})$ and $r_\infty(M^{\mu\nu})$ commute.

Therefore, in order to extend the previous discussion of finite-dimensional representations to fields, the objects from the representation space can simply be viewed as fields $(\phi(x), \psi_L(x), \psi_R(x), \psi_D(x), V(x))$ whose arguments go to $\Lambda^{-1}x$ under transformations. In the following, all representations are taken to be field representations, but the field arguments will generally be omitted except when relevant.

## 3.3. Two-component Weyl spinor formalism and van der Waerden notation

### 3.3.1. Definition

The notation and conventions used in this section and the rest of the thesis are essentially the same as in [DHM10], which provides a vastly comprehensive review of Weyl and Dirac spinors as well as their correspondence to each other. Consequently, the formalism presented here was originally developed and compiled in that report.

Just like the four-vector formalism with raised and lowered Greek indices allows handling sums and products of Lorentz tensors of different types in a simple and unified way, a similar convention has been established for the left- and right-handed two-component Weyl spinors. This convention is called *van der Waerden notation* and uses dotted and undotted indices for the components of the different spinor kinds. Exactly like the four-vector index convention, it makes it obvious whether a theory is written in a manifestly Lorentz-invariant way.

Denoting a left-handed $(\frac{1}{2}, 0)$ Weyl spinor as $\psi$ and a right-handed $(0, \frac{1}{2})$ one as $\xi$, the basic foundation of van der Waerden notation is the definition that their components are denoted by $\psi_a$ and $\xi^{\dot{a}}$.[26] The transformations of objects with these

[26]The convention here actually differs in a small aspect (following [Sch15b]) from [DHM10] (which makes no difference to the following discussion, though). In (3.33), the right-handed part of the representations of the Lorentz group was defined to use the same representations $J_{(j)}$ of $\mathfrak{su}(2)$ as the left-handed part, leading to the form of $\rho_L$ given in (3.49) and (3.50). However, since these representations of SU(2) are real or pseudo-real, they are equivalent to their conjugates and differ only by a change of basis.

Thus, one could just as well define that $(0, 1/2)$ use the conjugate representation $J^*_{(j)}$, and that is precisely what is done in [DHM10]. This means that, instead, the spinors of $(0, 1/2)$ naturally have lowered indices $\xi_{\dot{a}}$ in their convention. This makes the fact that the representations are conjugates, $(j_L, j_R)^* = (j_R, j_L)$, manifest. On the other hand, one is more interested in the spinors with raised indices $\xi^{\dot{a}}$ because these are the objects appearing as the right-handed parts of Dirac spinors. Also, the definition of $\bar{\sigma}^{\mu\nu}$ naturally leads to (3.50), which is the transformation matrix for $\xi^{\dot{a}}$, not for $\xi_{\dot{a}}$. The net effect caused by this difference is that when talking about $(0, 1/2)$ without indices, $\xi^{\dot{a}}$ is exchanged for $\xi_{\dot{a}}$ and $(\rho_L^{-1})^\dagger$ for $\rho_L^*$. However, since one does not often refer to the representation itself explicitly, working instead with component notation, this distinction is mostly irrelevant.

indices are given by

$$\psi_a \mapsto (\rho_{\mathrm{L}})_a{}^b \psi_b = \exp\left(\frac{1}{2}(-\vec{\eta} - i\vec{\alpha}) \cdot \vec{\sigma}\right)_a{}^b \psi_b \tag{3.79}$$

$$\xi^{\dot{a}} \mapsto (\rho_{\mathrm{R}})^{\dot{a}}{}_{\dot{b}} \psi^{\dot{b}} = \exp\left(\frac{1}{2}(\vec{\eta} - i\vec{\alpha}) \cdot \vec{\sigma}\right)^{\dot{a}}{}_{\dot{b}} \xi^{\dot{b}} \tag{3.80}$$

From the observation that the two-dimensional Levi–Civita symbol $\varepsilon$ is invariant under $\mathrm{SL}(2, \mathbb{C})$, one can define the *spinor metric*

$$(\varepsilon^{ab}) = (\varepsilon^{\dot{a}\dot{b}}) = \varepsilon = i\sigma^2 = \begin{pmatrix} 0 & 1 \\ -1 & 0 \end{pmatrix} \tag{3.81}$$

$$(\varepsilon_{ab}) = (\varepsilon_{\dot{a}\dot{b}}) = \varepsilon^{-1} = -i\sigma^2 = \begin{pmatrix} 0 & -1 \\ 1 & 0 \end{pmatrix} \tag{3.82}$$

which can be used to raise and lower indices just like the Minkowski metric $\eta$ (see appendix C, chapter 4) to form invariant products:

$$\psi^a = \varepsilon^{ab}\psi_b \qquad \psi_a = \varepsilon_{ab}\psi^b \qquad \xi^{\dot{a}} = \varepsilon^{\dot{a}\dot{b}}\xi_{\dot{b}} \qquad \xi_{\dot{a}} = \varepsilon_{\dot{a}\dot{b}}\xi^{\dot{b}} \tag{3.83}$$

Using the identity $\sigma^2\sigma^i\sigma^2 = -(\sigma^i)^* = -(\sigma^i)^\mathsf{T}$, it can be shown that the remaining spinors with raised/lowered indices transform as[27]

$$\psi^a \mapsto (\varepsilon\rho_{\mathrm{L}}\varepsilon^{-1})^a{}_b \psi^b = \exp\left(\frac{1}{2}(\vec{\eta} + i\vec{\alpha}) \cdot \vec{\sigma}^\mathsf{T}\right)^a{}_b \psi^b \tag{3.84}$$

$$\xi_{\dot{a}} \mapsto (\varepsilon^{-1}\rho_{\mathrm{R}}\varepsilon)_{\dot{a}}{}^{\dot{b}} \psi_{\dot{b}} = \exp\left(\frac{1}{2}(-\vec{\eta} + i\vec{\alpha}) \cdot \vec{\sigma}^\mathsf{T}\right)_{\dot{a}}{}^{\dot{b}} \psi_{\dot{b}} \tag{3.85}$$

Thus, in total, it becomes clear that all the transformations can be expressed in terms of one matrix $\rho_{\mathrm{L}}$:[28]

$$\psi_a \mapsto (\rho_{\mathrm{L}})_a{}^b \psi_b \tag{3.86}$$

$$\psi^a \mapsto (\rho_{\mathrm{L}}^{-1^\mathsf{T}})^a{}_b \psi^b \tag{3.87}$$

[27] In this case, $\vec{\sigma}^\mathsf{T}$ means $\big((\sigma^i)^\mathsf{T}\big)$.

[28] This also reveals that the irreducible representations using $\rho_{\mathrm{L}}^{-1^\mathsf{T}}$ and $\rho_{\mathrm{L}}^{-1^\dagger}$ are equivalent to the ones using $\rho_{\mathrm{L}}$ and $\rho_{\mathrm{L}}^*$ because they are, respectively, simply related to each other by a basis transformation with $\varepsilon$ as the transformation matrix.

$$\tilde{\zeta}^{\dot{a}} \mapsto ({\rho_{\mathrm{L}}^{-1}}^{\dagger})^{\dot{a}}{}_{\dot{b}}\tilde{\zeta}^{\dot{b}} \tag{3.88}$$

$$\tilde{\zeta}_{\dot{a}} \mapsto (\rho_{\mathrm{L}}^{*})_{\dot{a}}{}^{\dot{b}}\tilde{\zeta}_{\dot{b}} \tag{3.89}$$

But this result immediately implies that everything can be expressed in terms of one kind of spinor because the conjugate of a left-handed spinor *is* a right-handed spinor and vice versa:[29]

$$\psi^{\dagger\dot{a}} = (\psi^{a})^{\dagger} \mapsto ({\rho_{\mathrm{L}}^{-1}}^{\dagger})^{\dot{a}}{}_{\dot{b}}\psi^{\dagger\dot{b}} \tag{3.90}$$

This is hardly surprising since, as has been asserted before, $(\frac{1}{2}, 0)$ and $(0, \frac{1}{2})$ are conjugate to each other. In light of this, the convention is generally that *all Weyl spinors are exclusively defined to be left-handed*. If a right-handed spinor is needed, the conjugate of the corresponding left-handed spinor will be used.

Another useful definition[30] to make is

$$\bar{\psi} = i\sigma^{2}\psi^{*} = \varepsilon\psi^{*} \tag{3.91}$$

(where complex conjugation means to form the adjoint of all components if $\psi$ is a field). This is sometimes called a "charge conjugation" operation for Weyl spinors. It differs from $\psi^{\dagger}$ only in that it "naturally" has a raised dotted index. On the component level, there is no difference between the different versions because the position of the index specifies whether the $\varepsilon$ metric must be used:

$$\bar{\psi}^{\dot{a}} = \psi^{*\dot{a}} = \psi^{\dagger\dot{a}} \tag{3.92}$$

Going forward, the notation using $\bar{\psi}$ will be used. It should be duly noted, however, that this notation is *completely unrelated* to the bar operation $\bar{\psi}_{\mathrm{D}} = \psi_{\mathrm{D}}^{\dagger}\beta$ on Dirac spinors!

As hinted at before, the spinor metric can be used to define invariant products

$$\psi\chi = \psi^{a}\chi_{a} = \varepsilon^{ab}\psi_{b}\chi_{a} \qquad \bar{\psi}\bar{\chi} = \bar{\psi}_{\dot{a}}\bar{\chi}^{\dot{a}} = \varepsilon_{\dot{a}\dot{b}}\bar{\psi}^{\dot{b}}\bar{\chi}^{\dot{a}} \tag{3.93}$$

which form Lorentz scalars due to the invariance of $\varepsilon$ (see appendix C). This invariant product has the interesting property that it is commutative if the components of $\psi$ and $\chi$ *anti*-commute, whereas $\psi\chi = -\chi\psi$ for commuting components (cf. section 4.1.2).

[29] Conjugate Weyl spinors are conventionally written as $\psi^{\dagger}$. As far as components and indices are concerned, the adjoint has the same effect as complex conjugation ($\psi^{\dagger\dot{a}} = \psi^{*\dot{a}}$).

[30] going beyond [DHM10]

These definitions establish a larger pattern: *Whenever two adjacent indices are summed from top to bottom (undotted)* $^a{}_a$ *or from bottom to top (dotted)* $_{\dot a}{}^{\dot a}$*, they can be omitted.* This convention is valid in general (for example, also for matrix multiplication). However, if this condition is not fulfilled, the indices *cannot* be omitted because the resulting expression would be different from the original one.

The index structure of the $\sigma$ matrices can be inferred from the definitions made so far:

$$(\sigma^\mu)_{a\dot b} \qquad (\bar\sigma^\mu)^{\dot a b} \qquad (\sigma^{\mu\nu})_a{}^b \qquad (\bar\sigma^{\mu\nu})^{\dot a}{}_{\dot b} \tag{3.94}$$

Also note that changing unrelated indices (such as Lorentz indices) has no effect on Weyl spinor indices.

Finally, the bar notation for Weyl spinors, which has been introduced piecemeal so far via the $\sigma$ matrices ($\sigma^\mu$, $\bar\sigma^\mu$, $\sigma^{\mu\nu}$, $\bar\sigma^{\mu\nu}$) and the conjugate spinors $\bar\psi$, could be viewed as a bigger picture: One could define a "Weyl bar operation" which maps any $\sigma$ matrix and spinor to their "barred" version, with double bars giving the unbarred object again. In this way, taking the bar consistently exchanges left- and right-handed spinors, including their transformation matrices with $\bar\rho_\mathrm{L} = \rho_\mathrm{L}^*$, $\bar\rho_\mathrm{L}^* = \bar{\bar\rho}_\mathrm{L} = \rho_\mathrm{L}$.

### 3.3.2. Correspondence to the Dirac bispinor formalism

Due to the widespread use of the four-component formalism using Dirac spinors, it is often necessary to translate between the two conventions. As seen in section 3.2.3, the correspondence is natural in the Weyl basis.

A Dirac spinor field $\psi_\mathrm{D}$ in the Weyl basis can be constructed from two left-handed Weyl spinor fields $\psi$ and $\chi$ via

$$\psi_\mathrm{D}(x) = \begin{pmatrix} \psi(x) \\ \bar\chi(x) \end{pmatrix} \tag{3.95}$$

In anticipation of the application to gauge theories, $\psi$ and $\chi$ must have the same mass and opposite charge – otherwise, nothing gauge-invariant can be constructed from this field. Applying Dirac and charge conjugation yields all variants of the Dirac spinor:

$$\begin{aligned} \psi_\mathrm{D} &= \begin{pmatrix} \psi \\ \bar\chi \end{pmatrix} = \begin{pmatrix} (\psi_a) \\ (\bar\chi^{\dot a}) \end{pmatrix} & \bar\psi_\mathrm{D} &= \begin{pmatrix} (\chi^a) & (\bar\psi_{\dot a}) \end{pmatrix} \\ \psi_\mathrm{D}^\mathrm{c} &= \begin{pmatrix} \chi \\ \bar\psi \end{pmatrix} = \begin{pmatrix} (\chi_a) \\ (\bar\psi^{\dot a}) \end{pmatrix} & \overline{\psi_\mathrm{D}^\mathrm{c}} &= \begin{pmatrix} (\psi^a) & (\bar\chi_{\dot a}) \end{pmatrix} \end{aligned} \tag{3.96}$$

In particular, it becomes clear that charge conjugation is simply the exchange of the two Weyl spinors making up the Dirac spinor. Majorana spinors are simply a special case of such Dirac spinors, with the charge conjugation condition

$$\psi_{\mathrm{M}}^{\mathrm{c}} = \psi_{\mathrm{M}} \Rightarrow \psi = \chi \tag{3.97}$$

The most important part is the translation of Dirac bilinears, because these are used to form Lorentz-invariant terms in the Lagrangian. These are also listed, for example, in [DHM10; Sem09].[31] With (3.54) and (3.96), these are now straightforward to compute. For two Dirac spinors $\Psi_1, \Psi_2$ with $\Psi_i = (\psi_i, \bar{\chi}_i)$, the scalar, pseudo-scalar ($\gamma_5$), vector ($\gamma^\mu$) and pseudo-vector ($\gamma^\mu\gamma_5$) bilinears correspond as follows:

$$\left.\begin{aligned} \bar{\Psi}_1 P_{\mathrm{L}} \Psi_2 &= \chi_1 \psi_2 \\ \bar{\Psi}_1 P_{\mathrm{R}} \Psi_2 &= \bar{\psi}_1 \bar{\chi}_2 \end{aligned}\right\} \bar{\Psi}_1 \Psi_2 = \chi_1 \psi_2 + \bar{\psi}_1 \bar{\chi}_2 \tag{3.98}$$

$$\left.\begin{aligned} \bar{\Psi}_1 \gamma_5 P_{\mathrm{L}} \Psi_2 &= -\chi_1 \psi_2 \\ \bar{\Psi}_1 \gamma_5 P_{\mathrm{R}} \Psi_2 &= \bar{\psi}_1 \bar{\chi}_2 \end{aligned}\right\} \bar{\Psi}_1 \gamma_5 \Psi_2 = -\chi_1 \psi_2 + \bar{\psi}_1 \bar{\chi}_2 \tag{3.99}$$

$$\left.\begin{aligned} \bar{\Psi}_1 \gamma^\mu P_{\mathrm{L}} \Psi_2 &= \bar{\psi}_1 \bar{\sigma}^\mu \psi_2 \\ \bar{\Psi}_1 \gamma^\mu P_{\mathrm{R}} \Psi_2 &= \chi_1 \sigma^\mu \bar{\chi}_2 \end{aligned}\right\} \bar{\Psi}_1 \gamma^\mu \Psi_2 = \bar{\psi}_1 \bar{\sigma}^\mu \psi_2 + \chi_1 \sigma^\mu \bar{\chi}_2 \tag{3.100}$$

$$\left.\begin{aligned} \bar{\Psi}_1 \gamma^\mu \gamma_5 P_{\mathrm{L}} \Psi_2 &= -\bar{\psi}_1 \bar{\sigma}^\mu \psi_2 \\ \bar{\Psi}_1 \gamma^\mu \gamma_5 P_{\mathrm{R}} \Psi_2 &= \chi_1 \sigma^\mu \bar{\chi}_2 \end{aligned}\right\} \bar{\Psi}_1 \gamma^\mu \gamma_5 \Psi_2 = -\bar{\psi}_1 \bar{\sigma}^\mu \psi_2 + \chi_1 \sigma^\mu \bar{\chi}_2 \tag{3.101}$$

One can take note that replacing $\Psi_i$ by $\Psi_i^{\mathrm{c}}$ results in the exchange of $\psi_i$ and $\chi_i$, while inserting $\gamma_5$ simply gives a minus sign in the product involving $P_{\mathrm{L}}$. The complete list of all combinations can be found in appendix D.

In particular, one can take note that the product of the two Weyl spinors $\psi_1$, $\psi_2$ is not recovered using the combination $\bar{\Psi}_{1,\mathrm{L}} \Psi_{2,\mathrm{L}}$ since

$$\bar{\Psi}_{1,\mathrm{L}} \Psi_{2,\mathrm{L}} = \bar{\Psi}_{1,\mathrm{R}} \Psi_{2,\mathrm{R}} = 0 \tag{3.102}$$

Instead, charge conjugation is necessary to obtain the product of two left-handed or right-handed components:

$$\overline{\Psi^{\mathrm{c}}_{1,\mathrm{L}}} \Psi_{2,\mathrm{L}} = \psi_1 \psi_2 \qquad \overline{\Psi^{\mathrm{c}}_{1,\mathrm{R}}} \Psi_{2,\mathrm{R}} = \chi_1 \chi_2 \tag{3.103}$$

[31] One should be careful, however, because the identities given on the second and sixth line of [Sem09, p. 15] are incorrect – they should read $\bar{\xi}_1 \bar{\eta}_2 = \bar{\psi}_1 P_{\mathrm{R}} \psi_2$ and $\bar{\eta}_1 \bar{\xi}_2 = \bar{\psi}_1^{\mathrm{c}} P_{\mathrm{R}} \psi_2^{\mathrm{c}}$.

This already hints at the difference between "Dirac mass terms", which have the form (3.102), and "Majorana mass terms", which have the form (3.103): For Dirac fermions, both left- and right-handed components are needed, while for Majorana fermions, a single Weyl spinor is sufficient. However, due to the appearance of charge conjugation in (3.103), a fermion which consists of only one Weyl spinor but is also massive must necessarily be neutral.

One can also emulate Weyl spinors within the bispinor formalism by using Dirac spinors which only appear together with a projector, e. g. $P_\mathrm{L}\psi_\mathrm{D}$. Equivalently, one directly define a projected Dirac spinor (whose upper or lower components are 0). Since such a spinor contains nothing but a single Weyl spinor, the equivalence to a "proper" Weyl spinor is obvious. In fact, this is already necessary in the Standard Model because the neutrinos, being massless and therefore Weyl fermions, do not have a right-handed counterpart. The neutrino field $\nu$ could be defined as follows:

$$\begin{aligned} \nu &= \begin{pmatrix} \nu_\mathrm{L} \\ 0 \end{pmatrix} & \bar{\nu} &= \begin{pmatrix} 0 & (\bar{\nu}_{\mathrm{L}\dot{a}}) \end{pmatrix} \\ \nu^\mathrm{c} &= \begin{pmatrix} 0 \\ \bar{\nu}_\mathrm{L} \end{pmatrix} & \overline{\nu^\mathrm{c}} &= \begin{pmatrix} (\nu_\mathrm{L}^a) & 0 \end{pmatrix} \end{aligned} \tag{3.104}$$

## 3.4. The Standard Model

This section will provide a brief overview of the defining characteristics of the Standard Model of particle physics. Even by itself, the Standard Model contains a vast amount of mathematical intricacies and phenomenology, which (of course) cannot all be discussed here.

The Standard Model of particle physics and most of its extensions are (relativistic) *gauge quantum field theories*. That means that the Standard Model is:

- A *quantum field theory* (QFT). Such theories incorporate both the special theory of relativity (SRT) and quantum mechanics (QM).
  - They are called field theories because the variables or "degrees of freedom" appearing in their equations of motion (and, accordingly, in the corresponding Lagrangian and Hamiltonian) are not simply numbers, but fields, i. e. functions of spacetime $\phi(x)$. This allows the description of a variable number of particles, which is necessary for relativistic physics due to the existence of antiparticles: The total number of particles can change through absorption, emission, decay, annihilation and pair production.

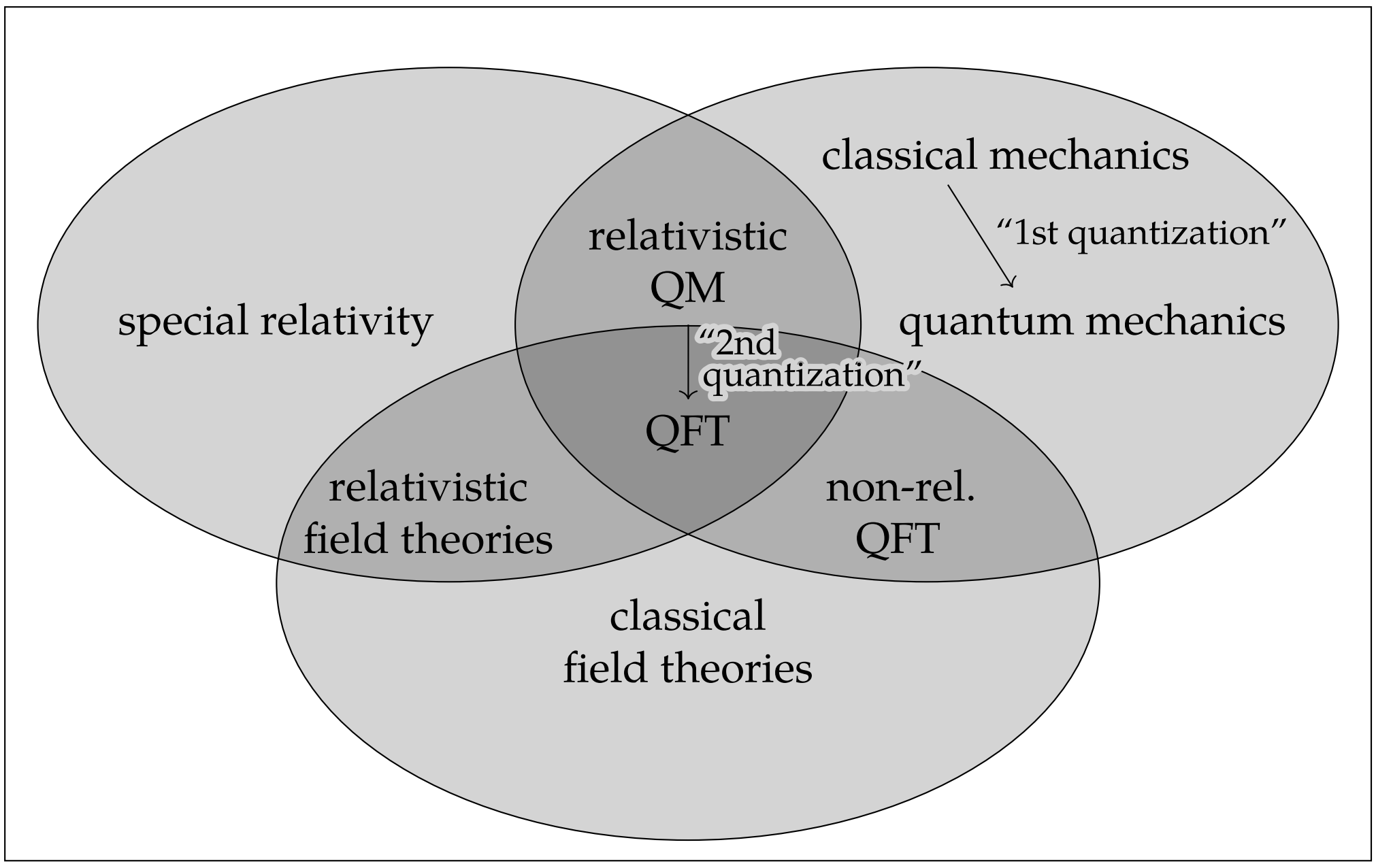

**Figure 3.1.:** Illustration of the different types of theories resulting from the combination of special relativity, quantum mechanics and field theories.

- They incorporate special relativity through Lorentz-invariant[32] quantities and equations.
- They incorporate quantum mechanics in that the fields $\phi$ are not ordinary (scalar-valued) functions, but operator-valued functions of spacetime. Their values $\phi(x)$ are quantum-mechanical operators acting on a Hilbert space whose elements represent the quantum-mechanical states. The passage from "single-particle" quantum mechanics ("wave functions yield complex numbers and represent quantum-mechanical states") to quantum field theory ("wave functions yield operators *acting on* quantum-mechanical states") is also called *second quantization* or *canonical quantization*.

- A *gauge theory*. Such theories are based on the principle of *gauge invariance*, which is the invariance under local[33] symmetry transformations called *gauge transformations*. A gauge symmetry is also called an "internal symmetry" in

[32] Technically, the theory should be invariant under the Poincaré group, which extends the Lorentz group by spacetime *translations*. However, since translations commute and thus form an abelian group, they are not difficult to handle in addition to the Lorentz group.

[33] i. e. depending on the spacetime point $x$

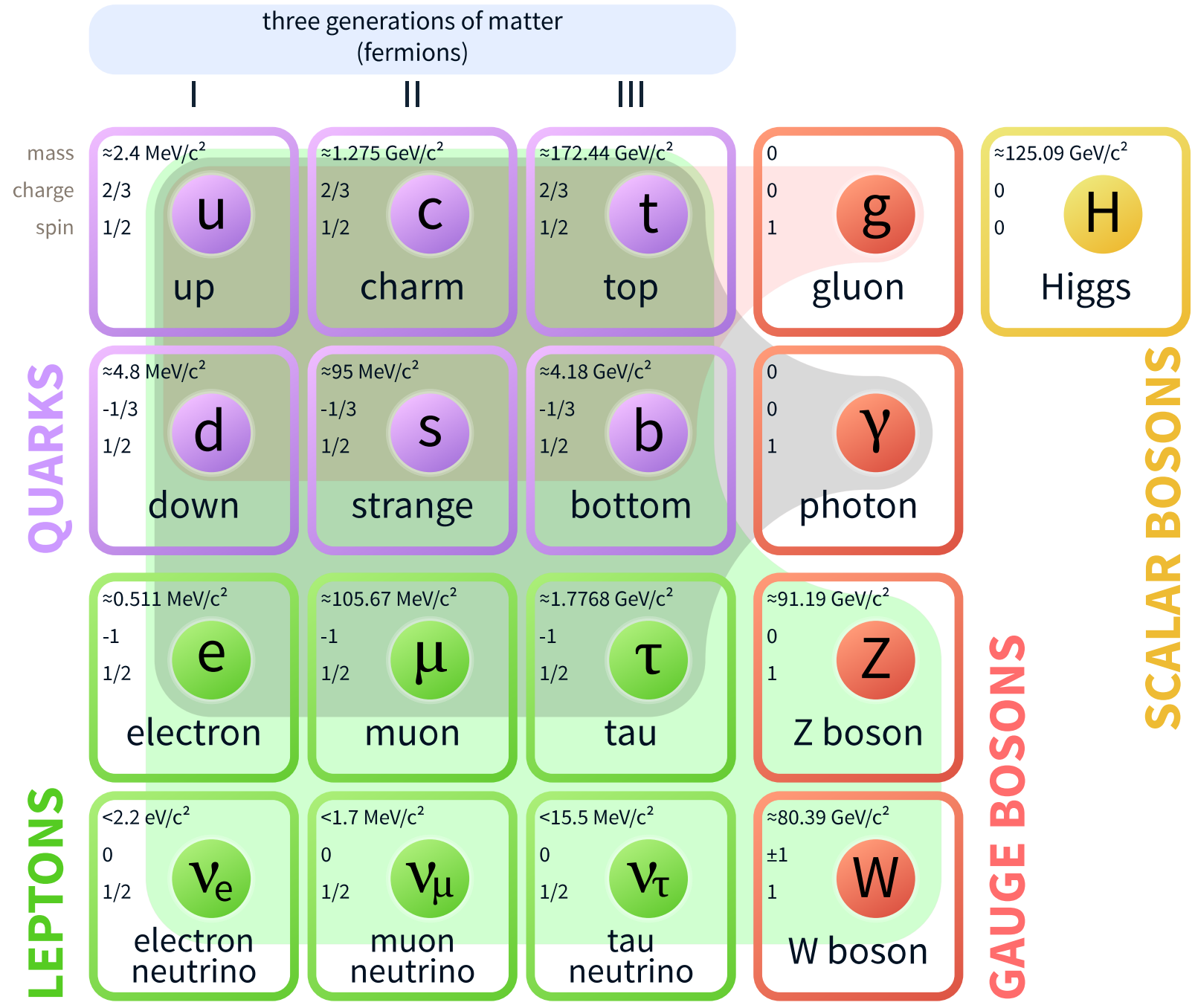

**Figure 3.2.:** A schematic overview of the particles and interactions of the Standard Model of particle physics. Based on `https://commons.wikimedia.org/wiki/File:Standard_Model_of_Elementary_Particles.svg` (retrieved 2018-05-07).

contrast with the "external symmetry" given by the symmetry group of space-time.

An overview of the different kinds of theories resulting from combinations of special relativity, quantum mechanics and field theories is shown in fig. 3.1.[34]

The Standard Model's gauge group is $SU(3) \times SU(2) \times U(1)$. The different subgroups correspond to the three different fundamental interactions (strong, weak and electromagnetic) described by the Standard Model, whose interplay with all the known elementary particles is illustrated schematically in fig. 3.2. $SU(3)$ corresponds to *quantum chromodynamics* (QCD), which describes the strong interaction of quarks and gluons. $SU(2) \times U(1)$ corresponds to the electroweak interaction, which is the unification of both the weak and the electromagnetic interaction at high energies. Through the Higgs mechanism, this group undergoes spontaneous symmetry break-ing at low energies, giving mass to the particles and splitting into the weak and electromagnetic interactions.

[34] A note on relativistic quantum mechanics: Relativistic QM treats the relativistic wave equations (Klein–Gordon equation, Dirac equation) as equations for single-particle wave functions. However, at the same time, these relativistic equations always imply the existence of antiparticles, contradicting the assumption of a fixed number of particles. Due to this, relativistic quantum mechanics exhibits a number of inconsistencies, necessitating the formulation of field theories.

It should be stressed that the Standard Model of particle physics has been and still is extremely successful in describing high-precision experimental data and even predicting new phenomena before their first observation. Currently, there is no data from collider experiments up to the TeV scale which contradicts its predictions. There *are*, however, a number of unexplained phenomena not incorporated into the Standard Model, two of which are the subject of this thesis.

### 3.4.1. Definition of the theory

In order to construct a gauge theory, one must specify the gauge groups and the fields of the theory – in particular, which representation of the Lorentz group and the gauge group it transforms under. The representation of the Lorentz group determines the spin of the field's particles, and thus in particular whether they are bosons or fermions, while the representation of the gauge group determines how many different particles correspond to a field (via the dimension of the representation space) and what charges (which are conserved quantities due to the gauge symmetry) they carry.

Subsequently, it would also be necessary to specify the Lagrangian density

$$\mathcal{L}\big(\phi_i(x), \partial_\mu \phi_i(x)\big) \tag{3.105}$$

to fully define the theory. However, empirically, the principle seems to be that any term allowed by symmetry (Lorentz- and gauge-invariant) that could appear in the Lagrangian must necessarily be present for a full description of nature. Taken to its conclusion, this means that it is sufficient to specify the gauge groups (which already imply the gauge bosons) and the additional fields in order to obtain a complete definition of the theory. The Lagrangian can then simply be derived by enumerating all the allowed (invariant) terms.

Correspondingly, the Standard Model can be defined using table 3.1. The column "generations" means that the theory contains that many identical copies of the given field. The only difference is in the parameters related to each copy. The spin gives the intrinsic angular momentum of the field's particles and is related to its Lorentz representation. The columns SU(3), SU(2) and U(1) show what representation of the gauge groups a field transforms as. The $n$-dimensional irreducible representations of SU($N$) are conventionally labeled as $\mathbf{n}$, while $\bar{\mathbf{n}}$ is the conjugate representation of $\mathbf{n}$. Finally, the Standard Model also has some global symmetries: *lepton number* and

**Table 3.1.:** The field content of the Standard Model of particle physics with assignment of "quantum numbers" (representations of the symmetry groups). The columns $L$ and $B$ show the charges under the "accidental" global symmetries of lepton number and baryon number. $G$, $W$ and $B$ are the gauge boson fields. Symbols like $e_\mathrm{R}^\mathrm{c}$ are simply names for left-handed spinors and do not imply the use of an operation on a field "$e_\mathrm{R}$".

| Field | Generations | Spin | Lorentz rep. | SU(3) | SU(2) | U(1) | $L$ | $B$ |
|---|---|---|---|---|---|---|---|---|
| $G$ | — | 1 | $(1/2, 1/2)$ | **8** | **1** | 0 | 0 | 0 |
| $W$ | — | 1 | $(1/2, 1/2)$ | **1** | **3** | 0 | 0 | 0 |
| $B$ | — | 1 | $(1/2, 1/2)$ | **1** | **1** | 0 | 0 | 0 |
| $L$ | 3 | $1/2$ | $(1/2, 0)$ | **1** | **2** | $-1$ | 1 | 0 |
| $Q$ | 3 | $1/2$ | $(1/2, 0)$ | **3** | **2** | $1/3$ | 0 | $1/3$ |
| $e_\mathrm{R}^\mathrm{c}$ | 3 | $1/2$ | $(1/2, 0)$ | **1** | **1** | 2 | $-1$ | 0 |
| $u_\mathrm{R}^\mathrm{c}$ | 3 | $1/2$ | $(1/2, 0)$ | $\bar{\mathbf{3}}$ | **1** | $-4/3$ | 0 | $-1/3$ |
| $d_\mathrm{R}^\mathrm{c}$ | 3 | $1/2$ | $(1/2, 0)$ | $\bar{\mathbf{3}}$ | **1** | $2/3$ | 0 | $-1/3$ |
| $H$ | 1 | 0 | $(0, 0)$ | **1** | **2** | 1 | 0 | 0 |

*baryon number*. They are called "accidental" because they were not postulated in the beginning, but follow from the specific combination of defined fields and symmetry groups. Using the "charge" assignments of table 3.1, the Standard Model is invariant under appropriate global (constant, independent of $x$) U(1) transformations.

The components of the Standard Model's SU(2) multiplets conventionally have the following names:

$$L = \begin{pmatrix} \nu_\mathrm{L} \\ e_\mathrm{L} \end{pmatrix} \qquad Q = \begin{pmatrix} u_\mathrm{L} \\ d_\mathrm{L} \end{pmatrix} \qquad H = \begin{pmatrix} H^+ \\ H^0 \end{pmatrix} \tag{3.106}$$

Moreover, the different generations of fields are also often assigned individual names, as shown in fig. 3.2:

$$L_1 = \begin{pmatrix} \nu_{e\mathrm{L}} \\ e_\mathrm{L} \end{pmatrix} \qquad L_2 = \begin{pmatrix} \nu_{\mu\mathrm{L}} \\ \mu_\mathrm{L} \end{pmatrix} \qquad L_3 = \begin{pmatrix} \nu_{\tau\mathrm{L}} \\ \tau_\mathrm{L} \end{pmatrix} \tag{3.107}$$

$$(e_\mathrm{R}^\mathrm{c})_1 = e_\mathrm{R}^\mathrm{c} \qquad (e_\mathrm{R}^\mathrm{c})_2 = \mu_\mathrm{R}^\mathrm{c} \qquad (e_\mathrm{R}^\mathrm{c})_3 = \tau_\mathrm{R}^\mathrm{c} \tag{3.108}$$

$$Q_1 = \begin{pmatrix} u_\mathrm{L} \\ d_\mathrm{L} \end{pmatrix} \qquad Q_2 = \begin{pmatrix} c_\mathrm{L} \\ s_\mathrm{L} \end{pmatrix} \qquad Q_3 = \begin{pmatrix} t_\mathrm{L} \\ b_\mathrm{L} \end{pmatrix} \tag{3.109}$$

$$(u_\mathrm{R}^\mathrm{c})_1 = u_\mathrm{R}^\mathrm{c} \qquad (u_\mathrm{R}^\mathrm{c})_2 = c_\mathrm{R}^\mathrm{c} \qquad (u_\mathrm{R}^\mathrm{c})_3 = t_\mathrm{R}^\mathrm{c} \tag{3.110}$$

$$(d^{\mathrm{c}}_{\mathrm{R}})_1 = d^{\mathrm{c}}_{\mathrm{R}} \qquad (d^{\mathrm{c}}_{\mathrm{R}})_2 = s^{\mathrm{c}}_{\mathrm{R}} \qquad (d^{\mathrm{c}}_{\mathrm{R}})_3 = b^{\mathrm{c}}_{\mathrm{R}} \tag{3.111}$$

Unfortunately, the names for the first generation and those for all generations as a whole are ambiguous. For this reason, these names for the different generations will not be used. The symbols $e_{\mathrm{L}}$, $e^{\mathrm{c}}_{\mathrm{R}}$, $u_{\mathrm{L}}$, $u^{\mathrm{c}}_{\mathrm{R}}$, $d_{\mathrm{L}}$ and $d^{\mathrm{c}}_{\mathrm{R}}$ will always refer to the "vector" of all generations as a whole, with individual generations denoted by an index.

### 3.4.2. The Lagrangian

A concise overview of the Standard Model's Lagrangian can be found (for example) in [Shi15]. It is given by[35,36]

$$\begin{aligned}
\mathcal{L}_{\mathrm{SM}} = & \underbrace{-\frac{1}{2}\operatorname{Tr}\left(G_{\mu\nu}G^{\mu\nu}\right) - \frac{1}{2}\operatorname{Tr}\left(W_{\mu\nu}W^{\mu\nu}\right) - \frac{1}{4}B_{\mu\nu}B^{\mu\nu}}_{\text{gauge kinetic terms}} \\
& + \underbrace{\sum_{\text{fermions } \psi} i\bar{\psi}\bar{\sigma}^{\mu}D_{\mu}\psi}_{\text{fermion kinetic terms}} + \underbrace{(D_{\mu}H)^{\dagger}(D^{\mu}H)}_{\text{Higgs kinetic term}} \\
& \underbrace{- Y_u^{ij}(H\cdot Q_i)(u^{\mathrm{c}}_{\mathrm{R}})_j - Y_d^{ij}(H^{\dagger}Q_i)(d^{\mathrm{c}}_{\mathrm{R}})_j - Y_e^{ij}(H^{\dagger}L_i)(e^{\mathrm{c}}_{\mathrm{R}})_j + \text{H.c.}}_{\text{Yukawa terms}} \\
& \underbrace{+ \mu^2 H^{\dagger}H - \lambda(H^{\dagger}H)^2}_{\text{Higgs potential}}
\end{aligned} \tag{3.112}$$

where $G_{\mu\nu}$, $W_{\mu\nu}$ and $B_{\mu\nu}$ are the field strength tensors

$$G_{\mu\nu} = \partial_{\mu}G_{\nu} - \partial_{\nu}G_{\mu} - ig_3\left[G_{\mu}, G_{\nu}\right] \tag{3.113}$$

$$W_{\mu\nu} = \partial_{\mu}W_{\nu} - \partial_{\nu}W_{\mu} - ig_2\left[W_{\mu}, W_{\nu}\right] \tag{3.114}$$

$$B_{\mu\nu} = \partial_{\mu}B_{\nu} - \partial_{\nu}B_{\mu} \tag{3.115}$$

with

$$G_{\mu} = G^a_{\mu}\frac{\lambda_a}{2} \qquad W_{\mu} = W^a_{\mu}\frac{\sigma_a}{2} \tag{3.116}$$

where $\lambda_a$ are the Gell-Mann matrices and $\sigma_a$ are the Pauli matrices, which are the generators of SU(3) and SU(2). The object $D_{\mu}$ is a central part of a gauge theory. It

---

[35] The Yukawa couplings $Y_u^{ij}$, $Y_d^{ij}$ and $Y_e^{ij}$ should not be confused with hypercharges, which are also denoted using the letter $Y$.

[36] Even though the notation is unambiguous, it should be noted that the traces appearing in (3.112) are over the SU($N$) generators, which may be unclear on first sight.

is defined in such a way that it transforms as

$$D_\mu \psi \mapsto U D_\mu \psi \tag{3.117}$$

even under *local* gauge transformations. For the Standard Model gauge group, it has the form

$$D_\mu = \partial_\mu - i g_3 G^a_\mu (T_{\mathrm{R}}^{(\mathrm{SU}(3))})_a - i g_2 W^a_\mu (T_{\mathrm{R}}^{(\mathrm{SU}(2))})_a - i g_1 B_\mu Y \tag{3.118}$$

where $(T_{\mathrm{R}})_a$ are the generators of the respective gauge group in the representation of the field which $D_\mu$ acts on. For the definition of the product of SU(2) doublets appearing in the up-type Yukawa terms, see chapter 4.

# 4 Component notation for representations of $\mathrm{SU}(2)$

A variety of index notations are in common use for working with multi-dimensional mathematical objects with tensorial structure. Combined with the Einstein summation convention, this allows one to keep track of various sub-structures in a compact way. Such sub-substructures are a constant companion in gauge theories like the Standard Model of particle physics, where the objects are representations of a $\mathrm{SU}(3) \times \mathrm{SU}(2) \times \mathrm{U}(1)$ gauge group in addition to a representation of the Lorentz group, so each object has many different kinds of components (i. e. indices) to worry about.

The principle behind these conventions is to define another set of objects which complement the existing vectors. When the components of these new objects are combined with those of a vector through the summation convention, this operation realizes a scalar product, so that the result of summing over two matching indices is always a scalar. Extending this to tensors, as long as all indices involved in an expression are matched appropriately, the result should be a scalar.

The most well-known and consistently applied index convention is probably that of four-vectors in special relativity (“upper and lower indices”), where a superscripted index denotes components of a “contravariant vector”, while a subscripted index denotes a “covariant vector” (dual vector or 1-form). Another example is van der Waerden notation (“dotted and undotted indices”) used with Weyl spinors, which was introduced in section 3.3. However, even though $\mathfrak{sl}(2,\mathbb{C}) = \mathfrak{su}(2) \oplus i\mathfrak{su}(2)$, the Lie algebra of $\mathrm{SL}(2,\mathbb{C})$, really only consists of two copies of $\mathfrak{su}(2)$, the Lie algebra of $\mathrm{SU}(2)$, there is no commonly used notation for $\mathrm{SU}(2)$ like there is with van der Waerden notation for $\mathrm{SL}(2,\mathbb{C})$. In fact, $\mathrm{SU}(2)$ and its properties are usually discussed only very tersely in context with the Standard Model. This chapter will introduce an analogous index notation as those mentioned above and give some details on $\mathrm{SU}(2)$ as a group.

The fundamental basis of these index notations is the existence of a scalar product,

or equivalently, a tensor of order two ("metric tensor") which is invariant with respect to the group transformations under consideration. The components of an $(n, 0)$-tensor, i. e. a tensor composed of $n$ "vector-like" objects, transforms "like a vector in each index". For a simple $(2, 0)$-tensor $T = T^{ab}\, \vec{e}_a \otimes \vec{e}_b$ and a transformation $M$, this has the form

$$T^{ab} \mapsto M^a{}_m M^b{}_n T^{mn}$$

and can equivalently be written as

$$T \mapsto (M \otimes M)(T^{ab}\, \vec{e}_a \otimes \vec{e}_b) = MTM^\mathsf{T}$$

in matrix form. Note that the transpose is on the right here, instead of on the left as in the expressions below. In the case where $M^\mathsf{T}$ is also an element of the group, both versions are equivalent.[1]

There are several cases in physics where this is relevant:

- $\mathrm{SO}(N) = \{\, O \in \mathrm{GL}(N, \mathbb{R}) \mid O^\mathsf{T} O = \mathbb{1} \wedge \det(O) = 1 \,\}$, the group of rotations in $N$ dimensions, leaves the identity matrix invariant (the same applies for $\mathrm{O}(N)$, of course):

  $$O^\mathsf{T} \mathbb{1} O = \mathbb{1} \quad \Leftrightarrow O_a{}^b O_c{}^d \delta_{bd} = \delta_{ac} \tag{4.1}$$

  For this group, the notation is trivial because the metric tensor is simply the identity. A priori, there is no need to introduce different kinds of indices since the scalar product is symmetric. The metric tensor does not change the "left" vector, so the components of vectors and dual vectors are the same:

  $$x_i = \delta_{ij} x^j = x^i \qquad x \cdot y = x^i \delta_{ij} y^j = x^i y^i = x_i y_i = x^i y_i = x_i y^i \tag{4.2}$$

- $\mathrm{SO}(1,3)^+ = \{\, \Lambda \in \mathrm{GL}(4, \mathbb{R}) \mid \Lambda^\mathsf{T} \eta \Lambda = \eta \wedge \det(\Lambda) = 1 \,\}$, the group of proper orthochronous Lorentz transformations, leaves the Minkowski metric tensor

[1] It should be noted that there is indeed a bit of tension here concerning the issue of which of the equations $MTM^\mathsf{T} = T$ or $M^\mathsf{T} TM = T$ should be used. For a tensor $T$, the transformation is invariably $T \mapsto MTM^\mathsf{T}$, as written above. However, the motivation stems not from the tensor itself, but the scalar product $T(v, w)$. On the other hand, if *this* is to be invariant, the form $M^\mathsf{T} TM = T$ is needed because the so-called "tensor" doesn't actually transform here. Instead, $v$ and $w$ transform simultaneously, which naturally leads to the form $v^\mathsf{T} T w \mapsto v^\mathsf{T} \underbrace{M^\mathsf{T} TM}\, w$. However, as mentioned here, in the common case where $M^\mathsf{T}$ is also a group element, both versions can be treated as equivalent. Also see appendix C.1 for more details.

invariant:

$$\Lambda^\mathsf{T}\eta\Lambda = \eta \quad \Leftrightarrow \Lambda_a{}^b\Lambda_c{}^d\eta_{bd} = \eta_{ac}. \tag{4.3}$$

This enables the well-known notation from special relativity:

$$x_\mu = \eta_{\mu\nu}x^\nu \qquad x \cdot y = x^\mu\eta_{\mu\nu}y^\nu \tag{4.4}$$

- $\mathrm{SL}(2, F) = \{ M \in \mathrm{GL}(2, F) \mid \det(M) = 1 \}$ (where $F$ is any field) leaves the Levi-Civita tensor invariant:

$$M^\mathsf{T}\varepsilon M = \varepsilon \quad \Leftrightarrow M_{ab}M_{cd}\varepsilon_{bd} = \varepsilon_{ac} \tag{4.5}$$

In particular, this includes subgroups, such as SO(2) and SU(2), which will be discussed in detail.

SU($N$) does also leave the standard complex scalar product invariant (as seen from its defining condition $U^\dagger U = \mathbb{1}$); however, since the complex scalar product is a sesquilinear form instead of a 2-form (which would be bilinear), it does not have a matrix (order-2 tensor) representation in the strictest sense. However, sesquilinear forms can be represented by matrices if one of its arguments[2] is taken to its complex conjugate before multiplying with the matrix. Thus, one can take note that if a transformation $M$ leaves an order-2 tensor $T$ invariant in the sense that

  - $M^\mathsf{T}TM = T$, there is an invariant bilinear form ("real scalar product") given by $(x, y) \mapsto x_iT_{ij}y^j$;
  - $M^\dagger TM = T$, there is an invariant sesquilinear form ("complex scalar product") given by $(x, y) \mapsto x_i^*T_{ij}y^j$.

SU(2) (and indeed U($N$) in general) therefore leaves the identity $\mathbb{1}$ invariant in the sesquilinear sense just like O($N$) leaves it invariant in the bilinear sense, so that the former has an invariant complex scalar product and the latter an invariant real scalar product (as shown above).

## 4.1. SU(2) doublets

As $\mathfrak{sl}(2, \mathbb{C})$ is essentially "two copies of $\mathfrak{su}(2)$", an index notation for SU(2) really boils down to a simplified version of van der Waerden notation, that is, with analogous

[2] in the physics convention: typically the first one

conventions and properties, but without the need for dotted indices (because there is only one "type" of transformation). As mentioned above, the common basis for these notations is the property of SL(2, ℂ) and SU(2) that, for a group element $g$ in the (2-dimensional) fundamental representation,

$$g^{\mathsf{T}} \varepsilon g = \varepsilon \tag{4.6}$$

### 4.1.1. Basic conventions and transformation of doublets

Thus, to begin with, consider the fundamental representation of SU(2). The group elements act as 2 × 2-matrices[3]

$$U = e^{-\frac{i}{2}\vec{\alpha}\cdot\vec{\sigma}} \tag{4.7}$$

on 2-component complex vectors $\chi$, the doublets (where $\sigma_i$ are the Pauli matrices[4]). Now, *define* the notation for the action of $U$ on $\chi$ as follows:

$$\chi \mapsto U\chi \Leftrightarrow \chi_a \mapsto U_a{}^b \chi_b \tag{4.8}$$

That is, the components[5] of the doublet vector $\chi$ are denoted as $\chi_a$ (with lower indices), and the components of the matrix $U$ are denoted as $U_a{}^b$. As one can see, the action of $U$ on a doublet $\chi$ with lower indices yields another object of the same type, as it should be.

The notation is defined in such a way that when adjacent indices are summed "from upper to lower index", the indices can be omitted and the resulting expression is equivalently valid in index-free notation (e. g. matrix multiplication, as shown above, or the scalar product, as defined below). If this rule is adhered to, objects with a lower index can be interpreted as column vectors, objects with an upper index can be interpreted as row vectors and the usual rules of matrix multiplication can be employed in index-free form.[6] The reason why upper and lower indices must be summed from left to right in this manner (as opposed to four-vector notation,

[3]As the basis of the Lie algebra $\mathfrak{su}(2)$, $\{ X_i = -{}^i\!/\!{}_2\, \sigma_i \}$ will be used as the basis so that the structure constants are $f_{ij}{}^k = \varepsilon_{ij}{}^k$, i. e. $\left[X_i, X_j\right] = \varepsilon_{ij}{}^k X_k$.

[4]The Pauli matrices seem to be some of the exceedingly rare objects for which there really exists only one convention about their precise definition.

[5]in a Cartesian, orthonormal basis

[6]Note that this convention, which is "inherited" from van der Waerden notation, is opposite to that for relativistic four-vectors, where vectors with an upper index are usually interpreted as a column vector and those with a lower index are interpreted as a row vector. As this SU(2) notation is intended to be analogous to van der Waerden notation, the same conventions are adopted here.

for example, where the order does not matter) is because the metric tensor $\varepsilon$ is not symmetric. The precise reasoning will become clear once the scalar product is defined.

### 4.1.2. Dual doublets and scalar product

Using the definitions

$$(\varepsilon^{ab}) = \varepsilon = i\sigma_2 = \begin{pmatrix} 0 & 1 \\ -1 & 0 \end{pmatrix} \tag{4.9}$$

$$(\varepsilon_{ab}) = \varepsilon^{-1} = -i\sigma_2 = \begin{pmatrix} 0 & -1 \\ 1 & 0 \end{pmatrix} \tag{4.10}$$

with

$$\varepsilon^{ab}\varepsilon_{bc} = \varepsilon_{cb}\varepsilon^{ba} = \delta^a_c$$

as in section 3.3, one can define the dual doublets, i. e. those with upper indices (denoted here in index-free form with a tilde):

$$\tilde{\chi} := \varepsilon\chi \Leftrightarrow \chi^a := \varepsilon^{ab}\chi_b \tag{4.11}$$

Similarly, because $(\varepsilon^{ab})$ and $(\varepsilon_{ab})$ are inverses, one can confirm that

$$\chi_a = \varepsilon_{ab}\chi^b \ (= \varepsilon_{ab}\varepsilon^{bc}\chi_c = \delta^c_a\chi_c = \chi_a) \tag{4.12}$$

The dual doublets induce a bilinear "scalar product"

$$\chi \cdot \xi = \chi^a\xi_a = \chi_a\varepsilon^{ba}\xi_b \ (= \tilde{\chi}\xi) \tag{4.13}$$

Note that, since $\varepsilon^{ab} \neq \varepsilon^{ba}$, in general one has

$$\chi \cdot \xi = \chi^a\xi_a \neq \chi_a\xi^a \tag{4.14}$$

and so the order of terms (i. e. sum from upper to lower index) in the scalar product is significant. For the special cases of commuting and anti-commuting components:

| Commuting: | Anti-commuting: | |
|---|---|---|
| $\chi \cdot \xi = \chi^a\xi_a = \xi_a\chi^a = -\xi^a\chi_a = -\xi \cdot \chi$ | $\chi \cdot \xi = \chi^a\xi_a = -\xi_a\chi^a = \xi^a\chi_a = \xi \cdot \chi$ | (4.15) |

so if the components anti-commute (as in the case of fermionic operators), the doublets "commute inside the scalar product", just like the product of Weyl spinors. Arbitrary expressions with balanced lower and upper indices still always produce a scalar, but depending on the order in which the indices appear, its specific value may differ.

### 4.1.3. Transformation of dual doublets

Now, the next step is to find out how the dual doublets transform. For this, it will be useful to know that

$$\varepsilon U \varepsilon^{-1} = (U^{-1})^{\mathsf{T}} \ (= (U^{\dagger})^{\mathsf{T}} = U^{*}) \tag{4.16}$$

This follows by rearranging the starting observation of $U^{\mathsf{T}} \varepsilon U = \varepsilon$.[7] Consequently, for the dual doublets:

$$\chi^a = \varepsilon^{ab} \chi_b \mapsto \varepsilon^{ab} U_b{}^c \chi_c = \varepsilon^{ab} U_b{}^c \varepsilon_{cd} \chi^d = \left((U^{-1})^{\mathsf{T}}\right)^a{}_d \chi^d = (U^{-1})_d{}^a \chi^d = \chi^d (U^{-1})_d{}^a \tag{4.17}$$

Transposition does what one would expect: It simply exchanges both indices of an order-2 tensor, without affecting their "vertical" position. So, a dual doublet transforms with the inverse group element of the original doublet. In index-free matrix notation (denoting a dual doublet as $(\chi^a) = \tilde{\chi}$ again):

$$\tilde{\chi} \mapsto \tilde{\chi} U^{-1} = \tilde{\chi} U^{\dagger} \tag{4.18}$$

This easily demonstrates that the scalar product indeed produces a scalar with respect to SU(2) transformations:

$$\chi \cdot \xi = \tilde{\chi} \xi \mapsto \tilde{\chi} \underbrace{U^{\dagger} U}_{=\mathbb{1}} \xi = \chi \cdot \xi \tag{4.19}$$

[7] As a side note, the relation $\varepsilon U \varepsilon^{-1} = (U^{-1})^{\mathsf{T}}$ is the reason that one kind of index is needed for SU(2) (so no dotted indices as for SL(2, $\mathbb{C}$) and no separate anti-fundamental representation as in SU(3), e. g. anti-quarks). This relation means that the fundamental representation (transforming with $U$) and the anti-fundamental representation (transforming with $U^{-1}$) are equivalent for SU(2), see definition 9. This is related to the fact that SU(2)'s fundamental representation is pseudo-real. (Technically, the relation only shows that $U$ and $(U^{-1})^{\mathsf{T}}$ are related by a basis transformation. However, every matrix is similar to its transpose, so there is matrix $A$ such that $A(U^{-1})^{\mathsf{T}} A^{-1} = U^{-1}$. Consequently, $A\varepsilon U (A\varepsilon)^{-1} = A(U^{-1})^{\mathsf{T}} A^{-1} = U^{-1}$, so $U$ is similar to $U^{-1}$ via the matrix $A\varepsilon$.) For SU(3), this is not the case.

**Table 4.1.:** Index structure of the different transformation matrices.

| Object | Index structure | Definition |
|---|---|---|
| $U$ | $U_a{}^b$ | explicit definition in (4.8) |
| $U^\mathsf{T}$ | $(U^\mathsf{T})^a{}_b = U_b{}^a$ | transposition explicitly defined as index exchange |
| $U^\dagger = U^{-1}$ | $(U^\dagger)_a{}^b = (U^*)^b{}_a$ | implicitly obtained from (4.17) |
| $U^*$ | $(U^*)^a{}_b = (U_b{}^a)^*$ | implicitly obtained from $U^* = (U^\dagger)^\mathsf{T}$ |

With the invariant scalar product in hand, the purpose of the notation with upper and lower indices can be generalized in the usual way: As long as an expression contains and equal number of upper and lower indices, it will be invariant under SU(2) transformations.

At this point, it is useful to keep track of the different index structures of all the different variations of the group element representatives $U$ which have been defined either explicitly or implicitly. Such an overview is given in table 4.1.

### 4.1.4. Adjoint doublets

Since this representation acts on complex vectors, there is another object that is commonly of interest: the adjoint of a doublet. Its transformation is easily derived from the transformation of the original doublet:

$$\chi^\dagger \mapsto (U\chi)^\dagger = \chi^\dagger U^\dagger \tag{4.20}$$

This demonstrates that the object $\chi^\dagger$ transforms in the same way as $\tilde{\chi}$. Correspondingly, its index structure must be the same:

$$(\chi^\dagger)^a \mapsto (\chi^\dagger)^b (U^\dagger)_b{}^a \tag{4.21}$$

So, as far as SU(2) transformations are concerned, either $\tilde{\chi}$ or $\chi^\dagger$ can be used with $\chi$ to form a scalar.

The correspondence of dual and adjoint doublet properties highlights how the two defining conditions of SU(2) lead to two ways of forming an SU(2) scalar, which is invariant under SU(2) transformations, from two SU(2) doublets $\chi$ and $\xi$:

1. $\tilde{\chi}\xi = \chi^a \xi_a$ from the condition $\det(U) = 1 \Rightarrow U^\mathsf{T}\varepsilon U = \varepsilon$ and

2. $\chi^\dagger\xi = (\chi^\dagger)^a \xi_a$ from the condition $U^\dagger \mathbb{1} U = \mathbb{1}$.

For reference, here are the explicit forms of all the doublet variants:

$$(\chi_a) = \begin{pmatrix} \chi_1 \\ \chi_2 \end{pmatrix} \qquad ((\chi^\dagger)^a) = \begin{pmatrix} \chi_1^* & \chi_2^* \end{pmatrix}$$

$$(\chi^a) = \begin{pmatrix} \chi_2 & -\chi_1 \end{pmatrix} \qquad ((\chi^\dagger)_a) = \begin{pmatrix} -\chi_2^* \\ \chi_1^* \end{pmatrix} \tag{4.22}$$

### 4.1.5. The question of transpose and conjugate doublets

At this point, one might wonder what the properties of $\chi^\mathsf{T}$ and $\chi^*$ are – after all, these too are common operations on complex vectors from linear algebra. In fact, the same question could already be asked in section 3.3. It turns out that transposition of doublets does not introduce anything interesting from the perspective of component notation. After all, the components of an arbitrary doublet $v$ and its transpose are the same: $v_a = (v^\mathsf{T})_a$, so on the level of components, there is no benefit in defining an object $v^\mathsf{T}$.

The purpose of the operation "transposition" on vectors is to interpret them as linear functionals. Within the framework of matrix notation, this means that they are written as row instead of column vectors, which allows the usual rules of matrix multiplication to yield the correct result. However, in the present conventions for index-free notation, the objects which are to be interpreted as "row vectors" have already been chosen: they are precisely those whose components carry an upper index. In order to interpret $\chi^\mathsf{T}$ as a row vector as well, it would have to carry an upper index. But since its components are the same as those of $\chi$, the rule that correct summation of upper and lower indices always produces an SU(2) scalar would break!

Moreover, looking at the transformation of $\chi^\mathsf{T}$ (for simplicity, writing all indices as ordinary subscripts for the moment):

$$\chi^\mathsf{T} \mapsto (U\chi)^\mathsf{T} = \chi^\mathsf{T} U^\mathsf{T} = \left((\chi^\mathsf{T})_b (U^\mathsf{T})_{ba}\right) = \left((\chi^\mathsf{T})_b U_{ab}\right) = \left(U_{ab} (\chi^\mathsf{T})_b\right) = (U_{ab}\chi_b) \tag{4.23}$$

The transformations of $\chi^\mathsf{T}$ and $\chi$ are identical on the component level! Thus, it seems that one should take the relation $(\chi^\mathsf{T})_a = \chi_a$ seriously and assign a lower index to $\chi^\mathsf{T}$, just as was done for $\chi$. However, it is not as simple as leaving it at that, because there is no way to reconcile the rule that indices can be omitted if a sum occurs from upper to lower index with the transformation $\chi^\mathsf{T} \mapsto \chi^\mathsf{T} U^\mathsf{T}$, which would read $(\chi^\mathsf{T})_a \mapsto (\chi^\mathsf{T})_b (U^\mathsf{T})^b{}_a$ in indices. Again, the concept of a transpose being a row vector (which requires it to be written on the left of the matrix in index-free notation) and that of objects with upper indices being row vectors clash. In light of this, the only conclusion is that there is no use for the concept of vector transposition in this notation. Transpose doublets could be included by introducing the convention that their indices must be summed from lower to upper index instead of the other way around, but this seems like a needless complication which yields little benefit.

As for conjugate doublets $\chi^*$, there is the same issue in relation to adjoint doublets as there is for transpose and "ordinary" doublets. Adjoint doublets are row vectors in index-free notation and due to their transformation properties, they naturally have an upper index, so the two concepts align. On the other hand, the components of conjugate doublets (which are the transpose of adjoint doublets, $(\chi^\dagger)^\mathsf{T} = \chi^*$) transform in the same way, but they are column vectors in index-free notation. In this sense, doublets $\chi$ and their adjoints $\chi^\dagger$ are the natural objects to use in this convention, while $\chi^\mathsf{T}$ and $\chi^*$ are equivalent, but cause undesirable complications in reconciling the notation with and without indices. Another issue arises when dealing with field multiplets, because then the operation of "complex conjugation" on a field would have to be defined.[8]

Even though there is no need for transposition from the point of view of components, when given a Lagrangian in index-free notation, the expressions may well contain transpose fields so that it becomes necessary to "translate" them to the present index convention. For example, if $H$ is the Higgs doublet of the Standard model (with hypercharge $Y = 1$) and $\Phi$ is another Lorentz scalar SU(2) doublet with $Y = -1$, the only gauge-invariant combination of fields is[9] $\tilde{H}\Phi = H^a\Phi_a$. Ordinarily, however, only $H$ and $H^\dagger$ are defined and neither has the desired properties of $\tilde{H}$ (upper index *and* $Y = 1$).

In this case, the object $H^\mathrm{c} = -i\sigma_2 H^*$ is sometimes defined. With the above, it is now revealed that $H^*$ has an upper index and the purpose of $-i\sigma_2 = \varepsilon^{-1}$ is to lower that index to obtain an object that transforms like $H$ ($H^\mathrm{c} \mapsto UH^\mathrm{c}$), but has $Y = -1$. In this convention, one would simply write $H^\mathrm{c}_a = (H^\dagger)_a$. However, due to the problems mentioned above, $H^\mathrm{c}$ is not defined as $-i\sigma_2 H^\dagger$ because that would make it a row vector in the standard convention, so the transformation matrix would have to act on

[8] The most natural thing to do would be to define

$$\psi(x)^* = (\psi_a(x))^* = \left([\psi_a(x)]^\dagger\right) = \left((\psi^\dagger)^a(x)\right) \tag{4.24}$$

but still, this would constitute additional effort.

[9] $H^a\Phi_a = -\Phi^a H_a$, so there is no need to include both.

it from the right instead of the left.

As the last missing piece, there is then the object $(H^c)^\dagger = H^\mathsf{T} i\sigma_2$. This has $H^\mathsf{T} i\sigma_2 = H^\mathsf{T}\varepsilon = (H_b \varepsilon^{ba}) = (-H^a)$, so $((H^c)^\dagger)^a = -H^a$. The additional minus sign may seem unexpected; it arises because with the previous definitions, indices are both raised and lowered from the left: $\tilde{\chi} = \varepsilon\chi$, $\chi = \varepsilon^{-1}\tilde{\chi}$, whereas applying the adjoint flips the order of $H$ and $\varepsilon$.[10] Using $(H^c)^\dagger = H^\mathsf{T} i\sigma_2$, the required invariant term can be written down as $(H^c)^\dagger \Phi = H^\mathsf{T} i\sigma_2 \Phi = -\tilde{H}\Phi$.

In general, when encountering transpose or conjugate fields, a correct translation procedure is to make the intuitive substitutions (where $H$ could be any doublet)

$$\begin{aligned} H^\mathsf{T} &\to H_a \\ H^* &\to (H^\dagger)^a \\ i\sigma_2 &\to \varepsilon^{ab} \\ -i\sigma_2 &\to \varepsilon_{ab} \end{aligned} \tag{4.25}$$

and to sum over adjacent indices as usual for matrix multiplication. This is not explained by the simple rule that indices can be omitted if summing from adjacent upper to lower index, but the result will nevertheless be correct. For the common cases explained in the example above, one can directly use:

$$\begin{aligned} H^\mathsf{T}(i\sigma_2) &\to -H^a \\ -i\sigma_2 H^* &\to (H^\dagger)_a \end{aligned} \tag{4.26}$$

## 4.2. SU(2) triplets

Triplet representations are special in SU(2). The reason is that the Lie algebra of SU($N = 2$) forms a $N^2 - 1 = 3$-dimensional vector space, which means that the adjoint representation (which is the representation where the vector space which is acted on by the matrices representing the group elements is the Lie algebra itself) is a triplet representation. This is why there are two commonly used bases for the triplet representation of SU(2):

- One stems from the usual procedure of selecting the eigenvectors of one of the generators as the basis of the vector space (i. e. choosing one of the generators to be diagonal) and deriving the matrix elements of the other generators in this basis from their algebraic properties. This is also called the “spherical basis”, where the generators $s_i$ have the following form:[11,12]

[10] For an alternative that avoids this issue, see [DHM10, pp. 11–13, footnote 5].

$$is_1 = \frac{1}{\sqrt{2}}\begin{pmatrix} 0 & -1 & 0 \\ -1 & 0 & 1 \\ 0 & 1 & 0 \end{pmatrix} \quad is_2 = \frac{1}{\sqrt{2}}\begin{pmatrix} 0 & i & 0 \\ -i & 0 & -i \\ 0 & i & 0 \end{pmatrix} \quad is_3 = \begin{pmatrix} 1 & 0 & 0 \\ 0 & 0 & 0 \\ 0 & 0 & -1 \end{pmatrix} \tag{4.28}$$

with, of course, $\left[s_i, s_j\right] = \varepsilon_{ij}{}^k s_k$.

- The other stems from the adjoint representation, where the vectors are the Lie algebra elements themselves. The group elements $g \in \mathrm{SU}(2)$ in this representation can be viewed in two different ways. The first way is that the group elements act on the elements of the Lie algebra not through ordinary matrix multiplication, but through the *adjoint action*, which is the linear map $X \mapsto gXg^{-1}$. For SU(2), the group elements $g$ remain the unitary $2 \times 2$ matrices $U$, and an element $X$ of the Lie algebra is $X = x^i e_i$, where the basis $\{e_i\}$ of the Lie algebra consists of the Pauli matrices $-{}^{i}/{}_{2}\,\sigma_i$, as before. Thus, the action of a group element on a general vector looks like

$$X \mapsto UXU^{-1} = -\frac{i}{2}Ux^i\sigma_i U^\dagger = -\frac{i}{2}e^{-\frac{i}{2}\vec{\alpha}\cdot\vec{\sigma}}\begin{pmatrix} z & x - iy \\ x + iy & -z \end{pmatrix} e^{\frac{i}{2}\vec{\alpha}\cdot\vec{\sigma}}. \tag{4.29}$$

On the other hand, the components of $X$ can be written as a 3-dimensional column vector. The representations of the group elements act on these column

[11] In the physicist's convention, where the generators are Hermitian (so that their eigenvalues are real), one would use the matrices $is_i$.

[12] Note that even specifying the commutation relations and letting $s_3$ be diagonal does not fully fix the matrices $s_i$. For example, in quantum mechanics, the operators for an angular momentum of $j = 1$ (which are also simply the generators of a 3-dimensional representation of SU(2)) are usually (e. g. in [Sch15b, pp. 51, 58 (but note the errata); Nol12, p. 22; Bar+15, pp. 915–916]) given as

$$is'_1 = \frac{1}{\sqrt{2}}\begin{pmatrix} 0 & 1 & 0 \\ 1 & 0 & 1 \\ 0 & 1 & 0 \end{pmatrix} \quad is'_2 = \frac{1}{\sqrt{2}}\begin{pmatrix} 0 & -i & 0 \\ i & 0 & -i \\ 0 & i & 0 \end{pmatrix} \quad is'_3 = \begin{pmatrix} 1 & 0 & 0 \\ 0 & 0 & 0 \\ 0 & 0 & -1 \end{pmatrix} \tag{4.27}$$

where $\left[s'_i, s'_j\right] = \varepsilon_{ij}{}^k s'_k$ is also satisfied. This freedom corresponds to the choice of the phase of the resulting vectors when acting with the ladder operators $s_\pm = s_1 \pm is_2$ on the eigenvectors of $s_3$. The definition of the $s'_i$ has the advantage that a ladder operator $is'_\pm$ acting on an eigenvector of $is'_3$ always results in another eigenvector multiplied by a real, positive number. However, this definition is not used here because it would not lead to the formula $T^\pm = {}^1/{}_{\sqrt{2}}\,(T^1 \mp iT^2)$ and thus to $(T^+)^\dagger = T^-$ for real $T^i$ below.

vectors through ordinary matrix multiplication, where the matrices encode the effect of the adjoint action. They are the $3 \times 3$ matrices given by $M(g) = e^{\vec{\alpha}\cdot\vec{l}}$, so the group action is

$$X \mapsto M(g)X = e^{\vec{\alpha}\cdot\vec{l}} \begin{pmatrix} x \\ y \\ z \end{pmatrix} \tag{4.30}$$

The generators $l_i$ in this representation are (up to a constant factor) composed of the structure constants: $(l_i)_{jk} = -\varepsilon_{ijk}$,

$$l_1 = \begin{pmatrix} 0 & 0 & 0 \\ 0 & 0 & -1 \\ 0 & 1 & 0 \end{pmatrix} \quad l_2 = \begin{pmatrix} 0 & 0 & 1 \\ 0 & 0 & 0 \\ -1 & 0 & 0 \end{pmatrix} \quad l_3 = \begin{pmatrix} 0 & -1 & 0 \\ 1 & 0 & 0 \\ 0 & 0 & 0 \end{pmatrix} \tag{4.31}$$

which fulfill $\left[l_i, l_j\right] = \varepsilon_{ij}{}^{k} l_k$ as required. This representation emphasizes the link between SU(2) and SO(3), as the $l_i$ are exactly the usual basis of $\mathfrak{so}(3)$. However, the exponential map of $\mathfrak{so}(3)$ does not generate SU(2), but SO(3). SU(2) is the universal covering group of SO(3), so they have the same Lie algebra ($\mathfrak{so}(3) \cong \mathfrak{su}(2)$), but SU(2) contains "more" elements (for the two elements $g$ and $-g$ with $g \in \mathrm{SU}(2)$, the adjoint action produces the same element $X'$).

Both of these options are just different possible bases; they span the same vector space.

### 4.2.1. Triplets from the adjoint representation

For the purposes of coupling doublets and triplets to once again produce scalars, the interpretation of viewing the triplet representation from the perspective of the adjoint representation (i. e. as $2 \times 2$ matrices instead of 3-component vectors) proves to be very useful. Because doublet and matrix-form triplet indices take on the same values, there is a natural way of coupling them. Moreover, the transformation behavior of a triplet $X$ is already known (the adjoint action) and one can determine its index structure through the role of its basis (the Pauli matrices) as the generators of the doublet representation. As noted in the beginning, the doublet representation's index structure is

$$U_a{}^b = \left(e^{-\frac{i}{2}\vec{\alpha}\cdot\vec{\sigma}}\right)_a{}^b = \left(e^{-\frac{i}{2}\alpha^i \sigma_i}\right)_a{}^b \tag{4.32}$$

The conclusion is that the Pauli matrices have the same index structure $(\sigma_i)_a{}^b$ and $(\sigma^i)_a{}^b$, which also directly translates to the triplets. This gives the triplet transformation in indices

$$X_a{}^b = -\frac{i}{2}x^i(\sigma_i)_a{}^b \mapsto U_a{}^j X_j{}^k (U^\dagger)_k{}^b = -\frac{i}{2}U_a{}^j x^i(\sigma_i)_j{}^k(U^\dagger)_k{}^b = (X')_a{}^b. \qquad (4.33)$$

Here, all the different index conventions neatly come together. The index $i = 1, 2, 3$ enumerating the Pauli matrices, which have already suggestively been written as a 3-dimensional "Pauli vector" $\vec{\sigma}$, is "canceled" via summation with the 3-dimensional vector components $x^i$.

With the known transformation behavior of triplets in the adjoint representation, it is also immediately clear how to produce scalars – for example, using a triplet $T$ and two doublets $\chi$ and $\xi$. The triplet has an upper and a lower index, so the doublets are needed in a form where one has an upper and one has a lower index to properly sum everything. Consequently, there are two ways of producing scalars again (plus their adjoints).

- Using $\tilde{\chi}$ and $\xi$:

$$\begin{aligned} &\tilde{\chi}T\xi \mapsto (\tilde{\chi}U^\dagger)(UTU^\dagger)(U\xi) = \tilde{\chi}T\xi \\ \Leftrightarrow\ &\chi^a T_a{}^b \xi_b \mapsto \left(\chi^a (U^\dagger)_a{}^b\right)\left(U_b{}^c T_c{}^d (U^\dagger)_d{}^e\right)\Big((U^\dagger)_e{}^f \xi_f\Big) = \chi^a T_a{}^b \xi_b \end{aligned} \qquad (4.34)$$

- Using $\chi^\dagger$ and $\xi$:

$$\begin{aligned} &\chi^\dagger T\xi \mapsto (\chi^\dagger U^\dagger)(UTU^\dagger)(U\xi) = \chi^\dagger T\xi \\ \Leftrightarrow\ &(\chi^\dagger)^a T_a{}^b \xi_b \mapsto \left((\chi^\dagger)^a (U^\dagger)_a{}^b\right)\left(U_b{}^c T_c{}^d (U^\dagger)_d{}^e\right)\Big((U^\dagger)_e{}^f \xi_f\Big) = (\chi^\dagger)^a T_a{}^b \xi_b \end{aligned} \qquad (4.35)$$

Another way of looking at this is that a triplet multiplied with a doublet produces another doublet (since $T_a{}^b \xi_b$ has exactly one lower index), which can then be combined with another doublet in the known ways.

### 4.2.2. Spherical basis

While the adjoint representation is very useful for coupling doublets and triplets, the spherical basis where $s_3$ is diagonal is still needed for a very important purpose. Since none of the generators $l_i$ are diagonal, there are no fixed eigenvalues associated with the components of a triplet in the adjoint representation. In the spherical basis $\{s_i\}$, on the other hand, $s_3$ is diagonal and the $s_i$ are thus written in a basis of eigenvectors of $s_3$, which can be labeled by its eigenvalues $-i, 0, i$. These eigenvalues are rather important in particle physics because they determine the charge $Q = {}^1\!/\!_2 Y + T_3$, where $T_3$ is the eigenvalue of $t_3 = is_3$.[13] The charges associated with a general triplet $T$ in the spherical basis are thus

$$T = \begin{pmatrix} T^{\frac{Y}{2}+1} \\ T^{\frac{Y}{2}} \\ T^{\frac{Y}{2}-1} \end{pmatrix} \overset{(Y=0)}{=} \begin{pmatrix} T^{+} \\ T^{0} \\ T^{-} \end{pmatrix} = T^{+}\vec{e}_{+} + T^{0}\vec{e}_{0} + T^{-}\vec{e}_{-} \tag{4.36}$$

where the charges $Q$ are in the superscripts. For example, a triplet with $Y = 0$ would lead to charged particles $T^+$, $T^0$ and $T^-$ (see above).

One possible matrix (up to a phase factor) effecting the change between these two bases is the unitary matrix

$$V = \frac{1}{\sqrt{2}} \begin{pmatrix} 1 & 0 & 1 \\ i & 0 & -i \\ 0 & \sqrt{2} & 0 \end{pmatrix} \qquad V^{\dagger} = V^{-1} = \frac{1}{\sqrt{2}} \begin{pmatrix} 1 & -i & 0 \\ 0 & 0 & \sqrt{2} \\ 1 & i & 0 \end{pmatrix}$$

which transforms the bases as follows:

$$\begin{pmatrix} \vec{e}_{+} & \vec{e}_{0} & \vec{e}_{-} \end{pmatrix} = \begin{pmatrix} \vec{e}_{1} & \vec{e}_{2} & \vec{e}_{3} \end{pmatrix} V$$
$$V^{-1} l_i V = s_i.$$

Using this basis transformation matrix, the change from the spherical basis to the basis of the $l_i$ can be performed to see what happens to the charged components:[14]

$$\begin{pmatrix} T^1 \\ T^2 \\ T^3 \end{pmatrix} = V \begin{pmatrix} T^{+} \\ T^{0} \\ T^{-} \end{pmatrix} = \frac{1}{\sqrt{2}} \begin{pmatrix} 1 & 0 & 1 \\ i & 0 & -i \\ 0 & \sqrt{2} & 0 \end{pmatrix} \begin{pmatrix} T^{+} \\ T^{0} \\ T^{-} \end{pmatrix} = \begin{pmatrix} \frac{1}{\sqrt{2}}(T^{+} + T^{-}) \\ \frac{i}{\sqrt{2}}(T^{+} - T^{-}) \\ T^{0} \end{pmatrix} \tag{4.37}$$

[13]The factor of $i$ arises from the interpretation of the generators as observables in quantum mechanics and quantum field theories, which must be Hermitian so that they have real eigenvalues. Thus, $t_3 = is_3$ must be used if the (naturally anti-Hermitian) generator is to be interpreted as an observable.

[14]$V$ is the basis transformation from the "Cartesian" to the spherical basis. To transform in the other direction, the matrix $V^{-1}$ is needed. However, coordinate vectors transform contravariantly, so the inverse is used again and $V$ is in fact the correct matrix to transform from the coordinates $T^+, T^0, T^-$ to $T^1, T^2, T^3$.

Inserting these 3-vector components into the 2 × 2 matrix representation yields

$$T = -\frac{i}{2} T^i \sigma_i = -\frac{i}{2} \begin{pmatrix} T^0 & \sqrt{2} T^+ \\ \sqrt{2} T^- & -T^0 \end{pmatrix} \tag{4.38}$$

It becomes apparent that the matrix entries of the triplet in 2 × 2 form do all have a well-defined charge, combining the advantages of both bases. Naturally, the triplet in matrix form is traceless, just like the Pauli matrices.

While the normalization used for the triplet $T$ above may be natural from the perspective of representation theory, it is rather impractical for the use in a Lagrangian. While the matrix form of the triplet makes formulating gauge-invariant interaction terms much easier, a small disadvantage is that it is not as obvious anymore how to write the (correctly-normalized) kinetic and mass terms. A more useful normalization for the triplet matrix is

$$T = \frac{1}{\sqrt{2}} T^i \sigma_i = \begin{pmatrix} \frac{1}{\sqrt{2}} T^0 & T^+ \\ T^- & -\frac{1}{\sqrt{2}} T^0 \end{pmatrix} \tag{4.39}$$

This is also the form that will be used from here on. Using this definition, a scalar triplet mass term can be written analogously to that for scalar singlets:

$$\text{T real: } \frac{1}{2} M^2 \operatorname{Tr}(T^2) = \frac{1}{2} M^2 (T^0)^2 + M^2 T^+ T^- \tag{4.40}$$

$$\text{T complex: } M^2 \operatorname{Tr}(T^\dagger T) = M^2 (T^0)^\dagger T^0 + M^2 (T^+)^\dagger T^+ + M^2 (T^-)^\dagger T^- \tag{4.41}$$

This results in the correct factor of $1/2$ for the real component $T^0$ in the real case, while all charged (and thus complex) scalars do not have this factor.

## 4.3. Relation to the Standard Model and its extensions

Within the Standard Model, the scalar product between doublets which has been defined here does not manifest itself very often. Since the gauge group of the Standard Model is SU(3) × SU(2) × U(1), many fields have a hypercharge $Y \neq 0$ in addition to being part of some representation of SU(3) and/or SU(2). Terms in the Lagrangian must be hypercharge-neutral, so this prevents most opportunities of combining two doublets using the $\varepsilon$ scalar product – the doublets would have to have opposite values of $Y$ for this to work. Thus, by far the most common combination is of the form $\chi^\dagger \xi$, as taking the conjugate inverts the sign of $Y$. However, even within the Standard model, there is one case where it is necessary to raise the index of a doublet *without* using the adjoint, because the hypercharges of the involved doublets are already opposite.

This case is the Higgs mechanism for the up-type quark masses. The masses of the charged leptons and the down-type quarks result from the terms[15]

$$Y^e H^\dagger L e^{\mathrm{c}}_{\mathrm{R}} + \mathrm{H.\,c.} \tag{4.42}$$

$$Y^d H^\dagger Q d^{\mathrm{c}}_{\mathrm{R}} + \mathrm{H.\,c.} \tag{4.43}$$

For the up-type quarks, however, this combination would not work – the resulting hypercharge would be $-2$ instead of 0. Instead, the following term is used (cf. section 4.1.5):

$$Y^u \underbrace{(-i\sigma_2 H^*)^\dagger}_{=H^{\mathsf{T}} i\sigma_2} Q u^{\mathrm{c}}_{\mathrm{R}} + \mathrm{H.\,c.} \tag{4.44}$$

As explained in section 4.1.5, with understanding of the transformation behavior of SU(2) doublets, it becomes clear that the terms $-i\sigma_2 H^*$ and $H^{\mathsf{T}} i\sigma_2$ are used to obtain $(H^\dagger)_a$ and $H^a$ (which do not have dedicated symbols in the standard notation), i. e. lowering or raising the index without changing $Y$.

In extensions of the Standard Model, however, the need to consider both ways of producing an SU(2) scalar from doublets arises much more commonly.

## 4.4. Higher-dimensional representations

In the case of SU(2), the fact that the adjoint representation is also the triplet representation has directly lead to the idea that a triplet can be represented as an object with two "doublet indices" – a second-order tensor. In fact, this is exactly how the Lorentz vector representation $(\frac{1}{2}, \frac{1}{2}) = (\frac{1}{2}, 0) \otimes (0, \frac{1}{2})$ in $1+3$ spacetime dimensions was obtained in section 3.2.5 – from a tensor product of two spin-1/2 representations. Analogously, the vector representation in 3 space dimensions, which is acted on by orthogonal rotation matrices, is nothing but the triplet representation of SU(2), which is the same as the defining representation of SO(3). It, too, can be constructed from a tensor product of two spin-1/2 representations in the corresponding spacetime – after all, the doublets **2** of SU(2) are simply the non-relativistic Pauli spinors, which are the spinors in three spatial dimensions.

[15] See table 3.1 for the definitions of the fields.

The general tensor product **2** ⊗ **2** is not exactly the same as the adjoint representation, however. Considering the Clebsch–Gordan decomposition of irreducible representations $\mathbf{n}_i$ of SU(2),[16]

$$\mathbf{n}_1 \otimes \mathbf{n}_2 = \bigoplus_{n=|n_1-n_2|+1}^{n_1+n_2-1} \mathbf{n} \tag{4.46}$$

(where $n$ increases by $\Delta n = 2$ for each step in the sum) shows that the general tensor product of two doublets gives the familiar addition of two spins $^1/_2$,

$$\mathbf{2} \otimes \mathbf{2} = \mathbf{3} \oplus \mathbf{1} \tag{4.47}$$

resulting in both a spin-1 and a spin-0 representation.

For two SU(2) doublets $D, E \in \mathbf{2}$, such a tensor product $T$ has the form[17]

$$T = (D_a) \otimes (E^b) = (D_a E^b) = (\varepsilon^{bc} D_a E_c) = \begin{pmatrix} D_1 E_2 & -D_1 E_1 \\ D_2 E_2 & -D_2 E_1 \end{pmatrix} \tag{4.48}$$

and it transforms as

$$T \mapsto (U_a{}^c D_c E^d (U^\dagger)_d{}^b) = U T U^\dagger \tag{4.49}$$

This matrix should contain both **3** and **1** as subspaces. And indeed, it is known that the trace of a matrix is invariant under a similarity transformation, which is exactly what its transformation behavior is. So the one-dimensional subspace **1** can be identified with the trace of this matrix,

$$\mathrm{Tr}(T) = D_1 E_2 - D_2 E_1 \tag{4.50}$$

And indeed, this is hardly a surprise, because the Clebsch–Gordan coefficients for $j_1 = j_2 = {}^1/_2$ say exactly that the singlet ($j = 0$) arising from the addition of two spins $^1/_2$ is given by the antisymmetric combination $D_1 E_2 - D_2 E_1$ of the two spinors. Thus, to extract the triplet part, these tensors should be restricted to the subspace of

[16]Using $n = 2j + 1$, (4.46) can also be written as

$$(\mathbf{2j_1} + \mathbf{1}) \otimes (\mathbf{2j_2} + \mathbf{1}) = \bigoplus_{j=|j_1-j_2|}^{j_1+j_2} (\mathbf{2j} + \mathbf{1}) \tag{4.45}$$

with $\Delta j = 1$.

[17]The specific form with one lowered and one raised index is used here for comparison with the triplet (4.38).

traceless tensors

$$T - \frac{\mathrm{Tr}(T)}{2} = \begin{pmatrix} \frac{D_1E_2+D_2E_1}{2} & -D_1E_1 \\ D_2E_2 & -\frac{D_1E_2+D_2E_1}{2} \end{pmatrix} \tag{4.51}$$

and again, the three symmetric combinations $D_1E_1$ $(m = 1)$, $D_2E_2$ $(m = -1)$ and $D_1E_2 + D_2E_1$ $(m = 0)$ demanded by the Clebsch–Gordan coefficients for the triplet $(j = 1)$ can be recognized. The tensor $T$ can thus be decomposed as

$$\underbrace{T_a{}^b}_{\mathbf{2}\otimes\mathbf{2}} = \underbrace{\left(T_a{}^b - \frac{\mathrm{Tr}(T)}{2}\delta_a^b\right)}_{\mathbf{3}} + \underbrace{\frac{\mathrm{Tr}(T)}{2}\delta_a^b}_{\mathbf{1}} \tag{4.52}$$

Now, identifying

$$\Delta^0 = \frac{D_1E_2 + D_2E_1}{2} \qquad \frac{\Delta^+}{\sqrt{2}} = -D_1E_1 \qquad \frac{\Delta^-}{\sqrt{2}} = D_2E_2 \tag{4.53}$$

(and demanding $(D_2E_2)^\dagger = -D_1E_1$ as well as that $D_1E_2 + D_2E_1$ be Hermitian) results in the previously obtained (real) triplet.

This procedure, exemplified above using the simplest case of the triplet, can be generalized to produce *any* finite-dimensional irreducible representation of SU(2). For example, a tensor made of three doublets is part of the representation

$$\mathbf{2}^{\otimes 3} = \mathbf{2}\otimes\mathbf{2}\otimes\mathbf{2} = (\mathbf{3}\oplus\mathbf{1})\otimes\mathbf{2} = \mathbf{4}\oplus\mathbf{2}\oplus\mathbf{2} \tag{4.54}$$

The highest-dimensional subspace of such a representation $\mathbf{2}^{\otimes(n-1)} = \mathbf{n} \oplus \ldots$ is of dimension $n$; conversely, an $n$-plet of SU(2) can also be represented using an order-$(n-1)$ tensor, i. e. an object with $n-1$ "doublet indices".

In short, every irreducible representation of SU(2) can be obtained through tensor products using only its defining representation, the doublet representation **2**. This is why the doublet representation of SU(2) is called its fundamental representation.

# 5 Theories of dark matter and massive neutrinos

## 5.1. Renormalizability

Throughout the history of physics, any theory formulated in a spacetime continuum has been plagued with infinite quantities to some extent. Even in classical mechanics and electromagnetism, quantities like a point particle's self-energy in its own potential or even the amount of energy contained within the potential generated by a point particle raised questions. However, such issues only became a major focus during the development of quantum field theory, where infinite quantities are pervasive even for the calculation of observables which can clearly be measured to be finite. The concepts of *regularization* and *renormalization* have evolved to extract testable predictions from quantum field theories even in the face of such infinities. A vast amount of literature is available on this subject as continuous progress has been made over the decades (see e. g. the Nobel Prize 1999). A modern treatment is available, for example, in [Sch14].

In essence, regularization is a method of parametrizing divergences in the calculation of quantities. For example, fig. 5.1 shows a simple loop correction which is,

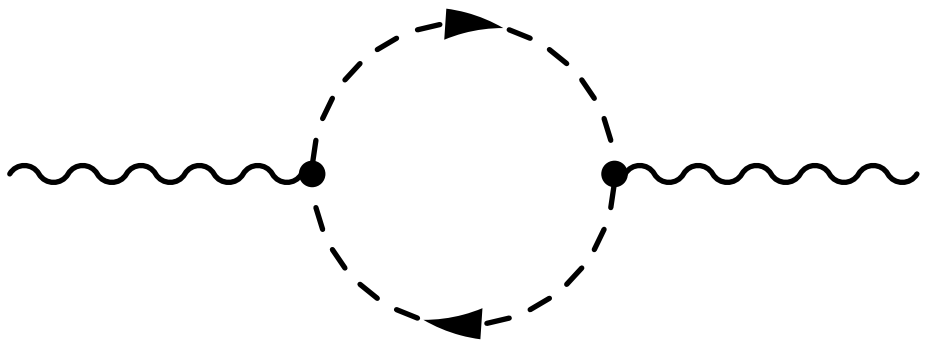

**Figure 5.1.:** Divergent vacuum polarization diagram in scalar quantum electrodynamics.

however, divergent. The value of the corresponding matrix element is

$$iM = e^2 \int_{\mathbb{R}^4} \frac{d^4k}{(2\pi)^4} \frac{2k^\mu - p^\mu}{(k-p)^2 - m^2 + i\varepsilon} \frac{2k^\nu - p^\nu}{k^2 - m^2 + i\varepsilon} \approx 4e^2 \int_{\mathbb{R}^4} \frac{d^4k}{(2\pi)^4} \frac{k^2}{k^4} \sim \int_0^\infty dk\, k = \infty \tag{5.1}$$

On the other hand, the value of this diagram is not directly measurable. Yet, it is an important mathematical step in calculating quantities which *are* measurable. A number of regularization procedures exist to avoid the appearance of the value "$\infty$", which cannot really be handled mathematically. One of the simplest ones simply introduces a cutoff $\Lambda$ for the integration such that

$$\int_0^\infty dk\, k \to \int_0^\Lambda dk\, k \sim \Lambda^2 \tag{5.2}$$

where the original integral is recovered for $\Lambda \to \infty$. Different divergent diagrams can now be summed while keeping track of the divergences through the parameter $\Lambda$.

However, $\Lambda$ is not a physical parameter. Renormalization is then used to ensure that any physical observables are indeed independent of this parameter. An important step in this procedure is the realization that there is no reason to believe that the parameters appearing in an interacting theory's Lagrangian, such as the coupling strength $e$ or a mass parameter $m$, are the same quantities as (in quantum electrodynamics) the electric charge or a particle's mass which can be physically measured. The parameters used in the Lagrangian are thus, generally, just as unphysical as the regularization parameter $\Lambda$. From a physical point of view, the divergence appearing in the diagram fig. 5.1 could be seen as an artifact of a specific method of computation and choice of parametrization, and not as a problem with any physical observable itself. Renormalization is the procedure that allows one to re-parametrize the results computed for physical observables (such as scattering cross sections) in terms of the actual physical parameters (charges, masses, ...) and to confirm that they do not depend on any unphysical ones.

However, not all theories can be renormalized. If the Lagrangian contains terms with operators whose mass dimension is too large, new Feynman diagrams with divergences will appear at each order, requiring an infinite number of renormalization parameters: The divergences cannot all be "removed" by re-parametrization of the Lagrangian. In general, this occurs if any operator in the Lagrangian has a mass dimension $> 4$. Thus, in order to construct a (potentially) renormalizable theory, only operators of mass dimension $\leq 4$ are allowed. This limitation restricts the possible Lorentz-invariant terms appearing in a renormalizable Lagrangian to just five different types, shown in table 5.1. A term's mass dimension can be worked out by recalling that, for any boson field $\phi$ and fermion field $\psi$,

$$[\phi] = \mathsf{M}^1 \qquad [\psi] = \mathsf{M}^{\frac{3}{2}} \tag{5.3}$$

**Table 5.1.:** Types of terms (without derivatives) which are allowed in a renormalizable theory for a set of bosons $\{\phi_i\}$ and fermions $\{\psi_i\}$.

| Term | Type $(N_s, N_f)$ | Mass dimension | Description |
|---|---|---|---|
| $\phi_i\phi_j$ | (2, 0) | $\mathsf{M}^2$ | bosonic mass term |
| $\psi_i\psi_j$ | (0, 2) | $\mathsf{M}^3$ | fermionic mass term |
| $\phi_i\phi_j\phi_k$ | (3, 0) | $\mathsf{M}^3$ | cubic interaction |
| $\phi_i\phi_j\phi_k\phi_m$ | (4, 0) | $\mathsf{M}^4$ | quartic interaction |
| $\phi_i\psi_j\psi_k$ | (2, 1) | $\mathsf{M}^4$ | Yukawa interaction |

Non-renormalizable theories may, at first glance, seem without much merit because an infinite number of parameters would be needed to get rid of all divergences. However, they can still be useful in the framework of *effective field theories*. Since the Lagrangian itself has mass dimension $[\mathcal{L}] = \mathsf{M}^4$, non-renormalizable terms are also those whose coupling parameters are inverse powers of a mass/energy scale $\Lambda$. If a non-renormalizable theory is now interpreted as one which is simply a low-energy approximation of another unknown theory, whose effects only become relevant beyond the energy scale $\Lambda$, it can be very useful to describe physics in a certain energy region. Higher-order terms are suppressed at low energies because inverse powers of the large scale $\Lambda$ are small. This can be thought of by imagining a large energy scale as a small distance scale: From far away, an interaction may appear to be fundamental, and only by "zooming in" can the sub-structure be revealed. Historically, this has happened many times in physics (e. g. with the discovery of the quarks or, as a prime example of a non-renormalizable theory, with Fermi's four-fermion coupling as a description of the weak interaction). In this thesis, though, only renormalizable theories are considered because the theories under consideration here are, after all, proposals of what such new-scale physics could look like.

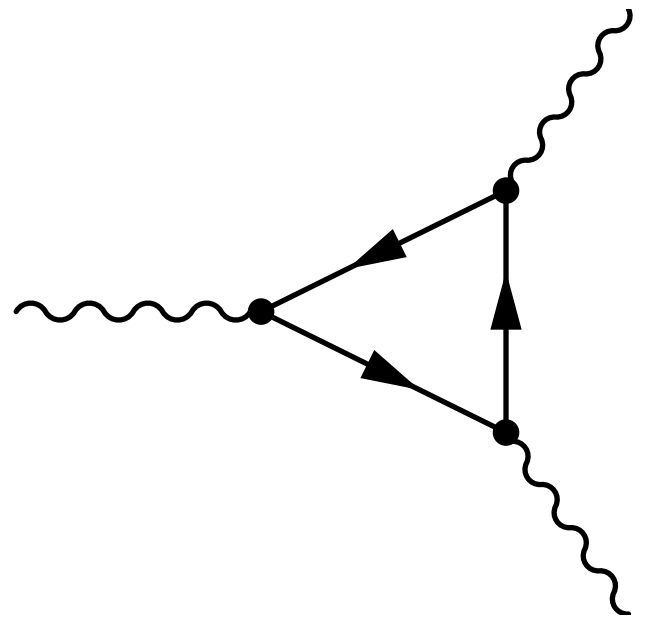

**Figure 5.2.:** Triangle diagram demonstrating gauge anomalies.

## 5.2. Gauge anomalies

Perhaps surprisingly, many features ascribed to quantum field theories can already be found in classical (relativistic) field theories. In fact, the only "quantum" aspect of QFT is the "promotion" of the classical field values to operators on a Hilbert space.The entire remaining discussion of gauge theories in chapter 3 can be done equivalently in a classical theory, treating boson fields as ordinary (commuting) $c$-numbers and fermions as (anti-commuting) Grassmann $a$-numbers. On the level of Feynman diagrams, the truly novel effects appear through loop diagrams, which are absent from classical theories. It is these features which truly cause classically unknown phenomena. Tree-level diagrams, on the other hand, can be used for computations both in QFT and in classical theories [Hel].

One of these phenomena is the appearance of so-called "anomalies". They occur if a symmetry of the classical theory is not valid anymore for the corresponding quantum theory. For global symmetries of the Lagrangian, this is not necessarily a problem, but an anomaly of the gauge group leads to an inconsistent theory. If the theory is not actually invariant under gauge transformations, the currents coupling to the gauge bosons are not conserved and unphysical states with negative probability (such as extra polarizations of the massless photon) can be produced.

In terms of Feynman diagrams, gauge anomalies manifest themselves in so-called triangle diagrams (shown in fig. 5.2) and can fully and exactly be treated at the level of one-loop diagrams. Calculating such triangle diagrams, which can be interpreted as the expectation value of an operator related to the conserved current, it becomes apparent that for some theories, the current is not conserved after all:

$$\langle B|\partial_\mu J^\mu|B\rangle \neq 0 \tag{5.4}$$

**Table 5.2.:** Anomaly constraints for a gauge theory with the Standard Model gauge group. For SU(3), it is assumed that all fermions are either in the representation $\mathbf{1}_3$ or $\mathbf{3}_3$, while for SU(2), $\mathbf{1}_2$, $\mathbf{2}_2$ and $\mathbf{3}_2$ are considered. The subscripts to the representations indicate which group they belong to. grav stands for anomalies caused by a hypothetical graviton. Based on [Sch14, ch. 30].

| Anomaly | Constraint |
|---|---|
| $SU(3)^3$ | only for chiral QCD |
| $SU(2)^3$ | none (always vanishes) |
| $U(1)^3$ | $\sum_{\psi} Y_{\psi}^3 = 0$ |
| $SU(3)^2 \times U(1)$ | $\sum_{\psi \in \mathbf{3}_3} Y_{\psi} = 0$ |
| $SU(2)^2 \times U(1)$ | $\sum_{\psi \in \mathbf{2}_2} Y_{\psi} + 4 \sum_{\psi \in \mathbf{3}_2} Y_{\psi} = 0$ |
| $SU(N) \times U(1)^2$ | none (always vanishes) |
| $SU(3) \times SU(2) \times U(1)$ | none (always vanishes) |
| $\text{grav}^2 \times SU(N)$ | none (always vanishes) |
| $\text{grav}^2 \times U(1)$ | $\sum_{\psi} Y_{\psi} = 0$ |
| Witten SU(2) | even number of SU(2) fermion doublets |

(where $B$ is a gauge boson), in direct contradiction with the assumption that the theory is invariant under the corresponding gauge group.

In general, the anomaly contribution is proportional to the symmetric tensor

$$d_R^{abc} = 2\,\mathrm{Tr}\big(T_R^a\{T_R^b, T_R^c\}\big) = 2A(R)\,\mathrm{Tr}\big(T_\mathrm{F}^a\{T_\mathrm{F}^b, T_\mathrm{F}^c\}\big) \tag{5.5}$$

where $T_R^a$, $T_R^b$, $T_R^c$ are the generators of the group $a$, $b$ or $c$ in the representation $R$, $T_\mathrm{F}^a$, $T_\mathrm{F}^b$, $T_\mathrm{F}^c$ are the generators of the corresponding SU($N$) or U(1) group in the defining representation and the anomaly coefficient $A(R)$ is a number that depends on the representation $R$ (with $A(\mathrm{F}) = 1$ for the defining representation) [Sch14, ch. 30.4]. The sum of the $d_R^{abc}$ for all the fields of a theory must be zero for the anomalies to cancel.

The resulting conditions that a theory's fields must satisfy to be free of gauge anomalies is shown in table 5.2. The Standard Model fulfills all of these conditions and thus does not have any gauge anomalies.

In the Standard Model, all fermions are either singlets or in the defining represen-tation of the gauge groups, so $A(R)$ is always either 0 or 1. However, for the purposes of [RZY13] and this thesis, triplets of SU(2) are also considered, so deriving the constraint for this representation is not as simple. Looking at table 5.2, the only place where the specific representation of SU(2) is actually relevant is for the anomaly $\mathrm{SU}(2)^2 \times \mathrm{U}(1)$. The factor of 4 in the condition comes from considering (5.5) for both doublets and triplets of SU(2) (**2** and **3**):[1]

$$\begin{aligned} 2\,\mathrm{Tr}\big(Y\{T_{\mathbf{2}}^b, T_{\mathbf{2}}^c\}\big) &= 2\,\mathrm{Tr}\left(Y\left\{-\frac{i}{2}\sigma^b, -\frac{i}{2}\sigma^c\right\}\right) = -\delta^{bc}\,\mathrm{Tr}(Y) \quad (A(\mathbf{2}) = 1) \\ 2\,\mathrm{Tr}\big(Y\{T_{\mathbf{3}}^b, T_{\mathbf{3}}^c\}\big) &= 2\,\mathrm{Tr}\big(Y\{s^b, s^c\}\big) = -4\delta^{bc}\,\mathrm{Tr}(Y) \qquad\qquad \Rightarrow A(\mathbf{3}) = 4 \end{aligned} \tag{5.6}$$

using the matrices $s_a$ defined in (4.28) and the identities $\mathrm{Tr}\big(\big\{-\frac{i}{2}\sigma^b, -\frac{i}{2}\sigma^c\big\}\big) = -\frac{1}{2}\delta^{bc}$, $\mathrm{Tr}\big(\{s^b, s^c\}\big) = -2\delta^{bc}$.

One particular anomaly that does not arise in the same way as the others pertains only to SU(2) as a gauge group. It is the last entry in table 5.2 and was first described by Witten in [Wit82], whom it is also named after. Essentially, this anomaly occurs if a theory contains an odd number of fermions which are doublets under an SU(2) gauge group, rendering it mathematically inconsistent.

The constraints shown in table 5.2 mean that the fields' hypercharges must satisfy a number of nonlinear equations. Together with the Witten anomaly, the only way to comply with all of these conditions is often to simply make all new fermion fields beyond the Standard Model with non-zero hypercharge *vector-like*. A vector-like fermion is simply a Dirac fermion whose left- and right-handed parts are in the same representation of the gauge groups. Correspondingly, such fermions could also be called non-chiral because the gauge interactions do not distinguish between the differently-handed components. Viewing every Weyl spinor field as left-handed according to the convention, this means that a vector-like fermion $\psi_\mathrm{V}$ is of the form

$$\psi_\mathrm{V} = \begin{pmatrix} \psi_\mathrm{L} \\ \psi_\mathrm{R}^\mathrm{c} \end{pmatrix} \qquad \psi_\mathrm{L} \in R, \psi_\mathrm{R}^\mathrm{c} \in \bar{R} \tag{5.7}$$

where $R$ is a representation of the gauge group, i. e. $\psi_\mathrm{L}$ and $\psi_\mathrm{R}^\mathrm{c}$ belong to representations which are conjugate to each other. In particular, this means that they have opposite hypercharges. Using table 5.2, it is then immediately clear that adding such a vector-like fermion to an anomaly-free theory yields another anomaly-free theory.

[1] For comparison, see, for example, [EFT17, p. 12], where the conditions for anomaly cancellation with SU(2) triplets are also considered.

The name "vector-like" stems from the form of the conserved current which results such fermions. The part of the Lagrangian responsible for the "weak charged current" (involving the $W^\pm$ bosons) is $\mathcal{L}_{\mathrm{CC}} \sim (J^\mu W^+_\mu + \mathrm{H.c.})$. In the Standard Model, the fermions are chiral with respect to the SU(2) gauge group and, using the leptons as an example, the current thus takes the form

$$J^\mu \sim \bar{\nu}_{\mathrm{L}} \bar{\sigma}^\mu e_{\mathrm{L}} \sim \bar{\nu}_{\mathrm{D}} \gamma^\mu (1 - \gamma_5) e_{\mathrm{D}} = \underbrace{\bar{\nu}_{\mathrm{D}} \gamma^\mu e_{\mathrm{D}}}_{=J^\mu_{\mathrm{V}}} - \underbrace{\bar{\nu}_{\mathrm{D}} \gamma^\mu \gamma_5 e_{\mathrm{D}}}_{=J^\mu_{\mathrm{A}}} \tag{5.8}$$

where $\nu_{\mathrm{D}}$ and $e_{\mathrm{D}}$ are Dirac spinors with left- and right-handed components $\nu_{\mathrm{L}}$, $\overline{\nu^{\mathrm{c}}_{\mathrm{R}}}$ and $e_{\mathrm{L}}$, $\overline{e^{\mathrm{c}}_{\mathrm{R}}}$. $J^\mu_{\mathrm{V}}$ behaves like a Lorentz vector under parity transformations ($J^\mu_{\mathrm{V}} \mapsto -J^\mu_{\mathrm{V}}$), while $J^\mu_{\mathrm{A}}$ behaves like a pseudo-vector, also called axial vector ($J^\mu_{\mathrm{A}} \mapsto J^\mu_{\mathrm{A}}$). Accordingly, the weak interaction of the Standard Model is often said to have a $V - A$ structure. However, if $\nu_{\mathrm{D}}$ and $e_{\mathrm{D}}$ were vector-like fermions, the current would be

$$J^\mu \sim \bar{\nu}_{\mathrm{L}} \bar{\sigma}^\mu e_{\mathrm{L}} + \overline{\nu^{\mathrm{c}}_{\mathrm{R}}} \bar{\sigma}^\mu e^{\mathrm{c}}_{\mathrm{R}} = \underbrace{\bar{\nu}_{\mathrm{D}} \gamma^\mu e_{\mathrm{D}}}_{=J^\mu_{\mathrm{V}}} \tag{5.9}$$

so the conserved current as a whole would behave like a *vector*.

The cases where adjustments need to be made to cancel anomalies are explicitly mentioned in [RZY13], where any fermion with hypercharge which lacks an oppositely-charged "partner" is taken to be vector-like to cancel anomalies. Moreover, although SU(2) doublets without hypercharge do not generally appear in these theories due to their inability to form useful couplings (their components would have the "exotic" charges $\pm 1/2$), another reason (apart from hypercharge) why "unpaired" fermionic doublets must be made vector-like is the Witten anomaly.

## 5.3. Mixing and mass diagonalization

The "masses" of a field's particles arise from terms quadratic or bilinear in the fields in the Lagrangian. However, in general, it is not the case that these terms are "diagonal" in the originally defined fields, that is, there may be mixed terms. When this happens, the fields are said to mix and the mass matrix, which is the matrix of coefficients of these bilinear terms, must be diagonalized in order to obtain those states which have a well-defined mass.

Mixed mass terms have the following form:

$$\mathcal{L} = -\frac{1}{2}(M_\mathrm{R}^2)^{ij}\phi_i\phi_j - (M_\mathrm{C}^2)^{ij}\Phi_i^\dagger\Phi_j - \left(\frac{1}{2}M_\mathrm{M}^{ij}\chi_i\chi_j - M_\mathrm{D}^{ij}\psi_i^1\psi_j^2 + \mathrm{H.c.}\right) \tag{5.10}$$

with real scalars $\phi_i$, complex scalars $\Phi_i$, neutral Weyl spinors $\chi_i$ and two sets of charged Weyl spinors $\psi_i^1$, $\psi_i^2$. In general, there are four different kinds of mass terms (bilinears) which lead to the mixing of fields in different ways [DHM10; Sta08]:[2]

- Real scalars: The squared mass matrix is real and symmetric, so it can be diagonalized by an orthogonal matrix.

$$\phi' = O\phi, \quad (m_\mathrm{R}^2)^\mathrm{diag} = OM_\mathrm{R}^2O^\mathrm{T} \tag{5.11}$$

- Complex scalars: The squared mass matrix is Hermitian, so it can be diagonalized by a unitary matrix.

$$\Phi' = U\Phi, \quad (m_\mathrm{C}^2)^\mathrm{diag} = UM_\mathrm{C}^2U^\dagger \tag{5.12}$$

- Majorana mass terms ("symmetric"): The mass matrix is complex and symmetric, so it can be diagonalized by a unitary matrix.

$$\chi' = U\chi, \quad m_\mathrm{M}^\mathrm{diag} = U^*M_\mathrm{M}U^\dagger \tag{5.13}$$

- Dirac mass terms ("not symmetric"): The mass matrix is a general complex square matrix, so it can be diagonalized by two unitary matrices.

$$\psi'^1 = U\psi^1, \psi'^2 = V\psi^2, \quad m_\mathrm{D}^\mathrm{diag} = U^*M_\mathrm{D}V^\dagger \tag{5.14}$$

The mass terms are then diagonal in the primed fields.[3]

[2] For the fermions, the expressions in the four-component spinor formalism are usually slightly different: $m_\mathrm{M}^\mathrm{diag} = UM_\mathrm{M}U^\dagger$ and $m_\mathrm{D}^\mathrm{diag} = UM_\mathrm{D}V^\dagger$ usually appear (without the conjugates). This is because the basis transformation is defined slightly differently. In the Weyl spinor formalism, one naturally defines a new basis for the left-handed spinors $\chi$, $\psi^1$, $\psi^2$. On the other hand, in the Dirac spinor formalism, the relevant objects are $\psi_\mathrm{D,L} = P_\mathrm{L}\psi_\mathrm{D}$ and $\psi_\mathrm{D,R} = P_\mathrm{R}\psi_\mathrm{D}$. For $\psi_\mathrm{D,L}$, there is no difference, but $\psi_\mathrm{D,R}$ contains a *right-handed* Weyl spinor. These are obtained from left-handed ones through complex conjugation, explaining the appearance of $U^*$.

[3] The transformations are sometimes also defined in the opposite way, i. e. $\phi = O\phi'$, $\Phi = U\Phi'$, $\chi = U\chi'$, $\psi^1 = U\psi'^1$, $\psi^2 = V\psi'^2$. The mixing matrices are then the inverses compared to the previous definition.

**Table 5.3.:** Overview of the different types of fermions.

| Fermion type | Properties | Mathematical representation |
|---|---|---|
| Weyl fermion | massless, neutral or charged, fixed helicity | one Weyl spinor |
| Majorana fermion | massive, neutral | one Weyl spinor or one Majorana spinor |
| Dirac fermion | massive, charged | two Weyl spinors or one Dirac spinor |

While the situation is simple enough for scalars, fermions are more complicated because two Weyl spinors can make up one Dirac fermion, whose mass terms are actually not diagonal in the left- and right-handed components. Moreover, given an arbitrary collection of Weyl spinors, one may wonder how one can tell which of these will assemble to Dirac fermions and which to Majorana fermions. The mass diagonalization procedure for fermions is discussed at length in [DHM10], with the following important result [DHM10, p. 35]:

> "Given an arbitrary collection of two-component [...] $(\frac{1}{2}, 0)$ fermions, the distinction between Majorana and Dirac fermions depends on whether the Lagrangian is invariant under a global (or local) continuous symmetry group $G$, and the corresponding multiplet structure of the fermion fields. If no such continuous symmetry exist[s], then the fermion mass eigenstates will consist of Majorana fermions. If the Lagrangian is invariant under a symmetry group $G$, then the collection of two-component fermions will break up into a sum of multiplets that transform irreducibly under $G$. [...] If a multiplet transforms under a real representation of $G$, then the corresponding fermion mass eigenstates are Majorana fermions. If a multiplet transforms under a complex representation of $G$, then the corresponding fermion mass eigenstates are Dirac fermions."

In other words, Majorana fermions are neutral with respect to *any* symmetry of the Lagrangian, even global ones (like lepton number). The properties of the different kinds of fermions are summarized in table 5.3. For massless fermions, there is no particular reason to combine Weyl spinors into a Dirac spinor because there are no mass terms. In that case, the fermion is called a (massless) *Weyl fermion*. As will become apparent soon (section 5.6), this is why the neutrinos are Majorana fermions in the presence of the Weinberg operator, which breaks the lepton number symmetry.

As an application of these statements, consider the Standard Model. Since there is only one scalar field in the Standard Model (the Higgs field), there is no mixing in the scalar sector. The terms involving the leptons are

$$
\begin{aligned}
\mathcal{L}_{(\text{lepton})} &= i\bar{L}^i \bar{\sigma}^\mu D_\mu L_i + i(\bar{e}^{\text{c}}_{\text{R}})^i \bar{\sigma}^\mu D_\mu (e^{\text{c}}_{\text{R}})_i - (Y_e^{ij} H^\dagger L_i (e^{\text{c}}_{\text{R}})_j + \text{H.c.}) \\
&= i\bar{L}^i \left( \partial_\mu - \frac{i}{2} g_2 \begin{pmatrix} W^3_\mu & W^1_\mu - iW^2_\mu \\ W^1_\mu + iW^2_\mu & -W^3_\mu \end{pmatrix} - \frac{i}{2} g_1 Y_L B_\mu \right) \bar{\sigma}^\mu L_i \\
&\quad + i(\bar{e}^{\text{c}}_{\text{R}})^i \left( \partial_\mu - \frac{i}{2} g_1 Y_{e^{\text{c}}_{\text{R}}} B_\mu \right) (e^{\text{c}}_{\text{R}})_i - \left( Y_e^{ij} \begin{pmatrix} (\nu_{\text{L}})_i & (e_{\text{L}})_i \end{pmatrix} H (e^{\text{c}}_{\text{R}})_j + \text{H.c.} \right) \\
&= i\bar{L}^i \left( \partial_\mu - \frac{i}{2} g_2 \begin{pmatrix} W^3_\mu & \sqrt{2} W^+_\mu \\ \sqrt{2} W^-_\mu & -W^3_\mu \end{pmatrix} - \frac{i}{2} g_1 Y_L B_\mu \right) \bar{\sigma}^\mu L_i \\
&\quad + i(\bar{e}^{\text{c}}_{\text{R}})^i \left( \partial_\mu - \frac{i}{2} g_1 Y_{e^{\text{c}}_{\text{R}}} B_\mu \right) (e^{\text{c}}_{\text{R}})_i - \left( Y_e^{ij} \begin{pmatrix} (\nu_{\text{L}})_i & (e_{\text{L}})_i \end{pmatrix} H (e^{\text{c}}_{\text{R}})_j + \text{H.c.} \right) \\
&= i(\bar{\nu}_{\text{L}})^i \left( \partial_\mu - \frac{i}{2} (g_2 W^3_\mu + g_1 Y_L B_\mu) \right) \bar{\sigma}^\mu (\nu_{\text{L}})_i \\
&\quad + i(\bar{e}_{\text{L}})^i \left( \partial_\mu - \frac{i}{2} (-g_2 W^3_\mu + g_1 Y_L B_\mu) \right) \bar{\sigma}^\mu (e_{\text{L}})_i \\
&\quad - \left( \frac{1}{\sqrt{2}} g_2 (\bar{\nu}_{\text{L}})^i W^+_\mu \bar{\sigma}^\mu (e_{\text{L}})_i + \text{H.c.} \right) + i(\bar{e}^{\text{c}}_{\text{R}})^i \left( \partial_\mu - \frac{i}{2} g_1 Y_{e^{\text{c}}_{\text{R}}} B_\mu \right) (e^{\text{c}}_{\text{R}})_i \\
&\quad - \left( Y_e^{ij} \begin{pmatrix} (\nu_{\text{L}})_i & (e_{\text{L}})_i \end{pmatrix} H (e^{\text{c}}_{\text{R}})_j + \text{H.c.} \right)
\end{aligned}
\tag{5.15}
$$

with generation indices $i, j \in \{1, 2, 3\}$.

An SU(2) gauge (the so-called *unitary gauge*) can be chosen so that the Higgs doublet has the form

$$
H = \begin{pmatrix} 0 \\ \frac{v+h}{\sqrt{2}} \end{pmatrix}
$$

where $v$ is the *vacuum expectation value* (VEV) of $H$ (which is non-zero after electroweak symmetry breaking). With this, the Yukawa term takes the form

$$
-Y_e^{ij} H^\dagger \begin{pmatrix} (\nu_{\text{L}})_i & (e_{\text{L}})_i \end{pmatrix} (e^{\text{c}}_{\text{R}})_j + \text{H.c.} = -e_{\text{L}}^{\text{T}} M_e e^{\text{c}}_{\text{R}} - \frac{\sqrt{2}}{v} e_{\text{L}}^{\text{T}} M_e e^{\text{c}}_{\text{R}} h + \text{H.c.} \tag{5.16}
$$

with the charged lepton mass matrix

$$
M_e = \frac{v}{\sqrt{2}} Y_e \tag{5.17}
$$

In this basis, the three lepton generations are mixed in the Yukawa terms, but not in the kinetic terms. However, one can also switch to the basis of mass eigenstates by defining new fields $e'_{\mathrm{L}} = U_e^{\mathrm{L}} e_{\mathrm{L}}$, $\nu'_{\mathrm{L}} = U_\nu^{\mathrm{L}} \nu_{\mathrm{L}}$, $e_{\mathrm{R}}^{\mathrm{c}\,\prime} = U_e^{\mathrm{R}} e_{\mathrm{R}}^{\mathrm{c}}$, where the $U$ matrices are unitary $3 \times 3$ matrices with generation indices, and expressing the Lagrangian through these new fields. Since the matrices are unitary, it holds that

$$e_{\mathrm{L}} = U_e^{\mathrm{L}\dagger} e'_{\mathrm{L}} \tag{5.18}$$

$$\nu_{\mathrm{L}} = U_\nu^{\mathrm{L}\dagger} \nu'_{\mathrm{L}} \tag{5.19}$$

$$e_{\mathrm{R}}^{\mathrm{c}} = U_e^{\mathrm{R}\dagger} e_{\mathrm{R}}^{\mathrm{c}\,\prime} \tag{5.20}$$

Inserting this into the Lagrangian:

$$\begin{aligned}
\mathcal{L}_{(\mathrm{lepton})} = {} & i\nu_{\mathrm{L}}^{\prime\mathrm{T}} \cancel{U_\nu^{\mathrm{L}*}} \left( \partial_\mu - \frac{i}{2}(g_2 W_\mu^3 - g_1 Y_L B_\mu) \right) \bar{\sigma}^\mu \cancel{U_\nu^{\mathrm{L}\dagger}} \nu'_{\mathrm{L}} \\
& + i e_{\mathrm{L}}^{\prime\mathrm{T}} \cancel{U_e^{\mathrm{L}*}} \left( \partial_\mu - \frac{i}{2}(-g_2 W_\mu^3 + g_1 Y_L B_\mu) \right) \bar{\sigma}^\mu \cancel{U_{\mathrm{L}}^{e\dagger}} e'_{\mathrm{L}} \\
& - \left( \frac{1}{\sqrt{2}} g_2 \nu_{\mathrm{L}}^{\prime\mathrm{T}} \underbrace{U_{\mathrm{L}}^{\nu*} U_{\mathrm{L}}^{e\dagger}}_{=V_{\mathrm{PMNS}}} W_\mu^+ \bar{\sigma}^\mu e'_{\mathrm{L}} + \mathrm{H.c.} \right) + i e_{\mathrm{R}}^{\prime\mathrm{T}} \cancel{U_e^{\mathrm{R}}} \left( \partial_\mu - \frac{i}{2} g_1 Y_{e_{\mathrm{R}}} B_\mu \right) \cancel{U_{\mathrm{R}}^{e\dagger}} e'_{\mathrm{R}} \\
& - e_{\mathrm{L}}^{\prime\mathrm{T}} \underbrace{U_{\mathrm{L}}^{e*} M^e U_{\mathrm{R}}^{e\dagger}}_{=\mathrm{diag}(m_e, m_\mu, m_\tau)} e'_{\mathrm{R}} + \frac{\sqrt{2}}{v} e_{\mathrm{L}}^{\prime\mathrm{T}} \underbrace{U_{\mathrm{L}}^{e*} M^e U_{\mathrm{R}}^{e\dagger}}_{=\mathrm{diag}(m_e, m_\mu, m_\tau)} e'_{\mathrm{R}} h + \mathrm{H.c.}
\end{aligned} \tag{5.21}$$

One can see that the $U$ matrices simply cancel in most terms.

Since the primed fields are supposed to form the mass eigenstates, $U_e^{\mathrm{L}}$ and $U_e^{\mathrm{R}}$ are chosen such that $M_e$ becomes diagonal (see above), which means that there is no mixing between generations in the Yukawa terms in this basis. After this choice, the only places which still have mixing between generations are the terms involving $W_\mu^\pm$, the weak charged currents. This is where the PMNS matrix

$$V_{\mathrm{PMNS}} = U_\nu^{\mathrm{L}*} U_e^{\mathrm{L}\dagger} \tag{5.22}$$

appears. For the quarks, the procedure is completely analogous (except that there is also a right-handed up quark) and one would combine the mixing matrices in these terms into the CKM matrix

$$V_{\mathrm{CKM}} = U_u^{\mathrm{L}*} U_d^{\mathrm{L}\dagger} \tag{5.23}$$

Since the neutrinos are massless in the Standard Model, they do not have Yukawa

(mass) terms. Correspondingly, $U_\nu^\mathrm{L}$ can be chosen completely arbitrarily and there would still be no mixing between neutrino generations. Choosing $U_\nu^\mathrm{L} = U_e^{\mathrm{L}*}$, the mixing matrices in the weak charged current simply cancel ($V_\mathrm{PMNS} = \mathbb{1}$) and there is no mixing between lepton generations of the mass eigenstates at all.

## 5.4. Redundant terms in the Lagrangian

This section will demonstrate that once all possible renormalizable combinations have been added to a Lagrangian, it is often the case that not all of these terms are actually needed in the sense that they have an independent parameter. As an example for motivation, consider [Far09], which presents an implementation of the model T1-1-A with a parameter of $\alpha = 0$. The Lagrangian given there is (notation adapted to [RZY13]):

$$\begin{aligned}\mathcal{L} = \mathcal{L}_\mathrm{SM} + \mathcal{L}_\mathrm{kin} - m_{\phi'}^2 \phi'^\dagger \phi' - \frac{m_\mathrm{s}^2}{2}\varphi^2 - (m_{\varphi\phi'}\varphi H^\mathrm{T} i\sigma^2 \phi' + \mathrm{H.\,c.})\\ - \lambda_1 |H^\mathrm{T} i\sigma^2\phi'|^2 - \left(\frac{\lambda_2}{2}(H^\mathrm{T} i\sigma^2\phi')^2 + \mathrm{H.\,c.}\right) - \lambda_3\varphi^2 H^\dagger H - \lambda_4(\phi'^\dagger\phi')(H^\dagger H)\\ - \frac{\lambda_1'}{2}(\phi'^\dagger\phi')^2 - \frac{\lambda_2'}{2}\varphi^4 - \lambda_3'\varphi^2(\phi'^\dagger\phi')\\ - (g\psi(\phi'^\dagger L) + \mathrm{H.\,c.}^4) - \frac{M}{2}(\psi\psi + \mathrm{H.\,c.}) \end{aligned} \tag{5.24}$$

At first, it seems like there is a missing invariant term which could still be added to the Lagrangian:

$$\tilde{\mathcal{L}} = -\tilde{\lambda}(H^\dagger\phi')(\phi'^\dagger H) \tag{5.25}$$

However, on closer inspection, it becomes apparent that this term is redundant with other terms in the Lagrangian and can thus be omitted. This can be seen as follows: The quartic terms involving both $H$ and $\phi'$ are

$$\begin{aligned}-\lambda_1|H^\mathrm{T} i\sigma^2\phi'|^2 &= -\lambda_1\left|\begin{pmatrix}H^+ & H^0\end{pmatrix}\begin{pmatrix}0 & 1\\ -1 & 0\end{pmatrix}\begin{pmatrix}\phi'^0\\ \phi'^-\end{pmatrix}\right|^2 = -\lambda_1|H^+\phi'^- - H^0\phi'^0|^2\\ &= \underbrace{-\lambda_1|H^+\phi'^-|^2 - \lambda_1|H^0\phi'^0|^2}_{\boxed{1}}\ \underbrace{+\lambda_1\big(H^-\phi'^+H^0\phi'^0 + \mathrm{H.\,c.}\big)}_{\boxed{2}}\end{aligned} \tag{5.26}$$

[4] Curiously, the Hermitian conjugate which is necessary here is missing from the Lagrangians given in [Far09; FH10a; Far11].

$$-\frac{\lambda_2}{2}(H^{\mathsf{T}} i\sigma^2 \phi')^2 = -\frac{\lambda_2}{2}\left(\begin{pmatrix} H^+ & H^0 \end{pmatrix}\begin{pmatrix} 0 & 1 \\ -1 & 0 \end{pmatrix}\begin{pmatrix} \phi'^0 \\ \phi'^- \end{pmatrix}\right)^2 = -\lambda_2\big(H^+\phi'^- - H^0\phi'^0\big)^2$$
$$= -\frac{\lambda_2}{2}(H^+\phi'^-)^2 - \frac{\lambda_2}{2}(H^0\phi'^0)^2 + \lambda_2 H^+\phi'^- H^0\phi'^0 \tag{5.27}$$

$$-\left(\frac{\lambda_2}{2}(H^{\mathsf{T}} i\sigma^2\phi')^2\right)^\dagger = -\frac{\lambda_2}{2}(H^-\phi'^+)^2 - \frac{\lambda_2}{2}((H^0)^\dagger(\phi'^0)^\dagger)^2 + \lambda_2 H^-\phi'^+(H^0)^\dagger(\phi'^0)^\dagger \tag{5.28}$$

$$-\lambda_4(\phi'^\dagger\phi')(H^\dagger H) = -\lambda_4\big(|\phi'^-|^2 + |\phi'^0|^2\big)\big(|H^+|^2 + |H^0|^2\big)$$
$$= \underbrace{-\lambda_4|H^+\phi'^-|^2 - \lambda_4|H^0\phi'^0|^2}_{\boxed{1}}\ \underbrace{-\lambda_4|H^+\phi'^0|^2 - \lambda_4|\phi'^-H^0|^2}_{\boxed{3}} \tag{5.29}$$

$$-\tilde\lambda(H^\dagger\phi')(\phi'^\dagger H) = -\tilde\lambda\big|H^-\phi'^0 + (H^0)^\dagger\phi'^-\big|^2$$
$$= \underbrace{-\tilde\lambda|H^+\phi'^0|^2 - \tilde\lambda|\phi'^-H^0|^2}_{\boxed{3}}\ \underbrace{-\tilde\lambda\big(H^-\phi'^+H^0\phi'^0 + \mathrm{H.c.}\big)}_{\boxed{2}} \tag{5.30}$$

Thus, all parts of $\tilde{\mathcal{L}}$ are already contained within the $\lambda_1$ and $\lambda_4$ terms. Defining a set of new parameters

$$\tilde\lambda_1 = \lambda_1 - \tilde\lambda \tag{5.31}$$
$$\tilde\lambda_4 = \lambda_4 + \tilde\lambda \tag{5.32}$$

results in

$$-\lambda_1|H^{\mathsf{T}} i\sigma^2\phi'|^2 - \lambda_4(\phi'^\dagger\phi')(H^\dagger H) - \tilde\lambda(H^\dagger\phi')(\phi'^\dagger H)$$
$$= \underbrace{-(\tilde\lambda_4 + \tilde\lambda_1)\big(|H^+\phi'^-|^2 + |H^0\phi'^0|^2\big)}_{\boxed{1}}\ \underbrace{+\tilde\lambda_1\big(H^-\phi'^+H^0\phi'^0 + \mathrm{H.c.}\big)}_{\boxed{2}}$$
$$\underbrace{-\tilde\lambda_4\big(|H^+\phi'^0|^2 + |\phi'^-H^0|^2\big)}_{\boxed{3}} \tag{5.33}$$

compared to

$$-\lambda_1|H^{\mathsf{T}} i\sigma^2\phi'|^2 - \lambda_4(\phi'^\dagger\phi')(H^\dagger H)$$
$$= \underbrace{-(\lambda_4 + \lambda_1)\big(|H^+\phi'^-|^2 + |H^0\phi'^0|^2\big)}_{\boxed{1}}\ \underbrace{+\lambda_1\big(H^-\phi'^+H^0\phi'^0 + \mathrm{H.c.}\big)}_{\boxed{2}}$$
$$\underbrace{-\lambda_4\big(|H^+\phi'^0|^2 + |\phi'^-H^0|^2\big)}_{\boxed{3}} \tag{5.34}$$

This illustrates that two equivalent free parameters – either $\lambda_1$ and $\lambda_4$ or $\tilde{\lambda}_1$ and $\tilde{\lambda}_4$ – are sufficient to describe the entirety of the parameter space, irrespective of the presence or absence of $\tilde{\mathcal{L}}$. Consequently, the parameter $\tilde{\lambda}$ is redundant with $\lambda_1$ and $\lambda_4$ and the term can be omitted without loss of generality.

Building on this example, one can take note of the following result:

**Theorem 1.** *Given a Lagrangian $\mathcal{L} = \sum_i \lambda_i A_i + R$ with parameters $\lambda_i$ and remaining terms $R$, if there is an identity of the form*

$$\sum_i A_i = 0$$

*then one of the terms $\lambda_i A_i$ can be omitted from the Lagrangian (equivalently, one can set $\lambda_i = 0$) without any loss of generality.*

*Proof.* Without loss of generality, let the $n$th term be the one to be eliminated with $\tilde{\lambda} = -\lambda_n$, $B = A_n$, rewriting $\sum_i^n A_i = 0$ as $\sum_i^{n-1} A_i = -B$. Then, $\mathcal{L}$ can be written as

$$\mathcal{L} = \sum_i^{n-1} \lambda_i A_i + \tilde{\lambda} B + R = \sum_i^{n-1} \lambda_i A_i + \tilde{\lambda} \sum_i^{n-1} A_i + R = \sum_i^{n-1} (\lambda_i + \tilde{\lambda}) A_i + R$$

Defining $\tilde{\lambda}_i = \lambda_i + \tilde{\lambda}$ where *both $\tilde{\lambda}_i$ and $\lambda_i$ are completely free parameters*, $\mathcal{L}$ takes the form

$$\mathcal{L} = \sum_i^{n-1} \tilde{\lambda}_i A_i + R$$

which is the same as before except that one of the terms $(\lambda_n A_n)$ has been eliminated. □

### 5.4.1. SU(2) doublets

**Identity 1.** *For any two scalar* SU(2) *doublets $D_1$ and $D_2$, the identity*

$$|D_1|^2 |D_2|^2 = |D_1 \cdot D_2|^2 + |D_1^\dagger D_2|^2 \tag{5.35}$$

*holds. By theorem 1, this implies that:*

- *For $D_1 \neq D_2$, only* two *of the three terms*
  1. $|D_1|^2 |D_2|^2$
  2. $|D_1^T i\sigma^2 D_2|^2 = |D_1 \cdot D_2|^2$
  3. $|D_1^\dagger D_2|^2$

*are relevant to the parameter space.*

- *For $D_1 = D_2 = D$, the second term is zero since $D \cdot D = -D \cdot D = 0$ for any bosonic doublet, and thus $|D|^2|D|^2 = |D^\dagger D|^2 = (D^\dagger D)^2$, so there is only* one *relevant term in this case.*

*Proof.* See appendix C. □

The discussion can be limited to scalar (or, in general, bosonic) fields here because all quartic terms involving fermions are non-renormalizable.

### 5.4.2. SU(2) triplets

[LG17] lists a number of identities for a SU(2) doublet $\phi$ and a SU(2) triplet $\Delta$ with arbitrary hypercharges. In an analogous procedure to that presented in section 5.4.1, parameters can be eliminated so that only a subset of these terms must be included in the Lagrangian.

The trilinear terms (at least those involving only triplets) are dealt with easily:

**Identity 2.** *For any scalar* SU(2) *triplets $\Delta_i = \Delta_{ij}\sigma_j$, the following equation holds:*

$$\mathrm{Tr}(\Delta_i \Delta_j \Delta_k) = 0 \quad \text{if at least two of the } i, j, k \text{ are equal.} \tag{5.36}$$

*Proof.* See appendix C. □

Now for the quartic terms. Again, only scalar/bosonic fields need to be considered as anything else yields non-renormalizable terms.

**Identity 3.** *For any scalar* SU(2) *doublets $D_1$, $D_2$ and triplets $\Delta_1$, $\Delta_2$, the identities*

1. $\mathrm{Tr}(\Delta_1^2 \Delta_2^2) = \frac{1}{2} \mathrm{Tr}(\Delta_1^2)\, \mathrm{Tr}(\Delta_2^2)$
2. $\mathrm{Tr}(\Delta_1^2 \Delta_2^2) + \mathrm{Tr}((\Delta_1 \Delta_2)^2) = \mathrm{Tr}(\Delta_1 \Delta_2)^2$
3. $D_1^\dagger \Delta_1 \Delta_2 D_2 + D_1^\dagger \Delta_2 \Delta_1 D_2 = D_1^\dagger \{\Delta_1, \Delta_2\} D_2 = D_1^\dagger D_2\, \mathrm{Tr}(\Delta_1 \Delta_2)$

*hold.*[5] *By theorem 1, this implies that*

- *Only* two *of the three terms*

[5]These identities are equally valid if written without any conjugates in the doublets by substituting $D_1' = D_1^\dagger$. $D_1^\dagger$ is only used to emphasize that the doublet on the left appears with a raised index. They also hold for a single complex triplet $\Delta$, setting $\Delta_1 = \Delta^\dagger$, $\Delta_2 = \Delta$.

1. $\mathrm{Tr}\big(\Delta_1^2\Delta_2^2\big) = \frac{1}{2}\,\mathrm{Tr}\big(\Delta_1^2\big)\,\mathrm{Tr}\big(\Delta_2^2\big)$
2. $\mathrm{Tr}\big((\Delta_1\Delta_2)^2\big)$
3. $\mathrm{Tr}(\Delta_1\Delta_2)^2$

*are relevant to the parameter space. If* $\Delta_1 = \Delta_2 = \Delta$, *it holds that* $\mathrm{Tr}(\Delta^2)^2 = 2\,\mathrm{Tr}(\Delta^4)$, *so there is only* one *relevant term in this case.*

- *For appropriate values of the hypercharges (otherwise, gauge invariance would be violated), only* two *of the three terms*

  1. $D_1^\dagger\Delta_1\Delta_2 D_2 + \mathrm{H.c.}$
  2. $D_1^\dagger\Delta_2\Delta_1 D_2 + \mathrm{H.c.}$
  3. $D_1^\dagger D_2\,\mathrm{Tr}(\Delta_1\Delta_2) + \mathrm{H.c.}$

  *are relevant to the parameter space. If* $\Delta_1 = \Delta_2 = \Delta$, *it holds that* $D_1^\dagger\Delta^2 D_2 = D_1^\dagger D_2\,\mathrm{Tr}(\Delta^2)$, *so there is only* one *relevant term in this case.*

*Proof.* See appendix C. □

**Corollary 1.** *For a single scalar* SU(2) *triplet* $\Delta$, *only the following two quartic terms must be included in a fully general Lagrangian:*

- $\mathrm{Tr}\big(\Delta^\dagger\Delta\big)^2$ *and*
- $\mathrm{Tr}\big((\Delta^\dagger\Delta)^2\big)$.

*If* $\Delta$ *is a real triplet* ($\Delta^\dagger = \Delta$), *it holds that* $\mathrm{Tr}(\Delta^2)^2 = 2\,\mathrm{Tr}(\Delta^4)$, *so only* one *term must be included in this case.*

*Proof.* See appendix C. □

## 5.5. Dark matter candidates

In principle, it is actually quite simple for a theory to accommodate WIMP dark matter. In fact, even though neutrinos are not WIMPs, they, too, make up a fraction of the dark matter content of the universe, albeit a small one. Hence, when thinking about WIMP dark matter candidates, one is essentially looking for something with the properties of a "heavy neutrino".

Any particle designated as a suitable dark matter candidate must have the following properties:

**Stability:** Since there is so much dark matter left in the universe today, it must be mostly stable against decay with respect a timescale like the age of the universe. On the other hand, depending on the production mechanism of the dark matter, it cannot be *too* stable (in the sense that its interactions cannot be arbitrarily weak) either. For example, with the freeze-out mechanism, a sufficient amount of (co-)annihilation is necessary to avoid an over-abundance of dark matter in the current universe.

**Electrical and color neutrality:** The requirement of zero electromagnetic charge is already in the name *dark matter* – any unknown charged particle would be detectable, even from afar, through its electromagnetic interactions. Quark-like dark matter (*strongly interacting massive particles*, SIMPs) is likewise difficult to imagine for the same reason that only color-neutral states can be observed in practice (confinement). Even if a heavy new quark were "stable", it would immediately be subject to violent QCD reactions, producing hadronic showers. Therefore, the only possible objects with QCD interactions which could act as dark matter are composite so that they are, as a whole, neutral. Such ideas have been pursued, for example, with the proposal of *quark nugget dark matter* [Zhi17; ARS99], although in general, any kind of strongly interacting dark matter is heavily constrained and thus usually omitted from consideration [CFS06; Sta+90].

In any case, this already leads well outside the WIMP paradigm which is the focus in this thesis. As the name suggests, WIMP dark matter can at most participate in weak interactions.

**Cold dark matter:** As discussed in chapter 2, dark matter must be cold to fit the observations. Otherwise, the evolution of the universe would be inconsistent with the observed structure formation. In particular, neutrinos cannot be the main component of dark matter because they are hot dark matter due to their small masses. Such hot matter would make it more difficult for clumps to form, "smearing out" the resulting structures.

However, the list given above only states the absolute minimum requirements for a WIMP dark matter candidate. As searches for dark matter continue, the constraints put on such candidates through direct detection, indirect detection and collider experiments grows. Especially relevant for the models considered here is the fact that SU(2) multiplets with non-zero hypercharge are already mostly ruled out as dark matter candidates by direct detection constraints, even if they have neutral components [CFS06; RZY13]. In fact, [RZY13] excludes all multiplets with non-zero hypercharge except scalar doublets as candidates on grounds of direct detection. The problem is that the scattering cross section through $Z$ bosons increases with hypercharge since the $Z$ boson is a linear combination of the $W_3$ and $B$ bosons, whose corresponding "charges" are $T_3$ and $Y$. A non-zero value of the hypercharge generally makes the neutral component's $Z$ coupling too strong, putting its cross section for elastic scattering on a nucleus above the limits established by direct detection experiments.

**Table 5.4.:** The field content of the Higgs portal dark matter model.

| Field | Generations | Spin | Lorentz rep. | SU(3) | SU(2) | U(1) | $\mathbb{Z}_2$ |
|---|---|---|---|---|---|---|---|
| SM fields (table 3.1) + ... | | | | | | | 1 |
| $S$ | 1 | 0 | (0,0) | **1** | **1** | 0 | −1 |

It turns out that the addition of a suitable dark matter candidate on its own to the Standard Model is actually rather straightforward. Arguably the simplest way to do it is by adding just a single new scalar field to the Standard Model, a neutral singlet $S$ (see table 5.4). The most general renormalizable Lagrangian with such a field $S$ is

$$\mathcal{L}_{\text{HP}} = \mathcal{L}_{\text{SM}} + \frac{1}{2}(\partial^\mu S)(\partial_\mu S) - \frac{1}{2}M_S^2 S^2 - \lambda_1 (H^\dagger H) S - \lambda_2 S^3 - \lambda_3 (H^\dagger H) S^2 - \lambda_4 S^4 \quad (5.37)$$

This is also called *Higgs portal dark matter* [Ple17] because the only interaction connecting the particles of the Standard Model with the "dark scalar" $S$ is through the Higgs boson, which acts as a "portal" to the "dark sector". However, the term $\lambda_1(H^\dagger H)S$ is problematic because it would allow a single $S$ particle to decay via Higgs bosons. It can be *stabilized* by postulating a new global symmetry which enforces $\lambda_1 = 0$.

The simplest possible symmetry group is the "parity group" $\mathbb{Z}_2$, under which a field $\phi$ transforms as $\phi \mapsto \pm\phi$. Indeed, most theories with dark matter candidates make use of this group (or a similar one) to ensure that the dark matter is stable [CFS06; Ple17]. The minimal dark matter models with radiative neutrino masses studied in this work are no different (see section 5.8). $S$ is now required to be odd ($S \mapsto -S$) under $\mathbb{Z}_2$ parity, while all the other fields are even ($\phi \mapsto \phi$). This assignment forbids the terms $\lambda_1(H^\dagger H)S$ and $\lambda_2 S^3$, requiring $\lambda_1 = \lambda_2 = 0$ and hence ensuring stability, as desired. (5.37) then becomes

$$\mathcal{L}_{\mathrm{HP}} = \mathcal{L}_{\mathrm{SM}} + \frac{1}{2}(\partial^\mu S)(\partial_\mu S) - \frac{1}{2}M_S^2 S^2 - \lambda_3 (H^\dagger H) S^2 - \lambda_4 S^4 \tag{5.38}$$

What is perhaps remarkable is that dark matter could, potentially, be explained in such a strikingly simple way.

## 5.6. The $d = 5$ Weinberg operator

This section now shifts the focus to the second main feature of the models which are the subject of this thesis: neutrino masses. There are many ways in which neutrinos could become massive, but as mentioned before, the problem is that their masses have experimentally been determined to be suspiciously tiny. The Standard Model Higgs mechanism for Dirac neutrinos offers no explanation for the enormous suppression compared to the other fermions.

An alternative to neutrinos as massive Dirac fermions would be Majorana neutrinos. As of yet, the question whether neutrinos are Dirac or Majorana fermions is still open. From an effective field theory point of view, the lowest-order operator that would bring about massive Majorana neutrinos in the Standard Model is the (non-renormalizable) Weinberg operator of mass dimension 5:[6]

$$O_{\mathrm{W}} \sim \frac{1}{\Lambda}(H^\dagger L)(H^\dagger L) + \mathrm{H.\,c.} \tag{5.39}$$

where $\Lambda$ is a parameter of mass dimension 1. This operator is not invariant under the global U(1) lepton number symmetry of the Standard Model (it "violates lepton number"), and consequently, the neutrinos are no longer charged under any continuous symmetry in its presence. Moreover, after electroweak symmetry breaking, it produces mass terms for the neutrinos, illustrated by the diagram in fig. 5.3.[7] As discussed in section 5.3, this means that they will become massive Majorana neutrinos after the mass diagonalization procedure.

[6]Such operators were introduced by Weinberg in [Wei79], although with more emphasis on baryon number (instead of lepton number) violation.

[7]This diagram may look unusual at first, with the appearance of the VEV $\langle H^0 \rangle$. Such diagrams are primarily supposed to represent terms in the Lagrangian which end up as mass terms for the neutrinos. In this sense, they are not to be understood as tools for computation like ordinary Feynman diagrams, but only for visualization.

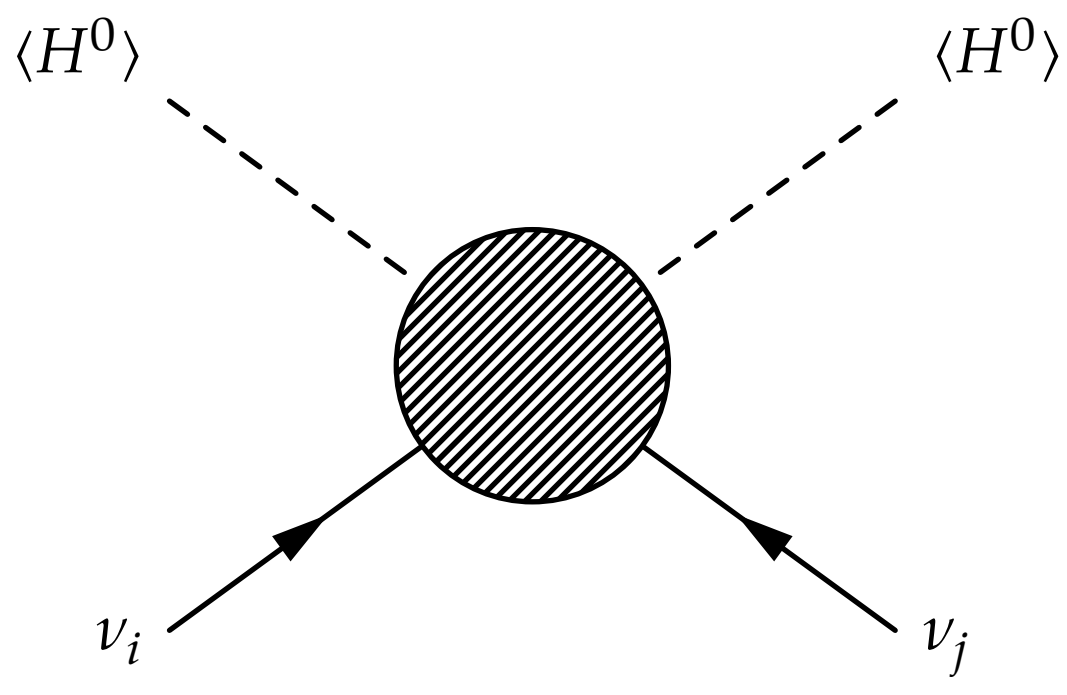

**Figure 5.3.:** Illustration of how neutrino masses are generated with the Weinberg operator after electroweak symmetry breaking.

By viewing (5.39) as an effective operator, one postulates the existence of new fields at the scale $\Lambda$ which, through their interactions, have the effect of this operator at lower energy scales. The goal is then to find which renormalizable theories manifest themselves through the Weinberg operator at low energies; these are its so-called *ultraviolet (UV) completions* (because they are valid for arbitrarily high energies) or *realizations*. By adding appropriate fields to the Standard Model, the "blob" in fig. 5.3 can be filled in with renormalizable interactions of these new fields. (5.39) then directly shows why the neutrino masses will be naturally small: They are suppressed by the large scale $\Lambda^{-1}$, which roughly corresponds to the masses of the heavy new particles.

The following sections will present the realizations of the Weinberg operator for neutrino masses appearing at zero and one loops. This progression corresponds to both how these models have been studied chronologically and how much attention they have received, with the well-known seesaw mechanism (section 5.7.1) having been thoroughly explored whereas many of the radiative models (section 5.8) remain completely unstudied so far.

## 5.7. Seesaw mechanisms at tree level

As shown by Ma in [Ma98], there are only three unique realizations of the Weinberg operator that bring about neutrino masses without loop corrections. These realizations are called the *seesaw mechanisms* of type I, II and III. They earned their names due to the suppression by the inverse mass scale $\Lambda$ in (5.39): The more this scale (the masses of the new particles) go up, the more the new interactions will be suppressed and thus, the more the neutrino masses will go down – just like on a seesaw. These seesaw mechanisms are well-known and studied (especially type I); a discussion is available, for example, in [Pat14].

**Table 5.5.:** The field content of the seesaw type I extension of the Standard Model.

| Field | Generations | Spin | Lorentz rep. | SU(3) | SU(2) | U(1) |
|---|---|---|---|---|---|---|
| SM fields (table 3.1) + ... | | | | | | |
| $N$ | $n$ | 1/2 | (1/2, 0) | **1** | **1** | 0 |

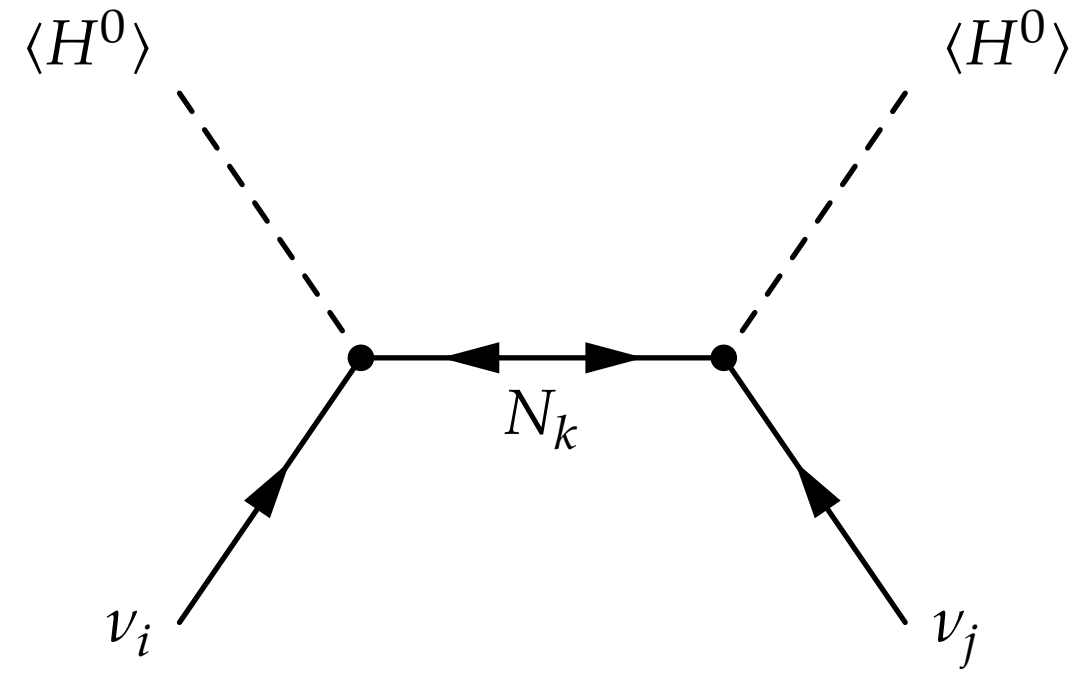

**Figure 5.4.:** Illustration of how neutrino masses are generated with the seesaw mechanism of type I.

### 5.7.1. Seesaw type I

The first and simplest possibility is to add $n$ generations of a neutral fermion singlet $N$ to the Standard Model (see table 5.5). When the term "seesaw mechanism" is used without a specified type, this variant is usually what is meant since it is the most well-known and well-studied one. As visible in table 5.5, the fermions $N_i$ are uncharged with respect to all gauge groups of the Standard Model and thus do not participate in gauge interactions at all; the only possible interaction of these fermions is through a Yukawa term with the Higgs boson. This is why they are often called *sterile neutrinos*.

The essence of the seesaw mechanism is not, however, to simply add right-handed fermion singlets for the neutrinos, as is done for all the other fermions, and have them obtain mass through the Higgs mechanism. Rather, unnaturally small values for the Yukawa couplings can be avoided because the neutral $N_i$ allow for ordinary mass terms in the Lagrangians. The mixing between the $N_i$ and the neutrinos can then be used to show that if the former have large masses, the latter become light due to the form of the mixing matrix.

The most general renormalizable Lagrangian for the type I seesaw mechanism is[8]

$$\begin{aligned}\mathcal{L}_{\mathrm{I}} &= \mathcal{L}_{\mathrm{SM}} + i\bar{N}^i \bar{\sigma}^\mu \partial_\mu N_i - \left(\frac{1}{2}(M_N)^{ij} N_i N_j + \mathrm{H.c.}\right) - \left((Y_N)_{ij}(HL_i)N_j + \mathrm{H.c.}\right) \\ &= \mathcal{L}_{\mathrm{SM}} + i\bar{N}^i \bar{\sigma}^\mu \partial_\mu N_i - \left(\frac{1}{2} N M_N N + \mathrm{H.c.}\right) - \left((H^0 \nu_{\mathrm{L}} - H^+ e_{\mathrm{L}}) Y_N N + \mathrm{H.c.}\right)\end{aligned} \tag{5.40}$$

where $M_N$ is the $n \times n$ mass matrix for the $N_i$ and $Y_N$ can be thought of as a $3 \times n$ matrix of Yukawa couplings. Figure 5.4 schematically shows how neutrino masses arise with this model. It should be stressed, however, that this is *not* a "real" Feynman diagram, but rather an illustration of the terms in the Lagrangian. In ordinary Feynman diagrams, the Higgs VEV, which acts here as a "mass insertion", would be left out and one would not use the mass eigenstates instead of the fields defined before electroweak symmetry breaking (which oscillate during propagation).

After electroweak symmetry breaking, the combined mass term for the neutral fermions takes the form

$$\begin{aligned}\mathcal{L}_{\mathrm{I}}^{(\mathrm{mass})} &= -\frac{1}{2}\begin{pmatrix}\nu_{\mathrm{L}} & N\end{pmatrix}\begin{pmatrix}0 & \frac{v}{2\sqrt{2}} Y_N \\ \frac{v}{2\sqrt{2}} Y_N^{\mathrm{T}} & M_N\end{pmatrix}\begin{pmatrix}\nu_{\mathrm{L}} \\ N\end{pmatrix} + \mathrm{H.c.} \\ &= -\frac{1}{2}\begin{pmatrix}\nu_{\mathrm{L}} & N\end{pmatrix}\begin{pmatrix}M_{\mathrm{L}} & M_{\mathrm{D}}^{\mathrm{T}} \\ M_{\mathrm{D}} & M_N\end{pmatrix}\begin{pmatrix}\nu_{\mathrm{L}} \\ N\end{pmatrix} + \mathrm{H.c.}\end{aligned} \tag{5.41}$$

where $M_{\mathrm{L}} = 0$ because there is no gauge-invariant mass term for the neutrinos $\nu_{\mathrm{L}}$. $M_{\mathrm{D}}$ arises from the "Dirac mass term" involving $\nu_{\mathrm{L}}$ and $N$ (although the appearance of this term does not imply that any of the fermions are Dirac fermions) while $M_N$ comes from the "Majorana mass term" for $N$.

Assuming that the eigenvalues of $M_N^\dagger M_N$ (which are the mass scales introduced by $N$) are much larger than the components of $M_{\mathrm{D}}$, diagonalization of the mass matrix in (5.41) will yield $n$ heavy and 3 light neutral fermions [Pat14]. The three light fermions are then simply the massive, but light neutrinos observed in experiments.

[8] A product like $(HL_i)N_j$ may seem a bit unclear if one is not used to the notation defined in section 3.3 and chapter 4. The scalar product of both spinors and SU(2) doublets appears here. In full, the term could be written $(HL_i)N_j = \varepsilon^{ab}\varepsilon^{\alpha\beta} H_b (L_i)_{a\beta} (N_j)_\alpha$, where $i, j$ are generation indices, $a, b$ are SU(2) indices and $\alpha, \beta$ are Weyl spinor indices. As this demonstrates, however, it becomes quite unwieldy to keep track of the different kinds of indices.

**Table 5.6.:** The field content of the seesaw type II extension of the Standard Model.

| Field | Generations | Spin | Lorentz rep. | SU(3) | SU(2) | U(1) |
|---|---|---|---|---|---|---|
| SM fields (table 3.1) + ... | | | | | | |
| $\Delta$ | 1 | 0 | (0, 0) | **1** | **3** | 2 |

For example, in the simplest case $n = 1$, the mass matrix has the form

$$\begin{pmatrix} M_\mathrm{L} & M_\mathrm{D}^\mathrm{T} \\ M_\mathrm{D} & M_N \end{pmatrix} = \begin{pmatrix} 0 & 0 & 0 & (M_\mathrm{D})_1 \\ 0 & 0 & 0 & (M_\mathrm{D})_2 \\ 0 & 0 & 0 & (M_\mathrm{D})_3 \\ (M_\mathrm{D})_1 & (M_\mathrm{D})_2 & (M_\mathrm{D})_3 & M_N \end{pmatrix} \tag{5.42}$$

which has the eigenvalues

$$m_{1,2} = \frac{M_N \pm \sqrt{M_N^2 + 4(M_\mathrm{D})^i(M_\mathrm{D})_i}}{2} \qquad m_3 = m_4 = 0 \tag{5.43}$$

In the limit $(M_\mathrm{D})_i \ll M_N$, the non-zero eigenvalues can be approximated as

$$m_1 \approx M_N \qquad m_2 \approx -\frac{(M_\mathrm{D})^i(M_\mathrm{D})_i}{M_N} \tag{5.44}$$

As predicted, there is one heavy mass ($m_1$) and three light ones ($m_2, m_3, m_4$), two of which are zero. Since at least two neutrino masses are observed to be non-zero, more than one generation of the heavy singlets $N_i$ are necessary. The neutrino masses are then suppressed by negative powers of the square roots of the eigenvalues of $M_N^\dagger M_N$, as seen in (5.44). This demonstrates the eponymous "seesaw" behavior.

In addition to neutrino masses, the seesaw model of type I actually also has dark matter candidates with its heavy neutral fermions. The stability of these fermions depends on the mixing angles, which need to be small in order to ensure that the particles mostly behave like the original sterile neutrinos so that they have a sufficiently long lifetime. However, at least three generations of $N$ are needed to accommodate dark matter and even then, the possible mass region is severely limited [LLT14; AS05].

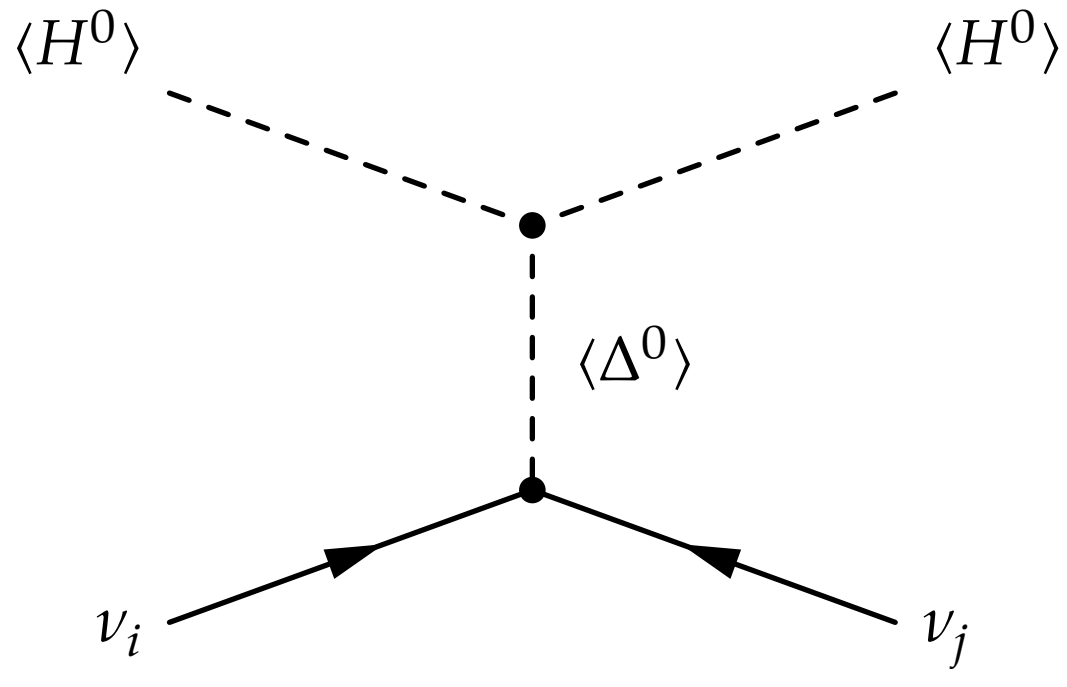

**Figure 5.5.:** Illustration of how neutrino masses are generated with the seesaw mechanism of type II.

### 5.7.2. Seesaw type II

The second type of seesaw mechanism uses a charged scalar triplet $\Delta$ with components $\Delta^{++}$, $\Delta^{+}$ and $\Delta^{0}$ (see table 5.6). This is also called a *Higgs triplet* because it can fulfill a similar role to the Higgs doublet of the Standard Model, and accordingly, the seesaw type II model is sometimes also called the Higgs triplet model (for example, see [KY12]). As shown in chapter 4, such a triplet can most conveniently be written as

$$\Delta = \frac{1}{\sqrt{2}}\Delta^i\sigma_i = \begin{pmatrix} \frac{1}{\sqrt{2}}\Delta^+ & \Delta^{++} \\ \Delta^0 & -\frac{1}{\sqrt{2}}\Delta^+ \end{pmatrix} \tag{5.45}$$

The most general renormalizable Lagrangian for the type II seesaw mechanism is

$$\begin{aligned}\mathcal{L}_{\mathrm{II}} = \mathcal{L}_{\mathrm{SM}} &+ \mathrm{Tr}\big((D^\mu\Delta)^\dagger(D_\mu\Delta)\big) - M_\Delta^2\,\mathrm{Tr}\big(\Delta^\dagger\Delta\big) - (\mu_\Delta H\Delta^\dagger H + \mathrm{H.c.}) \\ &- \lambda_1\,\mathrm{Tr}\big((\Delta^\dagger\Delta)^2\big) - \lambda_2\,\mathrm{Tr}\big(\Delta^\dagger\Delta\big)^2 - \lambda_3 H^\dagger\Delta^\dagger\Delta H - \lambda_4\,\mathrm{Tr}\big(\Delta^\dagger\Delta\big)(H^\dagger H) \\ &- \big((Y_\Delta)_{ij}L_i\Delta L_j + \mathrm{H.c.}\big)\end{aligned} \tag{5.46}$$

where terms like $\mathrm{Tr}\big((\Delta^\dagger)^2\Delta^2\big)$ and $H^\dagger\Delta\Delta^\dagger H$ have already been eliminated using identity 3. Here, the relevant term with regard to neutrino masses is the Yukawa term.

In this model, the neutrinos do acquire a mass through a Higgs mechanism, but one induced by the triplet $\Delta$ instead of a doublet, as shown in fig. 5.5. With a Higgs doublet, it is not possible to include a Yukawa coupling between two Lepton doublets (which generates Majorana mass terms for the neutrinos), whereas an appropriately-charged triplet enables such a coupling. The neutral component of the triplet can obtain a VEV just like the neutral component of the Higgs doublet:

$$\left\langle \Delta^0 \right\rangle = \frac{v_\Delta}{\sqrt{2}} \qquad \langle \Delta \rangle = \begin{pmatrix} 0 & 0 \\ \frac{v_\Delta}{\sqrt{2}} & 0 \end{pmatrix} \tag{5.47}$$

After electroweak symmetry breaking, the Yukawa term thus takes the form

$$\begin{aligned} \mathcal{L}_{\mathrm{II}}^{(Y,\langle\Delta\rangle)} &= -\left( (Y_\Delta)_{ij} \begin{pmatrix} (e_\mathrm{L})_i & -(\nu_\mathrm{L})_i \end{pmatrix} \begin{pmatrix} 0 & 0 \\ \frac{v_\Delta}{\sqrt{2}} & 0 \end{pmatrix} \begin{pmatrix} (\nu_\mathrm{L})_j \\ (e_\mathrm{L})_j \end{pmatrix} + \mathrm{H.c.} \right) \\ &= -\left( -(Y_\Delta)_{ij} \frac{v_\Delta}{\sqrt{2}} (\nu_\mathrm{L})_i (\nu_\mathrm{L})_j + \mathrm{H.c.} \right) = -\left( \frac{1}{2} \nu_\mathrm{L}^\mathrm{T} M_\nu \nu_\mathrm{L} + \mathrm{H.c.} \right) \end{aligned} \tag{5.48}$$

with the neutrino mass matrix $M_\nu = -\sqrt{2} v_\Delta Y_\Delta$.

The crucial point is now that $v_\Delta$ is necessarily small if $\Delta$ is heavy. The masses of the electroweak gauge bosons arise from the kinetic term of the Higgs boson and are thus (without loop corrections) directly proportional to the VEV $v$. However, the additional Higgs triplet $\Delta$ modifies the relationship between these masses through the addition of a second VEV. The experimental value of the so-called $\rho$ parameter

$$\rho = \frac{m_W^2}{m_Z^2 \cos(\theta_\mathrm{w})^2} \tag{5.49}$$

(with the weak mixing angle $\cos(\theta_\mathrm{w}) = \frac{g}{\sqrt{g^2+g'^2}}$), which is (without loop corrections) exactly 1 in the Standard Model, can be used to measure the deviation from the Standard Model Higgs mechanism. For the type II seesaw mechanism, this parameter is

$$\rho_{\mathrm{II}} = \frac{1 + 2\frac{v_\Delta^2}{v^2}}{1 + 4\frac{v_\Delta^2}{v^2}} \tag{5.50}$$

Since $\rho$ has been determined by electroweak precision data as $\rho = 1.000\,37 \pm 0.000\,23$ [PDG16], the ratio $\frac{v_\Delta}{v}$ is severely constrained:

$$\left| \frac{v_\Delta}{v} \right| \ll 1 \tag{5.51}$$

Using this constraint, minimization of the scalar potential yields

$$v_\Delta \approx \frac{\sqrt{2} \mu_\Delta v^2}{2 M_\Delta^2 + (\lambda_3 + \lambda_4) v^2} \overset{M_\Delta^2 \gg (\lambda_3+\lambda_4) v^2}{\approx} \frac{\mu_\Delta v^2}{\sqrt{2} M_\Delta^2} \tag{5.52}$$

**Table 5.7.:** The field content of the seesaw type III extension of the Standard Model.

| Field | Generations | Spin | Lorentz rep. | SU(3) | SU(2) | U(1) |
|---|---|---|---|---|---|---|
| SM fields (table 3.1) + ... | | | | | | |
| $\Sigma$ | $n$ | $1/2$ | $(1/2, 0)$ | **1** | **3** | 0 |

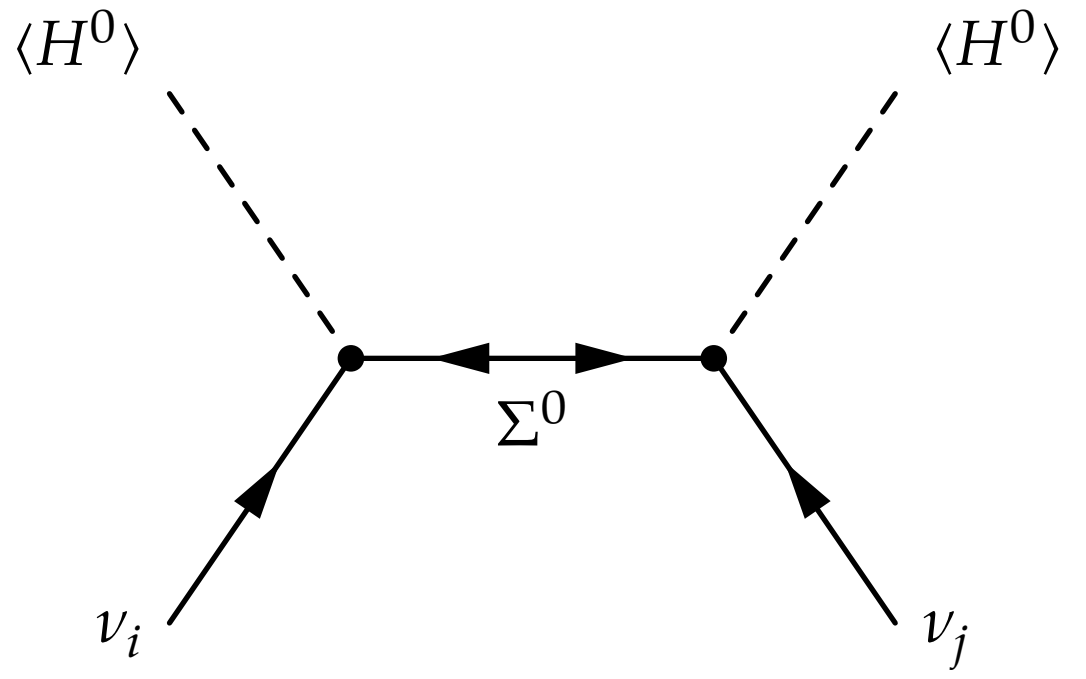

**Figure 5.6.:** Illustration of how neutrino masses are generated with the seesaw mechanism of type III (note that this is essentially the same diagram as fig. 5.4).

for the triplet VEV $v_\Delta$ [KY12; Pat14]. Again, the neutrino masses (which are proportional to $v_\Delta$) are inversely proportional to the squared mass of the new heavy particle, producing a "seesaw" effect.

The seesaw model of type II does contain a heavy neutral particle $\Delta^0$, but it is not a dark matter candidate due to its numerous interactions, which prevent stability. Even if it were stable, triplets with non-zero hypercharge are excluded by direct detection experiments (see section 5.5).

### 5.7.3. Seesaw type III

The final unique seesaw model with neutrino masses "at tree level" adds $n$ generations of a hypercharge-neutral fermion triplet $\Sigma$ with components $\Sigma^+$, $\Sigma^0$ and $\Sigma^-$ (see table 5.7). As far as neutrinos are concerned, there is very little difference between this variant and the seesaw mechanism of type I from section 5.7.1 – the fermion singlet is simply replaced by a triplet. The neutral components $\Sigma^0$ once again behave like sterile neutrinos due to their lack of interactions.

The most general renormalizable Lagrangian for the type III seesaw mechanism

is[9]

$$\begin{aligned}
\mathcal{L}_{\text{III}} &= \mathcal{L}_{\text{SM}} + i\,\text{Tr}_{\text{SU}(2)}\left(\bar{\Sigma}^i\bar{\sigma}^\mu\partial_\mu\Sigma_i\right) - \left(\frac{1}{2}(M_\Sigma)^{ij}\,\text{Tr}_{\text{SU}(2)}(\Sigma_i\Sigma_j) + \text{H.c.}\right) \\
&\quad - \left((Y_\Sigma)^{ij}L_i\Sigma_j H + \text{H.c.}\right) \\
&= \mathcal{L}_{\text{SM}} + i\,\text{Tr}_{\text{SU}(2)}\left(\bar{\Sigma}^i\bar{\sigma}^\mu\partial_\mu\Sigma_i\right) - \left(\frac{1}{2}\,\text{Tr}\left(\Sigma^{\text{T}_\text{g}}M_\Sigma\Sigma\right) + \text{H.c.}\right) \\
&\quad - \left(e_\text{L}^\text{T}Y_\Sigma\left(\frac{1}{\sqrt{2}}\Sigma^0H^+ + \Sigma^+H^0\right) - \nu_\text{L}^\text{T}Y_\Sigma\Sigma^-H^+ + \frac{1}{\sqrt{2}}\nu_\text{L}^\text{T}Y_\Sigma\Sigma^0H^0 + \text{H.c.}\right)
\end{aligned} \tag{5.53}$$

where (similarly to type I) $M_\Sigma$ is the $n \times n$ mass matrix for the $\Sigma_i$ and $Y_\Sigma$ is the $3 \times n$ matrix of Yukawa couplings. Figure 5.6, which is essentially identical to fig. 5.4, shows how neutrino masses arise.

After electroweak symmetry breaking, the combined mass term for the neutral fermions takes the form

$$\mathcal{L}_{\text{III}}^{(\text{mass},0)} = -\frac{1}{2}\begin{pmatrix}\nu_\text{L} & \Sigma^0\end{pmatrix}\begin{pmatrix}0 & \frac{v}{4}Y_\Sigma \\ \frac{v}{4}Y_\Sigma^\text{T} & M_\Sigma\end{pmatrix}\begin{pmatrix}\nu_\text{L} \\ \Sigma^0\end{pmatrix} + \text{H.c.} \tag{5.54}$$

In comparison with (5.41), the neutral mass matrix has exactly the same form as in the type I seesaw mechanism. The block elements of the mass matrix are now

$$M_\text{L} = 0 \qquad M_\text{D} = \frac{v}{4}Y_\Sigma^\text{T} \qquad M_N = M_\Sigma \tag{5.55}$$

The main difference between type I and type III is that the triplet also introduces charged fermions, which mix with the Standard Model charged leptons just like the new neutral fermions mix with the neutrinos. Accordingly, assuming again that the new mass eigenvalues of $M_\Sigma^\dagger M_\Sigma$ are large, the mass diagonalization for type III yields $n$ heavy and 3 light singly-charged fermions in addition to $n$ heavy and 3 light neutral ones [Pat14]. As in type I, the heavy neutral fermions are a potential dark matter candidate.

[9] Unfortunately, a triplet with several generations is both a matrix in "SU(2) space" and a vector in "generation space". To avoid any kind of ambiguity, the transpose has been marked $^{\text{T}_\text{g}}$ to make clear that it only applies to generation space. It should *not* be taken to mean that the triplet matrix is transposed!

**Table 5.8.:** The field content of the radiative seesaw model [Ma06].

| Field | Generations | Spin | Lorentz rep. | SU(3) | SU(2) | U(1) | $\mathbb{Z}_2$ |
|---|---|---|---|---|---|---|---|
| SM fields (table 3.1) + ... | | | | | | | 1 |
| $N$ | $n$ | $1/2$ | $(1/2, 0)$ | **1** | **1** | 0 | −1 |
| $\eta$ | 1 | 0 | $(0, 0)$ | **1** | **2** | 1 | −1 |

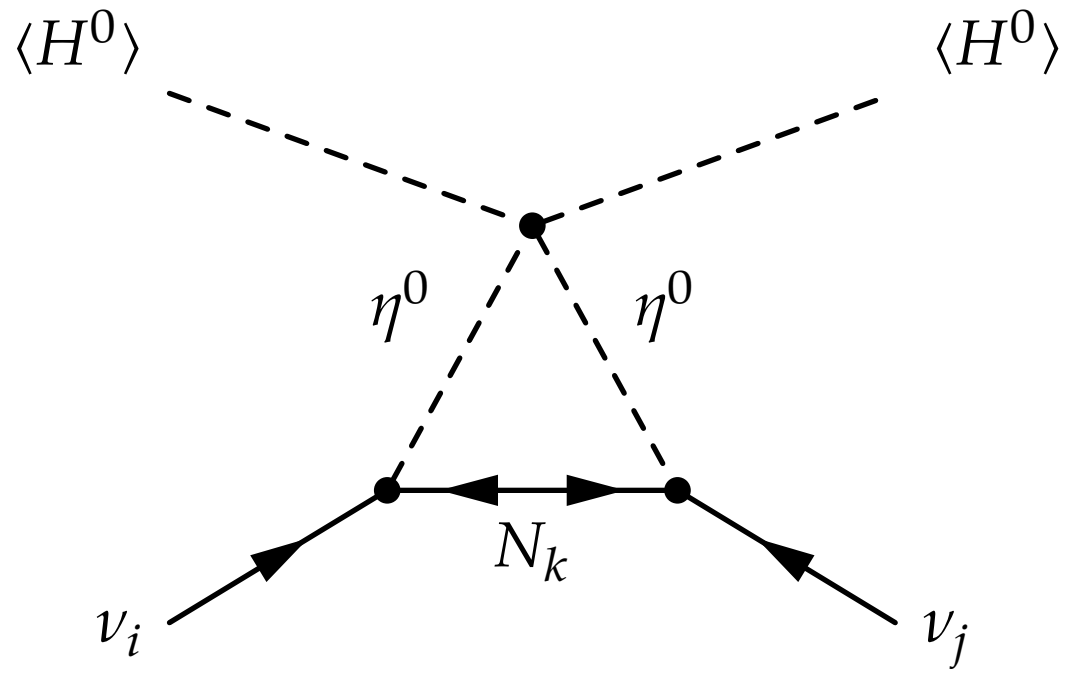

**Figure 5.7.:** Illustration of how neutrino masses are generated in the radiative seesaw model.

## 5.8. Radiative neutrino masses

Beyond the tree-level seesaw mechanisms, neutrino masses from the Weinberg operator can only be generated with radiative corrections (loops). This means that neutrinos only become massive when loop corrections to the masses are taken into account – without them, they would be massless. Such models with radiative neutrino masses are the primary subject of this thesis, as the title suggests. Their investigation has only started rather recently with the introduction of the *radiative seesaw model*, also called the *scotogenic model*, by Ma in [Ma06]. The most attractive feature of these models, which has motivated their study from the very beginning, is that the fields that must be introduced to implement the Weinberg operator in a renormalizable theory very often contain potential dark matter candidates.[10] In other words, the same mechanism which is responsible for the neutrino masses also takes care of the dark matter problem. In this way, both phenomena could be explained by the same principle.

### 5.8.1. The radiative seesaw model

The radiative seesaw model was not only the first step for this class of models, but it is also one of the simplest. As shown in table 5.8, it only adds a charged scalar doublet $\eta$ in addition to the fermion singlets $N_i$ (familiar from the type I seesaw mechanism) to the Standard model. Critically, however, a $\mathbb{Z}_2$ symmetry just like with Higgs portal dark matter (section 5.5) is postulated under which the new fields are odd, while the Standard Model ones are even. It is not spontaneously broken, either, so $\eta$ does not acquire a VEV. This $\mathbb{Z}_2$ symmetry means that every vertex must contain an even number of the new fields, having two effects: Firstly, it prevents decays of a single odd particle which proceeds only via Standard Model fields, stabilizing the *lightest odd particle* (LOP); secondly, it prevents the neutrinos from obtaining a mass via the type I seesaw mechanism with the fermion singlets.

Consequently, the most general renormalizable Lagrangian for the radiative seesaw model is

$$\begin{aligned}\mathcal{L}_{\text{RS}} = \mathcal{L}_{\text{SM}} &+ (D^\mu \eta)^\dagger (D_\mu \eta) + i N^i \bar{\sigma}^\mu \partial_\mu N_i - M_\eta^2 \eta^\dagger \eta - \left(\frac{1}{2} M_N^{ij} N_i N_j + \text{H.c.}\right) \\ &- \lambda_1 (H^\dagger H)(\eta^\dagger \eta) - \lambda_2 (H^\dagger \eta)(\eta^\dagger H) - \lambda_3 (\eta^\dagger \eta)^2 - \left(\lambda_4 (H^\dagger \eta)^2 + \text{H.c.}\right) \\ &- \left(Y^{ij} (\eta L_i) N_j + \text{H.c.}\right)\end{aligned} \tag{5.56}$$

This has two important consequences: If the LOP is neutral, it is an excellent dark matter candidate,[11] and the masses of the neutrinos only appear through radiative corrections with at least one loop, as shown in fig. 5.7. Small masses are ensured both because loop corrections are suppressed by the couplings and because the new fields, entering with their propagators in the loop, are heavy.

### 5.8.2. General one-loop realizations of the Weinberg operator

At this point, one might ask whether it is possible to find *all* realizations of the Weinberg operator at the level of one loop. Perhaps there is only a limited number of unique possibilities as with the tree-level seesaw mechanisms? Exactly this was done in [Bon+12], classifying all topologies of diagrams inducing neutrino masses.[12]

[10] The word "scotogenic" (from the Greek word σκοτος) could be defined as "created from darkness".

[11] In supersymmetry, dark matter candidates appear in exactly the same way: The new (SUSY) fields are odd under a discrete symmetry called $R$-parity, which leads to the existence of a *lightest supersymmetric particle* (LSP), which (if neutral) is the dark matter candidate.

[12] A similar work treating the realizations at the two-loop level appeared later with [Ari+15].

Subsequently, [RZY13] explicitly classified and listed all possible models with the following properties:[13]

- Up to four fields are added to the Standard Model.
- The new fields are odd under a global $\mathbb{Z}_2$ symmetry, while the Standard Model fields are even.
- The new fields are singlets of SU(3).
- The new fields are singlets, doublets or triplets of SU(2).

It turns out that these conditions encompass *all* realizations of the Weinberg operator at one loop if only SU(2) singlets, doublets and triplets are considered. Further, these models are analyzed in [RZY13] to see whether they contain viable dark matter candidates, resulting in *35 non-equivalent models that can simultaneously account for dark matter and neutrino masses*. These are the minimal dark matter models with radiative neutrino masses. For example, the radiative seesaw model introduced in the previous section is the model T3-B with $\alpha = -1$ in this classification.

As before in figs. 5.4 to 5.7, fig. 5.8 shows how neutrino masses arise in these models, where $\phi$, $\phi'$, $\varphi$, $\psi$, $\psi'$ and $\Psi$ are the new SU(2) multiplets. It becomes clear that neutrino masses in these radiative models are only possible if both scalars and fermions are added to the Standard Model.

Since the $\mathbb{Z}_2$ symmetry is such a central component of these models, a more detailed explanation is in order. $\mathbb{Z}_2$ is the cyclic group of order 2. It is sometimes called the "parity group" or "parity symmetry" because it comprises two elements behaving like parity transformations (both are their own inverse). All cyclic groups $\mathbb{Z}_n$ are finite abelian groups with $n$ elements. As all groups, their elements have an operation that can be thought of as some kind of multiplication (because it is associative). For abelian groups, this operation is also commutative (like multiplication of ordinary numbers), but this also allows one to think about an abelian group operation as addition, which is also both associative and commutative.[14]

[13] A study similar to [Bon+12] and [RZY13], but dealing with the equivalent of the mass dimension 5 Weinberg operator for *Dirac neutrinos* (an operator of mass dimension 6) was recently published with [YD17]. There, the realizations of this operator at tree and one-loop level as well as models with potential dark matter candidates were classified. Another work [CHH17] classifies realizations of neutrino masses from an operator of mass dimension 7 (the next-order operator generating Majorana neutrino masses after the Weinberg operator) at one loop.

[14] In fact, objects which have an addition (and an associative multiplication) operation are called rings and are nothing but abelian groups with respect to their addition operation.

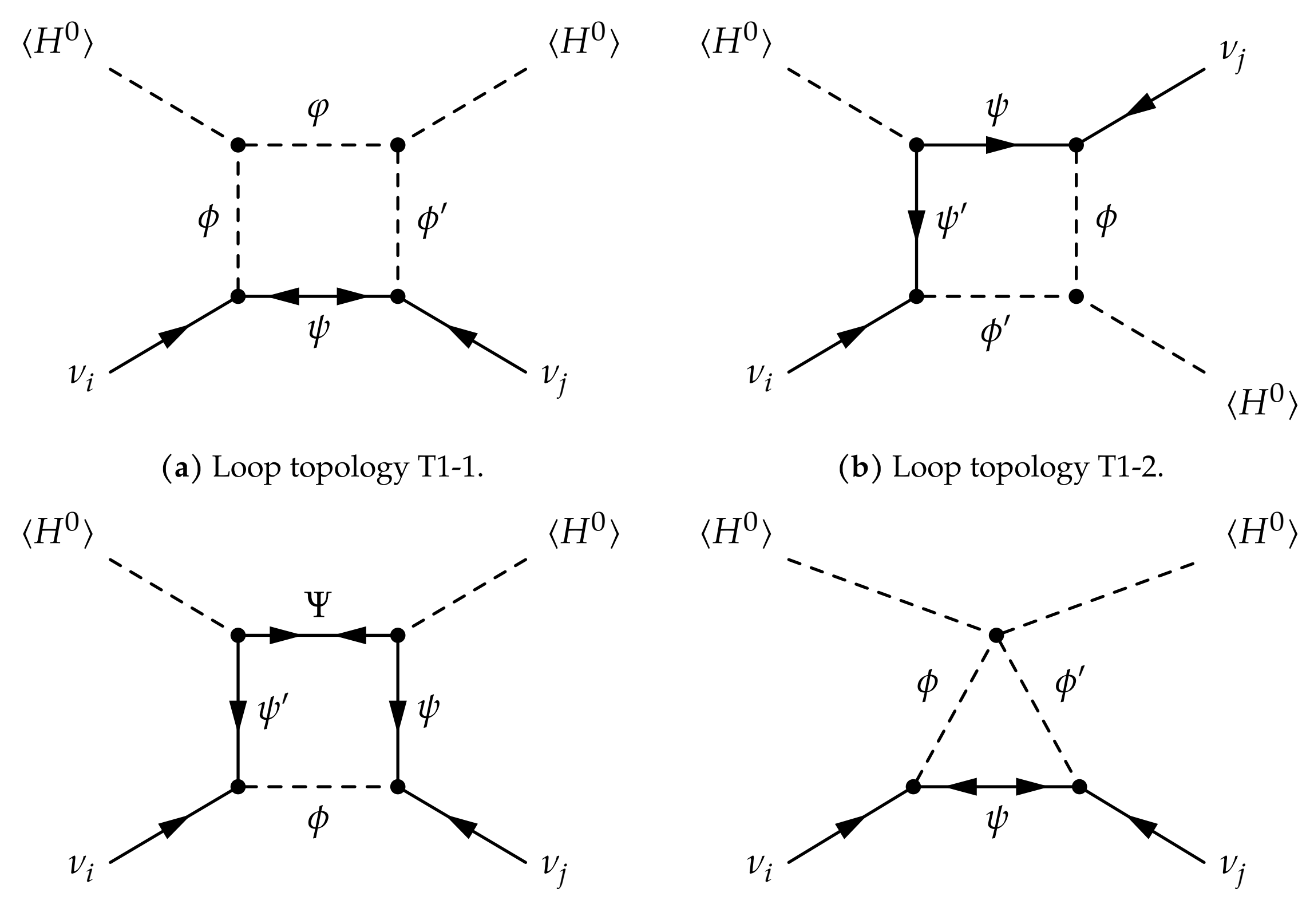

(**a**) Loop topology T1-1.

(**b**) Loop topology T1-2.

(**c**) Loop topology T1-3.

(**d**) Loop topology T3.

**Figure 5.8.:** Illustration of how neutrino masses are generated in minimal dark matter models with radiative neutrino masses. The symbols $\phi$, $\phi'$, $\varphi$, $\psi$, $\psi'$ and $\Psi$ mean that an appropriately-charged component of these multiplets would appear at the given location in a loop diagram.

This allows two different (but isomorphic) interpretations of the group $\mathbb{Z}_n$: The first is the group of integers with addition modulo $n$, which is also the origin of the notation $\mathbb{Z}_n = \mathbb{Z}/n\mathbb{Z}$. From this point of view, $(\mathbb{Z}_n, +)$ is the group

$$\mathbb{Z}_n = \{\, m \in \mathbb{N} \mid m < n \,\} \qquad + : \mathbb{Z}_n \to \mathbb{Z}_n, (a,b) \mapsto a + b \mod n \tag{5.57}$$

For $\mathbb{Z}_2$, this means that the group consists of the elements 0 and 1 with $1 + 1 \mod 2 = 0$. However, in general, groups are more like invertible matrices with matrix multiplication, and from this perspective, $\mathbb{Z}_n$ is more natural as a subgroup of U(1) with multiplication of complex numbers. Then, $(\mathbb{Z}_n, \cdot)$ is the group

$$\mathbb{Z}_n = \left\{ e^{2i\frac{m}{n}\pi} \,\middle|\, m \in \mathbb{Z} \right\} \qquad \cdot : \mathbb{Z}_n \to \mathbb{Z}_n, (a,b) \mapsto a \cdot b \tag{5.58}$$

The equivalence is easily seen: Due to the properties of the exponential function, products of exponentials become sums of their exponents, and since these are purely imaginary, their addition is effectively modulo $2\pi$. Moreover, since $|a| = 1\ \forall a \in \mathrm{U}(1)$, the absolute value does not change with multiplications. For $\mathbb{Z}_2$, this means that the group consists of the elements $-1$ and $1$ with $(-1) \cdot (-1) = 1$. Taking the analogy with $\mathrm{U}(1)$ further, one could think of $\mathbb{Z}_n$ as the group generated by discrete charges, with the numbers the exponent as the allowed charge values. For example, $\mathbb{Z}_2$ allows for charges of 0 and $\pm\pi$.

The perspective of $\mathbb{Z}_2$ as a subgroup of $\mathrm{U}(1)$ is the one that is most useful for quantum field theories because all other symmetry groups also act through multiplication, making this way much easier to implement in a compatible manner.

The function of the $\mathbb{Z}_2$ symmetry is twofold – the same as in the radiative seesaw model. It creates a stable LOP which can serve as a dark matter candidate and it ensures that neutrino masses only appear at loop level. As mentioned before, almost all current models of dark matter require some form of additional symmetry like this to fulfill the stability requirement. In this way, the introduction of such a symmetry may seem ad hoc and arbitrary, simply a means to achieve a desired result without further motivation. However, not least due to its simplicity as a group, it is not hard to imagine that this symmetry is merely a remnant of a larger, spontaneously broken symmetry. Its origin could be an extended gauge group, such as in *grand unified theories* (GUT) [KKR10; FH10b] or theories with an additional "dark" $\mathrm{U}(1)$ gauge group (producing "dark photons" or a $Z'$ boson) which is spontaneously broken to $\mathbb{Z}_2$ [ASV15], or a flavor symmetry of the lepton sector [Hir+10].

Considering the requirements of a dark matter candidate from section 5.5, it now becomes clear how such minimal models with radiative neutrino masses satisfy these conditions:

**Stability:** This is achieved through the global $\mathbb{Z}_2$ symmetry, allowing for one or several possible dark matter candidates (depending on the model). For a given set of parameters, the lightest odd particle (LOP) can be dark matter.

**Electrical & color neutrality:** Only color singlets are considered and the hypercharges can be assigned so that there is at least one new neutral particle.

**Cold dark matter:** The new particles are assumed to have weak-scale masses, in agreement with cold dark matter.

The motivation for considering these models is of course their ability to explain both dark matter and neutrino masses with a common mechanism, treating them not as separate problems, but related phenomena. As has become apparent from the previous discussion of the $\mathbb{Z}_2$ symmetry, however, the strategy with minimal models is quite different from previous attempts like grand unified theories or supersymmetry. The latter take a top-down approach, postulating a grand new symmetry or principle which, when studied in detail, is able to resolve a large number of problems and questions in particle physics. The former, though, represent a bottom-up approach, trying to explain only a small number of phenomena at a time in a minimal way. The focus here is not what deeper principle may lead to these minimal extensions in the end, but, starting from the Standard Model, what the essential building blocks are to solve individual problems. Of course, there should also be some motivation considering how (for example) new symmetries like the $\mathbb{Z}_2$ originate, but the details are left open as long as there is no evidence to be able to tell one way or another. In this sense, minimal models are effective theories in some (e. g. low-energy) limit.

## 5.9. Experimental constraints

As shown in chapter 2, there are many active efforts to search for dark matter and the exact properties of neutrinos are under continuous investigation as well. Moreover, since basically the whole field of particle physics is frantically looking for any hint as to how the known-to-be-incomplete Standard Model should be extended, any proposed extension of the Standard Model must face and satisfy numerous experimental constraints on observables that would be modified by the introduction of new fields.

At this point, the most relevant constraints that a new model must satisfy (for a given set of parameters!), both on a theoretical level and with respect to experimental data, shall be summarized.

- Gauge anomalies must cancel, as explained in section 5.2.
- There can be no new charged stable particles, as these would have long been detected.
- The dark matter relic density $\Omega h^2$ must be below the observed value, which is $0.1186 \pm 0.0020$ (2.4). Ideally, of course, the model should be able to match the observed value in order to explain the entire dark matter content of the universe. Otherwise, the model only describes a component of the dark matter, with the remaining parts still unexplained.

- The neutrino masses arising with the model must match the measurements of neutrino oscillations and the bound on the sum of the neutrino masses.
- Any new dark matter candidate must satisfy the bounds set by direct detection and indirect detection searches.
- If a model violates the Standard Model's global symmetries (such as lepton number) like the models considered here, it must satisfy the constraints from experiments on *lepton number violation* (LNV), such as neutrinoless double $\beta$ decay, and *lepton flavor violation* (LFV), such as the process $\mu \to e\gamma$.
- The properties of the Higgs boson must be in agreement with experiment. In particular, loop corrections can contribute quite significantly to the Higgs boson's mass, which needs to be taken into account to maintain the correct value of 125 GeV.

## 5.10. State of the art in studies of minimal models

As mentioned before and noted in [RZY13], most of the minimal dark matter models with radiative neutrino masses have not yet been studied in the literature. A detailed survey of the scientific literature on this subject has resulted in table 5.9, detailing which of the models have received research attention.

**Table 5.9.:** Survey of the literature on the viable minimal dark matter models with radiative neutrino masses presented in [RZY13].

| Model | $\alpha$ | References | Comments |
|---|---|---|---|
| T1-1-A | 0 | [Far09], [FH10a], [Far11], [Weg17] | $^{c,d,f}$ The Lagrangian given in [Far09; FH10a; Far11] is missing a term of the form $\bar{L}\Phi N$ |
| | $\pm 2$ | none | |
| T1-1-B, T1-1-C, T1-1-D, T1-1-F, T1-1-G and T1-1-H have not been studied in the literature so far | | | |
| T1-2-A | $-2$ | [Lon15], [Lon+16] | "inert Zee model"; $Y_\phi = 1$ (instead of $-1$) |
| | 0 | [EKY18] | |

**Table 5.9.:** (continued)

| Model | $\alpha$ | References | Comments |
|---|---|---|---|
| T1-2-B | −2 | none | |
| | 0 | none | |
| T1-2-D | −1 | none | |
| | 1 | none | |
| T1-2-F | −1 | [BLZ17] | "doublet–triplet DM model"; $Y_{\phi'} = 1$ (instead of −1) |
| | 1 | none | |
| T1-3-A | 0 | [FMP14], [Res+15a], [Res+15b], [Esc+18] | [d,e] "scotogenic inverse seesaw model" |
| T1-3-B, T1-3-C, T1-3-D, T1-3-F, T1-3-G and T1-3-H have not been studied in the literature so far | | | |
| T3-A | −2 | [LM13] (model D), [BPR14] | $Y_{\phi'} = 2$ (instead of −2) |
| | 0 | [FPS10], [Far11], [LM13] (model C), [OO16] | "AMEND model" ($Y_\psi = -1$ instead of 1 in [OO16]) |
| T3-B | −1 | [Ma06], [Ham+09], [KS12], [HI13], [Kla+13], [VY15], [IYZ16], [Lin+16] and many others | [b,f] "radiative seesaw model", "scotogenic model" |
| | −3, 1 | [AKY11], [LM13] (model A) | [f] |
| T3-C | −1 | [MS09], [Cha15], [Pah+16] | [b] "radiative seesaw with triplet fermion", "radiative type III seesaw (RSIII)" |
| | −3, 1 | [LM13] (model B) | [a] |
| T3-E | −2, 0 | [LM13] (model E) | [a] |

[a] field content defined and general properties discussed, but not studied in any detail
[b] only 2 fields
[c] only 3 fields
[d] real scalar singlet
[e] multiple generations of scalar singlet
[f] multiple generations of fermion singlet

There are a few things to note in table 5.9. For one, an overwhelming amount of works focus on the topology T3, to the extent that almost all viable models in this category have received at least some amount of research. The reason seems obvious: These models are the simplest, requiring at most three new fields. The models T3-B ($\alpha = -1$) and T3-C ($\alpha = -1$) can even be implemented with just two independent fields [RZY13]. Of these, the radiative seesaw model [Ma06] stands out with, by far, the largest amount of publications as the prototypical radiative model of dark matter and massive neutrinos.

However, beyond the T3 models, the amount of studies rapidly declines. There are only five models of the remaining topologies which have received any attention and most of the time, work on a particular model is mainly driven by a single author. For four of these models, the authors have chosen the "A" version of a given topology. Again, the explanation is simple: These are the models with the lowest-dimensional SU(2) multiplets and thus with the lowest amount of new field components. For example, none of these four models contains SU(2) triplets. This makes them not only easier to handle, but also reducing the amount of "unnecessary" degrees of freedom. Conspicuously, [BLZ17] with T1-2-F ($\alpha = -1$) is an exception to this rule going in the opposite direction: It contains many large SU(2) multiplets (only doublets and triplets, no singlets). This model was selected for its specific features, such as being more "general" than others and with all multiplets containing a neutral component as well as contributing to the neutrino masses.

To summarize, research on these models has been working itself up in terms of complexity, mirroring the progression from the seesaw mechanisms at tree level (simpler) to loop level. The T3 models have mostly been studied, while T1-1, T1-2 and T1-3 are almost completely unexplored.

# 6 Automatic generation of Lagrangians for minimal dark matter models

One of the goals of this thesis is to study the minimal models introduced in [RZY13] in a general way. A missing piece in [RZY13] are the Lagrangians for the individual models, which only specifies the field content for each model without giving any explicit Lagrangians at all. While it is a manageable task for someone experienced in building particle physics models to construct the Lagrangian for such a model manually, this process can be error-prone and time-consuming. Moreover, since the symmetry groups and the kinds of representations used in these models are limited and fixed, it is quite feasible to exhaustively list all the terms that could potentially appear in such a Lagrangian in full generality.

## 6.1. Program description: `minimal-lagrangians`

While dealing with the Lagrangians and the possible terms they can contain as part of the work on this thesis, a `Python` program started falling into place which is now capable of generating the Lagrangians for these models fully automatically, with different output formats. This program is called *`minimal-lagrangians`* and will be described in this chapter. It is currently available at [May18] upon request.

The program `minimal-lagrangians` allows one to specify the field content of an extension of the Standard Model of particle physics and, using this information, generates the most general renormalizable Lagrangian that describes such a model. As the program was written for the study of minimal dark matter models with radiative neutrino masses, it can handle fields beyond the Standard Model with the following properties:

- scalar or fermion fields[1]
- SU(3) singlets
- SU(2) singlets, doublets or triplets
- arbitrary hypercharge
- charged under a global $\mathbb{Z}_2$ symmetry (although the same code handling invariance under $\mathbb{Z}_2$ could be used for a general global U(1) symmetry)

The output generated by `minimal-lagrangians` for all the models given in [RZY13] is listed in appendix E. The program requires `Python` 3 and has been tested with `Python` $\geq$ 3.4. Apart from that, there are no external dependencies.

The program only prints the potential involving at least one new (i. e. non-SM) field, that is, the kinetic terms (which always have the same form) and the Standard Model Lagrangian are omitted. The models are not checked for anomalies (tools like `SARAH` can be used for this purpose), although the Witten SU(2) anomaly will be avoided by introducing vector-like fermions if necessary. The new models can be defined in a file `data.py`. A model can be added as an entry to the list in the following form:

```
BSMModel('<model_name>', (
    # list of fields
    # type        name    SU(2) rep.  hypercharge
    # for a scalar field, e.g. a scalar doublet with hypercharge 1:
    ScalarField ('S',    2,          Y=1),
    # for a fermion field, e.g. a fermion singlet with hypercharge 0:
    FermionField('F',    1,          Y=0),
    # …
    )
    # optional: parameter values for different hypercharge assignments (offsets), e.g.
    , (0, 2, …)
),
```

The program uses the convention where the hypercharge $Y$ is normalized such that the electric charge $Q$ is

$$Q = T_3 + \frac{Y}{2} \tag{6.1}$$

[1] Fermions are always defined in terms of Weyl spinors.

It should be noted that `minimal-lagrangians` automatically treats neutral scalars as real. For example, the model T1-3-B, which is studied for $\alpha = 0$ in chapter 7, is defined as

```
BSMModel('T1-3-B', (
    FermionField("Ψ",  1, Y= 0),
    FermionField("ψ'", 2, Y= 1),
    ScalarField ("ϕ",  3, Y= 0),
    FermionField("ψ",  2, Y=-1),
), (0, 2)), # α = -2 is equivalent to α = 2
            # |α| = 2 excluded by direct detection
```

Information on how to run the program on the command line can be obtained with `lagrangians.py -h`:

```
usage: lagrangians.py [-h] [--format {LaTeX,SARAH,plain}]
                      [--omit-self-interaction]
                      model [parameter α]

A Python program to generate the Lagrangians for minimal dark matter
 ↪  models
with radiative neutrino masses

positional arguments:
  model                 name of the model whose Lagrangian is to be
   ↪  generated
  parameter α           value of the model parameter α (determines
                        hypercharges of the fields)

optional arguments:
  -h, --help            show this help message and exit
  --format {LaTeX,SARAH,plain}
                        output format for the generated Lagrangian
                         ↪  (default:
                        plain)
  --omit-self-interaction
                        omit pure self-interactions of the new fields in
                         ↪  the
                        Lagrangian, that is, output only interaction
                         ↪  terms
                        which involve both SM and new fields (default:
                         ↪  output
                        all terms)
```

There is a comprehensive test suite, which can be run using `test.py`. Among other checks, this tests whether the program produces the correct Lagrangian for the following models:

- T1-1-A with $\alpha = 0$, as given in [Far09], which presents an implementation of this model.
- The simplified dark matter models given in [CS14]:
  - the singlet–doublet fermion model (SDF, "model A");
  - the singlet–doublet scalar model (SDS, "model B");
  - the singlet–triplet scalar model (STS, "model C").
- The seesaw mechanism type II (cf. section 5.7.2), also called the Higgs triplet model (see e. g. [KY12]).

Additionally, it has been verified manually that the generated output is correct for a variety of different models, in particular for the seesaw mechanisms of type I and II, the model T1-2-A ($\alpha = 0$) studied in [EKY18] and the model T1-3-B ($\alpha = 0$), which is the subject of chapter 7.

To get an impression of the LaTeX output capabilities of `minimal-lagrangians`, refer to appendix E. However, this is not the only implemented output format. By default, the program will output the Lagrangian in plain text to the command-line terminal for a clearer and more compact presentation which does not require a LaTeX processor. This output format makes heavy use of Unicode for optimal readability. For example, the command `lagrangians.py T1-1-A 0` prints the Lagrangian for the model T1-1-A with $\alpha = 0$ from [RZY13] in the following form:

```
- M_φ'² φ'^† φ' - ½ M_φ² φ²
- (λ₁ (H φ') φ + H.c.)
- λ₂ (H^† H) (φ'^† φ') - λ₃ (H^† φ'^†) (H φ') - λ₄ (H^† φ') (φ'^† H) -
 ↪  λ₅ (φ'^† φ')² - λ₆ (H^† H) φ² - λ₇ (φ'^† φ') φ²  - (λ₈ (H φ')² +
 ↪  H.c.) - λ₉ φ⁴
- (½ M_ψ ψ ψ + H.c.)
- (λ₁₀ (φ'^† L) ψ + H.c.)
```

As another example, running `lagrangians.py STS` prints the Lagrangian for model C (singlet–triplet scalar) from [CS14]:

```
- ½ M_T² Tr(T²) - ½ M_S² S²
- λ₁ (H^† H) Tr(T²) - λ₂ H^† T² H - λ₃ (H^† T H) S - λ₄ (H^† H) S² - λ₅
 ↪ Tr(T⁴) - λ₆ Tr(T³) S - λ₇ Tr(T²) S² - λ₈ S⁴
```

Although in this case, the term $\mathrm{Tr}(T^3)S$ is zero by identity 2 because there is only one generation of the scalar triplet.

Finally, the third output format supported by `minimal-lagrangians` allows one to generate model files for the tool `SARAH`, which can then be used to study the model in detail and subsequently generate code and model files for a large number of particle physics tools, such as `SPheno` and `micrOMEGAs`. In `SARAH`, one of the main tasks in implementing a model is specifying the Lagrangian, which has to be done manually. `minimal-lagrangians` mostly eliminates this step by generating the most general renormalizable Lagrangian automatically and creating all the files needed by `SARAH`. Together, these programs form a tool chain which, after specifying a model's field content, largely automates the programmatic implementation of the model's details and rapidly yields executable code to calculate physical observables (see section 6.3). An example making use of this tool chain is the analysis of the model T1-3-B ($\alpha = 0$) in chapter 7.

It should be noted that, as of yet, `minimal-lagrangians` does not (with some exceptions) remove terms using the identities described in section 5.4. It simply generates *all* possible invariant and renormalizable terms, without further simplifications.

## 6.2. Implementation

The way in which fields, terms and the Lagrangian as a whole are represented internally in `minimal-lagrangians` is very straightforward and simplistic. Fields are objects whose properties are their mathematical symbol and their quantum numbers,

- `type`: scalar or fermion;
- `su2_multiplicity`: the dimension of their representation under the gauge group SU(2) – the values 1, 2 and 3 (singlets, doublets and triplets) are supported;
- `hypercharge`: the charge under the gauge group U(1);

- `z2`: the charge ($\pm 1$) under the global $\mathbb{Z}_2$ symmetry.

Terms are then essentially lists of such field objects, with code in place to ensure a consistent order and grouping of fields within a term. A Lagrangian is then an (ordered) set of such terms.

An alternative to this very basic approach could have been to adapt a symbolic computation package with some additional rules for equivalence of terms, ordering and output formatting. However, for the Lagrangian of a minimal model, the only required operations are multiplication of fields and addition of terms, where the latter is even not really needed because none of the terms of the final Lagrangian can be simplified by addition. Consequently, a symbolic computation package would not have made the implementation much easier beyond providing the commutative property $\phi_1\phi_2 = \phi_2\phi_1$ in a product. On the other hand, adapting an existing library for a purpose such as this, which it was not designed for, would take a significant amount of effort. In a sense, `minimal-lagrangians` works at a higher level of abstraction – it does not think in terms of individual variables in a product, but really only cares about the terms as a whole. No operations are performed on these terms and the desired concept of "equality" of terms is not a precise one down to each constant factor, because these just affect the arbitrary definition of each term's coupling parameter. Such details are of no interest for the program's purpose, rather complicating the decision of whether two terms should be treated as equal. Only if integration with other symbolic computation tools is desired would it be useful to investigate whether employing such a package as a lower-level component is worthwhile.

The main component implementing the generation of all possible gauge- and Lorentz-invariant terms is the method `BSMModel.lagrangian`, together with the methods `Model.is_valid` and `Model.generate_terms`, which it uses to construct the Lagrangian. `BSMModel.lagrangian` works as follows: First, a list containing all of the model's fields and their adjoints is created. Then, all the possibilities of combining $n$ of these fields ($2 \le n \le 4$) are enumerated. For example, given two real fields $\phi_1$ and $\phi_2$, this list would contain the combinations $(\phi_1\phi_1\phi_1, \phi_1\phi_1\phi_2, \phi_1\phi_2\phi_2, \phi_2\phi_2\phi_2)$ for $n = 3$. Note that the order of the fields does not matter. However, only those combinations which can be used to form invariant terms at all are kept as candidates.

The check whether such a combination of fields can yield invariant terms is done by `Model.is_valid`. Having developed the index conventions for Weyl spinors and SU(2) multiplets in section 3.3 and chapter 4, it is now simple to determine whether this is the case: Every lowered index must appear in a sum with a raised index of the same kind. Since fermion fields have mass dimension $\mathsf{M}^{3/2}$, there can be at most two of them in a term anyway, so this reduces to a check that they are both either left-handed or conjugate left-handed (i. e. right-handed) spinors. In general, though, the number of indices must be even for each kind of index. For SU(2), an $n$-plet has $n - 1$ indices, so the sum $\sum_i (n_i - 1)$ must be an even number. For the abelian groups (U(1), $\mathbb{Z}_2$), the sum of each kind of charge must be zero: $\sum_i q_i = 0$. Alternatively, since the value of the "parity group" $\mathbb{Z}_2$ is usually given as $p_i = e^{iq_i} = \pm 1$ ("even/odd"), the product of all $\mathbb{Z}_2$ values must be one: $\prod_i p_i = 1$.

All the combinations of fields which are identified as potentially valid are then given to `Model.generate_terms`. This method contains the algorithm to determine, given an arbitrary collection of fields, what terms involving this specific combination must be added to the Lagrangian. As mentioned before, Lorentz invariance is easy to obtain since the combination of fields will either contain no or two Weyl spinors, and in both cases there is only one way to match up the spinor indices. For gauge and global invariance, SU(2) as the only non-abelian group is the only one which could couple the fields in non-trivial ways. Invariance under the abelian groups is automatically ensured because the charges were checked in the previous step. In any case, a term is either not or automatically invariant under abelian groups – there is only one way to couple the fields.

To recap, only three different representations of SU(2) are used in the considered models (singlets, doublets and triplets). Furthermore, the restriction to renormalizable Lagrangians limits the Lorentz structure of terms to the possibilities listed in table 5.1. Together, these assumptions ensure that there is only a finite number of different types of terms, so that the easiest way to enumerate all the possible terms for a given set $F$ of fields is to simply go through all the different cases. Denoting the singlets, doublets and triplets in $F$ by $S_i$, $D_i$ and $T_i$, the algorithm operates as follows:

1. If $F$ contains exactly one triplet and no doublets (i. e. the rest is singlets), the only possible term is zero due to $\mathrm{Tr}(\sigma^i) = 0$. No terms are added to the Lagrangian.

2. If $F$ contains exactly two identical scalar doublets and no triplets, the only possible terms are zero due because $D \cdot D = 0$ for any scalar doublet $D$. No terms are added to the Lagrangian.

3. If none of the previous conditions occur and if $F$ contains more than two fields (interaction terms):

a) If all the fields are singlets, there is only one distinct term for the given set of fields.

b) Else, if there is at least one singlet, there are the following cases:

    i. If there are two doublets and no triplets, there is only one distinct term for the given set of fields.

    ii. Else, if there are two doublets and one triplet, the term $(D_1 T_1 D_2) S_1$ is added to the Lagrangian.

    iii. Otherwise, there must be at least two triplets and no doublets. All permutations of the triplets are generated to form the trace of their product. Duplicates such as $\mathrm{Tr}(T_2 T_1)$, which is the same as $\mathrm{Tr}(T_1 T_2)$ by the cyclic property of the trace, are discarded. Using the remaining permutations, terms of the form $\mathrm{Tr}(T_1 T_2) S_1$ (2 triplets, 1 singlet), $\mathrm{Tr}(T_1 T_2) S_1 S_2$ (2 triplets, 2 singlets) or $\mathrm{Tr}(T_i T_j T_k) S_1$ (3 triplets, 1 singlet) are added to the Lagrangian.

c) Else, if all the fields are doublets, gauge invariance only allows a combination of exactly four doublets. All permutations of the doublets are generated and terms of the form $(D_{\pi(1)} \cdot D_{\pi(2)})(D_{\pi(3)} \cdot D_{\pi(4)})$ are added to the Lagrangian. Duplicate terms, such as $(D_2 \cdot D_1)(D_3 \cdot D_4)$ or $(D_3 \cdot D_4)(D_2 \cdot D_1)$, which are (up to a sign) the same as $(D_1 \cdot D_2)(D_3 \cdot D_4)$, are discarded. Moreover, terms containing a product like $(D \cdot D)$ are zero and discarded as well.

d) Else, if $F$ consists of three triplets, all permutations of the triplets are generated and terms of the form $\mathrm{Tr}(T_{\pi(1)} T_{\pi(2)} T_{\pi(3)})$ are added to the Lagrangian. Duplicates such as $\mathrm{Tr}(T_2 T_3 T_1)$, which is the same as $\mathrm{Tr}(T_1 T_2 T_3)$ by the cyclic property of the trace, are discarded.

e) Else, if $F$ consists of four triplets, all permutations of the triplets are generated and terms of the form $\mathrm{Tr}(T_{\pi(1)} T_{\pi(2)}) \mathrm{Tr}(T_{\pi(3)} T_{\pi(4)})$ as well as $\mathrm{Tr}(T_{\pi(1)} T_{\pi(2)} T_{\pi(3)} T_{\pi(4)})$ are added to the Lagrangian. Again, duplicates which are equal due to the cyclic property of the trace are discarded.

   Actually, if all four of the triplets are the same, a small "optimization" is performed: By corollary 1, only the terms $\mathrm{Tr}\big(T^\dagger T\big)^2$ and $\mathrm{Tr}\big((T^\dagger T)^2\big)$ (complex triplet) or $\mathrm{Tr}\big(T^4\big)$ (real triplet) are needed.

f) Otherwise: The combinations (3 doublets, $\leq 1$ triplet) and (1 doublet, 3 triplets) are not gauge-invariant. Thus, the only remaining combinations have 2 doublets and 1 or 2 triplets. For these, terms of the form $D_i T D_j$ (one triplet) or $D_i \operatorname{Tr}(T_a T_b) D_j$ and $(D_i \cdot D_j) \operatorname{Tr}(T_a T_b)$ (two triplets) are possible. In order to obtain all possibilities, all permutations of the doublets and triplets are formed separately and then the Cartesian product of these two sets is built. More concretely, depending on the number of triplets, this creates the lists

- one triplet: `[((`$D_1$`,`$D_2$`),(`$T$`,)),((`$D_2$`,`$D_1$`),(`$T$`,))]`;
- two triplets: `[((`$D_1$`,`$D_2$`),(`$T_1$`,`$T_2$`)),((`$D_1$`,`$D_2$`),(`$T_2$`,`$T_1$`)),`
  `((`$D_2$`,`$D_1$`),(`$T_1$`,`$T_2$`)),((`$D_2$`,`$D_1$`),(`$T_2$`,`$T_1$`))]`.

The fields from these lists are then used, in order, to generate the terms mentioned above. However, terms containing a product like $(D \cdot D)$ are zero and discarded.

4. Else, $F$ only contains two fields (mass terms). There is only one distinct non-zero term for the given set of fields.

## 6.3. Numerical analysis tool chain

As mentioned before, `minimal-lagrangians` adds another piece to a tool chain formed from existing particle physics code, allowing one to automate most of the model implementation starting just with a model's field content.

An illustration of how a model is implemented and analyzed using these tools is shown in fig. 6.1. At the beginning, the model's field content is defined in `data.py`, as explained above. `minimal-lagrangians` can then be used to generate model files for `SARAH` containing the most general renormalizable Lagrangian. Since `SARAH` has a great number of features, it is usually necessary or desirable to make some changes or additions to the generated model files. Documentation for `SARAH` is available, for example, in [Sta15]. In particular, the mixing of fields and Dirac spinors after electroweak symmetry breaking need to be defined, and if `SPheno` is to be used, the file `SPheno.m` must be written in order to define how the model parameters will be input to `SPheno` and to give some settings influencing the generated code. With this now complete set of model files, `SARAH` can be used to generate both the `SPheno` code and model files for `micrOMEGAs`.

At this point, the implementation of the model is complete. The generated code can be compiled and used to perform numerical calculations. `SPheno` takes a file in the *SUSY Les Houches Accord* (SLHA) format [Ska+04; All+09] as input. In this file, the input parameters are provided and a number of settings (for example, which file formats to output, whether some calculations should be disabled and what conventions are used) can be customized. `SARAH` provides a template for this input file. Running `SPheno` then produces a spectrum file, also in SLHA format. It contains the mass spectrum (including the mixing matrices) for the specified parameter points as well as lepton and quark flavor violation observables, some observables like $g-2$ and (if enabled) branching ratios for particle decays. In turn, this spectrum file can be used as an input to `micrOMEGAs`, which extracts the mass spectrum and uses it to perform the calculation of dark matter observables.

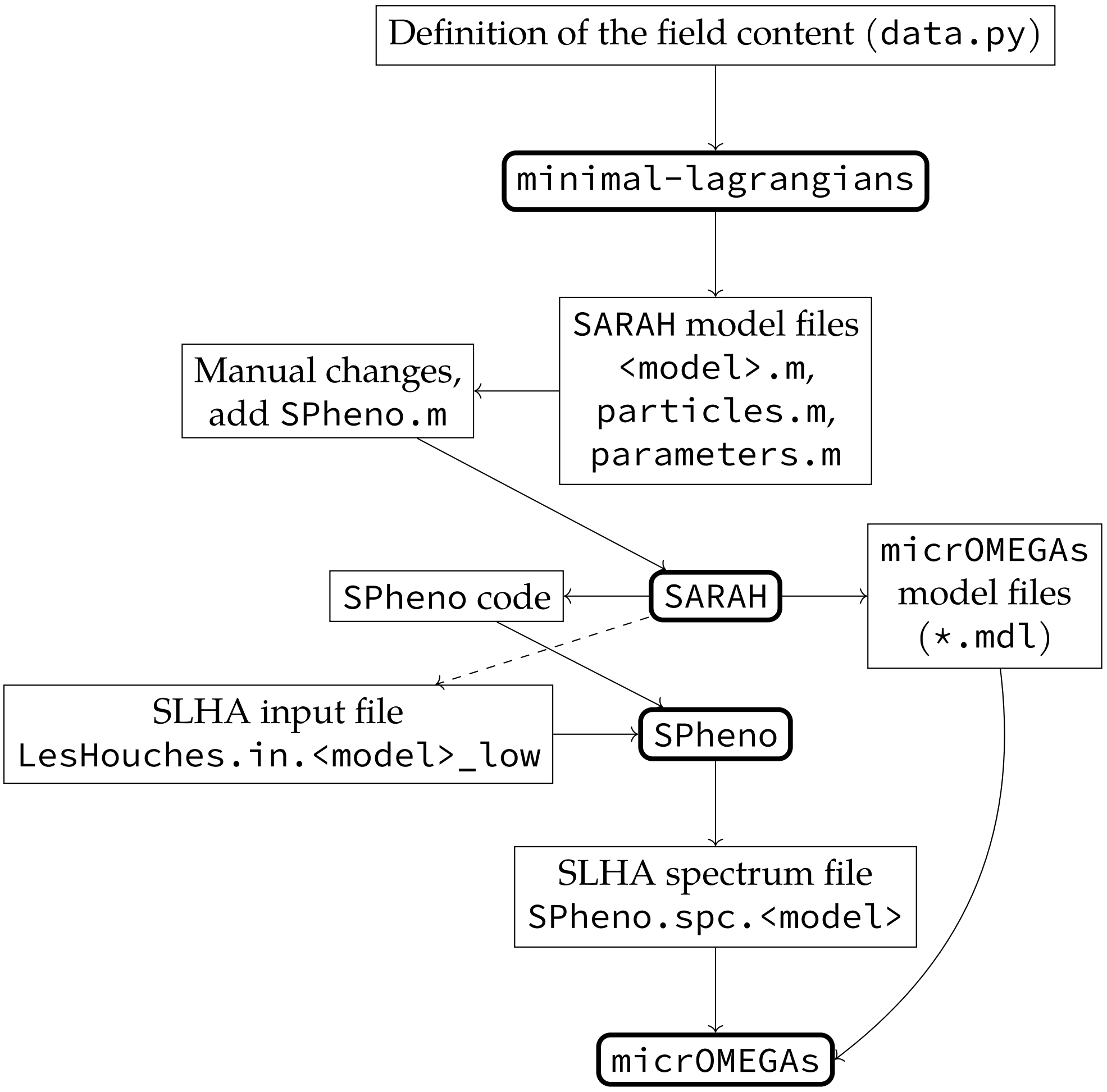

**Figure 6.1.:** Flowchart illustrating the procedure to implement a model and run the numerical code within the computational tool chain used in this work. Boxes with rounded corners and thick borders represent programs, while the others represent files.

# 7 The model T1-3-B ($\alpha = 0$)

So far, the dark matter models with radiative neutrino masses from [RZY13] have been discussed from a general perspective. This chapter, on the other hand, will make the step towards a concrete model, demonstrating how such a model can be explored in an exemplary way and how observables and predictions can be extracted using general-purpose computational tools. To this end, the model T1-3-B with $\alpha = 0$ was selected as the subject of closer investigation.

## 7.1. Motivation

With 35 viable models to choose from, the first question is, of course, why any particular model should be chosen. Perhaps there are some criteria that make some models more promising or interesting, setting themselves apart from the others? The first step in this decision-making process is table 5.9. While it may be worthwhile to study one model meticulously in full detail if it shows outstanding potential, the situation mostly comes down to the fact that presently, there is simply not enough data to make any of these models implausible. In light of this, once a model has been shown to be compatible with current observations and its parameter space has roughly been surveyed, showing how different observables behave, any further efforts run in danger of being speculation or duplicating previous results. Taking into account the additional fact that the majority of these models have not been studied at all makes the choice of a model without prior work an appealing one.

The next question is whether any of the unstudied options has particularly desirable features. One interesting aspect to consider is the problem of gauge coupling unification. Grand unified theories aim to extend the Standard Model in such a way that the gauge group of interactions is a simple group such as $SU(5)$ or $SO(10)$ instead of the Standard Model's $SU(3) \times SU(2) \times U(1)$. If this could be achieved, all three of the Standard Model's gauge interactions would originate from one truly fundamental interaction, which only splits into those of the Standard Model after symmetry breaking. However, this would imply that (at high energies) there is only one coupling parameter for the interactions of the GUT gauge bosons with the other fields. If this is the case, all three of the Standard Model's gauge couplings would have to meet at a certain point of their evolution with the renormalization scale, called the *GUT scale*. The Standard Model itself actually does not meet this criterion, with no single point of unification.

This is not entirely unexpected, though: The evolution of the gauge couplings depends on the numbers of the theory's boson and fermion fields. Hence, if there are any additional unknown fields, the couplings would be different, and inversely, adding fields to the Standard Model can have the effect of enabling a unification of couplings after all. For the models in question here, this point has been scrutinized in [Hag+16], identifying *ten minimal dark matter models with radiative neutrino masses which also allow for gauge coupling unificiation*. Of these, only the following five have no prior work: T1-1-D ($\alpha = -1$, $\alpha = 1$), T1-2-B ($\alpha = -2$, $\alpha = 0$), and T1-3-B ($\alpha = 0$).

Finally, as mentioned in the discussion of table 5.9, a model's simplicity is a contributing factor as well. Generally, models like those of the classes T1-1 and T1-2 with a larger number of scalars allow for a wider variety of interaction terms, whereas fermions are limited to mass terms and Yukawa terms (see table 5.1). To make matters worse, the number of terms tends to increase with larger representations of SU(2), such as triplets. As shown in appendix E, this leads to a veritable explosion of terms for models like T1-1-D, with both values of $\alpha$ giving more than 20 new couplings.[1] The situation is not much improved for T1-2-B, where both options still have at least 16 new interaction terms. Apart from the fact that such huge parameter spaces are difficult to handle, there is also the philosophical perspective that any model of reality should be as simple as possible, with as few arbitrary free parameters as feasible. In fact, a lot of criticism and a large motivation for more fundamental theories stem from the large number of free parameters and the perceived arbitrariness of the Standard Model.

In contrast, T1-3-B with $\alpha = 0$ introduces only six new interaction terms, one of which can be eliminated from the Lagrangian. As a result, this is the model which shall be studied below.

[1] This number of coupling parameters does not yet account for the fact that couplings of fields with several generations (such as the leptons) introduce a new parameter for each combination of the different generations, leading to an even larger parameter space. Also not accounted for are parameters which can be eliminated as shown in section 5.4.

**Table 7.1.:** The field content of the model T1-3-B with $\alpha = 0$.

| Field | Generations | Spin | Lorentz rep. | SU(3) | SU(2) | U(1) | $\mathbb{Z}_2$ |
|---|---|---|---|---|---|---|---|
| SM fields (table 3.1) + ... | | | | | | | 1 |
| $\Psi$ | 1 | $1/2$ | $(1/2, 0)$ | **1** | **1** | 0 | $-1$ |
| $\psi$ | 1 | $1/2$ | $(1/2, 0)$ | **1** | **2** | $-1$ | $-1$ |
| $\psi'$ | 1 | $1/2$ | $(1/2, 0)$ | **1** | **2** | 1 | $-1$ |
| $\phi$ | $n_s$ | 0 | $(0, 0)$ | **1** | **3** | 0 | $-1$ |

## 7.2. Definition

As usual, the model T1-3-B ($\alpha = 0$) is defined through its field content, which is given in table 7.1. The same names and hypercharge convention as in [RZY13] are used. In more detail, these are all the field components added to the Standard Model:

$$\Psi = \Psi^0 \qquad \psi = \begin{pmatrix} \psi^0 \\ \psi^- \end{pmatrix} \qquad \psi' = \begin{pmatrix} \psi'^+ \\ \psi'^0 \end{pmatrix} \qquad \phi_i = \begin{pmatrix} \frac{1}{\sqrt{2}}\phi_i^0 & \phi_i^+ \\ \phi_i^- & -\frac{1}{\sqrt{2}}\phi_i^0 \end{pmatrix} \tag{7.1}$$

where superscripts indicate the fields' electric charges. A priori, the number of generations of the scalar triplet $\phi$ is left open. It will become apparent later that in general, at least two generations are necessary. Furthermore, the scalar triplet is treated as real ($(\phi^0)^\dagger = \phi^0$, $(\phi^+)^\dagger = \phi^-$) since it has zero hypercharge. Taking such a scalar to be complex essentially only doubles its components, adding another charged scalar ($(\phi^+)^\dagger \neq \phi^-$) and a pseudo-scalar part of the neutral component ($\phi^0 = \phi^0_R + i\phi^0_I$).

$\Psi$ has the same quantum numbers as a hypothetical $\mathbb{Z}_2$-odd "right-handed neutrino", whereas $\psi$ can be viewed as a $\mathbb{Z}_2$-odd version of the lepton doublet. Together, $\psi$ and $\psi'$ form a field that is a $\mathbb{Z}_2$-odd analogy to a vector-like lepton doublet. Due to the appearance of this vector-like doublet, the model is anomaly-free without modification, as noted in [RZY13]. Further, all new SU(2) multiplets contain neutral components, which are all prospective dark matter candidates, allowing for triplet scalar or singlet–doublet fermion dark matter.

Taking the first step in the model analysis tool chain, `minimal-lagrangians` is used to generate the most general renormalizable Lagrangian for this model. The output for T1-3-B ($\alpha = 0$) is given in appendix E, (E.22). However, one of the terms in the scalar potential is redundant, as shown in identity 3. Consequently, one can set (for example) $\lambda_2 = 0$ in (E.22) without loss of generality.[2] Using this and explicitly inserting generation indices for the leptons and the triplet $\phi$, the most general renormalizable Lagrangian for this model is

$$\begin{aligned}\mathcal{L} = \mathcal{L}_{\mathrm{SM}} + \mathcal{L}_{\mathrm{kin}} - \frac{1}{2}(M_\phi^2)^{ij}\,\mathrm{Tr}\big(\phi_i\phi_j\big) - \left(\frac{1}{2}M_\Psi\Psi\Psi + \mathrm{H.c.}\right) - \big(M_{\psi\psi'}\psi\psi' + \mathrm{H.c.}\big)\\ - (\lambda_1)^{ij}(H^\dagger H)\,\mathrm{Tr}\big(\phi_i\phi_j\big) - (\lambda_3)^{ijkm}\,\mathrm{Tr}\big(\phi_i\phi_j\phi_k\phi_m\big)\\ - \big(\lambda_4(H^\dagger\psi')\Psi + \mathrm{H.c.}\big) - \big(\lambda_5(H\psi)\Psi + \mathrm{H.c.}\big) - \big((\lambda_6)^{ij}L_i\phi_j\psi' + \mathrm{H.c.}\big)\end{aligned} \tag{7.2}$$

The numbering of the parameters is kept (skipping $\lambda_2$) in order to stay in line with the output of `minimal-lagrangians`. For comparison, the full Lagrangian *without* the additional $\mathbb{Z}_2$ symmetry (also generated with `minimal-lagrangians`) would look like:[3]

$$\begin{aligned}\mathcal{L}' = \mathcal{L}_{\mathrm{SM}} + \mathcal{L}_{\mathrm{kin}} - \frac{1}{2}(M_\phi'^2)^{ij}\,\mathrm{Tr}\big(\phi_i\phi_j\big)\\ - \big(M'_{\psi\psi'}\psi\psi' + \mathrm{H.c.}\big) - \left(\frac{1}{2}M'_\Psi\Psi\Psi + \mathrm{H.c.}\right) - \big((M'_{L\psi'})^i L_i\psi' + \mathrm{H.c.}\big)\\ - (\lambda'_1)^i H^\dagger\phi_i H - (\lambda'_2)^{ijk}\,\mathrm{Tr}\big(\phi_i\phi_j\phi_k\big) - (\lambda'_3)^{ij}(H^\dagger H)\,\mathrm{Tr}\big(\phi_i\phi_j\big)\\ - (\lambda'_5)^{ijkm}\,\mathrm{Tr}\big(\phi_i\phi_j\phi_k\phi_m\big)\\ - \big(\lambda'_6(H^\dagger\psi')\Psi + \mathrm{H.c.}\big) - \big(\lambda'_7(H^\dagger\psi)e_R^c + \mathrm{H.c.}\big) - \big(\lambda'_8\psi'\phi\psi + \mathrm{H.c.}\big)\\ - \big(\lambda'_9(H\psi)\Psi + \mathrm{H.c.}\big) - \big((\lambda'_{10})^{ij}L_i\phi_j\psi' + \mathrm{H.c.}\big) - \big((\lambda'_{11})^i(HL_i)\Psi + \mathrm{H.c.}\big)\end{aligned} \tag{7.3}$$

Here, six new terms appear which are not present in (7.2): those with the parameters $M'_{L\psi'}$, $\lambda'_1$, $\lambda'_2$, $\lambda'_7$, $\lambda'_8$ and $\lambda'_{11}$.

This once again demonstrates the necessity of the stabilizing symmetry. The Lagrangian allows for decays of all the dark matter candidates to Standard Model particles: $\phi^0$ to Higgs bosons via the $\lambda'_1$ term and both $\psi^0$ and $\Psi$ to a Higgs boson and a neutrino via the $\lambda'_7$ and $\lambda'_{11}$ terms. The lack of a dedicated term for $\psi'^0$ does not save it, since it will inevitably mix with the other neutral fermions and thus be involved in their decay processes as well. In particular, the new fermion doublets can now mix with the lepton doublets through the $M'_{L\psi'}$ and $\lambda'_7$ terms. Concerning the neutrino masses, the $\lambda'_{11}$ term then leads to a seesaw type I contribution even without loop corrections (cf. (5.40)).

[2]In general, it is not possible to first eliminate redundant terms using the identities in section 5.4 and only then add generation indices to a Lagrangian. Adding additional fields affects the numbers of different kinds of terms differently, so the identities should generally only be applied to the full Lagrangian (including all fields). In this case, $(H^\dagger H)\,\mathrm{Tr}(\phi_i\phi_j)$ only has one additional term for $i \neq j$ (since $\mathrm{Tr}(\phi_i\phi_j) = \mathrm{Tr}(\phi_j\phi_i)$), while $H^\dagger\phi_i\phi_j H$ has two new terms for $i \neq j$. However, in this case, both choices are still equivalent if the parameters are real (no $CP$-violating phases), and keeping the term $(H^\dagger H)\,\mathrm{Tr}(\phi_i\phi_j)$ allows for comparison with previous work on similar models.

[3]Again, the term of the form $H^\dagger\phi_i\phi_j H$ has been omitted (explaining why there is no parameter $\lambda'_4$) and generation indices have been added.

## 7.3. Mixing

When the odd fields are restricted to just the fermion sector, the model T1-3-B ($\alpha = 0$) is nothing but the singlet–doublet fermion dark matter model studied, for example, in [Coh+12] and [CS14] (therein called "model A").[4]

After electroweak symmetry breaking, the following mass terms for the neutral fermions appear in the Lagrangian (7.2):

$$\mathcal{L}_{\mathrm{m},0} = -\frac{1}{2}M'_{\Psi}\Psi\Psi - M_{\psi\psi'}\psi^0\psi'^0 - \frac{\lambda_4 v}{\sqrt{2}}\psi'^0\Psi - \frac{\lambda_5 v}{\sqrt{2}}\psi^0\Psi + \mathrm{H.c.} \tag{7.4}$$

These mass terms are not diagonal in the neutral fermion fields $\Psi$, $\psi^0$, $\psi'^0$. In contrast, the new charged fermions do not mix because there is only one field for each charge value. Introducing a new basis for the fields in "generation space",

$$\chi = U_\chi \begin{pmatrix} \Psi \\ \psi^0 \\ \psi'^0 \end{pmatrix} \tag{7.5}$$

with a unitary mixing matrix $U_\chi$, the Lagrangian can be rewritten as

$$\mathcal{L}_{\mathrm{m},0} = -\frac{1}{2}\begin{pmatrix} \Psi & \psi^0 & \psi'^0 \end{pmatrix} M_0 \begin{pmatrix} \Psi \\ \psi^0 \\ \psi'^0 \end{pmatrix} + \mathrm{H.c.} = -\frac{1}{2}\chi^\mathsf{T} U_\chi^* M_0 U_\chi^\dagger \chi + \mathrm{H.c.} \tag{7.6}$$

The mixing matrix $U_\chi$ is chosen in such a way that it diagonalizes the mass matrix $M_0$ (see section 5.3):

$$U_\chi^* M_0 U_\chi^\dagger = \mathrm{diag}(m_{\chi_1}, m_{\chi_2}, m_{\chi_3}) \tag{7.7}$$

[4] For [Coh+12], the fields $\nu$ and $\nu^c$ correspond to the fields $\psi$ and $\psi'$ while the parameters $\lambda'$ and $\lambda$ are here called $\lambda_5$ and $\lambda_4$, respectively. For [CS14], $D_1$ and $D_2$ are the same as $\psi$ and $\psi'$ while the parameters $y_{D_1}$ and $y_{D_2}$ correspond to $\lambda_5$ and $\lambda_4$, respectively.

The Majorana fermions $\chi_i$ are then the mass eigenstates of the neutral fermions and thus correspond to the propagating physical particles.

The mass matrix itself can be read from the Lagrangian as

$$M_0 = \begin{pmatrix} M_\Psi & \frac{\lambda_5 v}{\sqrt{2}} & \frac{\lambda_4 v}{\sqrt{2}} \\ \frac{\lambda_5 v}{\sqrt{2}} & 0 & M_{\psi\psi'} \\ \frac{\lambda_4 v}{\sqrt{2}} & M_{\psi\psi'} & 0 \end{pmatrix} \tag{7.8}$$

and this is indeed the same as eq. (3) from [Coh+12] or eq. (10) from [CS14] once the different names of the parameters are substituted.[5] The analytical forms of the diagonal values $m_{\chi_i}$ are unfortunately quite unwieldy; they are given, for example, in [Res+15b, app. A].

For the scalar triplet, the fields only mix if there are several generations. For one generation, the "tree-level masses" (without radiative corrections) after electroweak symmetry breaking arise from the terms involving $M_\phi$ and $\lambda_1$:

$$m^2_{\phi^0} = m^2_{\phi^\pm} = M^2_\phi + \lambda_1 v^2 \tag{7.9}$$

Without loop corrections, the neutral and charged scalars are mass-degenerate. At one-loop level, though, a mass splitting

$$\Delta m_\phi = m_{\phi^\pm} - m_{\phi^0} = 166\,\text{MeV} \tag{7.10}$$

is introduced, making the neutral component lighter than the charged ones [CFS06].

For more than one generation, both $M_\phi$ and $\lambda_1$ become symmetric $n_\text{s} \times n_\text{s}$ matrices. The mass matrix is then

$$M^2_{\phi^0} = M^2_{\phi^\pm} = M^2_\phi + \lambda_1 v^2 \tag{7.11}$$

The neutral mass eigenstates $\eta^0_i$ are found by diagonalizing the scalar mass matrix using the orthogonal mixing matrix $O_\eta$:

$$\eta^0 = O_\eta \phi^0 \qquad O M^2_{\phi^0} O^\mathsf{T} = \text{diag}\Big( (m_{\eta^0_i}) \Big) \tag{7.12}$$

[5]Care must be taken with Yukawa terms like the $\lambda_5$ term. Since the SU(2) inner product has the property $H\psi = -\psi H$ if $H$ and $\psi$ commute, as they do in this case, a minus sign may be introduced to the parameter $\lambda_5$ depending on the order in which the product of $H$ and $\psi$ is written in the Lagrangian.

The next step in the tool chain, using the SARAH model files obtained through the output capabilities of minimal-lagrangians, is to specify the relevant information after electroweak symmetry breaking, namely the mixing of fields and (for further output to SPheno and micrOMEGAs) Dirac spinors. Mass diagonalization results in three neutral Majorana spinors $\chi_{\mathrm{M}i}$ and one negatively-charged Dirac spinor $\psi_{\mathrm{D}}$:

$$\chi_{\mathrm{M}i} = \begin{pmatrix} \chi_i \\ \bar{\chi}_i \end{pmatrix} \qquad \psi_{\mathrm{D}} = \begin{pmatrix} \psi^- \\ \overline{\psi'^+} \end{pmatrix} \tag{7.13}$$

The Dirac fermion $\psi_{\mathrm{D}}$ can be viewed as a $\mathbb{Z}_2$-odd equivalent of the electron.

## 7.4. Singlet–doublet fermion dark matter

Now that the SARAH model files are complete, the output of numerical code and model files for SPheno and micrOMEGAs can be performed and the numerical analysis can begin. First, the focus shall be on the singlet–doublet fermion (SDF) dark matter model which is contained within T1-3-B ($\alpha = 0$). If the scalar triplet is decoupled, that is, its mass is set to a large value like 1000 TeV and all its couplings are set to zero:

$$M_\phi = 1000\,\mathrm{TeV} \qquad \lambda_1 = \lambda_3 = \lambda_6 = 0 \tag{7.14}$$

it should be possible to treat the model as though this triplet did not exist for most purposes. It should cease to have any effect on observables like the dark matter relic density or the direct detection cross section, with either vanishing or at least tiny contributions which are numerically completely insignificant. The singlet–doublet fermion dark matter model would then be recovered.

As an example, figure 2 in [Coh+12] shows the dark matter relic density $\Omega h^2$ as well as the spin-dependent and -independent direct detection cross sections for scattering on a proton, $\sigma_{\mathrm{SI}}$ and $\sigma_{\mathrm{SD}}$, for a certain parameter point as a function of $\lambda_4$ (there called $\lambda'$). Using the output of SARAH, these results were checked within the model T1-3-B ($\alpha = 0$), employing the capabilities of SPheno to calculate the mass spectrum numerically for each parameter point and micrOMEGAs to compute the dark matter observables.[6]

In SPheno, no loop corrections to the masses were computed for this section (the calculations were disabled in the SLHA input file). There are several reasons for this approach. The first is consistency, since `micrOMEGAs` generally does not include radiative corrections for most computations. The second is easier comparison with the literature, where the particles' masses are generally calculated by `micrOMEGAs` itself and thus limited to tree level. Finally, when performing the same calculations including one-loop-corrected masses, there was simply not a big difference to the tree-level results in the cases considered here. It should be stressed again, however, that the use of `SPheno` does not only make it possible to include loop corrections for the masses in principle, but also to compute many additional observables like branching ratios from loop decays which are not handled by `micrOMEGAs`. Since a significant feature of this model are the radiative neutrino masses, the ability to include radiative corrections will also become very important in section 7.7.

[6] The versions of these programs which were used for this work are SARAH 4.12.3, SPheno 4.0.3 and micrOMEGAs 4.3.5, see section 7.6.

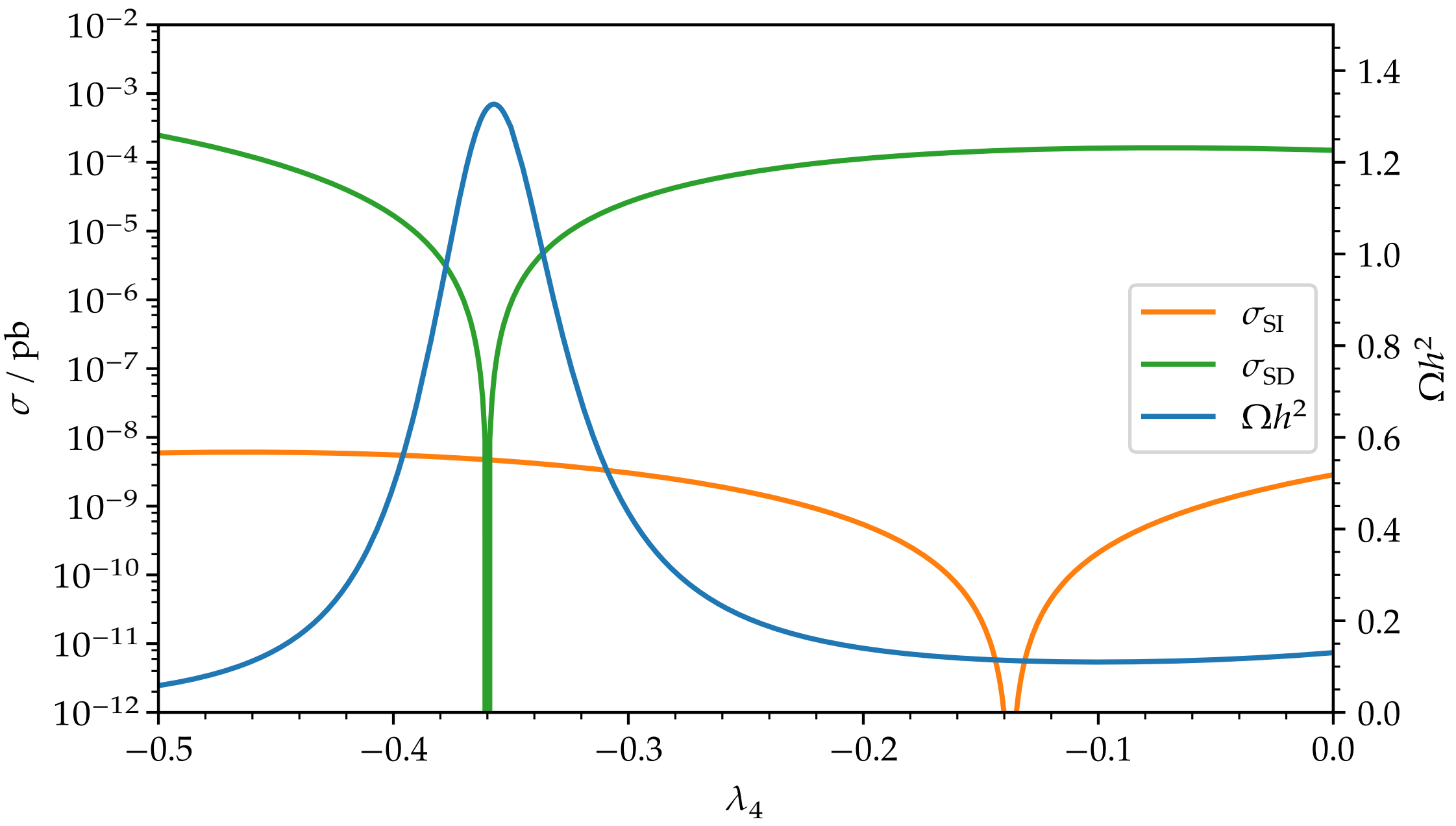

**Figure 7.1.:** The dark matter relic density $\Omega h^2$ and the spin-independent and spin-dependent cross sections for dark matter–proton scattering $\sigma_{\text{SI}}$ and $\sigma_{\text{SD}}$ as a function of $\lambda_4$ in the SDF limit. The parameter region shows examples of cancellation in the DM coupling to the Higgs boson ($\sigma_{\text{SI}}$) and the $Z$ boson ($\sigma_{\text{SD}}$). Used parameters: $M_\Psi = 200\,\text{GeV}$, $M_{\psi\psi'} = 300\,\text{GeV}$ and $\lambda_5 = 0.36$ (cf. [Coh+12, fig. 2]). The parameter $\lambda$ from the Higgs potential was set to $\lambda = 0.2612$ so that the Higgs mass is $m_h = \sqrt{\lambda} v = 125.08\,\text{GeV}$ at tree level (the measured value is $(125.09 \pm 0.24)\,\text{GeV}$ [PDG16]).

The result is shown in fig. 7.1, and indeed, it mostly agrees with [Coh+12]. However, there are some things to note. Firstly, even though it is difficult to tell because the logarithmic scale covers many orders of magnitude, the spin-independent cross section $\sigma_{\text{SI}}$ is smaller here by at least a factor of 2. The reason for this is that the version of `micrOMEGAs` used in [Coh+12] was `micrOMEGAs` 2.4. As stated in [Bél+17], later versions of `micrOMEGAs` use updated values of the nucleon form factors. The function `calcScalarQuarkFF` can be used to change these values.[7] A test using these form factors confirms that $\sigma_{\text{SI}}$ in fig. 7.1 increases, almost exactly matching [Coh+12].

Another thing to note is that the peak of the dark matter relic density in fig. 7.1 is somewhat larger than in [Coh+12]. Together with the slight remaining difference in $\sigma_{\text{SI}}$, this may partly be attributable to the different value of the Higgs mass used in [Coh+12]. The authors consider three different values for the mass of the Higgs boson (140 GeV, 200 GeV and 500 GeV), although they do not specify which one was used for figure 2. On the other hand, the implementation details of the `micrOMEGAs` model files and the main code can have an impact on the exact numerical results. Unfortunately, there is no way to know for sure without access to the original model files.

As for the physical interpretation, fig. 7.1 demonstrates that, depending on the mixing between the fermion doublets, the couplings to both the Higgs and $Z$ bosons, which directly impact the spin-independent and spin-dependent direct detection cross sections, can be suppressed by many orders of magnitude. This indicates that there are parameter regions which are "blind spots", evading direct detection experiments even for rather low masses. This is an important point to consider when giving exclusion limits for the WIMP mass in a given model or interpreting experimental results, which may not see dark matter interactions if they are only sensitive to either spin-independent or spin-dependent scattering.

Next, consider [CS14, fig. 5], which provides exclusion plots of the singlet–doublet fermion dark matter model for different ratios of the couplings $\lambda_4$ and $\lambda_5$ in the plane of the singlet and doublet mass parameters $M_\Psi$ and $M_{\psi\psi'}$.[8] Similar calculations were performed using the same setup as for fig. 7.1. Once again, radiative corrections were neglected [CS14, p. 6].

Figure 7.2 shows the results of the computations. A significant difference, however, is that the excluded regions shown here show the current limits set by the XENON1T experiment [XENON17] instead of the projections used in [CS14]. Similar to fig. 7.1,

[7] The form factors used in `micrOMEGAs` 2.4 and earlier versions are reproduced using `calcScalarQuarkFF(0.553, 18.9, 55., 243.5)` [Bél+17].

[8] In [CS14], the couplings $\lambda_4$ and $\lambda_5$ (there called $y_{D_2}$ and $y_{D_1}$) are re-parametrized as $\lambda_4 = y\sin(\theta)$, $\lambda_5 = y\cos(\theta)$.

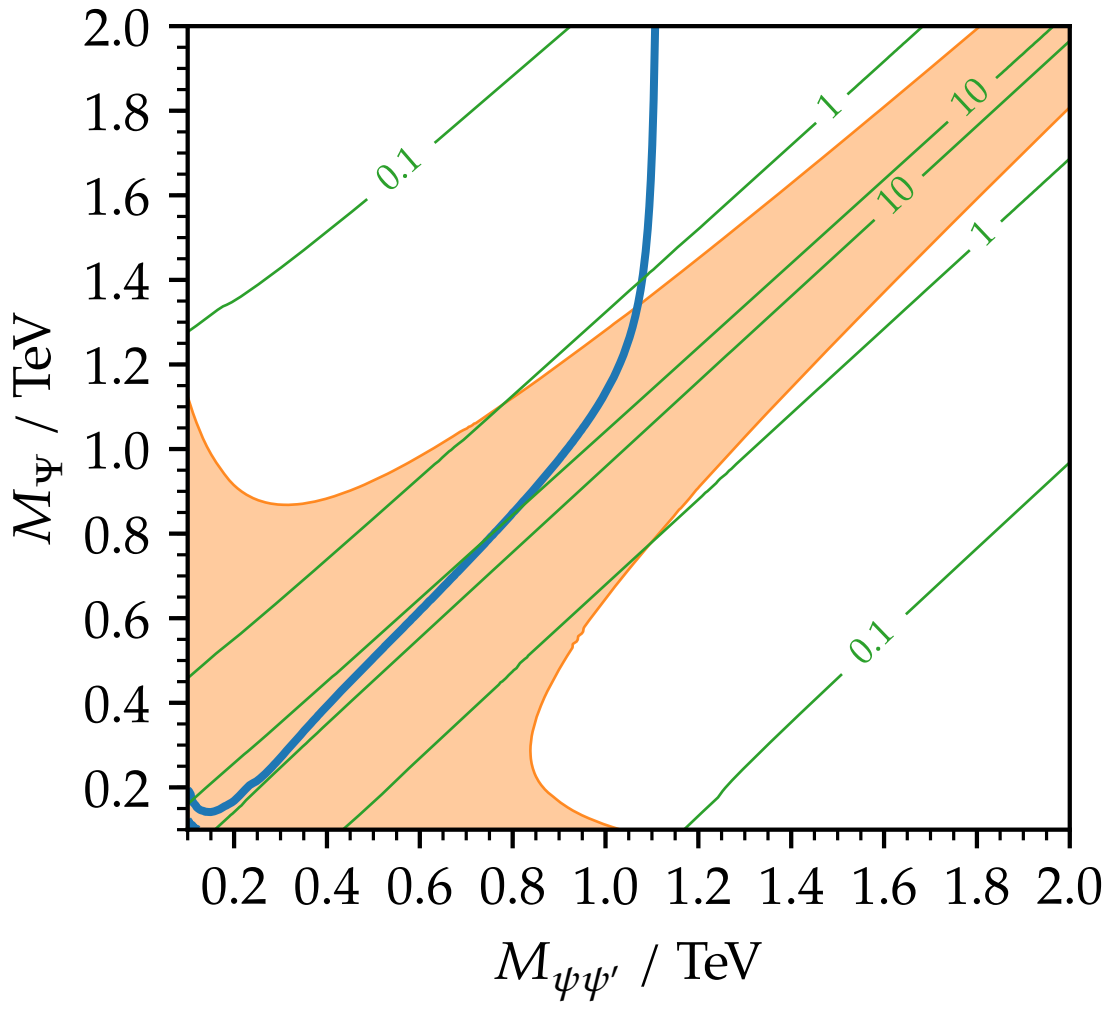

**(a)** $\tan(\theta) = \lambda_4/\lambda_5 = 2.$

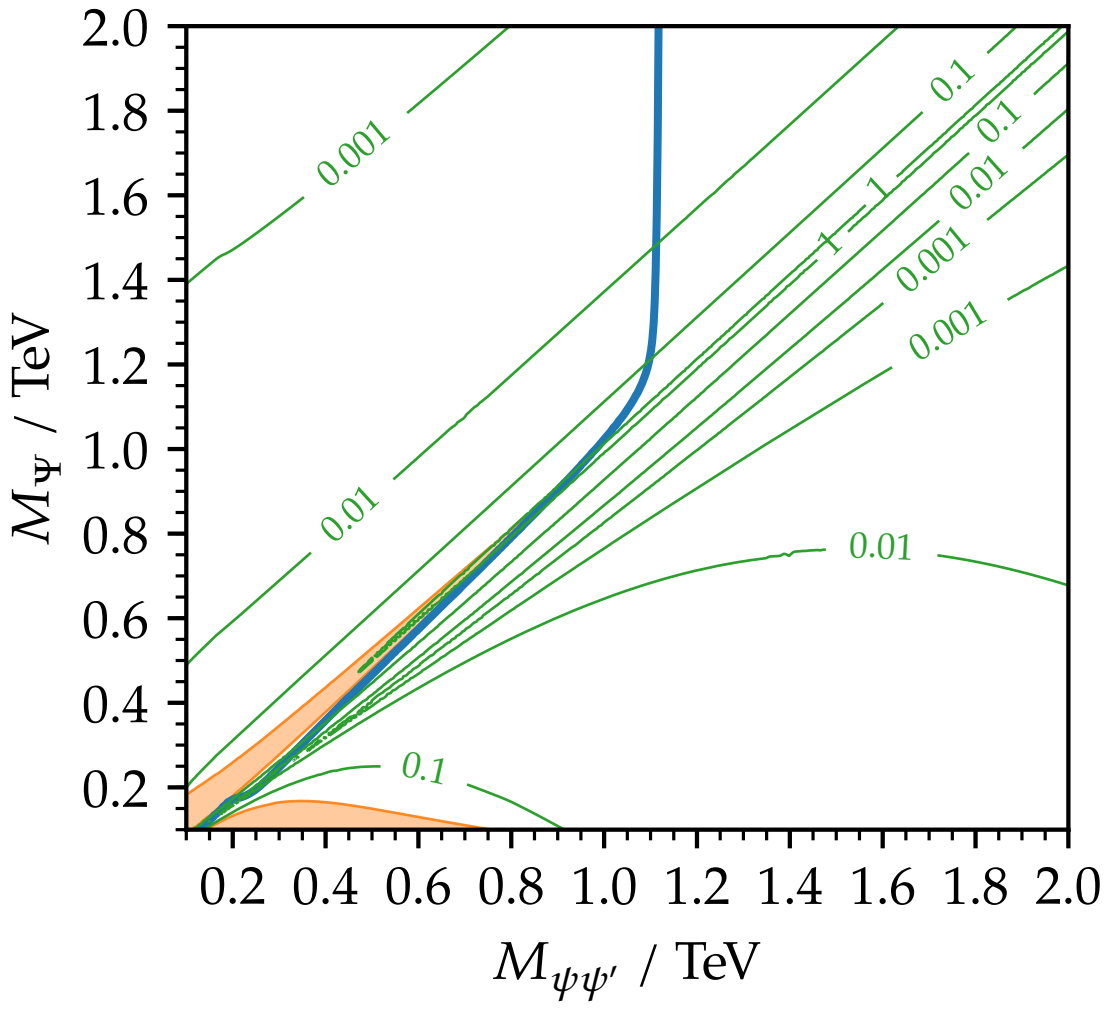

**(b)** $\tan(\theta) = \lambda_4/\lambda_5 = -2.$

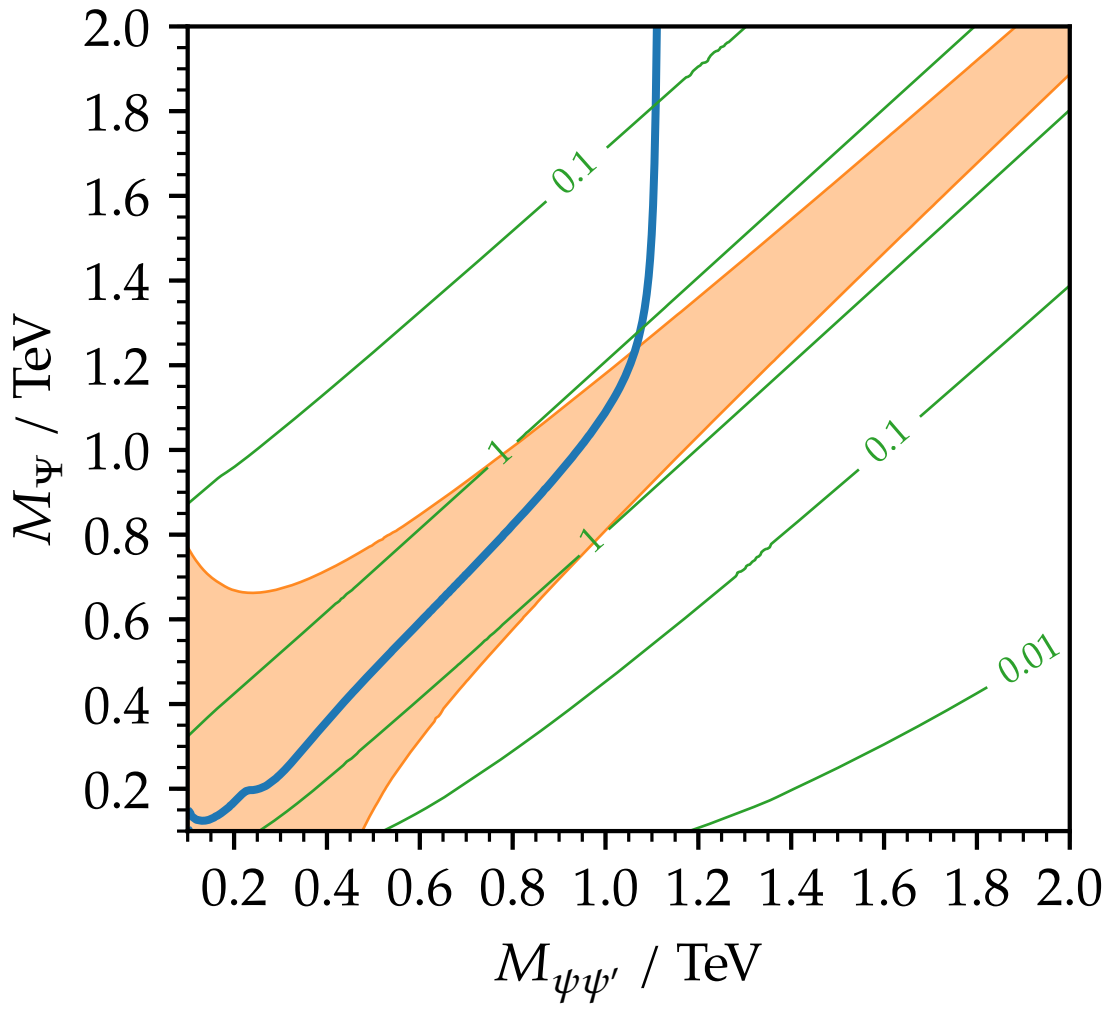

**(c)** $\tan(\theta) = \lambda_4/\lambda_5 = 10.$

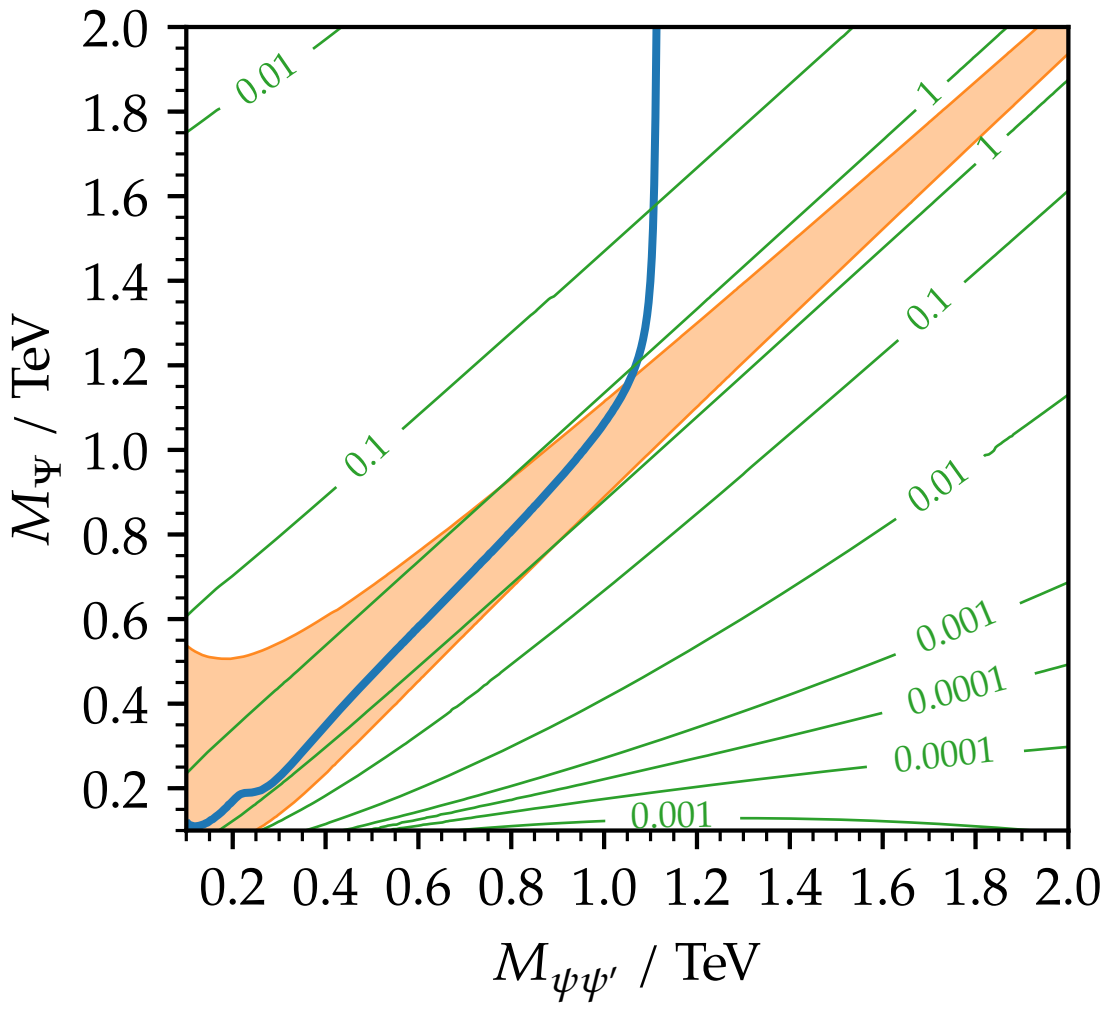

**(d)** $\tan(\theta) = \lambda_4/\lambda_5 = -10.$

**Figure 7.2.:** Contour plots of the dark matter relic density $\Omega h^2$ (blue) and the spin-independent dark matter–proton cross section $\sigma_{\mathrm{SI}}$ (green, in zb) on the $(M_{\psi\psi'}, M_\Psi)$ plane for $y = \sqrt{\lambda_4^2 + \lambda_5^2} = 0.3$ in the SDF limit. To the left of the blue line, $\Omega < \Omega_{\mathrm{obs}}$, while to the right, $\Omega > \Omega_{\mathrm{obs}}$ (cf. [CS14, fig. 5]). The gold shaded regions are excluded by current XENON1T bounds [XENON17]. (Higgs potential: $\lambda = 0.2612$.)

even though the qualitative behavior is the same, one can note a significant deviation in the magnitude of the spin-independent direct detection cross section $\sigma_{\text{SI}}$. As [CS14] used `micrOMEGAs` 2.4.5, the discrepancy is again due to the different nucleon form factors used in more recent versions of `micrOMEGAs`. However, calculations using the same form factors or even the same version of `micrOMEGAs` still showed slight deviations from the results displayed in [CS14, fig. 5], shifting the contour lines to the order of 100 GeV with respect to the mass axes.

Since the authors used `FeynRules` to generate the model files for `micrOMEGAs`, they were contacted in an attempt to see if the same results could be reproduced using their `FeynRules` model files. In that case, the cause of the deviation would have been that `SARAH` (used here) and `FeynRules` (used in [CS14]) generated differing `micrOMEGAs` model files and it would, perhaps, have been possible to identify a flaw in one of these two tools. Unfortunately, no reply was received from the authors. In this case, the lead of the different Higgs mass is not present either, because they used $m_h = 125.6\,\text{GeV}$ [CS14, p. 6]. On the other hand, the behavior of the relic density in fig. 7.2 is in excellent agreement with [CS14]. It seems that $\sigma_{\text{SI}}$ is rather sensitive to changes in the model implementation or recent updates, such as the nucleon form factors. The relic density and $\sigma_{\text{SD}}$, on the other hand, appear more stable, having shown little deviation between different numerical evaluations thus far.

Although the discrepancy concerning $\sigma_{\text{SI}}$ could, regrettably, not be resolved, the unexplained differences are small enough that the physical conclusions remain unchanged. For one, the same blind spot behavior as shown in fig. 7.1 can be seen here for $\tan(\theta) < 0$. These are also the regions which are less constrained by the XENON1T results in general.

Next, as stated in [CS14], it can be noticed that the "correct" dark matter density ($\Omega h^2 = \Omega_{\text{obs}} h^2 = 0.1186 \pm 0.0020$) is obtained for $M_{\psi\psi'} \approx M_\Psi$ up to a point of $M_\Psi \approx 1\,\text{TeV}$, beyond which $M_{\psi\psi'}$ becomes fixed to about 1 TeV. This is around the same value at which fermionic doublet dark matter yields the correct relic density as well. The same happens for all values of $\tan(\theta)$, indicating that dark matter annihilation is not primarily driven by processes involving the Higgs boson (since $\lambda_4$ and $\lambda_5$ are the dark matter couplings to the Higgs boson).

Finally, the regions with $M_{\psi\psi'} \approx M_\Psi < 1\,\text{TeV}$ are also those which are already excluded by XENON1T. In many cases, only pure doublet dark matter beyond 1 TeV is viable. However, the parameter space with $\tan(\theta) < 0$ is not nearly as constrained just yet, still allowing for masses of a few hundred GeV.

## 7.5. Triplet scalar dark matter

The analysis shall now turn to the other dark matter candidate of T1-3-B ($\alpha = 0$): triplet scalar dark matter. Once again, the model is studied first in the limit that the other fields (in this case, the fermions $\Psi$, $\psi$ and $\psi'$) are decoupled:

$$M_\Psi = M_{\psi\psi'} = 1000\,\text{TeV} \qquad \lambda_4 = \lambda_5 = \lambda_6 = 0 \tag{7.15}$$

leaving only the real scalar triplet $\phi$. Such a model with dark matter from a $\mathbb{Z}_2$-odd triplet is often called the *inert triplet model* (ITM) because the triplet does not participate in the Higgs mechanism due to its unbroken symmetry, contrary to the Higgs doublet.

Perhaps because an SU(2) triplet is a representation that is already beyond anything that is present in the Standard Model scalar and fermion field, or because a single triplet is additionally rather inflexible without any other fields to mix with (for example, model C in [CS14] is a singlet–triplet scalar model, not a pure triplet one), this model has not received as much research attention as, for example, the previously introduced singlet–doublet fermion one. In particular, the only really relevant parameters remaining in this limit are $M_\phi$ and $\lambda_1$. The quartic self-coupling $\lambda_3$ is hardly observable because it neither really affects the relic density (while it does allow for co-annihilation and thus conversion between different triplet components and generations, this generally does not make a significant difference in the end), nor does it contribute to direct or indirect detection processes. Nevertheless, there are a few studies of this model which can be found in the literature. They all deal with only a single generation of the scalar ($n_\text{s} = 1$).

The earliest detailed surveys that turned up in a survey of the literature were [AGN11a; AGN11b], which both present the same results. There, just this case of a real (neutral) $\mathbb{Z}_2$-odd scalar triplet is considered. [AGN11a, fig. 1] and [AGN11b, fig. 2] (which are identical) show the dark matter relic density as a function of the relevant parameters $M_\phi$ and $\lambda_1$.[9] The authors do not specify the value of $\lambda_3$ which was used due to its negligible impact. In the following, it will be taken to be $\lambda_3 = 0$.

Although it is not stated entirely clearly, the lack of discussion of radiative corrections, except for a reference to [CS09], indicates that they have not been taken into account for the calculations. Strangely, [CS09] is used to corroborate the claim that the neutral component of the scalar triplet becomes the LOP (and thus stable) after radiative corrections, with a mass splitting of $\Delta M_\phi = 166\,\text{MeV}$ between the neutral and charged component. However, [CS09] actually deals with fermionic quintuplets, mentioning other representations only in passing. A more fitting source would be [CFS06] by the same authors, which does show the same mass splitting also for scalar triplets.

[9] $\lambda_1$ is called $\lambda_3$ in [AGN11a; AGN11b].

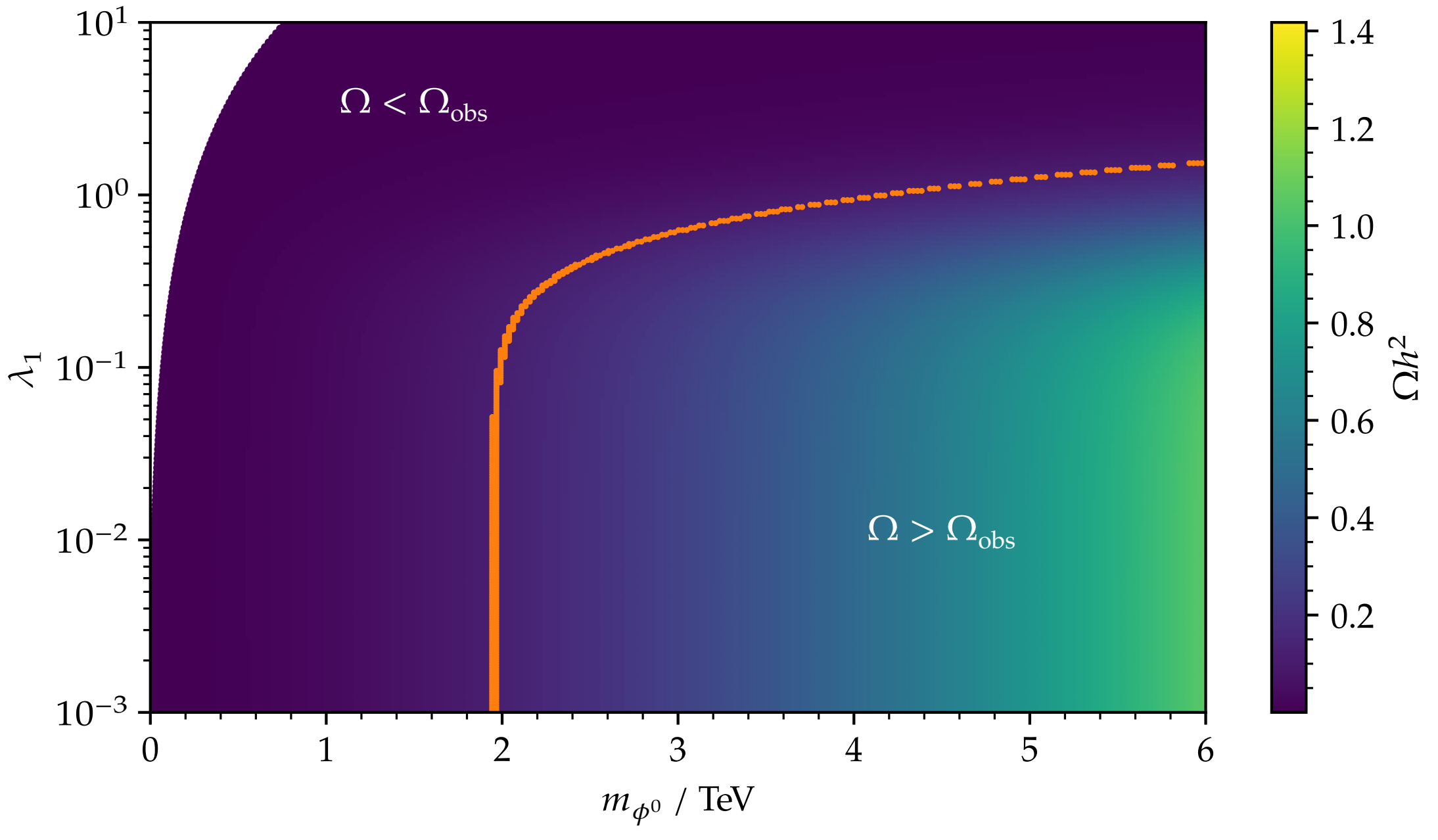

**Figure 7.3.:** The dark matter relic density $\Omega h^2$ as a function of the dark matter mass $m_{\phi^0}$ and $\lambda_1$ (with $\lambda_3 = 0$) in the ITM limit. The orange points show the region where $\Omega = \Omega_{\text{obs}}$ within the current uncertainty. (Higgs potential: $\lambda = 0.2612$.)

Consequently, loop corrections are once again disabled for the present calculations. It should be mentioned, however, that for some regions of the parameter space, the additional scalar can cause quite sizable corrections to the Higgs mass, which needs to be taken into account if the (at the time of [AGN11a; AGN11b] unknown) physical value is to be matched. Furthermore, if this variation of the Higgs mass is not adjusted for, the spin-independent cross section (fig. 7.5) behaves quite differently. The dark matter relic density, on the other hand (fig. 7.3), was insensitive to variations in the Higgs mass from radiative corrections.

Without further ado, the results for the relic density are shown in fig. 7.3. They clearly indicate a dark matter mass of around 2 TeV for the neutral component of the scalar triplet. Only for larger couplings $\lambda_1 \gtrsim 0.2$, when annihilation into Higgs bosons starts to ramp up, does the preferred mass shift to larger values, rapidly

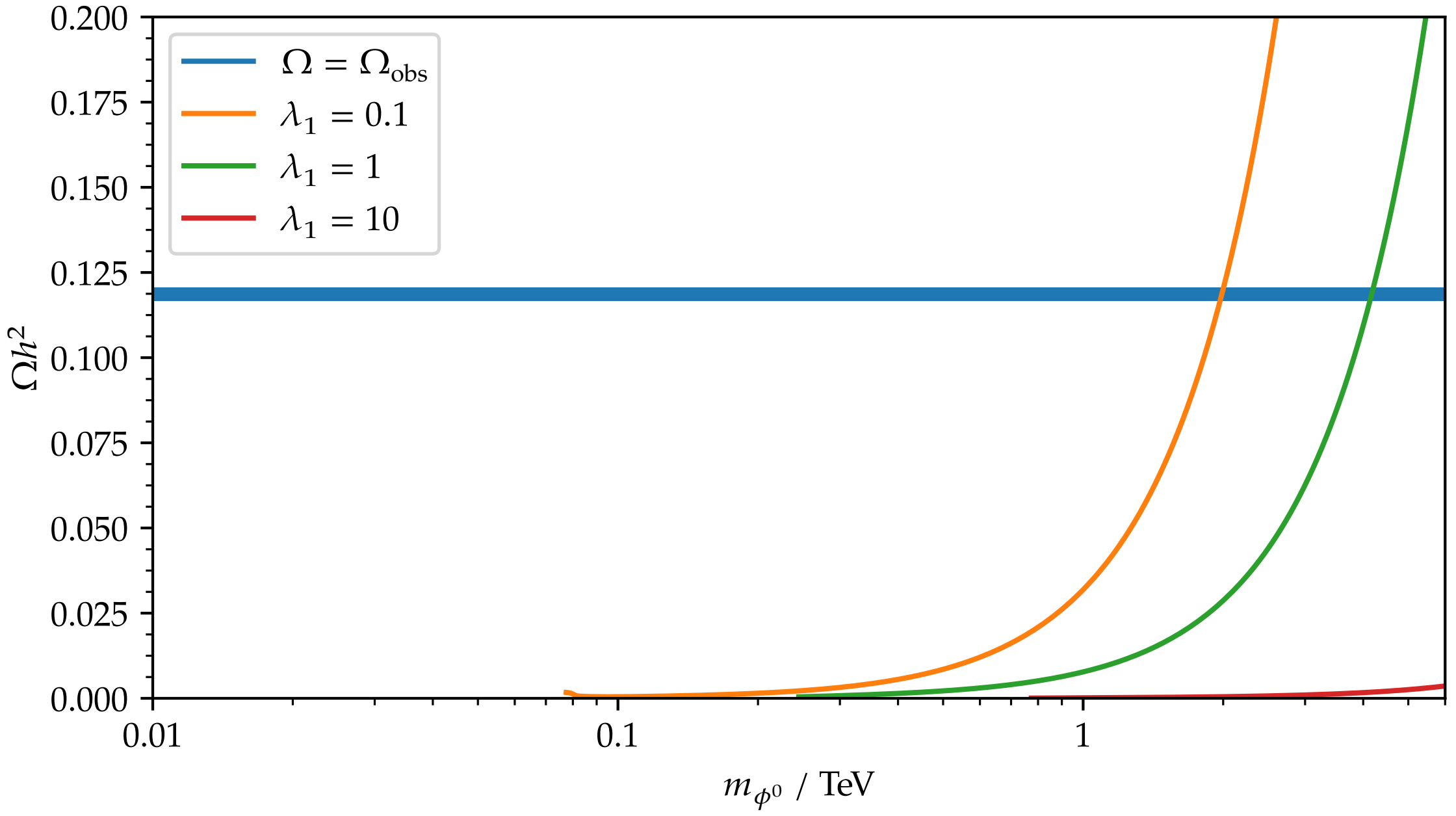

**Figure 7.4.:** The dark matter relic density $\Omega h^2$ as a function of the dark matter mass $m_{\phi^0}$ and for different values of $\lambda_1$ (with $\lambda_3 = 0$) in the ITM limit. The blue band shows the region where $\Omega = \Omega_{\text{obs}}$ within the current uncertainty. (Higgs potential: $\lambda = 0.2612$.)

reaching $m_{\phi^0} \approx 5\,\text{TeV}$ for $\lambda_1 \to 1$.

The problem now appears in the comparison with [AGN11a, fig. 1] and [AGN11b, fig. 2]. Although the qualitative behavior is the same, the data seem to be shifted by about 3.5 TeV on the mass axis. Accordingly, the authors claim a dark matter mass of 5.5 TeV compared to the 2 TeV obtained here. The value of the Higgs boson's mass is not available as an explanation this time since the authors use $m_h = 120\,\text{GeV}$, and in any case, no reasonable value of the Higgs mass could have an effect of this size.

In the search for an explanation of this major difference, one curious detail was discovered: The terms contributing to the triplet's mass are written in [AGN11a; AGN11b] as $M^2\,\text{Tr}(T^2) + \lambda_3 H^\dagger H\,\text{Tr}(T^2)$, and the masses of the triplet components are given in eq. 7 as $m_{T^0}^2 = m_{T^\pm}^2 = M^2 + \frac{1}{2}\lambda_3 v^2$. However, since this is a *real* triplet, a factor of $\frac{1}{2}$ needs to be taken out of the definition of $m_{T^0}^2$. The correct normalization of the mass is therefore $m_{T^0}'^2 = 2M^2 + \lambda_3 v^2 = 2m_{T^0}^2$. Depending whether the same expression was used in the numerical evaluation, it may be the case that the results in [AGN11a; AGN11b] have to be shifted by a factor of $\sqrt{2}$. Although this could be an improvement, there is still a significant difference between these results.

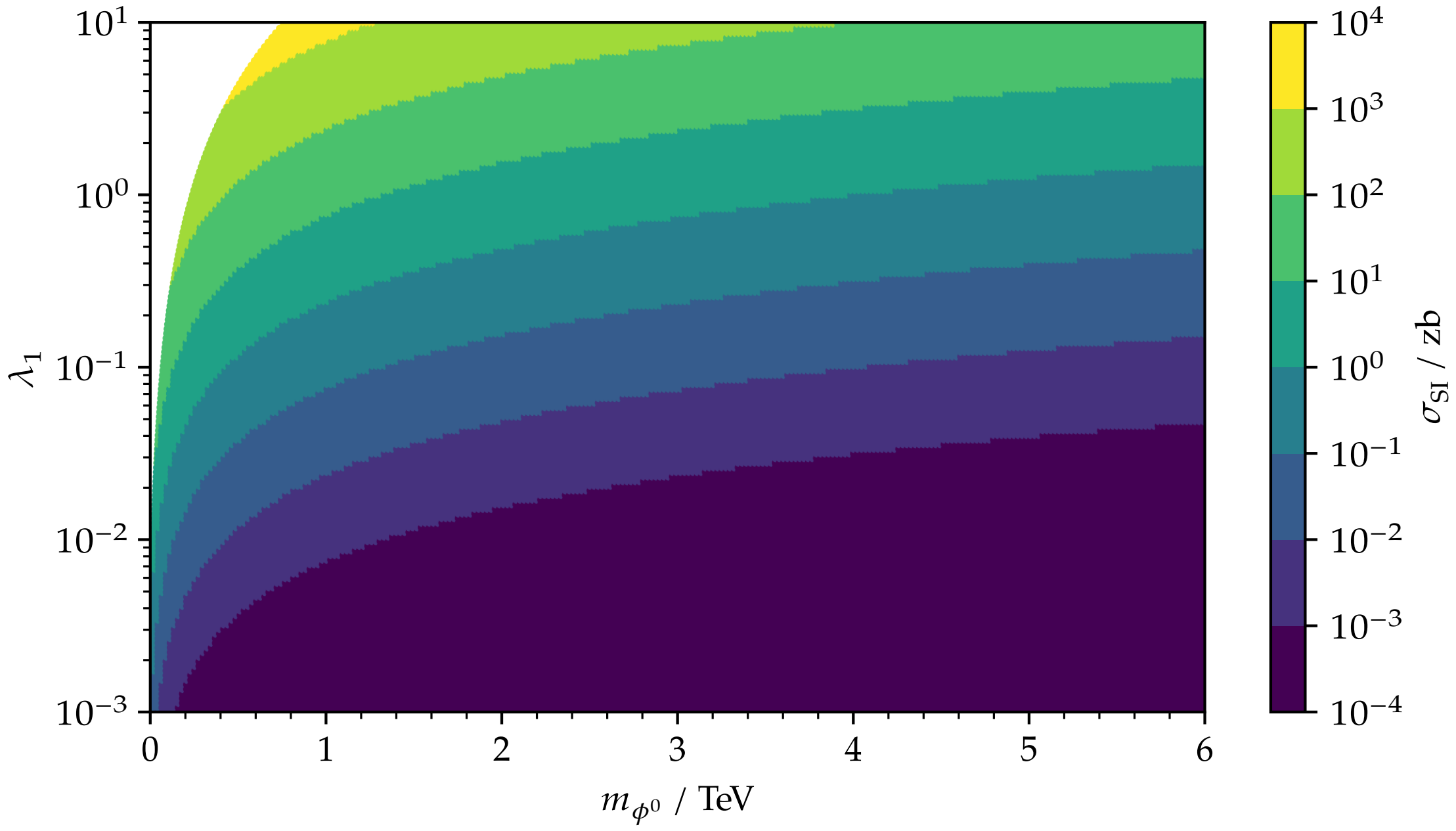

**Figure 7.5.:** The direct detection cross section $\sigma_{\mathrm{SI}}$ for scattering on the proton as a function of the dark matter mass $m_{\phi^0}$ and $\lambda_1$ (with $\lambda_3 = 0$) in the ITM limit. (Higgs potential: $\lambda = 0.2612$.)

Figure 7.4 shows a similar calculation as [AGN11b, fig. 1] for additional comparison. However, the conclusion is the same: For $\lambda_1 = 0.1$, the correct dark matter mass is at 2 TeV, shifting towards 5 TeV for $\lambda_1 \to 1$. [AGN11b], on the other hand, shows the same large value of >5 TeV for the correct relic density even for $\lambda_1 = 0.1$. Indeed, the figure only raises further questions: How is it possible that mass values all the way down to 1 GeV are shown? Since the tree-level mass $m_{\phi^0}$ is given by $m_{\phi^0} = \sqrt{M_\phi^2 + \lambda_1 v^2}$ (with $v \approx 246\,\mathrm{GeV}$), the minimum possible mass even for $M_\phi = 0$ and $\lambda_1 = 0.1$ is $m_{\phi^0} = 77.8\,\mathrm{GeV}$, which can clearly be seen in fig. 7.4. The suspicion arises that the plots may have been mislabeled, showing the parameter $M_\phi$ instead of the mass $m_{\phi^0}$ on the horizontal axis. Perhaps this is the case, and while this does shift the masses for larger values of $\lambda_1$, the impact is both comparatively small and would additionally have an effect opposite to the one observed here. That is, if the axes were actually showing $M_\phi$ in [AGN11a; AGN11b], it would be expected that the plots are shifted towards *lower* instead if higher masses.

Turning to the spin-independent direct detection cross section ([AGN11a, fig. 2] and [AGN11b, fig. 3]), the results are depicted in fig. 7.5. Mysteriously, they are in very good agreement with [AGN11a; AGN11b] (recalling $1\,\mathrm{b} = 10^{-24}\,\mathrm{cm}^2$, $1\,\mathrm{zb} =$ $10^{-45}$ cm$^2$). Considering the deviation from the previous results and the fact that the authors used `micrOMEGAs` 2.4, meaning that $\sigma_{\text{SI}}$ should be affected by the different nucleon form factors, the numerical agreement in this observable is quite puzzling.

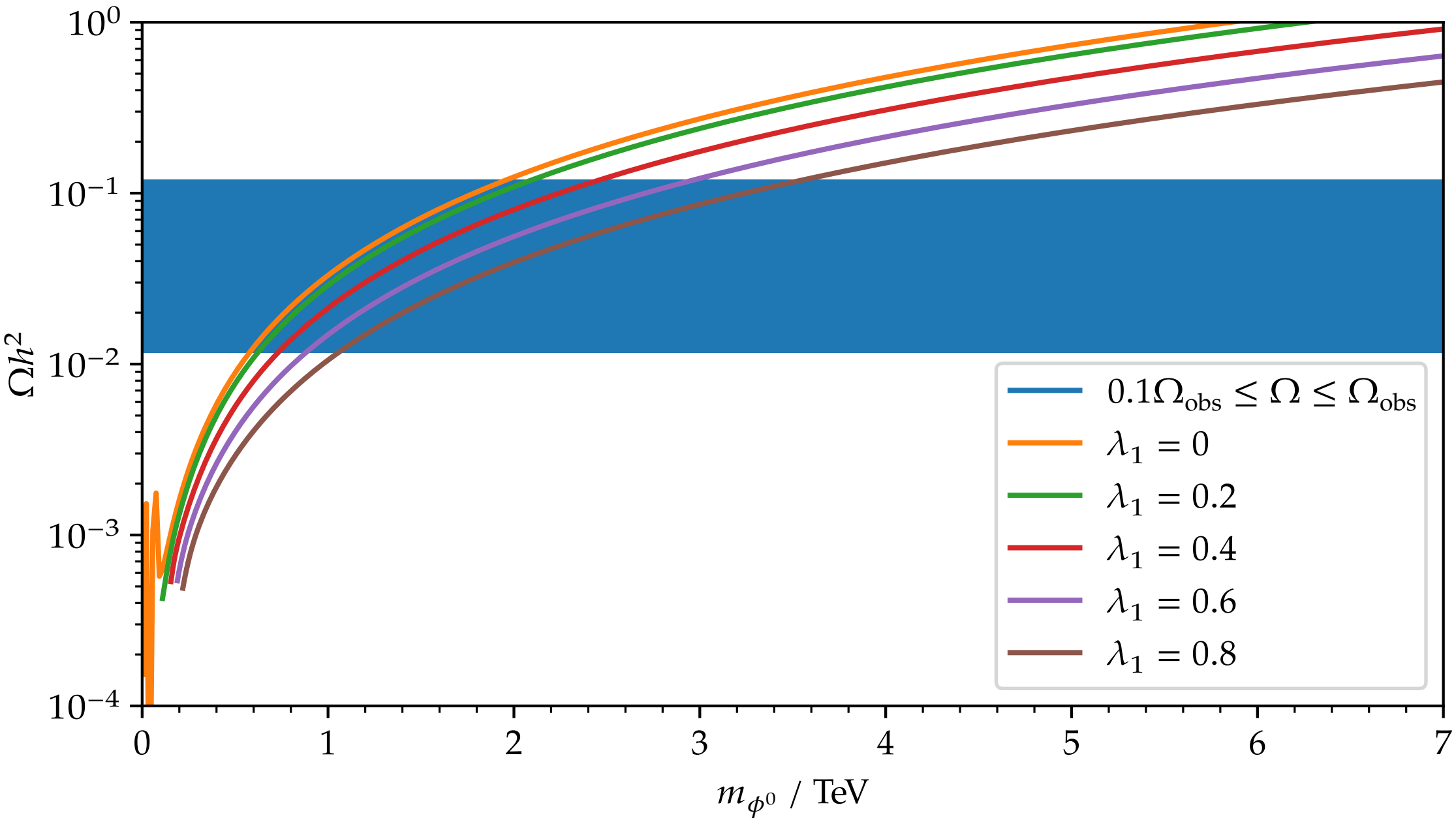

**Figure 7.6.:** The dark matter relic density $\Omega h^2$ as a function of the dark matter mass $m_{\phi^0}$ and for a range of values of $\lambda_1$ (with $\lambda_3 = 0$) in the ITM limit. The blue band shows the region where $0.1\Omega_{\text{obs}} \leq \Omega \leq \Omega_{\text{obs}}$, i. e. where the neutral triplet component contributes more than 10 % of the total dark matter density. (Higgs potential: $\lambda = 0.2612$.)

Further searching in the literature uncovered [YF15], where quite similar aspects of the inert triplet model were studied. Exactly the same definitions and conventions were used there (even including the erroneous factor of $\sqrt{2}$ in the mass), suggesting that a comparison should be straightforward. However, instead of shedding some much-needed light on the matter, this throws yet another value for the dark matter mass into the ring with $m_{\phi^0} = 7$ TeV. [YF15, fig. 2] is mostly the same as fig. 7.3, except that it uses a linear (instead of logarithmic) scale for $\lambda_1$ and also shows negative values of $\lambda_1$, which does not have any effect on the physical result ($\Omega h^2$ behaves the same for $\pm\lambda_1$).

[YF15, fig. 1] shows yet another plot of the relic density, this time on a logarithmic scale. But already on first sight, it is obvious that the figures disagree. Once again, the present results indicating a preferred mass of $m_{\phi^0} = 2$ TeV are in conflict with the claimed value of 7 TeV, which is also visible in their fig. 1. Figure 7.6 has a similar plot with different values of $\lambda_1$. Since the range of $\lambda_1$ used in the scatter plot [YF15, fig. 1] was not specified, it is difficult to do an exact comparison. What is clear once again, however, is that the observed relic density is reached for much lower masses of the WIMP $\phi^0$.

These differences are even more confusing because [AGN11a] is cited in [YF15]. However, despite the wildly different results obtained for the dark matter candidate's mass, this discrepancy is not mentioned at any point in the latter. The authors of both [AGN11a; AGN11b] and [YF15] were contacted in the hopes of clearing up these inconsistencies and requesting the model files used for `micrOMEGAs` and `FeynRules` (used by the latter). Using these model files, it would have been possible to run the same calculations again in order to confirm that they match the results given in the papers. The next step would then have been to analyze the differences in the model files in order to find the cause of these different outcomes. The authors replied, but unfortunately, the questions remained unresolved and none were able to produce any model files used in the publications.

Faced with this wholly unsatisfactory situation, further sources were sought out in order to settle the issue. [Kha18] is another publication about the inert triplet model which states a value for the dark matter mass. The relevant plot there is in figure 2. It should be noted that the parameter $\lambda_1$ (called $\lambda_3$ there) is scaled differently, with $\lambda_1 = 2\lambda_1^{(\text{theirs})}$. Thus, when "$\lambda_3 = 0.1$" is specified there, the corresponding value here is $\lambda_1 = 0.05$.

Figure 7.7 shows this work's comparison. And indeed, as stated in [Kha18], which gives a mass of $m_{\phi^0} \gtrsim 1.8\,\text{TeV}$, there is agreement between the results. On the flip side, this means that this is now the third different mass value reported in the literature, with all of them differing by at least 1.5 TeV. In addition, there is a small wrinkle concerning the value of $\lambda_1$. As mentioned, the corresponding value of $\lambda_1$ for the "$\lambda_3 = 0.1$" indicated in [Kha18] is 0.05, shown as a dashed green line in fig. 7.7. However, on comparing the two figures, one finds that the green line only reaches a mass as low as 55 GeV, while [Kha18, fig. 2] has data down to 30 GeV even though the lines should match. As explained in the discussion of fig. 7.4, a non-zero value of $\lambda_1$ induces a minimum value of $m_{\phi^0} = \sqrt{\lambda_1} v$ even for $M_\phi = 0$. In order to reach $m_{\phi^0} = 30\,\text{GeV}$, a value of at most $\lambda_1 = 0.015$ is possible. Since, as the figure shows, the relic density is insensitive to changes of $\lambda_1$ of this scale, it seems plausible that this is merely an oversight in [Kha18] and that in fact, a value of 0.01 (instead of 0.1) was used to create the figure.

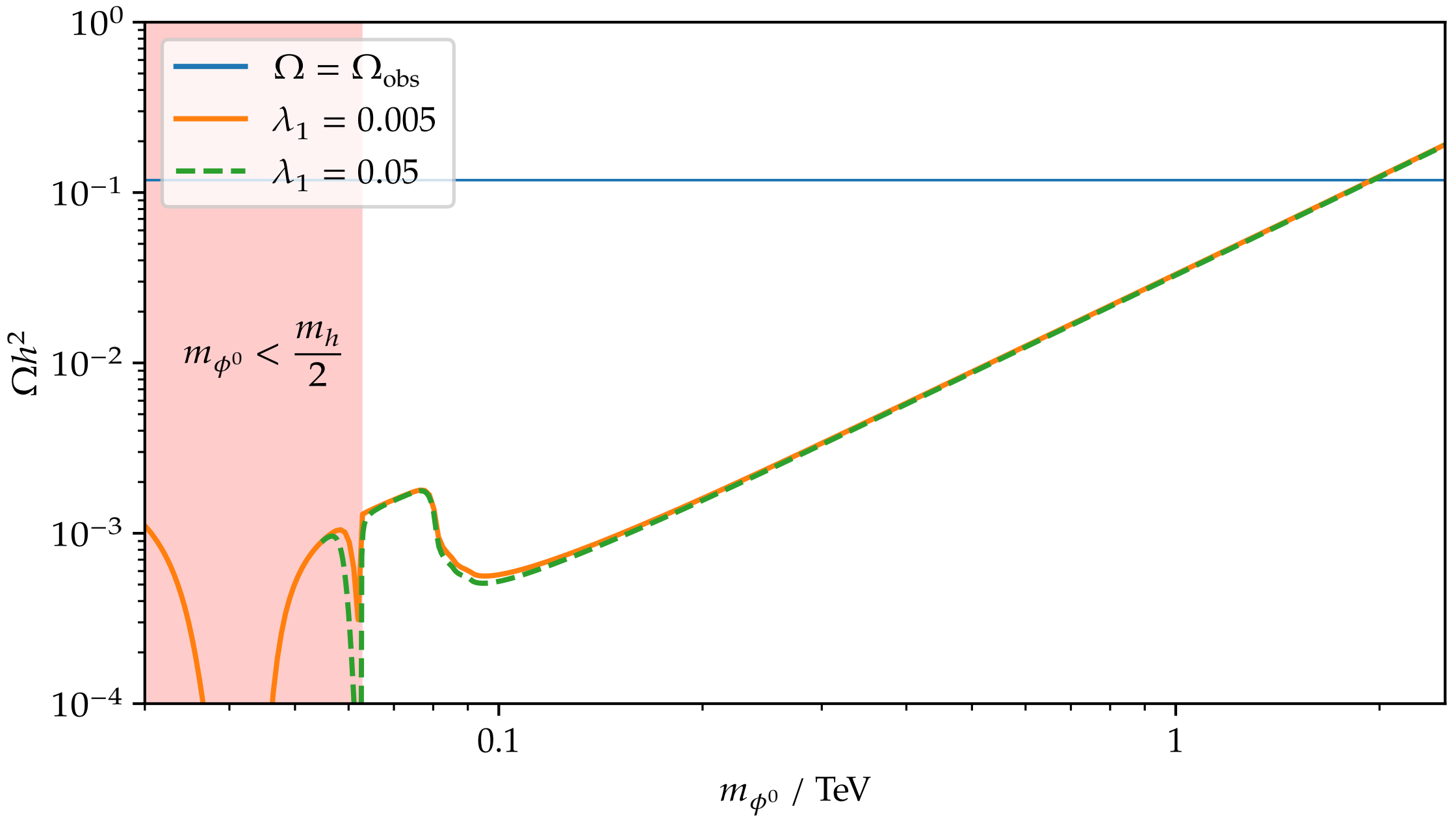

**Figure 7.7.:** The dark matter relic density $\Omega h^2$ as a function of the dark matter mass $m_{\phi^0}$ and some value of $\lambda_1$ (with $\lambda_3 = 0$) in the ITM limit. The blue band shows the region where $\Omega = \Omega_{obs}$ within the current uncertainty. (Higgs potential: $\lambda = 0.2612$.)

For low masses, there are some interesting features in fig. 7.7. Exactly as the mass $m_{\phi^0}$ reaches a value of half the Higgs mass, the relic density has a sharp minimum. A similar, but much wider minimum occurs for $40\,\text{GeV} \leq m_{\phi^0} \leq 45\,\text{GeV}$. These areas can clearly be identified with the resonances of the Higgs, $W$ and $Z$ bosons. At these mass values, annihilation to these bosons is greatly enhanced and the dark matter density is thus reduced drastically.

Confusingly, [Kha18] cites both [AGN11a] and [YF15] even though they found a WIMP mass which differs by 3.5 TeV and 5 TeV, respectively, from previous results. Again, no comment is made addressing these enormous discrepancies. The author was contacted concerning exactly this question, but unfortunately, there was no reply.

With [FB11], a final publication was found and compared to conclude the consideration of the ITM limit of T1-3-B ($\alpha = 0$). There, figure 5 shows one more view of the progression of the relic density with mass.

Figure 7.8 shows data which is analogous to [FB11, fig. 5]. Since the authors did not specify the used value of $\lambda_1$ (and, in any case, as seen before, it is negligible unless very large), stating only that it is "irrelevant", a value of $\lambda_1 = 0$ was assumed. Accounting for the updated value of the observed relic density in use here, which incorporates the findings of the Planck satellite [Planck16], the results agree once again with a WIMP mass of $m_{\phi^0} \approx 2$ TeV.

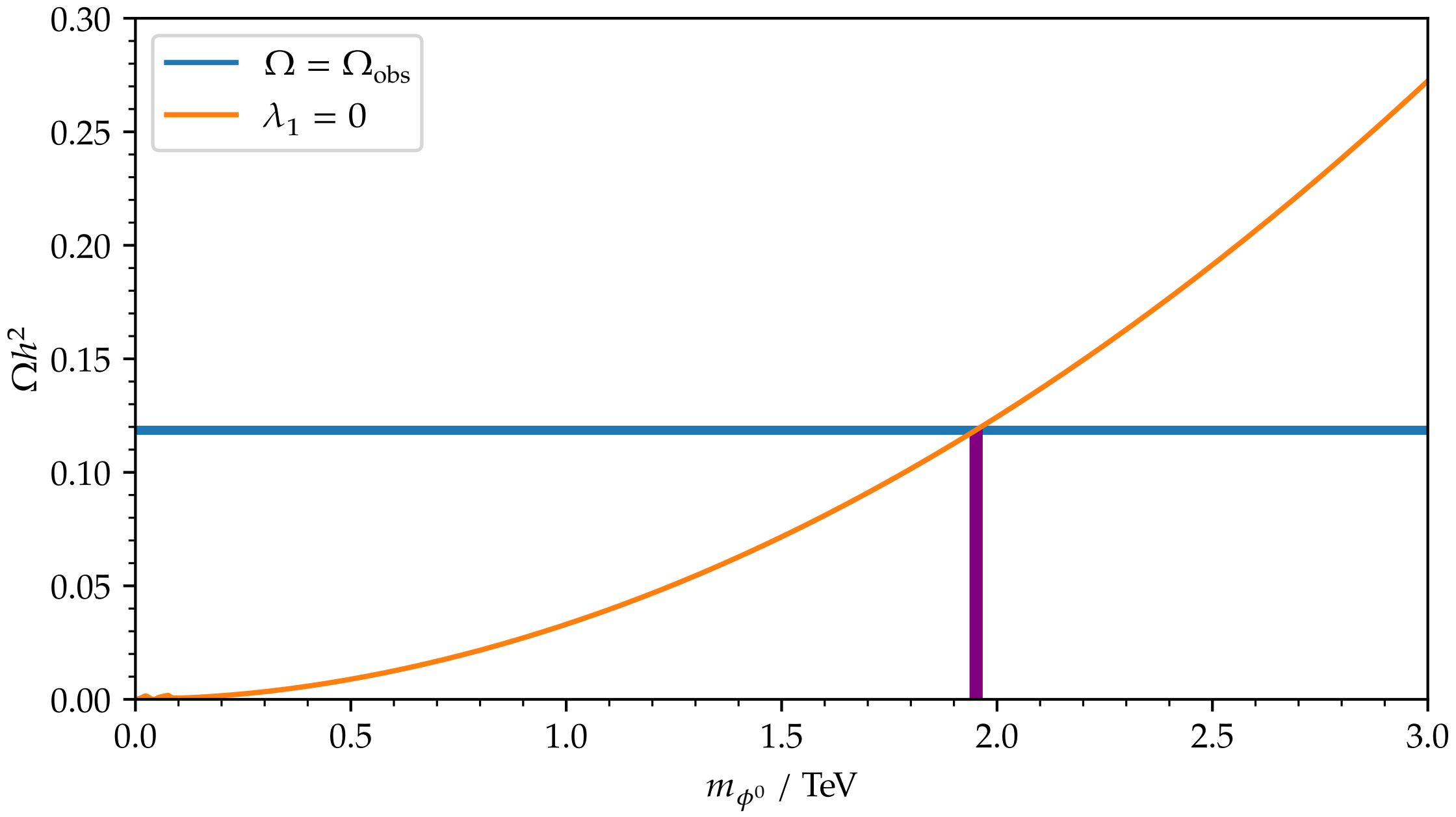

**Figure 7.8.:** The dark matter relic density $\Omega h^2$ as a function of the dark matter mass $m_{\phi^0}$ with $\lambda_1 = \lambda_3 = 0$ in the ITM limit. The blue band shows the region where $\Omega = \Omega_{obs}$ within the current uncertainty. (Higgs potential: $\lambda = 0.2612$.)

To summarize, the literature on the inert triplet model contains the following results:

- [AGN11a; AGN11b] with a WIMP mass of 5.5 TeV;
- [FB11] with a WIMP mass of $\approx 2$ TeV;
- [YF15] with a WIMP mass of 7 TeV;
- [Kha18] with a WIMP mass of 2 TeV.

Adding this work, again with 2 TeV, there seems little doubt that the results obtained here are correct, as they match the majority of publications. The results of [AGN11a; AGN11b] and [YF15], on the other hand, were never reproduced or confirmed independently. Regrettably, the root cause could not be identified in the end.

## 7.6. Note on the reproducibility of numerical results

Tools like `SARAH`, `SPheno` and `micrOMEGAs` are incredibly useful because they allow studying new models very rapidly, automating many repetitive steps that would be necessary when implementing a model "from scratch". However, a disadvantage is that they can cause one to lose sight of the lower-level conventions and implementation, which is often not discussed in detail in works using these higher-level tools; they are often treated as "black boxes" at this level of abstraction. Consequently, it can be very difficult, if not impossible, for someone else to reproduce results obtained in this manner.

As has become apparent in sections 7.4 and 7.5, even for the very simple models shown there, the literature contains some wildly conflicting results, with none of the authors being able to produce the model files or code used to obtain them. Contacting the authors and receiving either no reply or, arguably worse, learning that not even the authors themselves have access to the original files used in the computations was an unsettling experience. This is, of course, a distressing state of affairs, calling into question the very validity of the scientific results. In the interest of science, one hopes that most findings stand on firmer ground. Consequently, it is absolutely crucial for any work to specify exactly how numerical results were obtained, even though these computational tools can yield quick and easy results. In particular, any work should state precisely:

- Which tools were used to obtain any result.
- The *version numbers* of all tools that were used. Changes between different versions can be quite drastic in these codes and often contain numerous numerical corrections or updated experimental data.
- If the original, publicly available code was changed in any way, the corresponding *patches* must of course be included for there to be any chance for someone else to perform a valid comparison.
- The *input* that was used for these tools, such as *model files*, *SLHA files* and so on. Otherwise, as seen in section 7.4 and section 7.5, it is generally not possible to find the root cause of discrepancies between different authors.
- Ideally, system information such as the *compiler version*, *processor architecture* and *operating system*. Changes to compilers can especially cause significant differences, to the point of failure to compile at all.

- Finally, free tools and software should be used as much as possible so that anyone can verify the results.

Accordingly, this information for the present work shall be prominently summarized here:

- Tools: `SARAH`, `SPheno` and `micrOMEGAs` (see section 6.3).
- Version numbers: `SARAH` 4.12.3, `SPheno` 4.0.3 and `micrOMEGAs` 4.3.5.
- Modifications: A change had to be made to correct an error in `SARAH`'s `SPheno` output, see appendix F.
- Input files: See appendix G for the `SARAH` model files. The SLHA files used as input for `SPheno` were those generated by `SARAH` (the file with a name of the form `LesHouches.in.<model>_low`), with the following modifications:
  - Conventions (e. g. for the mixing matrices) were chosen to be compatible with `micrOMEGAs` ( `SPhenoInput` flags 50 and 77).
  - If radiative corrections were taken into account, `SPhenoInput` flags 55 (loop masses) and 77 (running SM parameters) were enabled.
  - The calculation of 3-body and loop decays as well as the output of a range of unused files and information in the spectrum file were disabled to improve the speed of computation.
  - Of course, the input values of the parameters were varied as specified in each case.

  The SLHA input files were processed using `pySLHA` [Buc15]. In `micrOMEGAs`, the flags `VWdecay = 0` and `VZdecay = 0` were set and the function for the calculation of the relic density, `darkOmega(&Xf, fast, Beps)`, was called with `fast = 1`, `Beps = 1e-5`.
- System information: `SARAH` was run on Mathematica 11.2.0.0. `C` and `Fortran` programs were compiled using `gcc/gfortran` version 7.2.0. The `Python` version used was mainly 3.6.3. All calculations were done on Linux 4.13.0 (`x86_64` processor).

## 7.7. Neutrino masses

### 7.7.1. Feynman rules for Weyl spinors

The most well-known version of Feynman diagrams and rules uses the four-component (Dirac bispinor) formalism. However, when a theory contains Majorana fermions, this formalism runs into problems because there is no unique "direction of fermion flow" that can be associated with these neutral fermions. Such ambiguities do not exist when using Feynman rules for Weyl spinors, as defined in [DHM10]. One of the main differences of this version of the Feynman rules compared to the four-component one is that the direction of the arrows on the line is not assigned depending on whether it corresponds to a particle or an anti-particle, but only signifies the handedness (left or right) of the spinors. To quote [DHM10, p. 39]:

> "In particular, the arrows indicate the spinor index structure, with fields of undotted indices flowing into any vertex and fields of dotted indices flowing out of any vertex."

For the purposes of calculating the radiative neutrino masses, only a small subset of these rules will be needed: The Yukawa vertices and the fermion propagators. In order to obtain them, however, the precise expression for the couplings must first be obtained from the Lagrangian (7.2). The Yukawa interactions enabling Majorana neutrino masses are given by the term

$$\begin{aligned}(\lambda_6)^{ij} L_i \phi_j \psi' + \text{H.c.} &= (\lambda_6)^{ij} \begin{pmatrix} e_\text{L} & -(\nu_\text{L})_i \end{pmatrix} \begin{pmatrix} \frac{\phi_j^0}{\sqrt{2}} & \phi_j^+ \\ \phi_j^- & -\frac{\phi_j^0}{\sqrt{2}} \end{pmatrix} \begin{pmatrix} \psi'^+ \\ \psi'^0 \end{pmatrix} + \text{H.c.} \\ &= (\lambda_6)^{ij} \left( (e_\text{L})_i \left( \frac{1}{\sqrt{2}} \phi_j^0 \psi'^+ + \phi_j^+ \psi'^0 \right) - (\nu_\text{L})_i \phi_j^- \psi'^+ + \frac{1}{\sqrt{2}} (\nu_\text{L})_i \phi_j^0 \psi'^0 \right) + \text{H.c.}\end{aligned} \tag{7.16}$$

Of these component terms, only the one with all fields neutral will contribute to the masses (see section 7.7.2).

However, it must also be considered that the propagating fields are not the interaction eigenstates $\phi^0$ and $\psi'^0$, but the mass eigenstates $\eta^0 = O_\eta \phi^0$ and $\chi = U_\chi(\Psi, \psi^0, \psi'^0)$. Accordingly, the expression above must be rewritten in terms of $\eta^0$

and $\chi$:

$$-\frac{1}{\sqrt{2}}(\lambda_6)^{ij}(\nu_{\mathrm{L}})_i\phi_j^0\psi'^0 + \mathrm{H.c.} = -\frac{1}{\sqrt{2}}(\lambda_6)^{ij}(U_\chi)^*_{k3}(O_\eta)_{lj}(\nu_{\mathrm{L}})_i\chi_k\eta_l^0 + \mathrm{H.c.} \tag{7.17}$$

The neutrino fields can remain as interaction eigenstates because they are massless before radiative corrections.

This leads to the following relevant Feynman rules (cf. [DHM10]):[10]

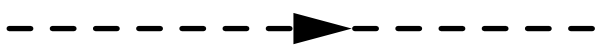

$$= \frac{i}{p^2 - m^2 + i\varepsilon} \tag{7.18}$$

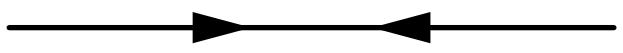

$$= \frac{im}{p^2 - m^2 + i\varepsilon}\delta^{\dot{a}}_{\dot{b}} \tag{7.19}$$

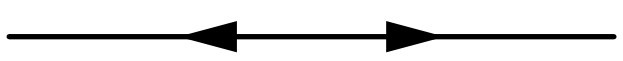

$$= \frac{im}{p^2 - m^2 + i\varepsilon}\delta^{b}_{a} \tag{7.20}$$

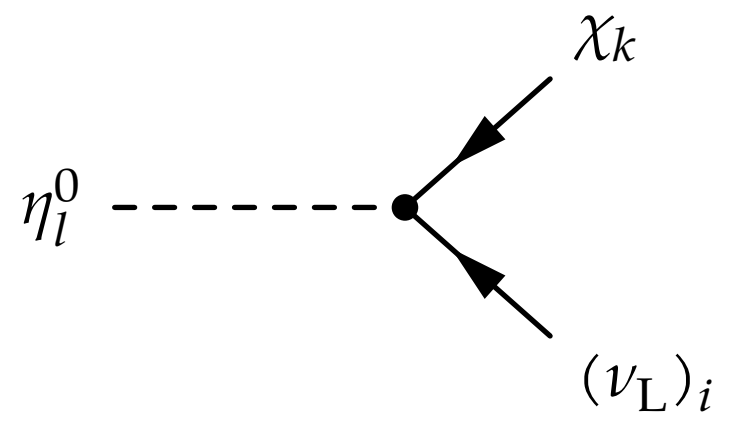

$$= -\frac{i}{\sqrt{2}}(\lambda_6)^{ij}(U_\chi)^*_{k3}(O_\eta)_{lj}\delta^b_a \tag{7.21}$$

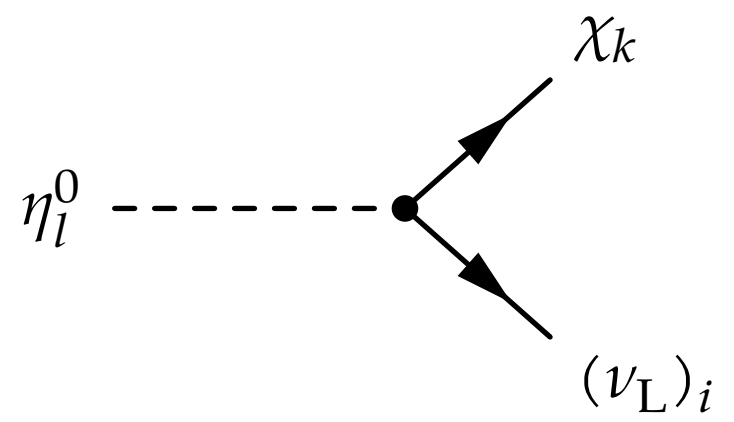

$$= -\frac{i}{\sqrt{2}}(\lambda_6^*)^{ij}(U_\chi)_{k3}(O_\eta)_{lj}\delta^{\dot{a}}_{\dot{b}} \tag{7.22}$$

with the field mass and momentum $m$ and $p$.

### 7.7.2. The neutrino mass matrix

The radiative corrections to the neutrino masses in T1-3-B ($\alpha = 0$) arise from the loop diagram shown in fig. 7.9. It should be noted that only the neutral fields contribute to the masses. Analogous diagrams involving charged fields instead only count towards the propagator correction because the fermion propagator exchanging the two Weyl components is not allowed in that case. Only the propagator preserving the direction of the arrow is possible. If this were not the case, then the neutrinos could already receive radiative masses in the Standard Model.

[10]The propagator for the scalar does not depend upon the convention for the fermions.

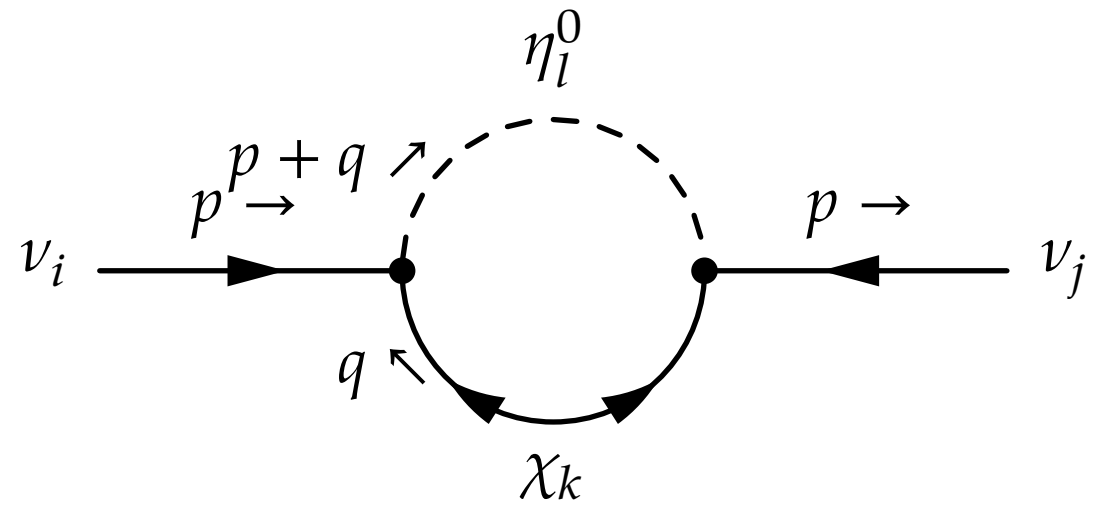

**Figure 7.9.:** Feynman diagram for the radiative contribution to the neutrino masses at the level of one loop.

Using the Feynman rules from the previous section, this diagram can be evaluated:

$$iM_{ij}^{kl} = \left( i\frac{1}{\sqrt{2}}(\lambda_6)^{im}(U_\chi)^*_{k3}(O_\eta)_{lm} \right) \int_{\mathbb{R}^4} \frac{1}{(2\pi)^4} \frac{im_{\chi_k}}{q^2 - m^2_{\chi_k} + i\varepsilon} \frac{i}{(p^2+q^2) - m^2_{\eta_l^0} + i\varepsilon} \mathrm{d}^4 q$$
$$\cdot \left( i\frac{1}{\sqrt{2}}(\lambda_6)^{jn}(U_\chi)^*_{k3}(O_\eta)_{ln} \right)$$

Identifying the loop integral with the scalar Passarino–Veltman integral $B_0$ yields [tV79]

$$= \frac{1}{2}\lambda_6^{im}\lambda_6^{jn}(O_\eta)_{ln}(O_\eta)_{lm}(U_\chi)^{*\,2}_{k3} m_{\chi_k}$$
$$\cdot \underbrace{\int_{\mathbb{R}^4} \frac{1}{(2\pi)^4} \frac{1}{\left(q^2 - m^2_{\chi_k} + i\varepsilon\right)\left((p^2+q^2) - m^2_{\eta_l^0} + i\varepsilon\right)} \mathrm{d}^4 q}_{=\frac{i}{16\pi^2}B_0\left(p^2, m^2_{\chi_k}, m^2_{\eta_l^0}\right)}$$
$$= \frac{i}{32\pi^2}\lambda_6^{im}\lambda_6^{jn}(O_\eta)_{ln}(O_\eta)_{lm}(U_\chi)^{*\,2}_{k3} m_{\chi_k} B_0(\underbrace{p^2}_{=0}, m^2_{\chi_k}, m^2_{\eta_l^0})$$
$$= \frac{i}{32\pi^2}\lambda_6^{im}\lambda_6^{jn}(O_\eta)_{ln}(O_\eta)_{lm}(U_\chi)^{*\,2}_{k3} m_{\chi_k} \tag{7.23}$$
$$\cdot \left( \Delta + 1 + \frac{m^2_{\chi_k}}{m^2_{\eta_l^0} - m^2_{\chi_k}} \ln\left(\frac{m^2_{\chi_k}}{\mu^2}\right) - \frac{m^2_{\eta_l^0}}{m^2_{\eta_l^0} - m^2_{\chi_k}} \ln\left(\frac{m^2_{\eta_l^0}}{\mu^2}\right) \right) \tag{7.24}$$

with $\Delta = \frac{1}{\varepsilon} - \gamma_\mathrm{E} + \ln(4\pi)$, Euler's constant $\gamma_\mathrm{E} = \lim_{n\to\infty} \sum_{k=1}^n \frac{1}{k} - \ln(n) \approx 0.5772$ and the renormalization scale $\mu$.

In order to obtain the total correction, all the neutral scalars and fermions must be summed over, yielding the neutrino mass matrix

$$
\begin{aligned}
(M_\nu)_{ij} &= \sum_{l=1}^{n_s}\sum_{k=1}^{3} M_{ij}^{kl} \\
&= \frac{1}{32\pi^2}\sum_{l=1}^{n_s}\lambda_6^{im}\lambda_6^{jn}(O_\eta)_{ln}(O_\eta)_{lm} \\
&\qquad \cdot \sum_{k=1}^{3}(U_\chi)_{k3}^{*\,2} m_{\chi_k}\left(\Delta + 1 + \frac{m_{\chi_k}^2}{m_{\eta_l^0}^2 - m_{\chi_k}^2}\ln\left(\frac{m_{\chi,k}^2}{\mu^2}\right) - \frac{m_{\eta_l^0}^2}{m_{\eta_l^0}^2 - m_{\chi_k}^2}\ln\left(\frac{m_{\eta_l^0}^2}{\mu^2}\right)\right) \\
&= \frac{1}{32\pi^2}\sum_{l=1}^{n_s}\lambda_6^{im}\lambda_6^{jn}(O_\eta)_{ln}(O_\eta)_{lm} \\
&\qquad \cdot \sum_{k=1}^{3}(U_\chi)_{k3}^{*\,2} m_{\chi_k}\left(\Delta + 1 + \frac{m_{\chi_k}^2}{m_{\eta_l^0}^2 - m_{\chi_k}^2}\ln\left(\frac{m_{\chi_k}^2}{m_{\eta_l^0}^2}\right) - \frac{m_{\eta_l^0}^2 - m_{\chi_k}^2}{m_{\eta_l^0}^2 - m_{\chi_k}^2}\ln\left(\frac{m_{\eta_l^0}^2}{\mu^2}\right)\right) \\
&= \frac{1}{32\pi^2}\sum_{l=1}^{n_s}\lambda_6^{im}\lambda_6^{jn}(O_\eta)_{ln}(O_\eta)_{lm}\left[\left(\Delta + 1 - \ln\left(\frac{m_{\eta_l^0}^2}{\mu^2}\right)\right)\sum_{k=1}^{3}(U_\chi)_{k3}^{*\,2} m_{\chi_k}\right. \\
&\qquad \left. + \sum_{k=1}^{3}(U_\chi)_{k3}^{*\,2} m_{\chi_k}\frac{m_{\chi_k}^2}{m_{\eta_l^0}^2 - m_{\chi_k}^2}\ln\left(\frac{m_{\chi_k}^2}{m_{\eta_l^0}^2}\right)\right]
\end{aligned}
$$

with $\sum_{k=1}^{3}(U_\chi)_{k3}^{*\,2} m_{\chi_k} = \sum_{k,k'=1}^{3}\left((U_\chi^{\mathrm{T}})_{3k}\,\mathrm{diag}(m_{\chi_1}, m_{\chi_2}, m_{\chi_3})_{kk'}(U_\chi)_{k'3}\right)^* = (M_0)_{33}^* = 0$:

$$
\begin{aligned}
&= \frac{1}{32\pi^2}\sum_{l=1}^{n_s}\lambda_6^{im}\lambda_6^{jn}(O_\eta)_{ln}(O_\eta)_{lm}\underbrace{\sum_{k=1}^{3}(U_\chi)_{k3}^{*\,2}\frac{m_{\chi_k}^3}{m_{\eta_l^0}^2 - m_{\chi_k}^2}\ln\left(\frac{m_{\chi_k}^2}{m_{\eta_l^0}^2}\right)}_{=A_l} \\
&= \frac{1}{32\pi^2}\sum_{l=1}^{n_s}A_l\lambda_6^{im}\lambda_6^{jn}(O_\eta)_{ln}(O_\eta)_{lm} \qquad (7.25)
\end{aligned}
$$

Any masses appearing in the expression are the masses without any corrections applied (tree level). As can be seen in this derivation, the divergences vanish in the neutrino mass matrix. This is, in fact, expected and required by the fact that T1-3-B ($\alpha = 0$) is renormalizable. The argument is that if there were any divergences, they would need to be canceled by counter terms during the renormalization procedure. However, there is no parameter for the neutrino masses because they do not exist at tree level. Consequently, if the radiative corrections to the neutrino masses were divergent, it would not be possible to cancel all divergences and the theory would not be renormalizable, contradicting the fact that it is, indeed, renormalizable.

For one generation of scalars, (7.25) simplifies to

$$(M_\nu)_{ij} = \frac{1}{32\pi^2}\lambda_6^i\lambda_6^j \sum_{k=1}^{3} (U_\chi)_{k3}^{*\,2} \frac{m_{\chi_k}^3}{m_{\phi^0}^2 - m_{\chi_k}^2} \ln\left(\frac{m_{\chi_k}^2}{m_{\phi^0}^2}\right) = \frac{A}{32\pi^2}\lambda_6^i\lambda_6^j \tag{7.26}$$

As discussed in section 5.3, the symmetric matrix $M_\nu$ can be diagonalized using a unitary matrix $U_\nu$, the neutrino mixing matrix, such that

$$\mathrm{diag}(m_{\nu_1}, m_{\nu_2}, m_{\nu_3}) = U_\nu^\mathsf{T} M_\nu U_\nu \tag{7.27}$$

$U_\nu$ is nothing but the Pontecorvo–Maki–Nakagawa–Sakata (PMNS) matrix.

Since the neutrino mass matrix is of the same form as for the model T1-3-A ($\alpha = 0$), it can be approximated at leading order in $\lambda_4, \lambda_5 \ll 1$ in the same way as in [Esc+18]. The mass matrix can then roughly be estimated for TeV-scale masses as

$$M_\nu \approx 100\,\mathrm{meV}\frac{M_\Psi}{1\,\mathrm{TeV}}\left(\frac{\lambda_6^{ij}\lambda_{4,5}}{10^{-5}}\right)^2 \tag{7.28}$$

showing that the product of the parameters $\lambda_{4,5}$ and $\lambda_6$ only has to be around $10^{-5}$ to obtain meV-scale neutrinos [Esc+18].

### 7.7.3. One generation of scalars

First, consider only one generation of the scalar triplet $\phi$. The neutrino mass matrix then has the form given in (7.26):

$$M_\nu = \frac{A}{32\pi^2}\begin{pmatrix} (\lambda_6^e)^2 & \lambda_6^e\lambda_6^\mu & \lambda_6^e\lambda_6^\tau \\ \lambda_6^\mu\lambda_6^\tau & (\lambda_6^\mu)^2 & \lambda_6^\mu\lambda_6^\tau \\ \lambda_6^e\lambda_6^\tau & \lambda_6^\mu\lambda_6^\tau & (\lambda_6^\tau)^2 \end{pmatrix} \tag{7.29}$$

where the components of $\lambda_6$ have been written as $\lambda_6^1 = \lambda_6^e, \lambda_6^2 = \lambda_6^\mu, \lambda_6^3 = \lambda_6^\tau$. It turns out, however, that a matrix of this form invariably leads to

$$U_\nu^\mathsf{T} M_\nu U_\nu = \mathrm{diag}(m_{\nu_1}, m_{\nu_2}, m_{\nu_3}) = \mathrm{diag}\left(0, 0, \frac{A}{32\pi^2}(\lambda_6)^i(\lambda_6)_i\right) \tag{7.30}$$

that is, only one of the neutrinos becomes massive.

Although this scenario of only one massive neutrino is not consistent with observations, it is nevertheless worthwhile to consider it as a consistency check. To this end, the neutrino mass $m_{\nu_3}$ was calculated in two different ways: First, the tool chain which has been tried and tested in the previous sections was used, this time enabling loop corrections, to obtain the neutrino masses for a given set of parameters using SPheno. Then, for comparison, the fermion mass matrix (7.8) was diagonalized numerically, yielding $m_{\chi_k}$ and $U_\chi$, allowing for the calculation of the neutrino mass matrix (7.26), which was in turn also diagonalized numerically. In the end, this procedure should of course give the same neutrino mass as SPheno.

Figure 7.10 shows the result of this comparison. As expected, the curve shows a parabola, indicating the quadratic dependence of $m_{\nu_3}$ on $\lambda_6$. In addition, when all of the $\lambda_6^i$ are equal, the mass reaches zero for $\lambda_6 = 0$. The displayed agreement between SPheno and the self-implemented computation is excellent. This is a great confirmation of the stated results.

### 7.7.4. Two generations of scalars

In order to confirm that T1-3-B ($\alpha = 0$) can accommodate the observed neutrino mass differences, the model was also implemented with two generations of the scalar triplet (see appendix G.2). There were some minor complications due to the fact that the quartic coupling in the term $\lambda_3^{ijkm} \operatorname{Tr}(\phi^i \phi^j \phi^k \phi^m)$ has four indices, which cannot directly be handled by SARAH. However, as stated before, $\lambda_3$ is mostly irrelevant to observables and thus simply omitted ($\lambda_3 = 0$) for this case.

For two scalar generations, the coupling $\lambda_6$ gets a second index running from 1 to 2, so it can be interpreted as $3 \times 2$ matrix

$$\lambda_6 = \begin{pmatrix} \lambda_6^{e1} & \lambda_6^{e2} \\ \lambda_6^{\mu 1} & \lambda_6^{\mu 2} \\ \lambda_6^{\tau 1} & \lambda_6^{\tau 2} \end{pmatrix} \tag{7.31}$$

For simplicity, assume that the scalars do not mix ($O_\eta = \mathbb{1}$). The neutrino mass

**Table 7.2.:** The input parameters used in fig. 7.10. All components of $\lambda_6$ were set to the same value. Note that $\lambda_3$ does not contribute to the neutrino masses.

| Parameter | Values |
|---|---|
| $M_\phi, M_\Psi, M_{\psi\psi'}$ | 1 TeV |
| $\lambda_1, \lambda_3$ | 0.1 |
| $\lambda_4, \lambda_5$ | 0.01 |
| $\lambda_6$ | $[-0.2, 0.2]$ |

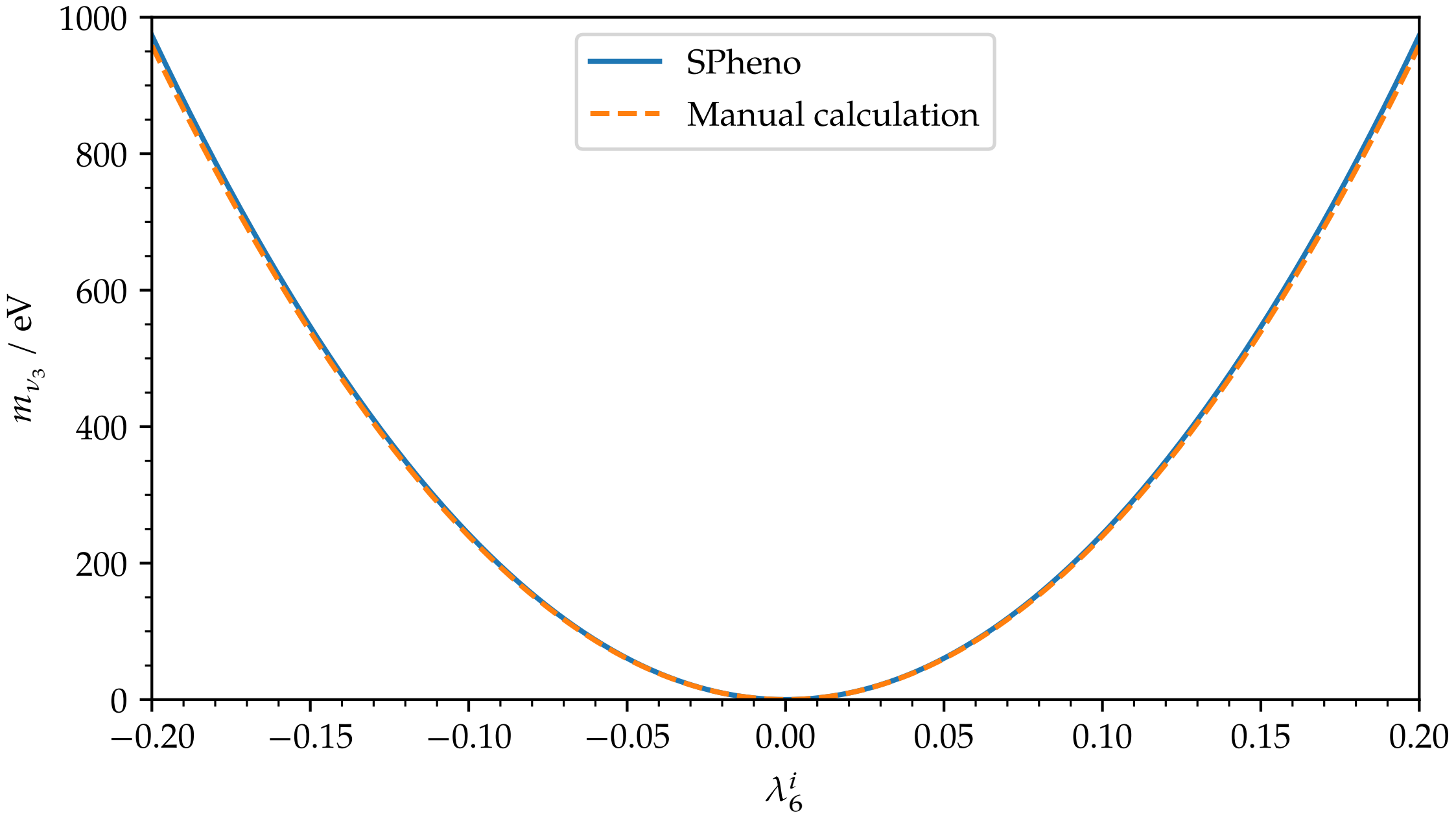

**Figure 7.10.:** The neutrino mass $m_{\nu_3}$ for the model T1-3-B ($\alpha = 0$) with one generation of scalar triplets as a function of $\lambda_6$, calculated both manually and using `SPheno`. The input parameters can be found in table 7.2.

matrix (7.25) then becomes

$$M_\nu = \frac{1}{32\pi^2} \sum_{l=1}^{n_s} A_l \lambda_6^{il} \lambda_6^{jl}$$
$$= \frac{1}{32\pi^2} \left( A_1 \begin{pmatrix} (\lambda_6^{e1})^2 & \lambda_6^{e1}\lambda_6^{\mu 1} & \lambda_6^{e1}\lambda_6^{\tau 1} \\ \lambda_6^{\mu 1}\lambda_6^{\tau 1} & (\lambda_6^{\mu 1})^2 & \lambda_6^{\mu 1}\lambda_6^{\tau 1} \\ \lambda_6^{e1}\lambda_6^{\tau 1} & \lambda_6^{\mu 1}\lambda_6^{\tau 1} & (\lambda_6^{\tau 1})^2 \end{pmatrix} + A_2 \begin{pmatrix} (\lambda_6^{e2})^2 & \lambda_6^{e2}\lambda_6^{\mu 2} & \lambda_6^{e2}\lambda_6^{\tau 2} \\ \lambda_6^{\mu 2}\lambda_6^{\tau 2} & (\lambda_6^{\mu 2})^2 & \lambda_6^{\mu 2}\lambda_6^{\tau 2} \\ \lambda_6^{e2}\lambda_6^{\mu 2} & \lambda_6^{\mu 2}\lambda_6^{\tau 2} & (\lambda_6^{\tau 2})^2 \end{pmatrix} \right) \tag{7.32}$$

In general, this matrix has

$$U_\nu^\mathsf{T} M_\nu U_\nu = \mathrm{diag}(m_{\nu_1}, m_{\nu_2}, m_{\nu_3})$$
$$= \frac{1}{64\pi^2} \mathrm{diag}\Big(0, \sum_{l=1}^{n_s} A_l (\lambda_6)^{il} (\lambda_6)_{il} - \sqrt{B},$$
$$\sum_{l=1}^{n_s} A_l (\lambda_6)^{il} (\lambda_6)_{il} + \sqrt{B}\Big) \tag{7.33}$$

upon diagonalization, where $B$ is a long polynomial in the elements $\lambda^{il}$ and the factors $A_l$. This shows that there is now a mass splitting $\Delta m_{32} = 2\sqrt{B}$ between $m_{\nu_3}$ and $m_{\nu_2}$. The lightest neutrino remains massless. For all three neutrinos to obtain a mass, three generations of the scalars would be required.

Assuming a normal hierarchy, the observed squared mass differences of the neutrinos are (cf. (2.8) and (2.9))

$$\Delta m_{21}^2 = (7.53 \pm 0.18) \times 10^{-5}\,\mathrm{eV}^2 \qquad |\Delta m_{32}^2| = (2.44 \pm 0.06) \times 10^{-3}\,\mathrm{eV}^2 \tag{7.34}$$

Even though only the mass differences are known, by enforcing $m_{\nu_1} = 0$, the other two neutrino masses are also fixed:

$$m_{\nu_2} = \sqrt{\Delta m_{21}^2} = 8.68\,\mathrm{meV} \qquad m_{\nu_3} = \sqrt{\Delta m_{32}^2 + m_{\nu_2}^2} = 50.2\,\mathrm{meV} \tag{7.35}$$

In order to give an example of how the model can match the experimental data, consider a simplified case reducing the parameters $\lambda_6^{il}$ to only three different ones. This is the minimum needed to obtain more than one non-zero diagonal value. Setting

$$\lambda^{i1} = a_1 \qquad \lambda^{i2} = a_2 \qquad \lambda^{3l} = b \qquad (i \neq 3) \tag{7.36}$$

the mass matrix becomes

$$M_\nu = \frac{1}{32\pi^2} \sum_{l=1}^{2} A_l \begin{pmatrix} a_l^2 & a_l^2 & a_l b \\ a_l^2 & a_l^2 & a_l b \\ a_l b & a_l b & b^2 \end{pmatrix} \tag{7.37}$$

Figure 7.11 then shows the two neutrino masses as a function of $a_2 = \lambda_6^{i2}$ with the input shown in table 7.3, including a point in agreement with the observed masses. The behavior of the masses is such that the larger $m_{\nu_3}$ still depends roughly quadratically on the couplings $\lambda_6$, while the smaller $m_{\nu_2}$ saturates for larger $\lambda_6$ with practically no variation after a certain point. Looking a the parameters used in table 7.3, it becomes apparent that it is not difficult to obtain the correct masses, even in this simplified case, by manually exploring the parameter space. There is no extreme fine-tuning of parameters and the couplings $\lambda_4$, $\lambda_5$ and $\lambda_6^{ij}$ are small, of the order $10^{-4}$ to $10^{-2}$, but not "unnaturally" tiny.

**Table 7.3.:** Input parameters used in fig. 7.11. Note that $\lambda_3$ does not contribute to the neutrino masses.

| Parameter | Values |
|---|---|
| $M_\phi$ | $\text{diag}(1.4\,\text{TeV}, 3\,\text{TeV})$ |
| $M_\Psi$ | $1\,\text{TeV}$ |
| $M_{\psi\psi'}$ | $1.5\,\text{TeV}$ |
| $\lambda_1, \lambda_3$ | $0$ |
| $\lambda_4, \lambda_5$ | $5 \times 10^{-3}$ |
| $\lambda_6^{i1}$ $(i \neq 3)$ | $3 \times 10^{-4}$ |
| $\lambda_6^{i2}$ $(i \neq 3)$ | $[10^{-3}, 10^{-2}]$ |
| $\lambda_6^{3l}$ | $3.05 \times 10^{-3}$ |

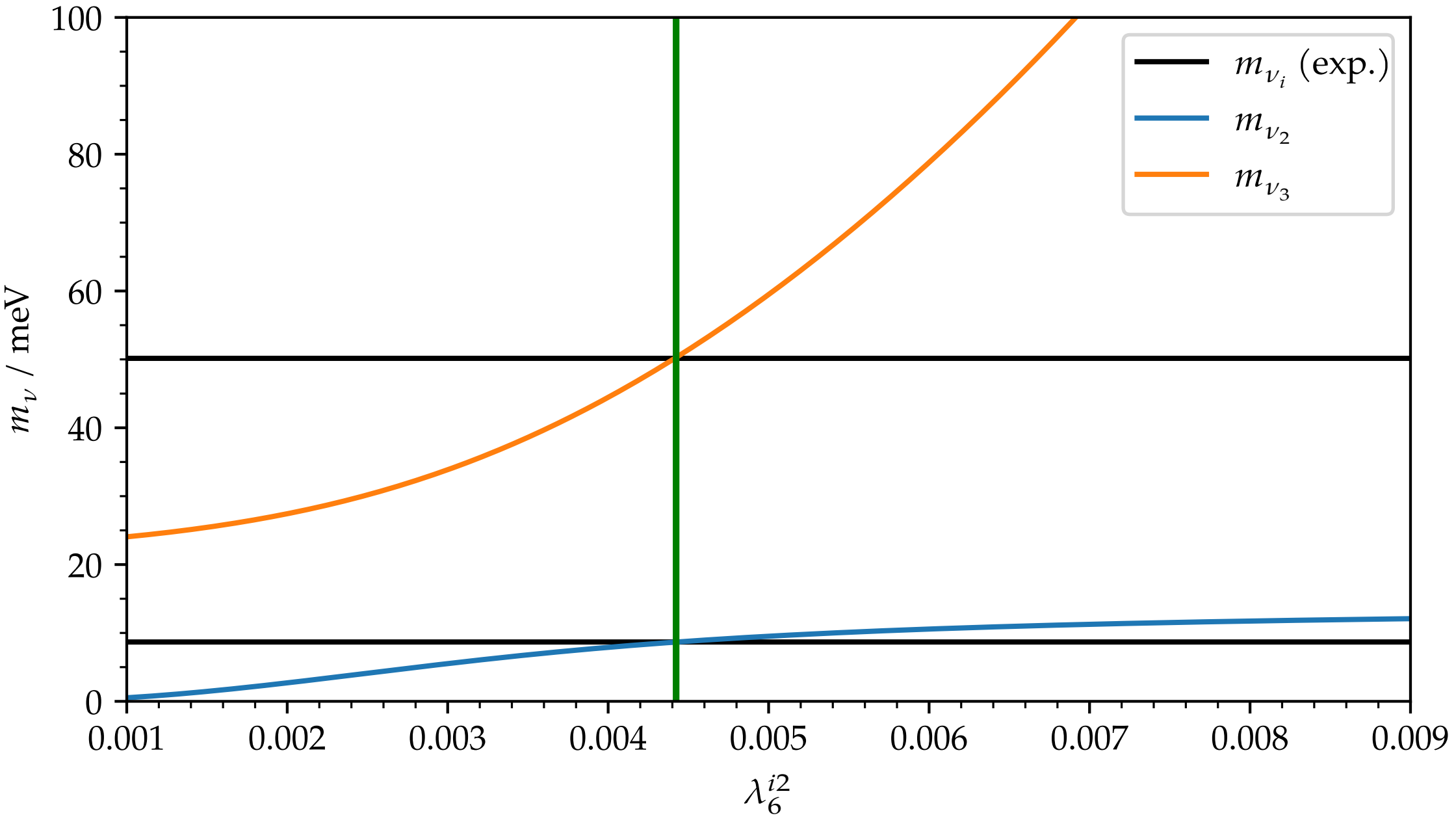

**Figure 7.11.:** The neutrino masses $m_{\nu_i}$ for the model T1-3-B ($\alpha = 0$) with two generations of scalar triplets as a function of $\lambda_6^{i2}$. The input parameters can be found in table 7.3. The experimentally determined neutrino masses for this scenario are shown in black. The correct values are reached for $\lambda_6^{i2} = 0.0044$.

## 7.8. Dark matter with massive neutrinos

Finally, to conclude this look at the properties of T1-3-B ($\alpha = 0$), all the previous cases shall be combined, allowing both scalar and fermion dark matter while also generating neutrino masses through loop corrections. For simplicity, the following will return to one scalar generation. The goal, at this point, is not to match all constraints simultaneously, but to show how the Yukawa coupling $\lambda_6$, which is responsible for neutrino masses, influences the dark matter relic density. It was not possible to see this in the previous cases (SDF, ITM) because $\lambda_6$ only appears when both fermions and scalars are present.

First, consider the dependence of the relic density on $\lambda_6$ for fixed masses, shown in fig. 7.12. For small values, coannihilation of the odd particles via leptons using the Yukawa coupling is not possible and other channels, such as annihilation to gauge bosons, dominate. As the Yukawa coupling increases, however, the new channel opens up and, as it gains significance, it reduces the relic density. However, this effect eventually saturates and for large $\lambda_6$, $\Omega h^2$ is again only weakly affected by changes to the coupling.

**Table 7.4.:** Input parameters used in fig. 7.12.

| Parameter | Fermion DM | Scalar DM |
|---|---|---|
| $M_\phi$ | 1.5 TeV | 2 TeV |
| $M_\Psi$ | 1.2 TeV | 2.2 TeV |
| $M_{\psi\psi'}$ | 1.2 TeV | 2.2 TeV |
| $\lambda_1, \lambda_3, \lambda_4, \lambda_5$ | $10^{-2}$ | $10^{-2}$ |
| $\lambda_6^i$ | $[0, 0.8]$ | $[0, 0.8]$ |

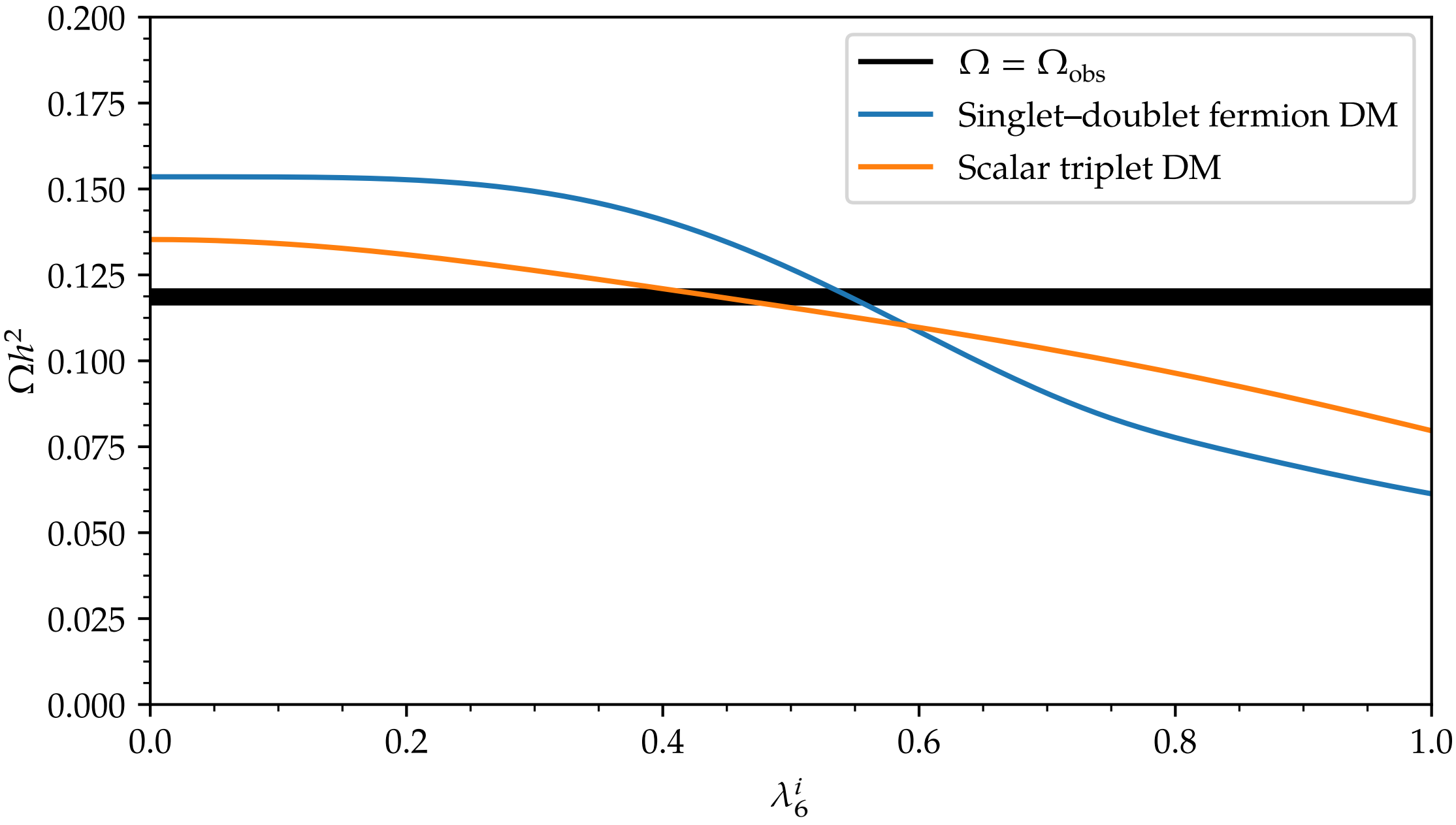

**Figure 7.12.:** The dark matter relic density $\Omega h^2$ as a function of the Yukawa coupling $\lambda_6$ for the cases of scalar ($m_{\phi^0} < m_{\chi_1}$) and fermion ($m_{\chi_1} < m_{\phi^0}$) dark matter. The input parameters for both cases can be found in table 7.4.

However, from the experimental point of view, one is most interested in how the observables vary with the mass $m_{\text{DM}}$ of the dark matter candidate. This is shown in fig. 7.13, demonstrating how different values of $\lambda_6$ affect the relic density. With the parameters used here, small changes in $\lambda_6$ do not noticeably impact $\Omega h^2$. Only for $\lambda_6^i > 0.1$ does this coupling become relevant.

If the dark matter is fermionic, $\lambda_6$ can influence the trend of $\Omega h^2$ quite significantly for large enough values. For $\lambda_6^i = 0.8$, the relic density even eventually flattens out instead rising continuously with increasing WIMP mass. Contrarily, the relic density of triplet dark matter is barely influenced by $\lambda_6^i$ at all. Even for large values, the main annihilation channels are processes to gauge bosons, such as $\phi^0\phi^0 \to W^+W^-$. Only for masses approaching the heavier fermions ($m_{\phi^0} > 2.5\,\text{TeV}$) do the Yukawa couplings effect any appreciable difference. This mirrors the results in fig. 7.12, where the relic density varies much slower in the scalar triplet case.

**Table 7.5.:** Input parameters used in fig. 7.13.

| Parameter | Fermion DM | Scalar DM |
|---|---|---|
| $M_\phi$ | 3 TeV | [0.5 TeV, 3.5 TeV] |
| $M_\Psi$ | [0.5 TeV, 2.8 TeV] | 3 TeV |
| $M_{\psi\psi'}$ | $= M_\Psi$ | 2.8 TeV |
| $\lambda_1, \lambda_3, \lambda_4, \lambda_5$ | $10^{-2}$ | $10^{-2}$ |
| $\lambda_6^i$ | 0.1, 0.5 and 0.8 | 0.1, 0.5 and 0.8 |

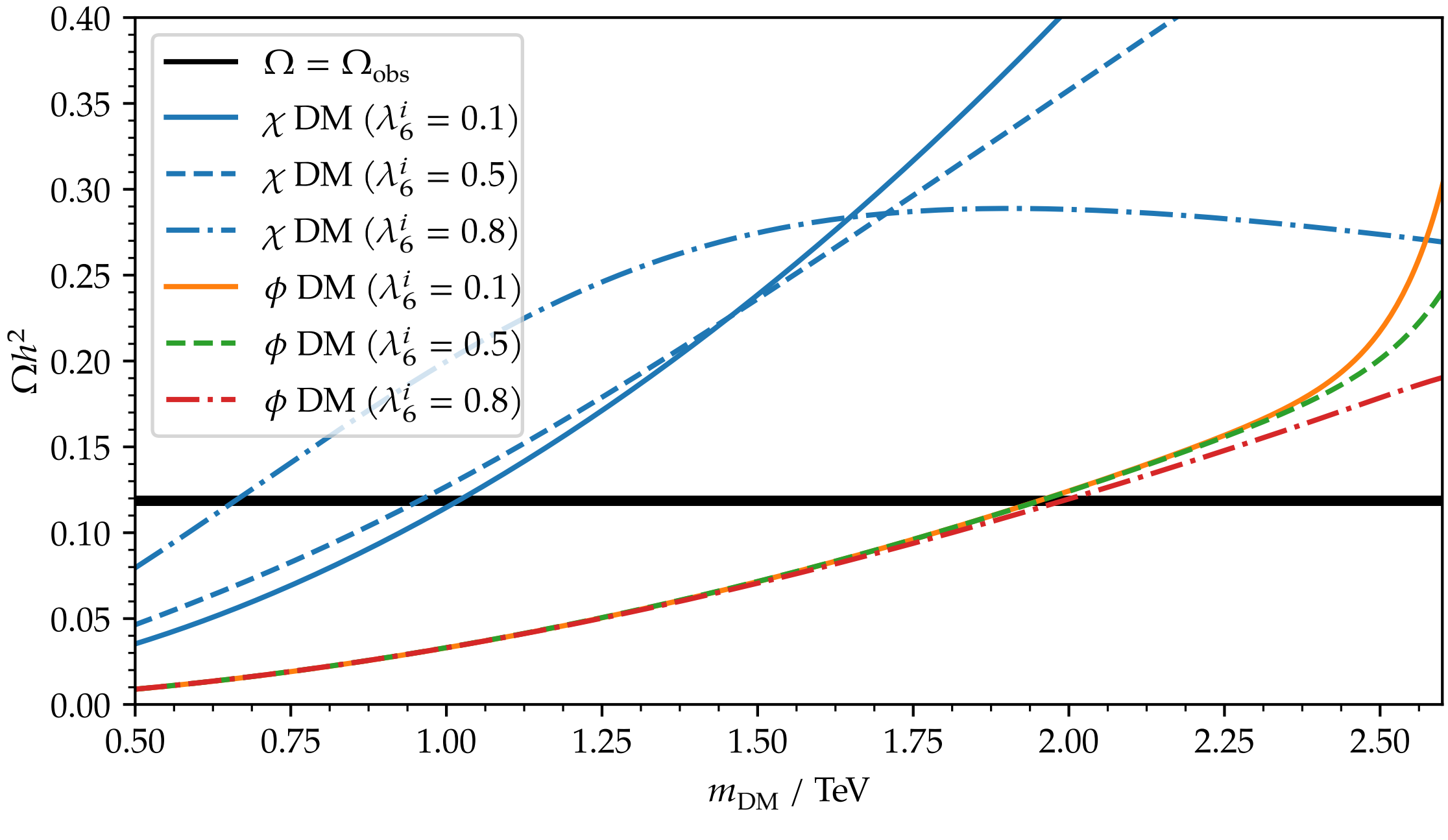

**Figure 7.13.:** The dark matter relic density $\Omega h^2$ as a function of the WIMP mass for the cases of scalar ($m_{\phi^0} < m_{\chi_1}$) and fermion ($m_{\chi_1} < m_{\phi^0}$) dark matter. Different values of the Yukawa coupling $\lambda_6$ are displayed. The input parameters for both cases of DM can be found in table 7.5.

# 8 Conclusion and outlook

In this thesis, the minimal dark matter models with radiative neutrino masses introduced in [RZY13] were examined both in general and, using the model T1-3-B ($\alpha = 0$) as a representative, in more detail. In the process, it was both shown how such models are built and a numerical analysis tool chain was established, automating as many steps as possible from the very beginning (deriving the Lagrangian from a given field content) to the very end (calculating observables such as the dark matter relic density, the direct detection cross section and the neutrino masses and mixing matrices).

First, the mathematical groundwork was laid out, describing how to deal with fermions (chapter 3) and $\mathrm{SU}(2)$ as a gauge group (chapter 4). As an important result, a convention for representations of $\mathrm{SU}(2)$ was established which, in this form, is not found in previous works. Further, in chapter 5, aside from reviewing the foundations of dark matter and neutrino mass models, methods for working with Lagrangians (sections 5.3 and 5.4.1) were presented and an overview of previously-studied models compiled, serving as a guideline for future research. All these methods and results are invaluable when engaging in model building, allowing one to swiftly and reliably treat the present class of models and many others beyond the Standard Model of particle physics on an analytical level. Even the Standard Model itself can become simpler to handle and interpret using these techniques.

Then, the capabilities and implementation of the program `minimal-lagrangians` were described in chapter 6. Its development is a major result of this thesis; it is of enormous use both for the construction of Lagrangians for new models as well their verification or quick surveys. It goes hand in hand with the previous results, making full use of the conventions established therein. The benefit of such automation must hardly be explained: Manual implementation of a model's Lagrangian, starting from the field content, is time-consuming and error-prone. Due to the large freedom with respect to how parameters can be defined (since arbitrary constant factors can be extracted), a consistent way of making such definitions greatly increases the ease of comparing different works. Furthermore, `minimal-lagrangians` can be applied to *any* model within the class it is designed to handle, extending its utility far beyond a one-time result that is only obtained for a single model.

Finally, the first steps in the analysis of the model T1-3-B ($\alpha = 0$) were taken in chapter 7. This work and its findings are completely novel since this model has not been studied at all before in the scientific literature. It was shown that the model can reproduce both the singlet–doublet fermion dark matter model and the inert triplet model in the appropriate limits. Moreover, the neutrino mass matrix was found and it was demonstrated that the model can accommodate current constraints on the neutrino masses. However, it was revealed that this is only possible with two generations of the scalar $SU(2)$ triplet. On the other hand, this is also the number of generations which allows this model to enable gauge coupling unification,[1] a potential hint towards its merit. Lastly, the behavior of the dark matter relic density as it depends on the parameter responsible for the generation of neutrino masses was investigated. In the course of this work, the foundations for a full analysis were laid. The model files and code used in this work are ready to be used for a more in-depth inspection.

Unfortunately, there were also some rather unedifying outcomes, discussed in section 7.6. On the one hand, automation, which was praised before, can be a great boost to research, not only in terms of efficiency, but also in correctness, as once there is a thoroughly-vetted general tool which performs the details of model implementation correctly, all users of this tool benefit from the same known-good basis. On the other hand, it must be mentioned again that treating such tools as black boxes without concern for their implementation and the details or limits of their capabilities carries great danger. It can be easy to lose sight of all the task that they really perform and what implications this brings. That is why it is of the utmost importance to ensure that others can reproduce the results by providing the model and input files the code was run with. When implementing numerical algorithms oneself, this is not even up for discussion, since others must be able to verify the results for them to have any worth. The simple consequence is that the aforementioned information cannot be omitted simply because someone else wrote the code. To do otherwise is to preclude any chances of determining the cause in the case of conflicting results, as the discrepancies could be due to anything from the use of different conventions to plain incorrect implementation.

As for the future, there are many different ways to build on the work in this thesis. For one, more features could be added to `minimal-lagrangians`. For example, the

[1] see [Hag+16]

identities shown in section 5.4 could be used to eliminate redundant terms from the generated Lagrangians or the output of `SARAH` model files could be automated even further (e. g. defining the mixing and Dirac spinors or generating the file `SPheno.m`). On the other hand, with more definite *applications* of the results in mind, `minimal-lagrangians` and the computational tool chain put to use in chapter 7 provide a great starting point for studying any of the 35 viable models from [RZY13].

As mentioned above, a very direct consequence prepared for by this thesis is the further study of T1-3-B ($\alpha = 0$). The immediate next step is to perform the Casas–Ibarra parametrization [CI01] in order to use the neutrino masses as input parameters instead of having them be the result of a particular choice of parameters. If the aim is to identify the areas of the parameters space compatible with *all* current constraints, inordinate amounts of time and computational power are wasted without this re-parametrization on points which have the wrong neutrino observables. Once this is done, observables like the dark matter relic density and direct detection constraints could be combined instead of looking at each individually. Finally, additional constraints can be taken into consideration, such as lepton flavor violation or the Higgs mass, which can have considerable corrections from the new fields. In fact, it is planned to continue this work in a future publication.

# A Notation and conventions

Throughout this work, the Einstein summation convention is employed unless noted otherwise, with a sum over any two raised and lowered indices. The position of the index does not have, a priori, any numerical meaning and is mostly used for the summation convention and to emphasize co- or contravariant transformation behavior. In the cases of Lorentz vectors, Weyl spinors and SU(2) multiplets, the index position does make a difference, however.

The signature of the metric tensor $\eta^{\mu\nu}$ is chosen to be $\eta = \mathrm{diag}(1, -1, -1, -1)$, i. e.

$$(\eta^{\mu\nu}) = (\eta_{\mu\nu}) = \begin{pmatrix} 1 & 0 & 0 & 0 \\ 0 & -1 & 0 & 0 \\ 0 & 0 & -1 & 0 \\ 0 & 0 & 0 & -1 \end{pmatrix} \tag{A.1}$$

With this, contravariant and covariant four-vectors are objects of the form

$$(A^\mu) = (A^0, \vec{A}) \quad \text{and} \quad (A_\mu) = (A_0, -\vec{A}) \tag{A.2}$$

respectively, where the symbols with the arrows represent three-component vectors. Accordingly, an inner product between two vectors $x, y$ in Minkowski space is defined as

$$x \cdot y = x^\mu y_\mu = x^\mu \eta_{\mu\nu} y^\mu \tag{A.3}$$

Index conventions for spinors and representations of SU(2) are defined in section 3.3 and chapter 4.

Physical quantities are given in natural units, i. e. velocities are expressed in units of the speed of light in vacuum $c$ and action or angular momenta are given in units of the reduced Planck constant $\hbar$. The notation employed is then

$$c = 1$$
$$\hbar = 1$$

with the physical dimension implicit in the seemingly dimensionless expressions on the right-hand side. The normalization for the hypercharge was chosen such that

$$Q = T_3 + \frac{Y}{2} \tag{A.4}$$

(Gell-Mann–Nishijima formula) where $T_3$ is an eigenvalue of the third SU(2) generator.

In Feynman diagrams, time is taken to be increasing towards the right on the horizontal axis while space increases upwards on the vertical axis. The usual conventions are followed: Fermions are represented by solid lines with arrows, scalar particles by dashed lines and electroweak gauge bosons by wavy lines.

There are different conventions about how the basis of the Lie algebra, its structure constants and the exponential map from the Lie algebra to the Lie group are defined. These arise from the interpretation of the generators as observables in quantum mechanics (in the case of SU(2): angular momentum), which must be Hermitian so that they have real eigenvalues. Consequently, a factor of $i$ is inserted both in the commutation relation and in the exponential map to make this work. In this thesis, the "mathematician's" convention without inserting any additional factors will be used, i. e.

- for a basis $\{X_i\}$ of the Lie algebra, the structure constants are the numbers $f^{ij}{}_k$ with $[X^i, X^j] = f^{ij}{}_k X^k$ and
- for any element $X$ of the Lie algebra, the corresponding group element is given by $g = e^X$.

# Common mathematical objects

B

This is a list of conventional names and symbols for commonly-used mathematical objects and operations. They are listed here for either as a reminder if they are always denoted in the same way or for completeness if they are not defined in the main text. When a symbol is not defined otherwise, its meaning is to be taken from this list. Many of these definitions can be found, for example, in [Hal15].

| Symbol | Description |
|---|---|
| $\mathbb{N}, \mathbb{Z}, \mathbb{R}, \mathbb{C}$ | the sets of natural numbers (including 0), integers, real numbers and complex numbers |
| $e, \pi, i$ | Euler's number, the number $\pi$ and the imaginary unit with $i^2 = -1$ |
| $\mathbb{1}, \mathbb{1}_n$ | the $n \times n$ identity matrix ($n$ can be omitted) with $\mathbb{1}^i_j = \delta^i_j$ |
| $\eta, \eta_{\mu\nu}$ | the Minkowski metric: $\eta = \mathrm{diag}(1, -1, -1, -1)$ |
| $\mathrm{diag}(A_1, \ldots, A_n)$ | the block matrix $A_1 \oplus \ldots \oplus A_n$; if the $A_i$ are square matrices, the result is block-diagonal; if the $A_i$ are numbers, this is the $n$-dimensional diagonal matrix with diagonal entries $A_i$, i. e. $\mathrm{diag}(A_1, \ldots, A_n)^i_j = A_i \delta^i_j$ (no sums) |
| $\delta^\mu_\nu$ | the Kronecker delta |
| $\varepsilon_{i_1 i_2 \ldots i_n}, \varepsilon^{i_1 i_2 \ldots i_n}$ | the $n$-dimensional Levi-Civita symbol |
| $\sigma^i, i \in \{1, 2, 3\}$ | the Pauli matrices $$\sigma^1 = \begin{pmatrix} 0 & 1 \\ 1 & 0 \end{pmatrix} \quad \sigma^2 = \begin{pmatrix} 0 & -i \\ i & 0 \end{pmatrix} \quad \sigma^3 = \begin{pmatrix} 1 & 0 \\ 0 & -1 \end{pmatrix}$$ |
| $\vec{\sigma}$ | the "Pauli vector" $\vec{\sigma} = (\sigma^1, \sigma^2, \sigma^3)^\mathsf{T}$ |
| $\sigma^\mu, \bar{\sigma}^\mu$ | the sigma matrices $(\sigma^\mu) = (\mathbb{1}_2, \vec{\sigma})$, $(\bar{\sigma}^\mu) = (\mathbb{1}_2, -\vec{\sigma})$ (sometimes called the generalized Pauli matrices) |
| $\gamma^\mu, \gamma_5$ | the Dirac matrices, which satisfy the Clifford algebra $\mathrm{Cl}_{1,3}(\mathbb{R})$: $\{\gamma^\mu, \gamma^\nu\} = 2\eta^{\mu\nu}\mathbb{1}_4$ with $\gamma_5 = i\gamma^0\gamma^1\gamma^2\gamma^3$ |

| Symbol | Description |
|---|---|
| $\partial_\mu$ | the partial derivative with respect to $x^\mu$: $\partial_\mu = \frac{\partial}{\partial x^\mu}$ |
| $D_\mu$ | the covariant derivative |
| $a \sim b$ | either "$a$ is proportional to $b$" or (if explicitly stated) "$a$ and $b$ are equivalent with respect to the equivalence relation $\sim$" |
| $\det(A)$ | the determinant of a matrix $A$ |
| $\mathrm{Tr}(A)$ | the trace of the matrix $A$: $\mathrm{Tr}(A) = A^i_i$ |
| $A^*$ | the complex conjugate of $A$ |
| $A^\mathsf{T}$ | the transpose of $A$ |
| $A^\dagger$ | the adjoint (Hermitian conjugate) of a linear operator, matrix or vector $A$ (for matrices or vectors, this is the same as the complex conjugate transpose; for numbers, this is simply the complex conjugate) |
| $[A, B]$ | commutator $[A, B] = AB - BA$ (or Lie bracket) of the linear operators $A$ and $B$ |
| $\{A, B\}$ | anticommutator $\{A, B\} = AB + BA$ of the linear operators $A$ and $B$ |
| $\dim(V)$ | the dimension of the vector space $V$ |
| $M(N, F)$ | the set of $N \times N$ matrices with entries from $F$ |
| $\mathbb{Z}_n$ | the cyclic group of order $n$, which is isomorphic to $\mathbb{Z}/n\mathbb{Z}$, the integers with addition modulo $n$ |
| $\mathrm{GL}(V)$ | the general linear group over a vector space $V$, i. e. the group of all bijective linear maps (invertible matrices) $V \to V$ |
| $\mathrm{GL}(N, F)$ | the general linear group of degree $N$ over the field $F$, which is the group of $N \times N$ invertible matrices with entries from $F$ |
| $\mathrm{SL}(N, F)$ | the special linear group of degree $N$ over the field $F$: $\mathrm{SL}(N, F) = \{M \in \mathrm{GL}(N, F) \mid \det(M) = 1\}$ |
| $\mathrm{SO}(N)$ | $\mathrm{SO}(N) = \{O \in \mathrm{GL}(N, \mathbb{R}) \mid O^\mathsf{T} O = \mathbb{1} \wedge \det(O) = 1\}$ |
| $\mathrm{O}(N)$ | the orthogonal group of degree $N$: $\mathrm{O}(N) = \{O \in \mathrm{GL}(N, \mathbb{R}) \mid O^\mathsf{T} O = \mathbb{1}\}$ |
| $\mathrm{SO}(N)$ | the special orthogonal group of degree $N$: $\mathrm{SO}(N) = \{O \in \mathrm{GL}(N, \mathbb{R}) \mid O^\mathsf{T} O = \mathbb{1} \wedge \det(O) = 1\}$ |
| $\mathrm{U}(N)$ | the unitary group of degree $N$: $\mathrm{U}(N) = \{U \in \mathrm{GL}(N, \mathbb{C}) \mid U^\dagger U = \mathbb{1}\}$ |

| Symbol | Description |
|---|---|
| $\mathrm{SU}(N)$ | the special unitary group of degree $N$: $\mathrm{SU}(N) = \{U \in \mathrm{GL}(N, \mathbb{C}) \mid U^\dagger U = \mathbb{1} \wedge \det(U) = 1\}$ |
| $\mathrm{O}(p,q)$ | the indefinite orthogonal group with signature $(p,q)$: $\mathrm{O}(p,q) = \{O \in \mathrm{GL}(p+q, \mathbb{R}) \mid O^\mathsf{T} g_{pq} O = g_{pq}\}$ where $g_{pq} = \mathbb{1}_p \oplus (-\mathbb{1}_q)$ |
| $\mathrm{SO}(p,q)$ | the indefinite special orthogonal group with signature $(p,q)$: $\mathrm{SO}(p,q) = \{O \in \mathrm{GL}(p+q, \mathbb{R}) \mid O^\mathsf{T} g_{pq} O = g_{pq} \wedge \det(O) = 1\}$ where $g_{pq} = \mathbb{1}_p \oplus (-\mathbb{1}_q)$ |
| $\mathrm{O}(1,3)$ | the Lorentz group: $\mathrm{O}(1,3) = \{\Lambda \in \mathrm{GL}(4, \mathbb{R}) \mid \Lambda^\mathsf{T} \eta \Lambda = \eta\}$ |
| $\mathrm{SO}(1,3)^+$ | the group of proper orthochronous Lorentz transformations: $\mathrm{SO}(1,3)^+ = \{\Lambda \in \mathrm{GL}(4, \mathbb{R}) \mid \Lambda^\mathsf{T} \eta \Lambda = \eta \wedge \det(\Lambda) = 1 \wedge \Lambda^0{}_0 > 0\}$ |
| $\mathfrak{gl}(V)$ | the Lie algebra of $\mathrm{GL}(V)$, which is the vector space of linear maps $V \to V$ (matrices) |
| $\mathfrak{gl}(N,F)$ | the Lie algebra of $\mathrm{GL}(N,F)$, which is the vector space of the set $M(N,F)$ ($N \times N$ matrices with entries from $F$) |
| $\mathfrak{sl}(N,F)$ | the Lie algebra of $\mathrm{SL}(N,F)$: $\mathfrak{sl}(N,F) = \{X \in \mathfrak{gl}(N,F) \mid \mathrm{Tr}(X) = 0\}$ |
| $\mathfrak{o}(N)$ | the Lie algebra of $\mathrm{O}(N)$: $\mathfrak{o}(N) = \{X \in \mathfrak{gl}(N, \mathbb{R}) \mid X^\mathsf{T} = -X\}$ |
| $\mathfrak{so}(N)$ | the Lie algebra of $\mathrm{SO}(N)$: $\mathfrak{so}(N) = \{X \in \mathfrak{gl}(N, \mathbb{R}) \mid X^\mathsf{T} = -X \wedge \mathrm{Tr}(X) = 0\}$ |
| $\mathfrak{u}(N)$ | the Lie algebra of $\mathrm{U}(N)$: $\mathfrak{u}(N) = \{X \in \mathfrak{gl}(N, \mathbb{C}) \mid X^\dagger = -X\}$ |
| $\mathfrak{su}(N)$ | the Lie algebra of $\mathrm{SU}(N)$: $\mathfrak{su}(N) = \{X \in \mathfrak{gl}(N, \mathbb{C}) \mid X^\dagger = -X \wedge \mathrm{Tr}(X) = 0\}$ |
| $\mathfrak{o}(p,q)$, $\mathfrak{so}(p,q)$ | the Lie algebra of $\mathrm{O}(p,q)$ and $\mathrm{SO}(p,q)$: $\mathfrak{o}(p,q) = \mathfrak{so}(p,q) = \{X \in \mathfrak{gl}(p+q, \mathbb{R}) \mid g_{pq} X^\mathsf{T} g_{pq} = -X\}$ (for the Lorentz group: $\mathfrak{o}(1,3) = \mathfrak{so}(1,3) = \{\lambda \in \mathfrak{gl}(4, \mathbb{R}) \mid \eta \lambda^\mathsf{T} \eta = -\lambda\}$) |
| $\mathbf{n}$, $\bar{\mathbf{n}}$ | the $n$-dimensional irreducible representations of $\mathrm{SU}(N)$ and their conjugates (the value of $N$ must be specified separately or as a subscript, i. e. $\mathbf{n}_N$) |
| $(j_\mathrm{L}, j_\mathrm{R})$ | the $(2j_\mathrm{L}+1)(2j_\mathrm{R}+1)$-dimensional irreducible representations of $\mathrm{SL}(2, \mathbb{C})$ |

**Table B.2.:** Classification of all simple Lie groups. Here, "rank" refers to the number of diagonal generators and "order" is the dimension of the Lie algebra, which is also the dimension of the group's corresponding manifold. Taken from [Kla17].

| Cartan name | Classical name | Rank | Order | Complex representations |
|---|---|---|---|---|
| $A_n (n \geq 1)$ | $\mathrm{SU}(n+1)$ | $n$ | $n(n+2)$ | $n \geq 2$ |
| $B_n (n \geq 2)$ | $\mathrm{SO}(2n+1)$ | $n$ | $n(2n+1)$ | No |
| $C_n (n \geq 3)$ | $\mathrm{Sp}(2n)$ | $n$ | $n(2n+1)$ | No |
| $D_n (n \geq 4)$ | $\mathrm{SO}(2n)$ | $n$ | $n(2n-1)$ | $n = 5, 7, 9, \ldots$ |
| $G_2$ | $G_2$ | 2 | 14 | No |
| $F_4$ | $F_4$ | 4 | 52 | No |
| $E_6$ | $E_6$ | 6 | 78 | Yes |
| $E_7$ | $E_7$ | 7 | 133 | No |
| $E_8$ | $E_8$ | 8 | 248 | No |

# Proofs

C

The commutation and anti-commutation relations of the Pauli matrices

$$\left[\sigma_i, \sigma_j\right] = 2i\varepsilon_{ijk}\sigma_k \tag{C.1}$$

$$\left\{\sigma_i, \sigma_j\right\} = 2\delta_{ij}\mathbb{1} \tag{C.2}$$

as well as the following well-known trace identities

$$\mathrm{Tr}(\sigma_i) = 0 \tag{C.3}$$

$$\mathrm{Tr}(\sigma_i\sigma_j) = 2\delta_{ij} \tag{C.4}$$

$$\mathrm{Tr}(\sigma_i\sigma_j\sigma_k) = 2i\varepsilon_{ijk} \tag{C.5}$$

$$\mathrm{Tr}(\sigma_i\sigma_j\sigma_k\sigma_m) = 2(\delta_{ij}\delta_{km} - \delta_{ik}\delta_{jm} + \delta_{im}\delta_{jk}) \tag{C.6}$$

will be used here. Using the properties (C.1) and (C.2), one can also show that

$$\sigma_i\sigma_j = \delta_{ij}\mathbb{1} + i\varepsilon_{ijk}\sigma_k \tag{C.7}$$

and

$$e^{i\alpha\vec{n}\cdot\vec{\sigma}} = \mathbb{1}\cos(\alpha) + i\vec{n}\cdot\vec{\sigma}\sin(\alpha) \tag{C.8}$$

with $\vec{n}^2 = 1$.

## C.1. Equivalence of invariant order-2 tensors and scalar products

Let $M$ be the transformation matrix, $T$ an order-2 tensor and $v, w$ vectors.

**Real scalar product** If $M^\mathsf{T}TM = T$ for an order-2 tensor $T$, then

$$T(v,w) = v^\mathsf{T}Tw \mapsto (v^\mathsf{T}\underbrace{M^\mathsf{T})T(M}_{=T}w) = v^\mathsf{T}Tw = T(v,w) \tag{C.9}$$

**Complex scalar product** (Same as for the real scalar product, but replace transpose with adjoint.) If $M^\dagger T M = T$ for an order-2 tensor $T$, then

$$T(v^*, w) = v^\dagger T w \mapsto (v^\dagger \underbrace{M^\dagger) T (M}_{=T} w) = v^\dagger T w = T(v^*, w) \tag{C.10}$$

## C.2. Invariance of the two-dimensional $\varepsilon$ symbol under $\mathrm{SL}(2, F)$

The condition $\det(M) = 1$ of $\mathrm{SL}(2, F)$ and its subgroups leads to the invariance of the two-dimensional Levi-Civita symbol under transformations from these groups. The determinant of a general Matrix $A$ can be expressed using the $n$-dimensional Levi-Civita symbol as follows:

$$\det(A) = \sum_{i_1,\ldots,i_n=1}^{n} \varepsilon^{i_1 \ldots i_n} A_{1i_1} \ldots A_{ni_n} \tag{C.11}$$

In the case of the $2 \times 2$ matrices $M$, this means

$$(M \varepsilon M^\mathsf{T})_{12} = \sum_{i,j=1}^{2} M_{1i} \varepsilon^{ij} (M^\mathsf{T})_{j2} = \sum_{i,j=1}^{2} \varepsilon^{ij} M_{1i} M_{2j} = \det(M) = 1 = \varepsilon^{12} \tag{C.12}$$

(and analogously for $(M \varepsilon M^\mathsf{T})_{21} = -\det(M) = -1 = \varepsilon^{21}$). For the matrix elements $(M \varepsilon M^\mathsf{T})_{11}$ and $(M \varepsilon M^\mathsf{T})_{22}$, the sum is symmetric with respect to exchanging $i$ and $j$, but $\varepsilon^{ij}$ is antisymmetric, so the result is $\varepsilon^{11} = \varepsilon^{22} = 0$. Thus,[1]

$$M \varepsilon M^\mathsf{T} = \varepsilon \tag{C.13}$$

or, using the element $M' = M^\mathsf{T} \in \mathrm{SL}(2, F)$ instead of $M$,

$$M'^\mathsf{T} \varepsilon M' = \varepsilon \tag{C.14}$$

For $\mathrm{SU}(2)$, there is also an alternative proof using (C.8) and the relation

$$\sigma^2 \sigma^i \sigma^2 = \varepsilon \sigma^i \varepsilon^{-1} = -(\sigma^i)^\mathsf{T} \tag{C.15}$$

[1] $\det(M^\mathsf{T}) = \det(M)$, so $M \in \mathrm{SL}(2, F) \Leftrightarrow M^\mathsf{T} \in \mathrm{SL}(2, F)$

Using these, one sees that for $U \in \mathrm{SU}(2)$,

$$\begin{aligned}
\varepsilon U \varepsilon^{-1} &= \varepsilon\, e^{-\frac{i}{2}\vec{\alpha}\cdot\vec{\sigma}}\, \varepsilon^{-1} \\
&= \varepsilon\Big(\mathbb{1}\cos\Big(-\frac{\alpha}{2}\Big) + i\vec{n}\cdot\vec{\sigma}\sin\Big(-\frac{\alpha}{2}\Big)\Big)\varepsilon^{-1} \\
&= \cos\Big(\frac{\alpha}{2}\Big) - in^i \underbrace{\varepsilon\sigma_i\varepsilon^{-1}}_{=-\sigma_i^{\mathsf{T}}} \sin\Big(\frac{\alpha}{2}\Big) = \cos\Big(\frac{\alpha}{2}\Big) + in^i\sigma_i^{\mathsf{T}}\sin\Big(\frac{\alpha}{2}\Big) \\
&= e^{\frac{i}{2}\alpha^i\sigma_i^{\mathsf{T}}} = \Big(e^{\frac{i}{2}\vec{\alpha}\cdot\vec{\sigma}}\Big)^{\mathsf{T}} = (U^{-1})^{\mathsf{T}} \qquad \text{(C.16)}
\end{aligned}$$

with $\vec{n} = \vec{\alpha}/a = -2\vec{\alpha}/\alpha$. This is the same relation which was obtained in section 4.1.3 and is hence equivalent to the relation $U^{\mathsf{T}}\varepsilon U = \varepsilon$.

## C.3. Proofs for section 5.4

*Proof of identity 1.*

$$\begin{aligned}
|D_1|^2|D_2|^2 &= (|D_{11}|^2 + |D_{12}|^2)(|D_{21}|^2 + |D_{22}|^2) \\
&= |D_{11}|^2|D_{21}|^2 + |D_{12}|^2|D_{21}|^2 + |D_{11}|^2|D_{22}|^2 + |D_{12}|^2|D_{22}|^2 \\
|D_1^{\mathsf{T}} i\sigma^2 D_2|^2 &= |D_{12}D_{21} - D_{11}D_{22}|^2 \\
&= |D_{12}D_{21}|^2 + |D_{11}D_{22}|^2 - D_{12}^\dagger D_{21}^\dagger D_{11}D_{22} - D_{11}^\dagger D_{22}^\dagger D_{12}D_{21} \\
|D_1^\dagger D_2|^2 &= |D_{11}^\dagger D_{21} + D_{12}^\dagger D_{22}|^2 \\
&= |D_{11}|^2|D_{21}|^2 + |D_{12}|^2|D_{21}|^2 + D_{12}^\dagger D_{21}^\dagger D_{11}D_{22} + D_{11}^\dagger D_{22}^\dagger D_{12}D_{21}
\end{aligned}$$

□

*Proof of identity 2.* Assume that $j = k$.

$$\begin{aligned}
\mathrm{Tr}(\Delta_i\Delta_j\Delta_k) = \mathrm{Tr}(\Delta_i\Delta_j\Delta_j) &= \Delta_i^l\Delta_j^m\Delta_j^n\,\mathrm{Tr}(\sigma_l\sigma_m\sigma_n) \\
&\overset{\text{(C.5)}}{=} 2i\Delta_i^l\Delta_j^m\Delta_j^n\varepsilon_{lmn} = 0
\end{aligned}$$

because $\Delta_j^m\Delta_j^n$ is even in the indices $m, n$ while $\varepsilon_{lmn}$ is odd, so their product is simply zero.

For the other cases ($i = j$ or $i = k$), the calculation is analogous, with an even factor multiplied by an odd factor (in $l, m$ or $l, n$, respectively). □

*Proof of identity 3.*

1.

$$\begin{aligned}
\mathrm{Tr}\big(\Delta_1^2\big)\,\mathrm{Tr}\big(\Delta_2^2\big) &= \mathrm{Tr}\big(\Delta_1^a\Delta_1^b\sigma_a\sigma_b\big)\,\mathrm{Tr}\big(\Delta_2^c\Delta_2^d\sigma_c\sigma_d\big)\\
&= \Delta_1^a\Delta_1^b\Delta_2^c\Delta_2^d\,\mathrm{Tr}(\sigma_a\sigma_b)\,\mathrm{Tr}(\sigma_c\sigma_d)\\
&\overset{(C.4)}{=} \Delta_1^a\Delta_1^b\Delta_2^c\Delta_2^d\cdot 2\delta_{ab}\cdot 2\delta_{cd}\\
&= 4\Delta_1^a(\Delta_1)_a\Delta_2^c(\Delta_2)_c\\
&= 4\big(\Delta_1^a\big)^2\big(\Delta_2^b\big)^2
\end{aligned}$$

$$\begin{aligned}
\mathrm{Tr}\big(\Delta_1^2\Delta_2^2\big) &= \mathrm{Tr}\big(\Delta_1^a\Delta_1^b\Delta_2^c\Delta_2^d\sigma_a\sigma_b\sigma_c\sigma_d\big)\\
&= \Delta_1^a\Delta_1^b\Delta_2^c\Delta_2^d\,\mathrm{Tr}(\sigma_a\sigma_b\sigma_c\sigma_d)\\
&\overset{(C.6)}{=} \Delta_1^a\Delta_1^b\Delta_2^c\Delta_2^d\cdot 2(\delta_{ab}\delta_{cd}-\delta_{ac}\delta_{bd}+\delta_{ad}\delta_{bc})\\
&= 2\Big(\big(\Delta_1^a\big)^2\big(\Delta_2^c\big)^2-\big(\Delta_1^a\Delta_2^b\big)^2+\big(\Delta_1^a\Delta_2^b\big)^2\Big)\\
&= 2\big(\Delta_1^a\big)^2\big(\Delta_2^b\big)^2\\
&= \frac{1}{2}\,\mathrm{Tr}\big(\Delta_1^2\big)\,\mathrm{Tr}\big(\Delta_2^2\big)
\end{aligned}$$

2.

$$\begin{aligned}
\mathrm{Tr}\big((\Delta_1\Delta_2)^2\big) &= \mathrm{Tr}\big(\Delta_1^a\Delta_2^b\Delta_1^c\Delta_2^d\sigma_a\sigma_b\sigma_c\sigma_d\big)\\
&= \Delta_1^a\Delta_2^b\Delta_1^c\Delta_2^d\,\mathrm{Tr}(\sigma_a\sigma_b\sigma_c\sigma_d)\\
&\overset{(C.6)}{=} \Delta_1^a\Delta_2^b\Delta_1^c\Delta_2^d\cdot 2(\delta_{ab}\delta_{cd}-\delta_{ac}\delta_{bd}+\delta_{ad}\delta_{bc})\\
&= 2\Big(\big(\Delta_1^a\Delta_2^c\big)^2-\big(\Delta_1^a\big)^2(\Delta_2^b)^2+\big(\Delta_1^a\Delta_2^c\big)^2\Big)\\
&= 4\big(\Delta_1^a\Delta_2^b\big)^2-2\big(\Delta_1^a\big)^2\big(\Delta_2^b\big)^2
\end{aligned}$$

$$\begin{aligned}
\mathrm{Tr}(\Delta_1\Delta_2)^2 &= \mathrm{Tr}\big(\Delta_1^a\Delta_2^b\sigma_a\sigma_b\big)^2\\
&= \big(\Delta_1^a\Delta_2^b\,\mathrm{Tr}(\sigma_a\sigma_b)\big)^2\\
&\overset{(C.4)}{=} \big(\Delta_1^a\Delta_2^b\cdot 2\delta_{ab}\big)^2\\
&= 4\big(\Delta_1^a\Delta_2^a\big)^2
\end{aligned}$$

$$\Rightarrow \operatorname{Tr}\left(\Delta_1^2\Delta_2^2\right) + \operatorname{Tr}\left((\Delta_1\Delta_2)^2\right) = 2\left(\Delta_1^a\right)^2\left(\Delta_2^b\right)^2 + 4\left(\Delta_1^a\Delta_2^b\right)^2 - 2\left(\Delta_1^a\right)^2\left(\Delta_2^b\right)^2$$
$$= 4\left(\Delta_1^a\Delta_2^b\right)^2$$
$$= \operatorname{Tr}(\Delta_1\Delta_2)^2$$

3.

$$D_1^\dagger\{\Delta_1, \Delta_2\}D_2 = D_1^\dagger\Delta_1^a\Delta_2^b\{\sigma_a, \sigma_b\}D_2$$
$$\overset{(C.2)}{=} D_1^\dagger\Delta_1^a\Delta_2^b \cdot 2\delta_{ab}\mathbb{1}D_2$$
$$\overset{(C.4)}{=} D_1^\dagger\Delta_1^a\Delta_2^b \operatorname{Tr}(\sigma_a\sigma_b)D_2$$
$$= D_1^\dagger \operatorname{Tr}(\Delta_1\Delta_2)D_2$$
$$= D_1^\dagger D_2 \operatorname{Tr}(\Delta_1\Delta_2)$$

□

*Proof of corollary 1.* Naively enumerating all possible combinations for quartic terms which could be formed using only one triplet Δ yields the following list:

(1) $\operatorname{Tr}\left(\Delta^\dagger\Delta\right)^2$

(2) $\operatorname{Tr}\left((\Delta^\dagger)^2\right)\operatorname{Tr}\left(\Delta^2\right)$

(3) $\operatorname{Tr}\left((\Delta^\dagger)^2\Delta^2\right)$

(3′) $\operatorname{Tr}\left(\Delta^2(\Delta^\dagger)^2\right)$

(3″) $\operatorname{Tr}\left(\Delta^\dagger\Delta^2\Delta^\dagger\right)$

(3‴) $\operatorname{Tr}\left(\Delta(\Delta^\dagger)^2\Delta\right)$

(4) $\operatorname{Tr}\left((\Delta^\dagger\Delta)^2\right)$

(4′) $\operatorname{Tr}\left((\Delta\Delta^\dagger)^2\right)$

Of these, all items labeled with the same number are equal by cyclicity of the trace. Further, (2) and (3) are equal up to a factor of two by identity 3. Finally, one of the three terms (1), (2) = (3) and (4) can be eliminated by identity 3 so that only two remain, for example (1) and (4). □

# D Fermion bilinears

This is the full list of bilinear spinor products forming Lorentz scalars, pseudo-scalars, vectors and pseudo-vectors, “translating” between the Dirac and Weyl spinor formalisms. These can be obtained directly from (3.54) and (3.96). $\Psi_1, \Psi_2$ are two Dirac spinors with components $\Psi_i = (\psi_i, \bar{\chi}_i)$ in the Weyl basis.

$$\left.\begin{aligned} \bar{\Psi}_1 P_\mathrm{L} \Psi_2 &= \chi_1 \psi_2 \\ \bar{\Psi}_1 P_\mathrm{R} \Psi_2 &= \bar{\psi}_1 \bar{\chi}_2 \end{aligned}\right\} \bar{\Psi}_1 \Psi_2 = \chi_1 \psi_2 + \bar{\psi}_1 \bar{\chi}_2 \tag{D.1}$$

$$\left.\begin{aligned} \overline{\Psi^c_1} P_\mathrm{L} \Psi_2 &= \psi_1 \psi_2 \\ \overline{\Psi^c_1} P_\mathrm{R} \Psi_2 &= \bar{\chi}_1 \bar{\chi}_2 \end{aligned}\right\} \overline{\Psi^c_1} \Psi_2 = \psi_1 \psi_2 + \bar{\chi}_1 \bar{\chi}_2 \tag{D.2}$$

$$\left.\begin{aligned} \bar{\Psi}_1 P_\mathrm{L} \Psi^c_2 &= \chi_1 \chi_2 \\ \bar{\Psi}_1 P_\mathrm{R} \Psi^c_2 &= \bar{\psi}_1 \bar{\psi}_2 \end{aligned}\right\} \bar{\Psi}_1 \Psi^c_2 = \chi_1 \chi_2 + \bar{\psi}_1 \bar{\psi}_2 \tag{D.3}$$

$$\left.\begin{aligned} \overline{\Psi^c_1} P_\mathrm{L} \Psi^c_2 &= \psi_1 \chi_2 \\ \overline{\Psi^c_1} P_\mathrm{R} \Psi^c_2 &= \bar{\chi}_1 \bar{\psi}_2 \end{aligned}\right\} \overline{\Psi^c_1} \Psi^c_2 = \psi_1 \chi_2 + \bar{\chi}_1 \bar{\psi}_2 \tag{D.4}$$

$$\left.\begin{aligned} \bar{\Psi}_1 \gamma_5 P_\mathrm{L} \Psi_2 &= -\chi_1 \psi_2 \\ \bar{\Psi}_1 \gamma_5 P_\mathrm{R} \Psi_2 &= \bar{\psi}_1 \bar{\chi}_2 \end{aligned}\right\} \bar{\Psi}_1 \gamma_5 \Psi_2 = -\chi_1 \psi_2 + \bar{\psi}_1 \bar{\chi}_2 \tag{D.5}$$

$$\left.\begin{aligned} \overline{\Psi^c_1} \gamma_5 P_\mathrm{L} \Psi_2 &= -\psi_1 \psi_2 \\ \overline{\Psi^c_1} \gamma_5 P_\mathrm{R} \Psi_2 &= \bar{\chi}_1 \bar{\chi}_2 \end{aligned}\right\} \overline{\Psi^c_1} \gamma_5 \Psi_2 = -\psi_1 \psi_2 + \bar{\chi}_1 \bar{\chi}_2 \tag{D.6}$$

$$\left.\begin{aligned} \bar{\Psi}_1 \gamma_5 P_\mathrm{L} \Psi^c_2 &= -\chi_1 \chi_2 \\ \bar{\Psi}_1 \gamma_5 P_\mathrm{R} \Psi^c_2 &= \bar{\psi}_1 \bar{\psi}_2 \end{aligned}\right\} \bar{\Psi}_1 \gamma_5 \Psi^c_2 = -\chi_1 \chi_2 + \bar{\psi}_1 \bar{\psi}_2 \tag{D.7}$$

$$\left.\begin{aligned} \overline{\Psi^c_1} \gamma_5 P_\mathrm{L} \Psi^c_2 &= -\psi_1 \chi_2 \\ \overline{\Psi^c_1} \gamma_5 P_\mathrm{R} \Psi^c_2 &= \bar{\chi}_1 \bar{\psi}_2 \end{aligned}\right\} \overline{\Psi^c_1} \gamma_5 \Psi^c_2 = -\psi_1 \chi_2 + \bar{\chi}_1 \bar{\psi}_2 \tag{D.8}$$

$$\left.\begin{aligned}\bar{\Psi}_1\gamma^\mu P_\mathrm{L}\Psi_2 &= \bar{\psi}_1\bar{\sigma}^\mu\psi_2\\ \bar{\Psi}_1\gamma^\mu P_\mathrm{R}\Psi_2 &= \chi_1\sigma^\mu\bar{\chi}_2\end{aligned}\right\}\ \bar{\Psi}_1\gamma^\mu\Psi_2 = \bar{\psi}_1\bar{\sigma}^\mu\psi_2 + \chi_1\sigma^\mu\bar{\chi}_2 \tag{D.9}$$

$$\left.\begin{aligned}\overline{\Psi^\mathrm{c}_1}\gamma^\mu P_\mathrm{L}\Psi_2 &= \bar{\chi}_1\bar{\sigma}^\mu\psi_2\\ \overline{\Psi^\mathrm{c}_1}\gamma^\mu P_\mathrm{R}\Psi_2 &= \psi_1\sigma^\mu\bar{\chi}_2\end{aligned}\right\}\ \overline{\Psi^\mathrm{c}_1}\gamma^\mu\Psi_2 = \bar{\chi}_1\bar{\sigma}^\mu\psi_2 + \psi_1\sigma^\mu\bar{\chi}_2 \tag{D.10}$$

$$\left.\begin{aligned}\bar{\Psi}_1\gamma^\mu P_\mathrm{L}\Psi^\mathrm{c}_2 &= \bar{\psi}_1\bar{\sigma}^\mu\chi_2\\ \bar{\Psi}_1\gamma^\mu P_\mathrm{R}\Psi^\mathrm{c}_2 &= \chi_1\sigma^\mu\bar{\psi}_2\end{aligned}\right\}\ \bar{\Psi}_1\gamma^\mu\Psi^\mathrm{c}_2 = \bar{\psi}_1\bar{\sigma}^\mu\chi_2 + \chi_1\sigma^\mu\bar{\psi}_2 \tag{D.11}$$

$$\left.\begin{aligned}\overline{\Psi^\mathrm{c}_1}\gamma^\mu P_\mathrm{L}\Psi^\mathrm{c}_2 &= \bar{\chi}_1\bar{\sigma}^\mu\chi_2\\ \overline{\Psi^\mathrm{c}_1}\gamma^\mu P_\mathrm{R}\Psi^\mathrm{c}_2 &= \psi_1\sigma^\mu\bar{\psi}_2\end{aligned}\right\}\ \overline{\Psi^\mathrm{c}_1}\gamma^\mu\Psi^\mathrm{c}_2 = \bar{\chi}_1\bar{\sigma}^\mu\chi_2 + \psi_1\sigma^\mu\bar{\psi}_2 \tag{D.12}$$

$$\left.\begin{aligned}\bar{\Psi}_1\gamma^\mu\gamma_5 P_\mathrm{L}\Psi_2 &= -\bar{\psi}_1\bar{\sigma}^\mu\psi_2\\ \bar{\Psi}_1\gamma^\mu\gamma_5 P_\mathrm{R}\Psi_2 &= \chi_1\sigma^\mu\bar{\chi}_2\end{aligned}\right\}\ \bar{\Psi}_1\gamma^\mu\gamma_5\Psi_2 = -\bar{\psi}_1\bar{\sigma}^\mu\psi_2 + \chi_1\sigma^\mu\bar{\chi}_2 \tag{D.13}$$

$$\left.\begin{aligned}\overline{\Psi^\mathrm{c}_1}\gamma^\mu\gamma_5 P_\mathrm{L}\Psi_2 &= -\bar{\chi}_1\bar{\sigma}^\mu\psi_2\\ \overline{\Psi^\mathrm{c}_1}\gamma^\mu\gamma_5 P_\mathrm{R}\Psi_2 &= \psi_1\sigma^\mu\bar{\chi}_2\end{aligned}\right\}\ \overline{\Psi^\mathrm{c}_1}\gamma^\mu\gamma_5\Psi_2 = -\bar{\chi}_1\bar{\sigma}^\mu\psi_2 + \psi_1\sigma^\mu\bar{\chi}_2 \tag{D.14}$$

$$\left.\begin{aligned}\bar{\Psi}_1\gamma^\mu\gamma_5 P_\mathrm{L}\Psi^\mathrm{c}_2 &= -\bar{\psi}_1\bar{\sigma}^\mu\chi_2\\ \bar{\Psi}_1\gamma^\mu\gamma_5 P_\mathrm{R}\Psi^\mathrm{c}_2 &= \chi_1\sigma^\mu\bar{\psi}_2\end{aligned}\right\}\ \bar{\Psi}_1\gamma^\mu\gamma_5\Psi^\mathrm{c}_2 = -\bar{\psi}_1\bar{\sigma}^\mu\chi_2 + \chi_1\sigma^\mu\bar{\psi}_2 \tag{D.15}$$

$$\left.\begin{aligned}\overline{\Psi^\mathrm{c}_1}\gamma^\mu\gamma_5 P_\mathrm{L}\Psi^\mathrm{c}_2 &= -\bar{\chi}_1\bar{\sigma}^\mu\chi_2\\ \overline{\Psi^\mathrm{c}_1}\gamma^\mu\gamma_5 P_\mathrm{R}\Psi^\mathrm{c}_2 &= \psi_1\sigma^\mu\bar{\psi}_2\end{aligned}\right\}\ \overline{\Psi^\mathrm{c}_1}\gamma^\mu\gamma_5\Psi^\mathrm{c}_2 = -\bar{\chi}_1\bar{\sigma}^\mu\chi_2 + \psi_1\sigma^\mu\bar{\psi}_2 \tag{D.16}$$

# E List of Lagrangians

Here, the Lagrangians for all of the 35 viable models given in [RZY13] are listed. They were generated by the program `minimal-lagrangians`, which was created as part of this thesis and is described in chapter 6. For the definitions of these models (in particular, the fields $\phi$, $\phi'$, $\varphi$, $\psi$, $\psi'$, $\Psi$ in each case) see [RZY13]. If numbered fermion fields ($\psi_1/\psi_2$, $\psi'_1/\psi'_2$ or $\Psi_1/\Psi_2$) appear, a fermion doublet had to be made vector-like to cancel the Witten SU(2) anomaly. Any mention of fields not defined there (such as $H$, $L$ or $e_{\mathrm{R}}^{\mathrm{c}}$) refers to the fields of the Standard Model defined in table 3.1. Generation indices for the fields are not written explicitly, but could trivially be inserted for the Standard Model fermions and any new fields which should appear with several generations.

### T1-1-A ($\alpha = 0$)

$$
\begin{aligned}
\mathcal{L}_{\text{T1-1-A }(\alpha = 0)} = {} & \mathcal{L}_{\text{SM}} + \mathcal{L}_{\text{kin}} - M_{\phi'}^2 \phi'^\dagger \phi' - \frac{1}{2} M_{\varphi}^2 \varphi^2 \\
& - (\lambda_1 (H\phi')\varphi + \text{H.c.}) \\
& - \lambda_2 (H^\dagger H)(\phi'^\dagger \phi') - \lambda_3 (H^\dagger \phi'^\dagger)(H\phi') - \lambda_4 (H^\dagger \phi')(\phi'^\dagger H) - \lambda_5 (\phi'^\dagger \phi')^2 \\
& \quad - \lambda_6 (H^\dagger H)\varphi^2 - \lambda_7 (\phi'^\dagger \phi')\varphi^2 - \left(\lambda_8 (H\phi')^2 + \text{H.c.}\right) - \lambda_9 \varphi^4 \\
& - \left(\frac{1}{2} M_\psi \psi\psi + \text{H.c.}\right) \\
& - \left(\lambda_{10} (\phi'^\dagger L)\psi + \text{H.c.}\right)
\end{aligned}
\tag{E.1}
$$

**T1-1-A ($\alpha = 2$)**

$$
\begin{aligned}
\mathcal{L}_{\text{T1-1-A }(\alpha\,=\,2)} = \mathcal{L}_{\text{SM}} + \mathcal{L}_{\text{kin}} - M_{\phi'}^2 {\phi'}^\dagger \phi' - M_\phi^2 \phi^\dagger \phi - M_\varphi^2 \varphi^\dagger \varphi \\
&- \left(\lambda_1 \varphi^\dagger (H\phi') + \text{H.c.}\right) - \left(\lambda_2 (\phi^\dagger H)\varphi + \text{H.c.}\right) \\
&- \lambda_3 (H^\dagger H)\left({\phi'}^\dagger \phi'\right) - \lambda_4 (H^\dagger H)(\phi^\dagger \phi) \\
&\quad - \left(\lambda_5 (H^\dagger \phi')^2 + \text{H.c.}\right) - \lambda_6 \left(H^\dagger {\phi'}^\dagger\right)(H\phi') - \lambda_7 (H^\dagger \phi')\left({\phi'}^\dagger H\right) \\
&\quad - \lambda_8 (H^\dagger \phi^\dagger)(H\phi) - \lambda_9 (H^\dagger \phi)(\phi^\dagger H) - \lambda_{10}\left({\phi'}^\dagger \phi'\right)^2 - \lambda_{11}\left({\phi'}^\dagger \phi'\right)(\phi^\dagger \phi) \\
&\quad - \lambda_{12}\left({\phi'}^\dagger \phi\right)(\phi^\dagger \phi') - \lambda_{13}\left(\phi^\dagger {\phi'}^\dagger\right)(\phi\phi') - \lambda_{14}(\phi^\dagger \phi)^2 - \lambda_{15}(H^\dagger H)\varphi^\dagger \varphi \\
&\quad - \lambda_{16}\left({\phi'}^\dagger \phi'\right)\varphi^\dagger \varphi - \left(\lambda_{17}\left(\phi^\dagger {\phi'}^\dagger\right)\varphi^2 + \text{H.c.}\right) - \lambda_{18}(\phi^\dagger \phi)\varphi^\dagger \varphi - \lambda_{19}{\varphi^\dagger}^2 \varphi^2 \\
&\quad - \left(\lambda_{20}(\phi^\dagger H)(H\phi') + \text{H.c.}\right) \\
&- \left(M_{\psi_1\psi_2}\psi_1\psi_2 + \text{H.c.}\right) \\
&- \left(\lambda_{21}\left({\phi'}^\dagger L\right)\psi_1 + \text{H.c.}\right) - \left(\lambda_{22}(L\phi)\psi_2 + \text{H.c.}\right)
\end{aligned}
\tag{E.2}
$$

**T1-1-B ($\alpha = 0$)**

$$
\begin{aligned}
\mathcal{L}_{\text{T1-1-B }(\alpha\,=\,0)} = \mathcal{L}_{\text{SM}} + \mathcal{L}_{\text{kin}} - M_{\phi'}^2 {\phi'}^\dagger \phi' - \frac{1}{2} M_\varphi^2 \varphi^2 \\
&- \left(\lambda_1 (H\phi')\varphi + \text{H.c.}\right) \\
&- \lambda_2 (H^\dagger H)\left({\phi'}^\dagger \phi'\right) - \lambda_3\left(H^\dagger {\phi'}^\dagger\right)(H\phi') - \lambda_4 (H^\dagger \phi')\left({\phi'}^\dagger H\right) - \lambda_5\left({\phi'}^\dagger \phi'\right)^2 \\
&\quad - \lambda_6 (H^\dagger H)\varphi^2 - \lambda_7\left({\phi'}^\dagger \phi'\right)\varphi^2 - \left(\lambda_8 (H\phi')^2 + \text{H.c.}\right) - \lambda_9 \varphi^4 \\
&- \left(\frac{1}{2} M_\psi \,\text{Tr}(\psi\psi) + \text{H.c.}\right) \\
&- \left(\lambda_{10}{\phi'}^\dagger \psi L + \text{H.c.}\right)
\end{aligned}
\tag{E.3}
$$

**T1-1-B ($\alpha = 2$)**

$$
\begin{aligned}
\mathcal{L}_{\text{T1-1-B}\,(\alpha\,=\,2)} = {} & \mathcal{L}_{\text{SM}} + \mathcal{L}_{\text{kin}} - M_{\phi'}^2 \phi'^\dagger \phi' - M_\phi^2 \phi^\dagger \phi - M_\varphi^2 \varphi^\dagger \varphi \\
& - \left(\lambda_1 \varphi^\dagger (H\phi') + \text{H.c.}\right) - \left(\lambda_2 (\phi^\dagger H)\varphi + \text{H.c.}\right) \\
& - \lambda_3 (H^\dagger H)\left(\phi'^\dagger \phi'\right) - \lambda_4 (H^\dagger H)(\phi^\dagger \phi) - \lambda_5 \left(H^\dagger \phi'^\dagger\right)(H\phi') - \lambda_6 (H^\dagger \phi')\left(\phi'^\dagger H\right) \\
& \quad - \lambda_7 (H^\dagger \phi^\dagger)(H\phi) - \lambda_8 (H^\dagger \phi)(\phi^\dagger H) - \left(\lambda_9 \left(\phi'^\dagger H\right)^2 + \text{H.c.}\right) \\
& \quad - \lambda_{10} \left(\phi'^\dagger \phi'\right)^2 - \lambda_{11} \left(\phi'^\dagger \phi'\right)(\phi^\dagger \phi) - \lambda_{12} \left(\phi'^\dagger \phi\right)(\phi^\dagger \phi') - \lambda_{13} \left(\phi^\dagger \phi'^\dagger\right)(\phi\phi') \\
& \quad - \lambda_{14} (\phi^\dagger \phi)^2 - \lambda_{15} (H^\dagger H)\varphi^\dagger \varphi - \lambda_{16} \left(\phi'^\dagger \phi'\right)\varphi^\dagger \varphi - \left(\lambda_{17} \left(\phi^\dagger \phi'^\dagger\right)\varphi^2 + \text{H.c.}\right) \\
& \quad - \lambda_{18} (\phi^\dagger \phi)\varphi^\dagger \varphi - \lambda_{19} \varphi^{\dagger 2} \varphi^2 - \left(\lambda_{20} (\phi^\dagger H)(H\phi') + \text{H.c.}\right) \\
& - \left(M_{\psi_1 \psi_2} \operatorname{Tr}(\psi_1 \psi_2) + \text{H.c.}\right) \\
& - \left(\lambda_{21} \phi'^\dagger \psi_1 L + \text{H.c.}\right) - \left(\lambda_{22} \phi \psi_2 L + \text{H.c.}\right)
\end{aligned}
\tag{E.4}
$$

**T1-1-C ($\alpha = 1$)**

$$
\begin{aligned}
\mathcal{L}_{\text{T1-1-C}\,(\alpha\,=\,1)} = {} & \mathcal{L}_{\text{SM}} + \mathcal{L}_{\text{kin}} - M_\varphi^2 \varphi^\dagger \varphi - M_\phi^2 \phi^\dagger \phi - \frac{1}{2} M_{\phi'}^2 \phi'^2 \\
& - \left(\lambda_1 (H^\dagger \varphi)\phi' + \text{H.c.}\right) - \left(\lambda_2 \phi^\dagger (H\varphi) + \text{H.c.}\right) \\
& - \lambda_3 (H^\dagger H)(\varphi^\dagger \varphi) - \lambda_4 (H^\dagger \varphi^\dagger)(H\varphi) - \lambda_5 (H^\dagger \varphi)(\varphi^\dagger H) - \left(\lambda_6 (\varphi^\dagger H)^2 + \text{H.c.}\right) \\
& \quad - \lambda_7 (\varphi^\dagger \varphi)^2 - \lambda_8 (H^\dagger H)\phi^\dagger \phi - \lambda_9 (\varphi^\dagger \varphi)\phi^\dagger \phi - \lambda_{10} \phi^{\dagger 2} \phi^2 - \lambda_{11} (H^\dagger H)\phi'^2 \\
& \quad - \lambda_{12} (\varphi^\dagger \varphi)\phi'^2 - \lambda_{13} \phi^\dagger \phi'^2 \phi - \lambda_{14} \phi'^4 \\
& - \left(M_{\psi_1 \psi_2} \psi_1 \psi_2 + \text{H.c.}\right) \\
& - \left(\lambda_{15} (\varphi^\dagger \psi_2) e_{\text{R}}^{\text{c}} + \text{H.c.}\right) - \left(\lambda_{16} (L\psi_1)\phi' + \text{H.c.}\right) - \left(\lambda_{17} (L\psi_2)\phi + \text{H.c.}\right)
\end{aligned}
\tag{E.5}
$$

**T1-1-D ($\alpha = -1$)**

$$
\begin{aligned}
\mathcal{L}_{\text{T1-1-D }(\alpha\,=\,-1)} = {} & \mathcal{L}_{\text{SM}} + \mathcal{L}_{\text{kin}} - M_{\varphi}^2 \varphi^\dagger \varphi - M_{\phi'}^2 {\phi'}^\dagger \phi' - \frac{1}{2} M_{\phi}^2 \operatorname{Tr}(\phi^2) \\
& - \big(\lambda_1 (\varphi^\dagger H) \phi' + \text{H.c.}\big) - (\lambda_2 H \phi \varphi + \text{H.c.}) \\
& - \lambda_3 (H^\dagger H)(\varphi^\dagger \varphi) - \lambda_4 (H^\dagger \varphi^\dagger)(H \varphi) - \lambda_5 (H^\dagger \varphi)(\varphi^\dagger H) - \lambda_6 (\varphi^\dagger \varphi)^2 \\
& \quad - \lambda_7 (H^\dagger H) {\phi'}^\dagger \phi' - \lambda_8 (\varphi^\dagger \varphi) {\phi'}^\dagger \phi' - \lambda_9 {\phi'}^{\dagger 2} {\phi'}^2 - \lambda_{10} (H^\dagger H) \operatorname{Tr}(\phi^2) \\
& \quad - \lambda_{11} H^\dagger \phi^2 H - \lambda_{12} (\varphi^\dagger \varphi) \operatorname{Tr}(\phi^2) - \lambda_{13} \varphi^\dagger \phi^2 \varphi - \lambda_{14} {\phi'}^\dagger \phi' \operatorname{Tr}(\phi^2) \\
& \quad - \Big(\lambda_{15} {\phi'}^\dagger (\varphi \phi \varphi) + \text{H.c.}\Big) - \lambda_{16} \operatorname{Tr}(\phi^4) - \Big(\lambda_{17} (H \varphi)^2 + \text{H.c.}\Big) \\
& \quad - (\lambda_{18} (H \phi H) \phi' + \text{H.c.}) \\
& - \Big(M_{\psi_1 \psi_2} \psi_1 \psi_2 + \text{H.c.}\Big) \\
& - \Big(\lambda_{19} {\phi'}^\dagger (L \psi_1) + \text{H.c.}\Big) - (\lambda_{20} \psi_2 \phi L + \text{H.c.}) - (\lambda_{21} (\varphi \psi_1) e_{\text{R}}^{\text{c}} + \text{H.c.})
\end{aligned}
\tag{E.6}
$$

**T1-1-D ($\alpha = 1$)**

$$
\begin{aligned}
\mathcal{L}_{\text{T1-1-D }(\alpha\,=\,1)} = {} & \mathcal{L}_{\text{SM}} + \mathcal{L}_{\text{kin}} - M_{\phi}^2 \operatorname{Tr}(\phi^\dagger \phi) - M_{\varphi}^2 \varphi^\dagger \varphi - \frac{1}{2} M_{\phi'}^2 {\phi'}^2 \\
& - (\lambda_1 H \phi^\dagger \varphi + \text{H.c.}) - \big(\lambda_2 (\varphi^\dagger H) \phi' + \text{H.c.}\big) \\
& - \lambda_3 \operatorname{Tr}\Big((\phi^\dagger \phi)^2\Big) - \lambda_4 \operatorname{Tr}(\phi^\dagger \phi)^2 - \lambda_5 H^\dagger \phi^\dagger \phi H - \lambda_6 \operatorname{Tr}(\phi^\dagger \phi)(H^\dagger H) \\
& \quad - \lambda_7 \operatorname{Tr}(\phi^\dagger \phi)(\varphi^\dagger \varphi) - \lambda_8 \varphi^\dagger \phi^\dagger \phi \varphi - \lambda_9 (H^\dagger H)(\varphi^\dagger \varphi) - \lambda_{10} (H^\dagger \varphi^\dagger)(H \varphi) \\
& \quad - \lambda_{11} (H^\dagger \varphi)(\varphi^\dagger H) - \Big(\lambda_{12} (\varphi^\dagger H)^2 + \text{H.c.}\Big) - \lambda_{13} (\varphi^\dagger \varphi)^2 \\
& \quad - (\lambda_{14} (H \phi^\dagger H) \phi' + \text{H.c.}) - \lambda_{15} \operatorname{Tr}(\phi^\dagger \phi) {\phi'}^2 - (\lambda_{16} (\varphi \phi^\dagger \varphi) \phi' + \text{H.c.}) \\
& \quad - \lambda_{17} (H^\dagger H) {\phi'}^2 - \lambda_{18} (\varphi^\dagger \varphi) {\phi'}^2 - \lambda_{19} {\phi'}^4 \\
& - \Big(M_{\psi_1 \psi_2} \psi_1 \psi_2 + \text{H.c.}\Big) \\
& - (\lambda_{20} (\varphi^\dagger \psi_2) e_{\text{R}}^{\text{c}} + \text{H.c.}) - (\lambda_{21} \psi_2 \phi L + \text{H.c.}) - (\lambda_{22} (L \psi_1) \phi' + \text{H.c.})
\end{aligned}
\tag{E.7}
$$

**T1-1-F ($\alpha = 1$)**

$$
\begin{aligned}
\mathcal{L}_{\text{T1-1-F }(\alpha\,=\,1)} = {} & \mathcal{L}_{\text{SM}} + \mathcal{L}_{\text{kin}} - M_\phi^2 \operatorname{Tr}(\phi^\dagger\phi) - M_\varphi^2 \varphi^\dagger\varphi - \frac{1}{2} M_{\phi'}^2 \operatorname{Tr}\big(\phi'^2\big) \\
& - (\lambda_1 H\phi^\dagger\varphi + \text{H.c.}) - (\lambda_2 \varphi^\dagger\phi' H + \text{H.c.}) \\
& - \lambda_3 \operatorname{Tr}\Big((\phi^\dagger\phi)^2\Big) - \lambda_4 \operatorname{Tr}(\phi^\dagger\phi)^2 - \lambda_5 H^\dagger\phi^\dagger\phi H - \lambda_6 \operatorname{Tr}(\phi^\dagger\phi)(H^\dagger H) \\
& \quad - \lambda_7 \operatorname{Tr}(\phi^\dagger\phi)(\varphi^\dagger\varphi) - \lambda_8 \varphi^\dagger\phi^\dagger\phi\varphi - \lambda_9 (H^\dagger H)(\varphi^\dagger\varphi) - \lambda_{10} (H^\dagger\varphi^\dagger)(H\varphi) \\
& \quad - \lambda_{11} (H^\dagger\varphi)(\varphi^\dagger H) - \Big(\lambda_{12} (\varphi^\dagger H)^2 + \text{H.c.}\Big) - \lambda_{13} (\varphi^\dagger\varphi)^2 - \lambda_{14} \operatorname{Tr}\big(\phi^\dagger\phi\phi'^2\big) \\
& \quad - \lambda_{15} \operatorname{Tr}(\phi^\dagger\phi'\phi\phi') - \lambda_{16} \operatorname{Tr}(\phi^\dagger\phi') \operatorname{Tr}(\phi\phi') - \lambda_{17} \operatorname{Tr}\big(\phi^\dagger\phi'^2\phi\big) \\
& \quad - \lambda_{18} \operatorname{Tr}(\phi^\dagger\phi) \operatorname{Tr}\big(\phi'^2\big) - (\lambda_{19} H\phi^\dagger\phi' H + \text{H.c.}) - \lambda_{20} (H^\dagger H) \operatorname{Tr}\big(\phi'^2\big) \\
& \quad - \lambda_{21} H^\dagger\phi'^2 H - (\lambda_{22} \varphi\phi^\dagger\phi'\varphi + \text{H.c.}) - \lambda_{23} (\varphi^\dagger\varphi) \operatorname{Tr}\big(\phi'^2\big) - \lambda_{24} \varphi^\dagger\phi'^2\varphi \\
& \quad - \lambda_{25} \operatorname{Tr}\big(\phi'^4\big) \\
& - \big(M_{\psi_1\psi_2} \psi_1\psi_2 + \text{H.c.}\big) \\
& - \big(\lambda_{26} (\varphi^\dagger\psi_2) e_{\text{R}}^{\text{c}} + \text{H.c.}\big) - (\lambda_{27} L\phi'\psi_1 + \text{H.c.}) - (\lambda_{28} \psi_2\phi L + \text{H.c.})
\end{aligned} \tag{E.8}
$$

**T1-1-G ($\alpha = 0$)**

$$
\begin{aligned}
\mathcal{L}_{\text{T1-1-G }(\alpha\,=\,0)} = {} & \mathcal{L}_{\text{SM}} + \mathcal{L}_{\text{kin}} - M_{\phi'}^2 \phi'^\dagger\phi' - \frac{1}{2} M_\varphi^2 \operatorname{Tr}(\varphi^2) \\
& - (\lambda_1 H\varphi\phi' + \text{H.c.}) \\
& - \lambda_2 (H^\dagger H)\big(\phi'^\dagger\phi'\big) - \lambda_3 \big(H^\dagger\phi'^\dagger\big)(H\phi') - \lambda_4 (H^\dagger\phi')\big(\phi'^\dagger H\big) - \lambda_5 \big(\phi'^\dagger\phi'\big)^2 \\
& \quad - \lambda_6 (H^\dagger H) \operatorname{Tr}(\varphi^2) - \lambda_7 H^\dagger\varphi^2 H - \lambda_8 \phi'^\dagger\varphi^2\phi' - \lambda_9 \big(\phi'^\dagger\phi'\big) \operatorname{Tr}(\varphi^2) \\
& \quad - \lambda_{10} \operatorname{Tr}(\varphi^4) - \big(\lambda_{11} (H\phi')^2 + \text{H.c.}\big) \\
& - \left(\frac{1}{2} M_\psi \psi\psi + \text{H.c.}\right) \\
& - \big(\lambda_{12} \big(\phi'^\dagger L\big)\psi + \text{H.c.}\big)
\end{aligned} \tag{E.9}
$$

**T1-1-G ($\alpha = 2$)**

$$\begin{aligned}
\mathcal{L}_{\text{T1-1-G }(\alpha\,=\,2)} = {} & \mathcal{L}_{\text{SM}} + \mathcal{L}_{\text{kin}} - M_\varphi^2 \operatorname{Tr}(\varphi^\dagger\varphi) - M_{\phi'}^2 \phi'^\dagger\phi' - M_\phi^2 \phi^\dagger\phi \\
& - (\lambda_1 H\varphi^\dagger\phi' + \text{H.c.}) - (\lambda_2 \phi^\dagger\varphi H + \text{H.c.}) \\
& - \lambda_3 \operatorname{Tr}\Big((\varphi^\dagger\varphi)^2\Big) - \lambda_4 \operatorname{Tr}(\varphi^\dagger\varphi)^2 - \lambda_5 H^\dagger\varphi^\dagger\varphi H - \lambda_6 \operatorname{Tr}(\varphi^\dagger\varphi)(H^\dagger H) \\
& \quad - \lambda_7 \operatorname{Tr}(\varphi^\dagger\varphi)(\phi'^\dagger\phi') - \lambda_8 \operatorname{Tr}(\varphi^\dagger\varphi)(\phi^\dagger\phi) - \Big(\lambda_9 \operatorname{Tr}\Big(\varphi^{\dagger 2}\Big)(\phi\phi') + \text{H.c.}\Big) \\
& \quad - \Big(\lambda_{10}\phi'\varphi^{\dagger 2}\phi + \text{H.c.}\Big) - \lambda_{11}\phi'^\dagger\varphi^\dagger\varphi\phi' - \lambda_{12}\phi^\dagger\varphi^\dagger\varphi\phi - \lambda_{13}(H^\dagger H)(\phi'^\dagger\phi') \\
& \quad - \lambda_{14}(H^\dagger H)(\phi^\dagger\phi) - \lambda_{15}(H^\dagger\phi'^\dagger)(H\phi') - \lambda_{16}(H^\dagger\phi')(\phi'^\dagger H) \\
& \quad - \lambda_{17}(H^\dagger\phi^\dagger)(H\phi) - \lambda_{18}(H^\dagger\phi)(\phi^\dagger H) - \Big(\lambda_{19}(\phi'^\dagger H)^2 + \text{H.c.}\Big) \\
& \quad - \lambda_{20}(\phi'^\dagger\phi')^2 - \lambda_{21}(\phi'^\dagger\phi')(\phi^\dagger\phi) - \lambda_{22}(\phi'^\dagger\phi)(\phi^\dagger\phi') - \lambda_{23}(\phi^\dagger\phi'^\dagger)(\phi\phi') \\
& \quad - \lambda_{24}(\phi^\dagger\phi)^2 - (\lambda_{25}(\phi^\dagger H)(H\phi') + \text{H.c.}) \\
& - \big(M_{\psi_1\psi_2}\psi_1\psi_2 + \text{H.c.}\big) \\
& - \big(\lambda_{26}(\phi'^\dagger L)\psi_1 + \text{H.c.}\big) - (\lambda_{27}(L\phi)\psi_2 + \text{H.c.}) \qquad \text{(E.10)}
\end{aligned}$$

**T1-1-H ($\alpha = 0$)**

$$\begin{aligned}
\mathcal{L}_{\text{T1-1-H }(\alpha\,=\,0)} = {} & \mathcal{L}_{\text{SM}} + \mathcal{L}_{\text{kin}} - M_{\phi'}^2 \phi'^\dagger\phi' - \frac{1}{2} M_\varphi^2 \operatorname{Tr}(\varphi^2) \\
& - (\lambda_1 H\varphi\phi' + \text{H.c.}) \\
& - \lambda_2(H^\dagger H)(\phi'^\dagger\phi') - \lambda_3(H^\dagger\phi'^\dagger)(H\phi') - \lambda_4(H^\dagger\phi')(\phi'^\dagger H) - \lambda_5(\phi'^\dagger\phi')^2 \\
& \quad - \lambda_6(H^\dagger H)\operatorname{Tr}(\varphi^2) - \lambda_7 H^\dagger\varphi^2 H - \lambda_8\phi'^\dagger\varphi^2\phi' - \lambda_9(\phi'^\dagger\phi')\operatorname{Tr}(\varphi^2) \\
& \quad - \lambda_{10}\operatorname{Tr}(\varphi^4) - \big(\lambda_{11}(H\phi')^2 + \text{H.c.}\big) \\
& - \Big(\frac{1}{2} M_\psi \operatorname{Tr}(\psi\psi) + \text{H.c.}\Big) \\
& - \big(\lambda_{12}\phi'^\dagger\psi L + \text{H.c.}\big) \qquad \text{(E.11)}
\end{aligned}$$

**T1-1-H ($\alpha = 2$)**

$$\begin{aligned}
\mathcal{L}_{\text{T1-1-H }(\alpha\,=\,2)} = {} & \mathcal{L}_{\text{SM}} + \mathcal{L}_{\text{kin}} - M_\varphi^2 \operatorname{Tr}(\varphi^\dagger \varphi) - M_{\phi'}^2 \phi'^\dagger \phi' - M_\phi^2 \phi^\dagger \phi \\
& - (\lambda_1 H \varphi^\dagger \phi' + \text{H.c.}) - (\lambda_2 \phi^\dagger \varphi H + \text{H.c.}) \\
& - \lambda_3 \operatorname{Tr}\Big((\varphi^\dagger\varphi)^2\Big) - \lambda_4 \operatorname{Tr}(\varphi^\dagger\varphi)^2 - \lambda_5 H^\dagger \varphi^\dagger \varphi H - \lambda_6 \operatorname{Tr}(\varphi^\dagger\varphi)(H^\dagger H) \\
& \quad - \lambda_7 \operatorname{Tr}(\varphi^\dagger\varphi)\big(\phi'^\dagger\phi'\big) - \lambda_8 \operatorname{Tr}(\varphi^\dagger\varphi)(\phi^\dagger\phi) - \Big(\lambda_9 \operatorname{Tr}\Big(\varphi^{\dagger 2}\Big)(\phi\phi') + \text{H.c.}\Big) \\
& \quad - \Big(\lambda_{10}\phi'\varphi^{\dagger 2}\phi + \text{H.c.}\Big) - \lambda_{11}\phi'^\dagger\varphi^\dagger\varphi\phi' - \lambda_{12}\phi^\dagger\varphi^\dagger\varphi\phi - \lambda_{13}(H^\dagger H)\big(\phi'^\dagger\phi'\big) \\
& \quad - \lambda_{14}(H^\dagger H)(\phi^\dagger\phi) - \lambda_{15}\big(H^\dagger\phi'^\dagger\big)(H\phi') - \lambda_{16}(H^\dagger\phi')\big(\phi'^\dagger H\big) \\
& \quad - \lambda_{17}(H^\dagger\phi^\dagger)(H\phi) - \lambda_{18}(H^\dagger\phi)(\phi^\dagger H) - \Big(\lambda_{19}\big(\phi'^\dagger H\big)^2 + \text{H.c.}\Big) \\
& \quad - \lambda_{20}\big(\phi'^\dagger\phi'\big)^2 - \lambda_{21}\big(\phi'^\dagger\phi'\big)(\phi^\dagger\phi) - \lambda_{22}\big(\phi'^\dagger\phi\big)(\phi^\dagger\phi') - \lambda_{23}\big(\phi^\dagger\phi'^\dagger\big)(\phi\phi') \\
& \quad - \lambda_{24}(\phi^\dagger\phi)^2 - (\lambda_{25}(\phi^\dagger H)(H\phi') + \text{H.c.}) \\
& - \big(M_{\psi_1\psi_2} \operatorname{Tr}(\psi_1\psi_2) + \text{H.c.}\big) \\
& - \big(\lambda_{26}\phi'^\dagger\psi_1 L + \text{H.c.}\big) - (\lambda_{27}\phi\psi_2 L + \text{H.c.})
\end{aligned} \tag{E.12}$$

**T1-2-A ($\alpha = -2$)**

$$\begin{aligned}
\mathcal{L}_{\text{T1-2-A }(\alpha\,=\,-2)} = {} & \mathcal{L}_{\text{SM}} + \mathcal{L}_{\text{kin}} - M_\phi^2 \phi^\dagger\phi - M_{\phi'}^2 \phi'^\dagger\phi' \\
& - (\lambda_1(\phi^\dagger H)\phi' + \text{H.c.}) \\
& - \lambda_2(H^\dagger H)(\phi^\dagger\phi) - \lambda_3(H^\dagger\phi^\dagger)(H\phi) - \lambda_4(H^\dagger\phi)(\phi^\dagger H) - \lambda_5(\phi^\dagger\phi)^2 \\
& \quad - \lambda_6(H^\dagger H)\phi'^\dagger\phi' - \lambda_7(\phi^\dagger\phi)\phi'^\dagger\phi' - \lambda_8\phi'^{\dagger 2}\phi'^2 - \big(\lambda_9(H\phi)^2 + \text{H.c.}\big) \\
& - \big(M_{\psi_1'\psi_2'}\psi_1'\psi_2' + \text{H.c.}\big) - \big(M_{\psi_1\psi_2}\psi_1\psi_2 + \text{H.c.}\big) \\
& - (\lambda_{10}(H^\dagger\psi_1')\psi_2 + \text{H.c.}) - \big(\lambda_{11}\phi'^\dagger(L\psi_1') + \text{H.c.}\big) - (\lambda_{12}(H\psi_2')\psi_1 + \text{H.c.}) \\
& \quad - (\lambda_{13}(L\phi)\psi_2 + \text{H.c.}) - (\lambda_{14}(\psi_1'\phi)e_{\text{R}}^{\text{c}} + \text{H.c.})
\end{aligned} \tag{E.13}$$

**T1-2-A ($\alpha = 0$)**

$$\begin{aligned}
\mathcal{L}_{\text{T1-2-A }(\alpha\,=\,0)} = {}& \mathcal{L}_{\text{SM}} + \mathcal{L}_{\text{kin}} - M_\phi^2 \phi^\dagger \phi - \frac{1}{2} M_{\phi'}^2 \phi'^2 \\
& - \left(\lambda_1 (H^\dagger \phi)\phi' + \text{H.c.}\right) \\
& - \lambda_2 (H^\dagger H)(\phi^\dagger \phi) - \lambda_3 (H^\dagger \phi^\dagger)(H\phi) - \lambda_4 (H^\dagger \phi)(\phi^\dagger H) - \left(\lambda_5 (\phi^\dagger H)^2 + \text{H.c.}\right) \\
& \quad - \lambda_6 (\phi^\dagger \phi)^2 - \lambda_7 (H^\dagger H)\phi'^2 - \lambda_8 (\phi^\dagger \phi)\phi'^2 - \lambda_9 \phi'^4 \\
& - \left(M_{\psi_1'\psi_2'} \psi_1' \psi_2' + \text{H.c.}\right) - \left(\frac{1}{2} M_\psi \psi\psi + \text{H.c.}\right) \\
& - \left(\lambda_{10} (H^\dagger \psi_1')\psi + \text{H.c.}\right) - \left(\lambda_{11} (\phi^\dagger \psi_2') e_{\text{R}}^{\text{c}} + \text{H.c.}\right) - \left(\lambda_{12} (H\psi_2')\psi + \text{H.c.}\right) \\
& \quad - \left(\lambda_{13} (L\psi_1')\phi' + \text{H.c.}\right) - \left(\lambda_{14} (L\phi)\psi + \text{H.c.}\right)
\end{aligned} \tag{E.14}$$

**T1-2-B ($\alpha = -2$)**

$$\begin{aligned}
\mathcal{L}_{\text{T1-2-B }(\alpha\,=\,-2)} = {}& \mathcal{L}_{\text{SM}} + \mathcal{L}_{\text{kin}} - M_{\phi'}^2 \operatorname{Tr}\left(\phi'^\dagger \phi'\right) - M_\phi^2 \phi^\dagger \phi \\
& - \left(\lambda_1 \phi^\dagger \phi' H + \text{H.c.}\right) \\
& - \lambda_2 \operatorname{Tr}\left(\left(\phi'^\dagger \phi'\right)^2\right) - \lambda_3 \operatorname{Tr}\left(\phi'^\dagger \phi'\right)^2 - \lambda_4 H^\dagger \phi'^\dagger \phi' H - \lambda_5 \operatorname{Tr}\left(\phi'^\dagger \phi'\right)(H^\dagger H) \\
& \quad - \lambda_6 \operatorname{Tr}\left(\phi'^\dagger \phi'\right)(\phi^\dagger \phi) - \lambda_7 \phi^\dagger \phi'^\dagger \phi' \phi - \lambda_8 (H^\dagger H)(\phi^\dagger \phi) - \lambda_9 (H^\dagger \phi^\dagger)(H\phi) \\
& \quad - \lambda_{10} (H^\dagger \phi)(\phi^\dagger H) - \lambda_{11} (\phi^\dagger \phi)^2 - \left(\lambda_{12} (H\phi)^2 + \text{H.c.}\right) \\
& - \left(M_{\psi_1'\psi_2'} \psi_1' \psi_2' + \text{H.c.}\right) - \left(M_{\psi_1\psi_2} \psi_1 \psi_2 + \text{H.c.}\right) \\
& - \left(\lambda_{13} (H^\dagger \psi_1')\psi_2 + \text{H.c.}\right) - \left(\lambda_{14} L \phi'^\dagger \psi_1' + \text{H.c.}\right) - \left(\lambda_{15} (H\psi_2')\psi_1 + \text{H.c.}\right) \\
& \quad - \left(\lambda_{16} (L\phi)\psi_2 + \text{H.c.}\right) - \left(\lambda_{17} (\psi_1' \phi) e_{\text{R}}^{\text{c}} + \text{H.c.}\right)
\end{aligned} \tag{E.15}$$

**T1-2-B ($\alpha = 0$)**

$$\begin{aligned}
\mathcal{L}_{\text{T1-2-B }(\alpha\,=\,0)} = {}& \mathcal{L}_{\text{SM}} + \mathcal{L}_{\text{kin}} - M_\phi^2 \phi^\dagger \phi - \frac{1}{2} M_{\phi'}^2 \operatorname{Tr}\left(\phi'^2\right) \\
& - \left(\lambda_1 \phi^\dagger \phi' H + \text{H.c.}\right) \\
& - \lambda_2 (H^\dagger H)(\phi^\dagger \phi) - \lambda_3 (H^\dagger \phi^\dagger)(H\phi) - \lambda_4 (H^\dagger \phi)(\phi^\dagger H) - \left(\lambda_5 (\phi^\dagger H)^2 + \text{H.c.}\right) \\
& \quad - \lambda_6 (\phi^\dagger \phi)^2 - \lambda_7 (H^\dagger H) \operatorname{Tr}\left(\phi'^2\right) - \lambda_8 H^\dagger \phi'^2 H - \lambda_9 \phi^\dagger \phi'^2 \phi \\
& \quad - \lambda_{10} (\phi^\dagger \phi) \operatorname{Tr}\left(\phi'^2\right) - \lambda_{11} \operatorname{Tr}\left(\phi'^4\right) \\
& - \left(M_{\psi_1'\psi_2'} \psi_1' \psi_2' + \text{H.c.}\right) - \left(\frac{1}{2} M_\psi \psi\psi + \text{H.c.}\right) \\
& - \left(\lambda_{12} (H^\dagger \psi_1')\psi + \text{H.c.}\right) - \left(\lambda_{13} (\phi^\dagger \psi_2') e_{\text{R}}^{\text{c}} + \text{H.c.}\right) - \left(\lambda_{14} (H\psi_2')\psi + \text{H.c.}\right) \\
& \quad - \left(\lambda_{15} L \phi' \psi_1' + \text{H.c.}\right) - \left(\lambda_{16} (L\phi)\psi + \text{H.c.}\right)
\end{aligned} \tag{E.16}$$

**T1-2-D ($\alpha = -1$)**

$$\begin{aligned}
\mathcal{L}_{\text{T1-2-D}\,(\alpha=-1)} = {} & \mathcal{L}_{\text{SM}} + \mathcal{L}_{\text{kin}} - M_{\phi'}^2 {\phi'}^\dagger \phi' - \frac{1}{2} M_\phi^2 \phi^2 \\
& - \left(\lambda_1 (H\phi')\phi + \text{H.c.}\right) \\
& - \lambda_2 \left(H^\dagger H\right)\left({\phi'}^\dagger \phi'\right) - \lambda_3 \left(H^\dagger {\phi'}^\dagger\right)(H\phi') - \lambda_4 \left(H^\dagger \phi'\right)\left({\phi'}^\dagger H\right) - \lambda_5 \left({\phi'}^\dagger \phi'\right)^2 \\
& \quad - \lambda_6 \left(H^\dagger H\right)\phi^2 - \lambda_7 \left({\phi'}^\dagger \phi'\right)\phi^2 - \left(\lambda_8 (H\phi')^2 + \text{H.c.}\right) - \lambda_9 \phi^4 \\
& - \left(\frac{1}{2} M_{\psi'} \operatorname{Tr}(\psi'\psi') + \text{H.c.}\right) - \left(M_{\psi_1\psi_2} \psi_1 \psi_2 + \text{H.c.}\right) \\
& - \left(\lambda_{10} H^\dagger \psi' \psi_2 + \text{H.c.}\right) - \left(\lambda_{11} {\phi'}^\dagger \psi' L + \text{H.c.}\right) - \left(\lambda_{12} H \psi' \psi_1 + \text{H.c.}\right) \\
& \quad - \left(\lambda_{13} (L\psi_2)\phi + \text{H.c.}\right) - \left(\lambda_{14} (\psi_1 \phi') e_{\text{R}}^{\text{c}} + \text{H.c.}\right)
\end{aligned} \tag{E.17}$$

**T1-2-D ($\alpha = 1$)**

$$\begin{aligned}
\mathcal{L}_{\text{T1-2-D}\,(\alpha=1)} = {} & \mathcal{L}_{\text{SM}} + \mathcal{L}_{\text{kin}} - M_{\phi'}^2 {\phi'}^\dagger \phi' - M_\phi^2 \phi^\dagger \phi \\
& - \left(\lambda_1 \phi^\dagger (H\phi') + \text{H.c.}\right) \\
& - \lambda_2 \left(H^\dagger H\right)\left({\phi'}^\dagger \phi'\right) - \lambda_3 \left(H^\dagger {\phi'}^\dagger\right)(H\phi') - \lambda_4 \left(H^\dagger \phi'\right)\left({\phi'}^\dagger H\right) \\
& \quad - \left(\lambda_5 \left({\phi'}^\dagger H\right)^2 + \text{H.c.}\right) - \lambda_6 \left({\phi'}^\dagger \phi'\right)^2 - \lambda_7 \left(H^\dagger H\right)\phi^\dagger \phi - \lambda_8 \left({\phi'}^\dagger \phi'\right)\phi^\dagger \phi \\
& \quad - \lambda_9 {\phi^\dagger}^2 \phi^2 \\
& - \left(M_{\psi_1'\psi_2'} \operatorname{Tr}(\psi_1' \psi_2') + \text{H.c.}\right) - \left(M_{\psi_1\psi_2} \psi_1 \psi_2 + \text{H.c.}\right) \\
& - \left(\lambda_{10} H^\dagger \psi_1' \psi_2 + \text{H.c.}\right) - \left(\lambda_{11} {\phi'}^\dagger \psi_1' L + \text{H.c.}\right) - \left(\lambda_{12} \left({\phi'}^\dagger \psi_2\right) e_{\text{R}}^{\text{c}} + \text{H.c.}\right) \\
& \quad - \left(\lambda_{13} H \psi_2' \psi_1 + \text{H.c.}\right) - \left(\lambda_{14} (L\psi_2)\phi + \text{H.c.}\right)
\end{aligned} \tag{E.18}$$

**T1-2-F ($\alpha = -1$)**

$$\begin{aligned}
\mathcal{L}_{\text{T1-2-F}\,(\alpha=-1)} = {} & \mathcal{L}_{\text{SM}} + \mathcal{L}_{\text{kin}} - M_{\phi'}^2 {\phi'}^\dagger \phi' - \frac{1}{2} M_\phi^2 \operatorname{Tr}(\phi^2) \\
& - \left(\lambda_1 H \phi \phi' + \text{H.c.}\right) \\
& - \lambda_2 \left(H^\dagger H\right)\left({\phi'}^\dagger \phi'\right) - \lambda_3 \left(H^\dagger {\phi'}^\dagger\right)(H\phi') - \lambda_4 \left(H^\dagger \phi'\right)\left({\phi'}^\dagger H\right) - \lambda_5 \left({\phi'}^\dagger \phi'\right)^2 \\
& \quad - \lambda_6 \left(H^\dagger H\right) \operatorname{Tr}(\phi^2) - \lambda_7 H^\dagger \phi^2 H - \lambda_8 \left({\phi'}^\dagger \phi'\right) \operatorname{Tr}(\phi^2) - \lambda_9 {\phi'}^\dagger \phi^2 \phi' \\
& \quad - \lambda_{10} \operatorname{Tr}(\phi^4) - \left(\lambda_{11} (H\phi')^2 + \text{H.c.}\right) \\
& - \left(\frac{1}{2} M_{\psi'} \operatorname{Tr}(\psi'\psi') + \text{H.c.}\right) - \left(M_{\psi_1\psi_2} \psi_1 \psi_2 + \text{H.c.}\right) \\
& - \left(\lambda_{12} H^\dagger \psi' \psi_2 + \text{H.c.}\right) - \left(\lambda_{13} {\phi'}^\dagger \psi' L + \text{H.c.}\right) - \left(\lambda_{14} H \psi' \psi_1 + \text{H.c.}\right) \\
& \quad - \left(\lambda_{15} \psi_2 \phi L + \text{H.c.}\right) - \left(\lambda_{16} (\psi_1 \phi') e_{\text{R}}^{\text{c}} + \text{H.c.}\right)
\end{aligned} \tag{E.19}$$

**T1-2-F ($\alpha = 1$)**

$$
\begin{aligned}
\mathcal{L}_{\text{T1-2-F }(\alpha\,=\,1)} &= \mathcal{L}_{\text{SM}} + \mathcal{L}_{\text{kin}} - M_\phi^2 \operatorname{Tr}\left(\phi^\dagger\phi\right) - M_{\phi'}^2 \phi'^\dagger\phi' \\
&\quad - \left(\lambda_1 H\phi^\dagger\phi' + \text{H.c.}\right) \\
&\quad - \lambda_2 \operatorname{Tr}\Big(\left(\phi^\dagger\phi\right)^2\Big) - \lambda_3 \operatorname{Tr}\left(\phi^\dagger\phi\right)^2 - \lambda_4 H^\dagger\phi^\dagger\phi H - \lambda_5 \operatorname{Tr}\left(\phi^\dagger\phi\right)\left(H^\dagger H\right) \\
&\qquad - \lambda_6 \operatorname{Tr}\left(\phi^\dagger\phi\right)\left(\phi'^\dagger\phi'\right) - \lambda_7 \phi'^\dagger\phi^\dagger\phi\phi' - \lambda_8\left(H^\dagger H\right)\left(\phi'^\dagger\phi'\right) \\
&\qquad - \lambda_9\left(H^\dagger\phi'^\dagger\right)\left(H\phi'\right) - \lambda_{10}\left(H^\dagger\phi'\right)\left(\phi'^\dagger H\right) - \Big(\lambda_{11}\left(\phi'^\dagger H\right)^2 + \text{H.c.}\Big) \\
&\qquad - \lambda_{12}\left(\phi'^\dagger\phi'\right)^2 \\
&\quad - \left(M_{\psi_1'\psi_2'} \operatorname{Tr}\left(\psi_1'\psi_2'\right) + \text{H.c.}\right) - \left(M_{\psi_1\psi_2}\psi_1\psi_2 + \text{H.c.}\right) \\
&\quad - \left(\lambda_{13} H^\dagger\psi_1'\psi_2 + \text{H.c.}\right) - \left(\lambda_{14}\phi'^\dagger\psi_1' L + \text{H.c.}\right) - \left(\lambda_{15}\left(\phi'^\dagger\psi_2\right)e_{\text{R}}^{\text{c}} + \text{H.c.}\right) \\
&\qquad - \left(\lambda_{16} H\psi_2'\psi_1 + \text{H.c.}\right) - \left(\lambda_{17}\psi_2\phi L + \text{H.c.}\right)
\end{aligned} \tag{E.20}
$$

**T1-3-A ($\alpha = 0$)**

$$
\begin{aligned}
\mathcal{L}_{\text{T1-3-A }(\alpha\,=\,0)} &= \mathcal{L}_{\text{SM}} + \mathcal{L}_{\text{kin}} - \frac{1}{2} M_\phi^2 \phi^2 \\
&\quad - \lambda_1\left(H^\dagger H\right)\phi^2 - \lambda_2\phi^4 \\
&\quad - \left(M_{\psi\psi'}\psi\psi' + \text{H.c.}\right) - \left(\frac{1}{2} M_\Psi \Psi\Psi + \text{H.c.}\right) \\
&\quad - \left(\lambda_3\left(H^\dagger\psi'\right)\Psi + \text{H.c.}\right) - \left(\lambda_4(H\psi)\Psi + \text{H.c.}\right) - \left(\lambda_5(L\psi')\phi + \text{H.c.}\right)
\end{aligned} \tag{E.21}
$$

**T1-3-B ($\alpha = 0$)**

$$
\begin{aligned}
\mathcal{L}_{\text{T1-3-B }(\alpha\,=\,0)} &= \mathcal{L}_{\text{SM}} + \mathcal{L}_{\text{kin}} - \frac{1}{2} M_\phi^2 \operatorname{Tr}\left(\phi^2\right) \\
&\quad - \lambda_1\left(H^\dagger H\right) \operatorname{Tr}\left(\phi^2\right) - \lambda_2 H^\dagger\phi^2 H - \lambda_3 \operatorname{Tr}\left(\phi^4\right) \\
&\quad - \left(M_{\psi\psi'}\psi\psi' + \text{H.c.}\right) - \left(\frac{1}{2} M_\Psi \Psi\Psi + \text{H.c.}\right) \\
&\quad - \left(\lambda_4\left(H^\dagger\psi'\right)\Psi + \text{H.c.}\right) - \left(\lambda_5(H\psi)\Psi + \text{H.c.}\right) - \left(\lambda_6 L\phi\psi' + \text{H.c.}\right)
\end{aligned} \tag{E.22}
$$

**T1-3-C ($\alpha = 1$)**

$$\begin{aligned}
\mathcal{L}_{\text{T1-3-C }(\alpha=1)} = {} & \mathcal{L}_{\text{SM}} + \mathcal{L}_{\text{kin}} - M_\phi^2 \phi^\dagger \phi \\
& - \lambda_1 (H^\dagger H)(\phi^\dagger \phi) - \lambda_2 (H^\dagger \phi^\dagger)(H\phi) - \lambda_3 (H^\dagger \phi)(\phi^\dagger H) - \left(\lambda_4 (\phi^\dagger H)^2 + \text{H.c.}\right) \\
& \quad - \lambda_5 (\phi^\dagger \phi)^2 \\
& - \left(M_{\Psi_1\Psi_2} \Psi_1 \Psi_2 + \text{H.c.}\right) - \left(\frac{1}{2} M_\psi \psi\psi + \text{H.c.}\right) - \left(M_{\psi_1'\psi_2'} \psi_1' \psi_2' + \text{H.c.}\right) \\
& - \left(\lambda_6 (H^\dagger \Psi_1)\psi + \text{H.c.}\right) - \left(\lambda_7 (H^\dagger \Psi_2)\psi_1' + \text{H.c.}\right) - \left(\lambda_8 (\phi^\dagger L)\psi_1' + \text{H.c.}\right) \\
& \quad - \left(\lambda_9 (\phi^\dagger \Psi_2) e_R^c + \text{H.c.}\right) - \left(\lambda_{10} (H\Psi_1)\psi_2' + \text{H.c.}\right) - \left(\lambda_{11} (H\Psi_2)\psi + \text{H.c.}\right) \\
& \quad - \left(\lambda_{12} (L\phi)\psi + \text{H.c.}\right)
\end{aligned} \tag{E.23}$$

**T1-3-D ($\alpha = -1$)**

$$\begin{aligned}
\mathcal{L}_{\text{T1-3-D }(\alpha=-1)} = {} & \mathcal{L}_{\text{SM}} + \mathcal{L}_{\text{kin}} - M_\phi^2 \phi^\dagger \phi \\
& - \lambda_1 (H^\dagger H)(\phi^\dagger \phi) - \lambda_2 (H^\dagger \phi^\dagger)(H\phi) - \lambda_3 (H^\dagger \phi)(\phi^\dagger H) - \lambda_4 (\phi^\dagger \phi)^2 \\
& \quad - \left(\lambda_5 (H\phi)^2 + \text{H.c.}\right) \\
& - \left(M_{\psi_1\psi_2} \operatorname{Tr}(\psi_1 \psi_2) + \text{H.c.}\right) - \left(M_{\Psi_1\Psi_2} \Psi_1 \Psi_2 + \text{H.c.}\right) - \left(\frac{1}{2} M_{\psi'} \psi' \psi' + \text{H.c.}\right) \\
& - \left(\lambda_6 H^\dagger \psi_2 \Psi_1 + \text{H.c.}\right) - \left(\lambda_7 (H^\dagger \Psi_2)\psi' + \text{H.c.}\right) - \left(\lambda_8 (\phi^\dagger L)\psi' + \text{H.c.}\right) \\
& \quad - \left(\lambda_9 H \psi_1 \Psi_2 + \text{H.c.}\right) - \left(\lambda_{10} (H\Psi_1)\psi' + \text{H.c.}\right) - \left(\lambda_{11} \phi \psi_2 L + \text{H.c.}\right) \\
& \quad - \left(\lambda_{12} (\Psi_1 \phi) e_R^c + \text{H.c.}\right)
\end{aligned} \tag{E.24}$$

**T1-3-D ($\alpha = 1$)**

$$\begin{aligned}
\mathcal{L}_{\text{T1-3-D }(\alpha=1)} = {} & \mathcal{L}_{\text{SM}} + \mathcal{L}_{\text{kin}} - M_\phi^2 \phi^\dagger \phi \\
& - \lambda_1 (H^\dagger H)(\phi^\dagger \phi) - \left(\lambda_2 (H^\dagger \phi)^2 + \text{H.c.}\right) - \lambda_3 (H^\dagger \phi^\dagger)(H\phi) - \lambda_4 (H^\dagger \phi)(\phi^\dagger H) \\
& \quad - \lambda_5 (\phi^\dagger \phi)^2 \\
& - \left(\frac{1}{2} M_\psi \operatorname{Tr}(\psi\psi) + \text{H.c.}\right) - \left(M_{\Psi_1\Psi_2} \Psi_1 \Psi_2 + \text{H.c.}\right) - \left(M_{\psi_1'\psi_2'} \psi_1' \psi_2' + \text{H.c.}\right) \\
& - \left(\lambda_6 H^\dagger \psi \Psi_1 + \text{H.c.}\right) - \left(\lambda_7 (H^\dagger \Psi_2)\psi_1' + \text{H.c.}\right) - \left(\lambda_8 (\phi^\dagger L)\psi_1' + \text{H.c.}\right) \\
& \quad - \left(\lambda_9 (\phi^\dagger \Psi_2) e_R^c + \text{H.c.}\right) - \left(\lambda_{10} H \psi \Psi_2 + \text{H.c.}\right) - \left(\lambda_{11} (H\Psi_1)\psi_2' + \text{H.c.}\right) \\
& \quad - \left(\lambda_{12} \phi \psi L + \text{H.c.}\right)
\end{aligned} \tag{E.25}$$

**T1-3-F ($\alpha = 1$)**

$$
\begin{aligned}
\mathcal{L}_{\text{T1-3-F}\,(\alpha\,=\,1)} = {} & \mathcal{L}_{\text{SM}} + \mathcal{L}_{\text{kin}} - M_\phi^2 \phi^\dagger \phi \\
& - \lambda_1 (H^\dagger H)(\phi^\dagger \phi) - \left(\lambda_2 (H^\dagger \phi)^2 + \text{H.c.}\right) - \lambda_3 (H^\dagger \phi^\dagger)(H\phi) - \lambda_4 (H^\dagger \phi)(\phi^\dagger H) \\
& \quad - \lambda_5 (\phi^\dagger \phi)^2 \\
& - \left(\frac{1}{2} M_\psi \operatorname{Tr}(\psi\psi) + \text{H.c.}\right) - \left(M_{\psi_1'\psi_2'} \operatorname{Tr}(\psi_1' \psi_2') + \text{H.c.}\right) \\
& \quad - \left(M_{\Psi_1\Psi_2} \Psi_1 \Psi_2 + \text{H.c.}\right) \\
& - (\lambda_6 H^\dagger \psi \Psi_1 + \text{H.c.}) - (\lambda_7 H^\dagger \psi_1' \Psi_2 + \text{H.c.}) - (\lambda_8 \phi^\dagger \psi_1' L + \text{H.c.}) \\
& \quad - (\lambda_9 (\phi^\dagger \Psi_2) e_{\text{R}}^{\text{c}} + \text{H.c.}) - (\lambda_{10} H \psi \Psi_2 + \text{H.c.}) - (\lambda_{11} H \psi_2' \Psi_1 + \text{H.c.}) \\
& \quad - (\lambda_{12} \phi \psi L + \text{H.c.})
\end{aligned} \tag{E.26}
$$

**T1-3-G ($\alpha = 0$)**

$$
\begin{aligned}
\mathcal{L}_{\text{T1-3-G}\,(\alpha\,=\,0)} = {} & \mathcal{L}_{\text{SM}} + \mathcal{L}_{\text{kin}} - \frac{1}{2} M_\phi^2 \phi^2 \\
& - \lambda_1 (H^\dagger H) \phi^2 - \lambda_2 \phi^4 \\
& - \left(\frac{1}{2} M_\Psi \operatorname{Tr}(\Psi\Psi) + \text{H.c.}\right) - \left(M_{\psi\psi'} \psi \psi' + \text{H.c.}\right) \\
& - \left(\lambda_3 H^\dagger \Psi \psi' + \text{H.c.}\right) - (\lambda_4 H \Psi \psi + \text{H.c.}) - (\lambda_5 (L\psi') \phi + \text{H.c.})
\end{aligned} \tag{E.27}
$$

**T1-3-H ($\alpha = 0$)**

$$
\begin{aligned}
\mathcal{L}_{\text{T1-3-H}\,(\alpha\,=\,0)} = {} & \mathcal{L}_{\text{SM}} + \mathcal{L}_{\text{kin}} - \frac{1}{2} M_\phi^2 \operatorname{Tr}(\phi^2) \\
& - \lambda_1 (H^\dagger H) \operatorname{Tr}(\phi^2) - \lambda_2 H^\dagger \phi^2 H - \lambda_3 \operatorname{Tr}(\phi^4) \\
& - \left(\frac{1}{2} M_\Psi \operatorname{Tr}(\Psi\Psi) + \text{H.c.}\right) - \left(M_{\psi\psi'} \psi \psi' + \text{H.c.}\right) \\
& - \left(\lambda_4 H^\dagger \Psi \psi' + \text{H.c.}\right) - (\lambda_5 H \Psi \psi + \text{H.c.}) - (\lambda_6 L \phi \psi' + \text{H.c.})
\end{aligned} \tag{E.28}
$$

**T3-A ($\alpha = -2$)**

$$
\begin{aligned}
\mathcal{L}_{\text{T3-A}\,(\alpha\,=\,-2)} = {} & \mathcal{L}_{\text{SM}} + \mathcal{L}_{\text{kin}} - M_{\phi'}^2 {\phi'}^\dagger \phi' - \frac{1}{2} M_\phi^2 \operatorname{Tr}(\phi^2) \\
& - \lambda_1 (H^\dagger H) {\phi'}^\dagger \phi' - \lambda_2 {\phi'}^{\dagger 2} {\phi'}^2 - \lambda_3 (H^\dagger H) \operatorname{Tr}(\phi^2) - \lambda_4 H^\dagger \phi^2 H \\
& \quad - \lambda_5 {\phi'}^\dagger \phi' \operatorname{Tr}(\phi^2) - \lambda_6 \operatorname{Tr}(\phi^4) - (\lambda_7 (H\phi H) \phi' + \text{H.c.}) \\
& - \left(M_{\psi_1\psi_2} \psi_1 \psi_2 + \text{H.c.}\right) \\
& - \left(\lambda_8 {\phi'}^\dagger (L\psi_1) + \text{H.c.}\right) - (\lambda_9 \psi_2 \phi L + \text{H.c.})
\end{aligned} \tag{E.29}
$$

**T3-A ($\alpha = 0$)**

$$\begin{aligned}\mathcal{L}_{\text{T3-A }(\alpha\,=\,0)} = {} & \mathcal{L}_{\text{SM}} + \mathcal{L}_{\text{kin}} - M_\phi^2 \operatorname{Tr}\left(\phi^\dagger\phi\right) - \frac{1}{2}M_{\phi'}^2\phi'^2 \\ & - \lambda_1 \operatorname{Tr}\left(\left(\phi^\dagger\phi\right)^2\right) - \lambda_2 \operatorname{Tr}\left(\phi^\dagger\phi\right)^2 - \lambda_3 H^\dagger\phi^\dagger\phi H - \lambda_4 \operatorname{Tr}\left(\phi^\dagger\phi\right)\left(H^\dagger H\right) \\ & \quad - \left(\lambda_5\left(H\phi^\dagger H\right)\phi' + \text{H.c.}\right) - \lambda_6 \operatorname{Tr}\left(\phi^\dagger\phi\right)\phi'^2 - \lambda_7\left(H^\dagger H\right)\phi'^2 - \lambda_8\phi'^4 \\ & - \left(M_{\psi_1\psi_2}\psi_1\psi_2 + \text{H.c.}\right) \\ & - \left(\lambda_9\psi_2\phi L + \text{H.c.}\right) - \left(\lambda_{10}(L\psi_1)\phi' + \text{H.c.}\right)\end{aligned} \tag{E.30}$$

**T3-B ($\alpha = -1$)**

$$\begin{aligned}\mathcal{L}_{\text{T3-B }(\alpha\,=\,-1)} = {} & \mathcal{L}_{\text{SM}} + \mathcal{L}_{\text{kin}} - M_{\phi'}^2\phi'^\dagger\phi' \\ & - \lambda_1\left(H^\dagger H\right)\left(\phi'^\dagger\phi'\right) - \lambda_2\left(H^\dagger\phi'^\dagger\right)\left(H\phi'\right) - \lambda_3\left(H^\dagger\phi'\right)\left(\phi'^\dagger H\right) - \lambda_4\left(\phi'^\dagger\phi'\right)^2 \\ & \quad - \left(\lambda_5\left(H\phi'\right)^2 + \text{H.c.}\right) \\ & - \left(\frac{1}{2}M_\psi\psi\psi + \text{H.c.}\right) \\ & - \left(\lambda_6\left(\phi'^\dagger L\right)\psi + \text{H.c.}\right)\end{aligned} \tag{E.31}$$

**T3-B ($\alpha = 1$)**

$$\begin{aligned}\mathcal{L}_{\text{T3-B }(\alpha\,=\,1)} = {} & \mathcal{L}_{\text{SM}} + \mathcal{L}_{\text{kin}} - M_{\phi'}^2\phi'^\dagger\phi' - M_\phi^2\phi^\dagger\phi \\ & - \lambda_1\left(H^\dagger H\right)\left(\phi'^\dagger\phi'\right) - \lambda_2\left(H^\dagger H\right)\left(\phi^\dagger\phi\right) - \left(\lambda_3\left(H^\dagger\phi'\right)^2 + \text{H.c.}\right) \\ & \quad - \lambda_4\left(H^\dagger\phi'^\dagger\right)\left(H\phi'\right) - \lambda_5\left(H^\dagger\phi'\right)\left(\phi'^\dagger H\right) - \lambda_6\left(H^\dagger\phi^\dagger\right)\left(H\phi\right) \\ & \quad - \lambda_7\left(H^\dagger\phi\right)\left(\phi^\dagger H\right) - \lambda_8\left(\phi'^\dagger\phi'\right)^2 - \lambda_9\left(\phi'^\dagger\phi'\right)\left(\phi^\dagger\phi\right) - \lambda_{10}\left(\phi'^\dagger\phi\right)\left(\phi^\dagger\phi'\right) \\ & \quad - \lambda_{11}\left(\phi^\dagger\phi'^\dagger\right)\left(\phi\phi'\right) - \lambda_{12}\left(\phi^\dagger\phi\right)^2 - \left(\lambda_{13}\left(\phi^\dagger H\right)\left(H\phi'\right) + \text{H.c.}\right) \\ & - \left(M_{\psi_1\psi_2}\psi_1\psi_2 + \text{H.c.}\right) \\ & - \left(\lambda_{14}\left(\phi'^\dagger L\right)\psi_1 + \text{H.c.}\right) - \left(\lambda_{15}(L\phi)\psi_2 + \text{H.c.}\right)\end{aligned} \tag{E.32}$$

**T3-C ($\alpha = -1$)**

$$\begin{aligned}\mathcal{L}_{\text{T3-C }(\alpha\,=\,-1)} = {} & \mathcal{L}_{\text{SM}} + \mathcal{L}_{\text{kin}} - M_{\phi'}^2\phi'^\dagger\phi' \\ & - \lambda_1\left(H^\dagger H\right)\left(\phi'^\dagger\phi'\right) - \lambda_2\left(H^\dagger\phi'^\dagger\right)\left(H\phi'\right) - \lambda_3\left(H^\dagger\phi'\right)\left(\phi'^\dagger H\right) - \lambda_4\left(\phi'^\dagger\phi'\right)^2 \\ & \quad - \left(\lambda_5\left(H\phi'\right)^2 + \text{H.c.}\right) \\ & - \left(\frac{1}{2}M_\psi \operatorname{Tr}(\psi\psi) + \text{H.c.}\right) \\ & - \left(\lambda_6\phi'^\dagger\psi L + \text{H.c.}\right)\end{aligned} \tag{E.33}$$

**T3-C ($\alpha = 1$)**

$$\begin{aligned}
\mathcal{L}_{\text{T3-C }(\alpha\,=\,1)} &= \mathcal{L}_{\text{SM}} + \mathcal{L}_{\text{kin}} - M_{\phi'}^2 \phi'^\dagger \phi' - M_\phi^2 \phi^\dagger \phi \\
&- \lambda_1 (H^\dagger H)\big(\phi'^\dagger \phi'\big) - \lambda_2 (H^\dagger H)(\phi^\dagger \phi) - \Big(\lambda_3 (H^\dagger \phi')^2 + \text{H.c.}\Big) \\
&\quad - \lambda_4 \big(H^\dagger \phi'^\dagger\big)(H\phi') - \lambda_5 (H^\dagger \phi')\big(\phi'^\dagger H\big) - \lambda_6 (H^\dagger \phi^\dagger)(H\phi) \\
&\quad - \lambda_7 (H^\dagger \phi)(\phi^\dagger H) - \lambda_8 \big(\phi'^\dagger \phi'\big)^2 - \lambda_9 \big(\phi'^\dagger \phi'\big)(\phi^\dagger \phi) - \lambda_{10} \big(\phi'^\dagger \phi\big)(\phi^\dagger \phi') \\
&\quad - \lambda_{11} \big(\phi^\dagger \phi'^\dagger\big)(\phi\phi') - \lambda_{12} (\phi^\dagger \phi)^2 - (\lambda_{13} (\phi^\dagger H)(H\phi') + \text{H.c.}) \\
&- \Big(M_{\psi_1\psi_2} \operatorname{Tr}(\psi_1 \psi_2) + \text{H.c.}\Big) \\
&- \Big(\lambda_{14} \phi'^\dagger \psi_1 L + \text{H.c.}\Big) - (\lambda_{15} \phi \psi_2 L + \text{H.c.})
\end{aligned} \tag{E.34}$$

**T3-E ($\alpha = 0$)**

$$\begin{aligned}
\mathcal{L}_{\text{T3-E }(\alpha\,=\,0)} &= \mathcal{L}_{\text{SM}} + \mathcal{L}_{\text{kin}} - M_\phi^2 \operatorname{Tr}(\phi^\dagger \phi) - \frac{1}{2} M_{\phi'}^2 \operatorname{Tr}\big(\phi'^2\big) \\
&- \lambda_1 \operatorname{Tr}\Big((\phi^\dagger \phi)^2\Big) - \lambda_2 \operatorname{Tr}(\phi^\dagger \phi)^2 - \lambda_3 H^\dagger \phi^\dagger \phi H - \lambda_4 \operatorname{Tr}(\phi^\dagger \phi)(H^\dagger H) \\
&\quad - \lambda_5 \operatorname{Tr}\big(\phi^\dagger \phi \phi'^2\big) - \lambda_6 \operatorname{Tr}(\phi^\dagger \phi' \phi \phi') - \lambda_7 \operatorname{Tr}(\phi^\dagger \phi') \operatorname{Tr}(\phi \phi') \\
&\quad - \lambda_8 \operatorname{Tr}\big(\phi^\dagger \phi'^2 \phi\big) - \lambda_9 \operatorname{Tr}(\phi^\dagger \phi) \operatorname{Tr}\big(\phi'^2\big) - (\lambda_{10} H \phi^\dagger \phi' H + \text{H.c.}) \\
&\quad - \lambda_{11} (H^\dagger H) \operatorname{Tr}\big(\phi'^2\big) - \lambda_{12} H^\dagger \phi'^2 H - \lambda_{13} \operatorname{Tr}\big(\phi'^4\big) \\
&- \Big(M_{\psi_1\psi_2} \psi_1 \psi_2 + \text{H.c.}\Big) \\
&- (\lambda_{14} L \phi' \psi_1 + \text{H.c.}) - (\lambda_{15} \psi_2 \phi L + \text{H.c.})
\end{aligned} \tag{E.35}$$

# F Patches for SARAH

Patches to the files SPhenoBoundaryEW.m and SPhenoLoopDecaysReal.m provided by SARAH author Florian Staub were applied to the code of SARAH version 4.12.3 used for this thesis. They correct some compatibility problems of the generated SPheno code when using the conventions required by micrOMEGAs (real mixing matrices). These patches are listed in the following. SARAH version 4.13.0, released 11th May 2018, incorporates these changes in an official release.

### SPhenoBoundaryEW.m

```
--- SARAH-4.12.3/Package/SPheno/SPhenoBoundaryEW.m    2017-10-14 15:09:06.000000000 +0200
+++ SARAH-4.12.3/Package/SPheno/SPhenoBoundaryEW.m    2018-03-06 13:34:58.380305423 +0100
@@ -232,7 +232,7 @@

 WriteString[sphenoSugra,"alphaQ = AlphaEw_T(alphaEW_MS,mudim,"];
 For[i=1,i<=Length[coupAlphaEWSB],
-WriteString[sphenoSugra,SPhenoForm[SPhenoMass[coupAlphaEWSB[[i,1]]]]];
+WriteString[sphenoSugra,"Abs("<>SPhenoForm[SPhenoMass[coupAlphaEWSB[[i,1]]]]<>")"];
 If[i!= Length[coupAlphaEWSB],
 WriteString[sphenoSugra,","];
 ];
@@ -250,7 +250,7 @@
 WriteString[sphenoSugra,"alpha3 = AlphaS_T(alphaS_MS,mudim,"];

 For[i=1,i<=Length[coupAlphaStrong],
-WriteString[sphenoSugra,SPhenoForm[SPhenoMass[coupAlphaStrong[[i,1]]]]];
+WriteString[sphenoSugra,"Abs("<>SPhenoForm[SPhenoMass[coupAlphaStrong[[i,1]]]]<>")"];
 If[i!= Length[coupAlphaStrong],
 WriteString[sphenoSugra,","];
 ];
@@ -2038,7 +2038,7 @@

```

```
 WriteString[sphenoSugra,"alphaMZ = AlphaEwDR(mZ,"];
 For[i=1,i<=Length[coupAlphaEWSB],
-WriteString[sphenoSugra,SPhenoForm[SPhenoMass[coupAlphaEWSB[[i,1]]]]];
+WriteString[sphenoSugra,"Abs("<>SPhenoForm[SPhenoMass[coupAlphaEWSB[[
↪  i,1]]]]<>")"];
 If[i!= Length[coupAlphaEWSB],
 WriteString[sphenoSugra,","];
 ];
@@ -2056,7 +2056,7 @@
 WriteString[sphenoSugra,"alpha3 = AlphaSDR(mZ,"];

 For[i=1,i<=Length[coupAlphaStrong],
-WriteString[sphenoSugra,SPhenoForm[SPhenoMass[coupAlphaStrong[[i,1]]]]];
+WriteString[sphenoSugra,"Abs("<>SPhenoForm[SPhenoMass[coupAlphaStrong[[
↪  i,1]]]]<>")"];
 If[i!= Length[coupAlphaStrong],
 WriteString[sphenoSugra,","];
 ];
```

**SPhenoLoopDecaysReal.m**

```
--- SARAH-4.12.3/Package/SPheno/SPhenoLoopDecaysReal.m    2017-06-29
↪  11:36:03.000000000 +0200
+++ SARAH-4.12.3/Package/SPheno/SPhenoLoopDecaysReal.m    2018-03-14
↪  14:39:22.987642799 +0100
@@ -198,7 +198,7 @@
 ];


-WriteString[outputfortran,"If (Mex1.gt.(Mex2+Mex3)) Then \n"];
+WriteString[outputfortran,"If (Abs(Mex1).gt.(Abs(Mex2)+Abs(Mex3))) Then
↪  \n"];
 Switch[type,
 FFV,

```

# G SARAH model files

As discussed in section 7.6, including all model files used with tools such as SARAH is crucial in allowing others to reproduce results. For this reason, the full SARAH model files used for chapter 7 are listed here.

## G.1. T1-3-B ($\alpha = 0$) with one triplet generation

**T1-3-B_0.m**

```
Model`Name = "T13Balpha0";
Model`NameLaTeX = "T1-3-B (\\alpha = 0)";
Model`Authors = "Simon May";
Model`Date = "2018-04-21";

(*-------------------------------------------*)
(*   particle content                        *)
(*-------------------------------------------*)

(* global symmetries *)
(* discrete 2 symmetry *)
Global[[1]] = {Z[2], Z2};

(* gauge groups *)
Gauge[[1]] = {B,   U[1], hypercharge, g1, False, 1};
Gauge[[2]] = {WB, SU[2], left,        g2, True,  1};
Gauge[[3]] = {G,  SU[3], color,       g3, False, 1};

(* matter fields *)
(*                    {name, gens, components, Y/2,  SU(2), SU(3), global} *)
(* Standard Model *)
FermionFields[[1]] = {q,    3,    {uL, dL},    1/6, 2,     3,     1};
FermionFields[[2]] = {l,    3,    {vL, eL},   -1/2, 2,     1,     1};
FermionFields[[3]] = {d,    3,    conj[dR],    1/3, 1,    -3,     1};
FermionFields[[4]] = {u,    3,    conj[uR],   -2/3, 1,    -3,     1};
FermionFields[[5]] = {e,    3,    conj[eR],      1, 1,     1,     1};

ScalarFields[[1]]  = {H,    1,    {Hp, H0},    1/2, 2,     1,     1};

(* new fields *)
FermionFields[[6]] = {CPsi, 1, CPsi0, 0, 1, 1, -1};
FermionFields[[7]] = {psip, 1, {psipp, psip0}, 1/2, 2, 1, -1};
FermionFields[[8]] = {psi, 1, {psi0, psim}, -1/2, 2, 1, -1};
```

```
ScalarFields[[2]]  = {phi, 1, {{phi0/Sqrt[2], phip}, {conj[phip], -phi0/Sqrt[2]}}, 0, 3, 1, -1};
RealScalars = {phi0};


(*----------------------------------------------*)
(*   DEFINITION                                 *)
(*----------------------------------------------*)
NameOfStates = {GaugeES, EWSB};

(* ----- before EWSB ----- *)
DEFINITION[GaugeES][LagrangianInput] = {
    {LagHC,      {AddHC -> True }},
    {LagNoHC,    {AddHC -> False}},
    {LagBSMHC,   {AddHC -> True }},
    {LagBSMNoHC, {AddHC -> False}}
};

(* Standard Model Lagrangian *)
LagHC      = -Yd conj[H].d.q - Ye conj[H].e.l - Yu u.q.H;
LagNoHC    = mu2 conj[H].H - 1/2 λ conj[H].H.conj[H].H;

(* Lagrangian involving the new fields *)
LagBSMNoHC = - 1/2 mφ2 phi.phi \
     - λ1 conj[H].H.phi.phi (*- λ2 conj[H].phi.phi.H*) - λ3 phi.phi.phi.phi;
LagBSMHC   = - 1/2 mΨ CPsi.CPsi - mψψp psi.psip \
     - λ4 conj[H].psip.CPsi - λ5 psi.H.CPsi - λ6 l.phi.psip;

(* ----- after EWSB ----- *)
(* gauge sector mixing *)
DEFINITION[EWSB][GaugeSector] = {
    {{VB,     VWB[3]}, {VP, VZ}, ZZ},
    {{VWB[1], VWB[2]}, {VWp, conj[VWp]}, ZW}
};

(* VEVs *)
DEFINITION[EWSB][VEVs] = {
    (* Standard Model Higgs VEV *)
    {H0,
        {v, 1/Sqrt[2]},
        {Ah, \[ImaginaryI]/Sqrt[2]},
        {hh, 1/Sqrt[2]}
    }
};

(* mixing *)
DEFINITION[EWSB][MatterSector] = {
    (* Standard Model mixing *)
    {{{dL}, {conj[dR]}}, {{DL, Vd}, {DR, Ud}}},
    {{{uL}, {conj[uR]}}, {{UL, Vu}, {UR, Uu}}},
    {{{eL}, {conj[eR]}}, {{EL, Ve}, {ER, Ue}}},
    (* mixing of new fields can be added here *)
    {{vL}, {VL, Uneu}},
    {{CPsi0, psi0, psip0}, {chi, Uferm}}
};
```

```
(* Dirac spinors *)
DEFINITION[EWSB][DiracSpinors] = {
    Fd -> {DL, conj[DR]},
    Fe -> {EL, conj[ER]},
    Fu -> {UL, conj[UR]},
    Fnu -> {VL, conj[VL]},
    (* new fields *)
    Fpsi -> {psim, conj[psipp]},
    Fchi -> {chi, conj[chi]}
};

(* phases, see http://stauby.de/sarah_wiki/index.php?title=Phases *)
DEFINITION[EWSB][Phases] = {
    {psim, phasepsi}
};
```

## particles.m

```
ParticleDefinitions[GaugeES] = {
    (* new fields *)
    {phip, {Description -> "BSM field ϕ+",
            LaTeX -> "\\phi^+",
            OutputName -> "phip",
            ElectricCharge -> 1
        }
    },
    {phi0, {Description -> "BSM field ϕ0",
            LaTeX -> "\\phi^0",
            OutputName -> "phi0",
            ElectricCharge -> 0
        }
    },

    {CPsi0, {Description -> "BSM field Ψ0",
            LaTeX -> "\\Psi^0",
            OutputName -> "CPsi0",
            ElectricCharge -> 0
        }
    },
    {psipp, {Description -> "BSM field ψ'+",
            LaTeX -> "\\psi'^+",
            OutputName -> "psipp",
            ElectricCharge -> 1
        }
    },
    {psip0, {Description -> "BSM field ψ'0",
            LaTeX -> "\\psi'^0",
            OutputName -> "psip0",
            ElectricCharge -> 0
        }
    },
    {psi0, {Description -> "BSM field ψ0",
            LaTeX -> "\\psi^0",
            OutputName -> "psi0",
```

```
            ElectricCharge -> 0
        }
    },
    {psim, {Description -> "BSM field ψ-",
            LaTeX -> "\\psi^-",
            OutputName -> "psim",
            ElectricCharge -> -1
        }
    },

    (* Standard Model *)
    {H0, {
            PDG -> {0},
            Width -> 0,
            Mass -> Automatic,
            FeynArtsNr -> 1,
            LaTeX -> "H^0",
            OutputName -> "H0"
        }
    },
    {Hp, {
            PDG -> {0},
            Width -> 0,
            Mass -> Automatic,
            FeynArtsNr -> 2,
            LaTeX -> "H^+",
            OutputName -> "Hp"
        }
    },

    {VB,  {Description -> "B-Boson"}},
    {VG,  {Description -> "Gluon"}},
    {VWB, {Description -> "W-Bosons"}},
    {gB,  {Description -> "B-Boson Ghost"}},
    {gG,  {Description -> "Gluon Ghost"}},
    {gWB, {Description -> "W-Boson Ghost"}}
};

ParticleDefinitions[EWSB] = {
    (* new fields *)
    (* to be completed manually *)
    {phip, {Description -> "BSM field ϕ+",
            LaTeX -> "\\phi^+",
            OutputName -> "php",
            PDG -> {901},
            FeynArtsNr -> 901,
            ElectricCharge -> 1
        }
    },
    {phi0, {Description -> "BSM field ϕ0",
            LaTeX -> "\\phi^0",
            OutputName -> "ph0",
            PDG -> {902},
            FeynArtsNr -> 902,
            ElectricCharge -> 0
```

```
        }
    },

    {Fpsi, {Description -> "BSM charged (-1) Dirac spinor Fψ",
            LaTeX -> "\\psi",
            OutputName -> "psi",
            PDG -> {903},
            FeynArtsNr -> 903,
            ElectricCharge -> -1
        }
    },
    {Fchi, {Description -> "BSM neutral Majorana spinor Fχ",
            LaTeX -> "\\chi",
            OutputName -> "ch",
            PDG -> {911, 912, 913},
            FeynArtsNr -> 911,
            ElectricCharge -> 0
        }
    },

    {Fnu,  {Description -> "Neutrinos",
            Mass -> LesHouches
        }
    },

    (* Standard Model *)
    {hh,   {Description -> "Higgs",
            PDG -> {25},
            PDG.IX -> {101000001}
        }
    },
    {Ah,   {Description -> "Pseudo-Scalar Higgs",
            PDG -> {0},
            PDG.IX -> {0},
            Mass -> {0},
            Width -> {0}
        }
    },
    {Hp,   {Description -> "Charged Higgs",
            PDG -> {0},
            PDG.IX -> {0},
            Width -> {0},
            Mass -> {0},
            LaTeX -> {"H^+", "H^-"},
            OutputName -> {"Hp", "Hm"},
            ElectricCharge -> 1
        }
    },

    {VP,   {Description -> "Photon"}},
    {VZ,   {Description -> "Z-Boson",
            Goldstone -> Ah
        }
    },
    {VG,   {Description -> "Gluon"}},
```

```
    {VWp,  {Description -> "W+ - Boson",
            Goldstone -> Hp
        }
    },
    {gP,   {Description -> "Photon Ghost"}},
    {gWp,  {Description -> "Positive W+ - Boson Ghost"}},
    {gWpC, {Description -> "Negative W+ - Boson Ghost"}},
    {gZ,   {Description -> "Z-Boson Ghost"}},
    {gG,   {Description -> "Gluon Ghost"}},

    {Fd,   {Description -> "Down-Quarks"}},
    {Fu,   {Description -> "Up-Quarks"}},
    {Fe,   {Description -> "Leptons"}}
};

WeylFermionAndIndermediate = {
    (* new fields *)
    {CPsi, {LaTeX -> "\\CPsi"}},
    {psip, {LaTeX -> "\\psi'"}},
    {phi, {LaTeX -> "\\phi"}},

    {chi, {LaTeX -> "\\chi"}},
    (* Standard Model *)
    {H, {
            PDG -> {0},
            Width -> 0,
            Mass -> Automatic,
            LaTeX -> "H",
            OutputName -> ""
        }
    },

    {dR,  {LaTeX -> "d_R"}},
    {eR,  {LaTeX -> "e_R"}},
    {l,   {LaTeX -> "l"}},
    {uR,  {LaTeX -> "u_R"}},
    {q,   {LaTeX -> "q"}},
    {eL,  {LaTeX -> "e_L"}},
    {dL,  {LaTeX -> "d_L"}},
    {uL,  {LaTeX -> "u_L"}},
    {vL,  {LaTeX -> "\\nu_L"}},

    {DR,  {LaTeX -> "D_R"}},
    {ER,  {LaTeX -> "E_R"}},
    {UR,  {LaTeX -> "U_R"}},
    {EL,  {LaTeX -> "E_L"}},
    {DL,  {LaTeX -> "D_L"}},
    {UL,  {LaTeX -> "U_L"}}
};
```

## parameters.m

```
ParameterDefinitions = {
    (* new parameters *)
    {mϕ2, {Description -> "BSM parameter mϕ^2",
```

```
            LaTeX -> "m_\\phi^2",
            OutputName -> "mphi2",
            LesHouches -> {"T13B", 11},
            Real -> True
        }
    },
    {mΨ, {Description -> "BSM parameter mΨ",
            LaTeX -> "m_\\Psi",
            OutputName -> "mCPsi",
            LesHouches -> {"T13B", 12},
            Real -> True (* not required (Real -> False: CP violation) *)
        }
    },
    {mψψp, {Description -> "BSM parameter mψψ'",
            LaTeX -> "m_{\\psi\\psi'}",
            OutputName -> "mpp",
            LesHouches -> {"T13B", 13},
            Real -> True (* not required (Real -> False: CP violation) *)
        }
    },
    {λ1, {Description -> "BSM parameter λ1",
            LaTeX -> "\\lambda_1",
            OutputName -> "lam1",
            LesHouches -> {"T13B", 21},
            Real -> True
        }
    },
    {λ2, {Description -> "BSM parameter λ2",
            LaTeX -> "\\lambda_2",
            OutputName -> "lam2",
            LesHouches -> {"T13B", 22},
            Real -> True
        }
    },
    {λ3, {Description -> "BSM parameter λ3",
            LaTeX -> "\\lambda_3",
            OutputName -> "lam3",
            LesHouches -> {"T13B", 23},
            Real -> True
        }
    },
    {λ4, {Description -> "BSM parameter λ4",
            LaTeX -> "\\lambda_4",
            OutputName -> "lam4",
            LesHouches -> {"T13B", 24},
            Real -> True (* not required (Real -> False: CP violation) *)
        }
    },
    {λ5, {Description -> "BSM parameter λ5",
            LaTeX -> "\\lambda_5",
            OutputName -> "lam5",
            LesHouches -> {"T13B", 25},
            Real -> True (* not required (Real -> False: CP violation) *)
        }
    },
```

```
{λ6, {Description -> "BSM parameter λ6",
        LaTeX -> "\\lambda_6",
        OutputName -> "lam6",
        LesHouches -> "LAM6",
        Real -> True (* not required (Real -> False: CP violation) *)
    }
},
{phasepsi, {Description -> "BSM phase for fermion component ψ-",
        LaTeX -> "p",
        OutputName -> "ppsi",
        LesHouches -> {"T13B", 30}
    }
},

{Uferm, {Description -> "Neutral fermion mixing matrix",
        LaTeX -> "U_\\chi",
        OutputName -> "Ufrm",
        LesHouches -> "CHIMIX"
    }
},

{Uneu, {Description -> "Neutrino mixing matrix",
        LaTeX -> "U_\\nu",
        OutputName -> "Uneu",
        LesHouches -> "NUMIX"
    }
},

(* Standard Model parameters *)
{g1,        {Description -> "Hypercharge-Coupling"}},
{g2,        {Description -> "Left-Coupling"}},
{g3,        {Description -> "Strong-Coupling"}},
{AlphaS,    {Description -> "Alpha Strong"}},
{e,         {Description -> "electric charge"}},

{Gf,        {Description -> "Fermi's constant"}},
{aEWinv,    {Description -> "inverse weak coupling constant at mZ"}},

{Yu,        {Description -> "Up-Yukawa-Coupling",
                DependenceNum -> Sqrt[2]/v * {
                    {Mass[Fu,1], 0, 0},
                    {0, Mass[Fu,2], 0},
                    {0, 0, Mass[Fu,3]}
                }
            }
},
{Yd,        {Description -> "Down-Yukawa-Coupling",
                DependenceNum -> Sqrt[2]/v * {
                    {Mass[Fd,1], 0, 0},
                    {0, Mass[Fd,2], 0},
                    {0, 0, Mass[Fd,3]}
                }
            }
},
{Ye,        {Description -> "Lepton-Yukawa-Coupling",
```

```
                    DependenceNum -> Sqrt[2]/v * {
                        {Mass[Fe,1], 0, 0},
                        {0, Mass[Fe,2], 0},
                        {0, 0, Mass[Fe,3]}
                    }
                }
    },


    {mu2,       {Description -> "SM Mu Parameter"}},
    {\[Lambda], {Description -> "SM Higgs Selfcouplings",
                    DependenceNum -> Mass[hh]^2/(v^2)
                }
    },
    {v,         {Description -> "EW-VEV",
                    DependenceNum -> Sqrt[4 * Mass[VWp]^2/(g2^2)],
                    DependenceSPheno -> None
                }
    },
    {mH2,       {Description -> "SM Higgs Mass Parameter"}},

    {ThetaW,    {Description -> "Weinberg-Angle",
                    DependenceNum -> ArcSin[Sqrt[1 - Mass[VWp]^2/Mass[VZ]^2]]
                }
    },

    {ZZ,        {Description -> "Photon-Z Mixing Matrix"}},
    {ZW,        {Description -> "W Mixing Matrix",
                    Dependence -> 1/Sqrt[2] {
                        {1, 1},
                        {\[ImaginaryI], -\[ImaginaryI]}
                    }
                }
    },

    {Vu,        {Description -> "Left-Up-Mixing-Matrix"}},
    {Vd,        {Description -> "Left-Down-Mixing-Matrix"}},
    {Uu,        {Description -> "Right-Up-Mixing-Matrix"}},
    {Ud,        {Description -> "Right-Down-Mixing-Matrix"}},
    {Ve,        {Description -> "Left-Lepton-Mixing-Matrix"}},
    {Ue,        {Description -> "Right-Lepton-Mixing-Matrix"}}
};
```

**SPheno.m**

```
OnlyLowEnergySPheno = True;
(* add Block TREELEVELUNITARITY to SPheno output spectrum file *)
(* AddTreeLevelUnitarityLimits = True; *)

MINPAR = {
    {1, lambdaInput},
    (*{2, mHInput},*)
    {11, mphiInput},
    {12, mCPsiInput},
    {13, mpsipsipInput},
```

```
    {21, lambda1Input},
    (*{22, lambda2Input},*)
    {23, lambda3Input},
    {24, lambda4Input},
    {25, lambda5Input}
};

BoundaryLowScaleInput = {
    {mϕ2, mphiInput^2},
    {mΨ, mCPsiInput},
    {mψψp, mpsipsipInput},
    {λ1, lambda1Input},
    (*{λ2, lambda2Input},*)
    {λ3, lambda3Input},
    {λ4, lambda4Input},
    {λ5, lambda5Input},
    {λ6, LHInput[λ6]},
    (* Standard Model *)
    {λ, lambdaInput}
    (*{λ, mHInput^2 / vSM^2}*)
};

DEFINITION[MatchingConditions] = {
    {Ye, YeSM},
    {Yd, YdSM},
    {Yu, YuSM},
    {g1, g1SM},
    {g2, g2SM},
    {g3, g3SM},
    {v, vSM}
};

DefaultInputValues = {
    mphiInput -> 500,
    mCPsiInput -> 610,
    mpsipsipInput -> 720,
    lambda1Input -> 0.011,
    (*lambda2Input -> 0.012,*)
    lambda3Input -> 0.013,
    lambda4Input -> 0.014,
    lambda5Input -> 0.015,
    λ6[1] -> 0.016,
    λ6[2] -> 0.016,
    λ6[3] -> 0.016,
    (* Standard Model *)
    (*mHInput -> 125.09,*) (* see PDG *)
    lambdaInput -> 0.2612 (* ≈ 125.09² / 244.74² *)
};

ParametersToSolveTadpoles = {mu2};

ListDecayParticles = {Fu, Fe, Fd, hh, Hp, phip, phi0, Fchi, Fpsi};
ListDecayParticles3B = {{Fu, "Fu.f90"}, {Fe, "Fe.f90"}, {Fd, "Fd.f90"}};
```

## G.2. T1-3-B ($\alpha = 0$) with two triplet generations

### T1-3-B_0_scalar2g.m

```
Model`Name = "T13Balpha02g";
Model`NameLaTeX = "T1-3-B (\\alpha = 0) with 2 scalar generations";
Model`Authors = "Simon May";
Model`Date = "2018-04-21";

(*---------------------------------------------*)
(*   particle content                          *)
(*---------------------------------------------*)

(* global symmetries *)
(* discrete 2 symmetry *)
Global[[1]] = {Z[2], Z2};

(* gauge groups *)
Gauge[[1]] = {B,   U[1], hypercharge, g1, False, 1};
Gauge[[2]] = {WB, SU[2], left,        g2, True,  1};
Gauge[[3]] = {G,  SU[3], color,       g3, False, 1};

(* matter fields *)
(*                   {name, gens, components, Y/2,  SU(2), SU(3), global} *)
(* Standard Model *)
FermionFields[[1]] = {q,    3,    {uL, dL},    1/6, 2,     3,     1};
FermionFields[[2]] = {l,    3,    {vL, eL},   -1/2, 2,     1,     1};
FermionFields[[3]] = {d,    3,    conj[dR],    1/3, 1,    -3,     1};
FermionFields[[4]] = {u,    3,    conj[uR],   -2/3, 1,    -3,     1};
FermionFields[[5]] = {e,    3,    conj[eR],      1, 1,     1,     1};

ScalarFields[[1]]  = {H,    1,    {Hp, H0},    1/2, 2,     1,     1};

(* new fields *)
FermionFields[[6]] = {CPsi, 1, CPsi0, 0, 1, 1, -1};
FermionFields[[7]] = {psip, 1, {psipp, psip0}, 1/2, 2, 1, -1};
FermionFields[[8]] = {psi, 1, {psi0, psim}, -1/2, 2, 1, -1};
ScalarFields[[2]]  = {phi, 2, {{phi0/Sqrt[2], phip}, {conj[phip], -phi0/Sqrt[2]}}, 0, 3, 1, -1};
RealScalars = {phi0};


(*------------------------------------------------*)
(*   DEFINITION                                   *)
(*------------------------------------------------*)
NameOfStates = {GaugeES, EWSB};

(* ----- before EWSB ----- *)
DEFINITION[GaugeES][LagrangianInput] = {
    {LagHC,      {AddHC -> True }},
    {LagNoHC,    {AddHC -> False}},
    {LagBSMHC,   {AddHC -> True }},
    {LagBSMNoHC, {AddHC -> False}}
};

```

```
(* Standard Model Lagrangian *)
LagHC      = -Yd conj[H].d.q - Ye conj[H].e.l - Yu u.q.H;
LagNoHC    = mu2 conj[H].H - 1/2 λ conj[H].H.conj[H].H;

(* Lagrangian involving the new fields *)
(* can't include λ3 term with four indices - see
http://stauby.de/sarah_userforum/viewtopic.php?f=4&t=402 *)
LagBSMNoHC = - 1/2 mϕ2 phi.phi \
     - λ1 conj[H].H.phi.phi (*- λ2 conj[H].phi.phi.H*) (*- λ3 phi.phi.phi.phi*);
LagBSMHC   = - 1/2 mΨ CPsi.CPsi - mψψp psi.psip \
     - λ4 conj[H].psip.CPsi - λ5 psi.H.CPsi - λ6 l.phi.psip;

(* ----- after EWSB ----- *)
(* gauge sector mixing *)
DEFINITION[EWSB][GaugeSector] = {
    {{VB,     VWB[3]}, {VP, VZ}, ZZ},
    {{VWB[1], VWB[2]}, {VWp, conj[VWp]}, ZW}
};

(* VEVs *)
DEFINITION[EWSB][VEVs] = {
    (* Standard Model Higgs VEV *)
    {H0,
        {v, 1/Sqrt[2]},
        {Ah, \[ImaginaryI]/Sqrt[2]},
        {hh, 1/Sqrt[2]}
    }
};

(* mixing *)
DEFINITION[EWSB][MatterSector] = {
    (* Standard Model mixing *)
    {{{dL}, {conj[dR]}}, {{DL, Vd}, {DR, Ud}}},
    {{{uL}, {conj[uR]}}, {{UL, Vu}, {UR, Uu}}},
    {{{eL}, {conj[eR]}}, {{EL, Ve}, {ER, Ue}}},
    (* mixing of new fields can be added here *)
    {{vL}, {VL, Uneu}},
    {{phip}, {etap, Uetp}},
    {{phi0}, {eta0, Uet0}},
    {{CPsi0, psi0, psip0}, {chi, Uferm}}
};

(* Dirac spinors *)
DEFINITION[EWSB][DiracSpinors] = {
    Fd -> {DL, conj[DR]},
    Fe -> {EL, conj[ER]},
    Fu -> {UL, conj[UR]},
    Fnu -> {VL, conj[VL]},
    (* new fields *)
    Fpsi -> {psim, conj[psipp]},
    Fchi -> {chi, conj[chi]}
};

(* phases, see http://stauby.de/sarah_wiki/index.php?title=Phases *)
DEFINITION[EWSB][Phases] = {
```

```
    {psim, phasepsi}
};
```

## particles.m

```
ParticleDefinitions[GaugeES] = {
    (* new fields *)
    {CPsi0, {Description -> "BSM field Ψ0",
            OutputName -> "CPsi0",
            LaTeX -> "\\Psi^0"
        }
    },
    {psipp, {Description -> "BSM field ψ'+",
            OutputName -> "psipp",
            LaTeX -> "\\psi'^+"
        }
    },
    {psip0, {Description -> "BSM field ψ'0",
            OutputName -> "psip0",
            LaTeX -> "\\psi'^0"
        }
    },
    {phip, {Description -> "BSM field ϕ+",
            OutputName -> "phip",
            LaTeX -> "\\phi^+",
            ElectricCharge -> 1
        }
    },
    {phi0, {Description -> "BSM field ϕ0",
            OutputName -> "phi0",
            LaTeX -> "\\phi^0",
            ElectricCharge -> 0
        }
    },
    {psi0, {Description -> "BSM field ψ0",
            OutputName -> "psi0",
            LaTeX -> "\\psi^0"
        }
    },
    {psim, {Description -> "BSM field ψ-",
            OutputName -> "psim",
            LaTeX -> "\\psi^-"
        }
    },

    (* Standard Model *)
    {H0, {
            PDG -> {0},
            Width -> 0,
            Mass -> Automatic,
            FeynArtsNr -> 1,
            LaTeX -> "H^0",
            OutputName -> "H0"
        }
    },
```

```
    {Hp, {
            PDG -> {0},
            Width -> 0,
            Mass -> Automatic,
            FeynArtsNr -> 2,
            LaTeX -> "H^+",
            OutputName -> "Hp"
        }
    },

    {VB,  {Description -> "B-Boson"}},
    {VG,  {Description -> "Gluon"}},
    {VWB, {Description -> "W-Bosons"}},
    {gB,  {Description -> "B-Boson Ghost"}},
    {gG,  {Description -> "Gluon Ghost"}},
    {gWB, {Description -> "W-Boson Ghost"}}
};

ParticleDefinitions[EWSB] = {
    (* new fields *)
    (* to be completed manually *)
    {etap, {Description -> "BSM field η+",
            OutputName -> "etp",
            LaTeX -> "\\eta^+",
            PDG -> {921, 922},
            FeynArtsNr -> 921,
            ElectricCharge -> 1
        }
    },
    {eta0, {Description -> "BSM neutral field η0",
            OutputName -> "et0",
            LaTeX -> "\\eta^0",
            PDG -> {931, 932},
            FeynArtsNr -> 931,
            ElectricCharge -> 0
        }
    },

    {Fpsi, {Description -> "BSM charged (-1) Dirac spinor Fψ",
            OutputName -> "psi",
            LaTeX -> "\\psi",
            PDG -> {903},
            FeynArtsNr -> 903,
            ElectricCharge -> -1
        }
    },
    {Fchi, {Description -> "BSM neutral Majorana spinor Fχ",
            OutputName -> "ch",
            LaTeX -> "\\chi",
            PDG -> {911, 912, 913},
            FeynArtsNr -> 911,
            ElectricCharge -> 0
        }
    },

```

```
    {Fnu,  {Description -> "Neutrinos",
            Mass -> LesHouches
        }
    },

    (* Standard Model *)
    {hh,   {Description -> "Higgs",
            PDG -> {25},
            PDG.IX -> {101000001}
        }
    },
    {Ah,   {Description -> "Pseudo-Scalar Higgs",
            PDG -> {0},
            PDG.IX -> {0},
            Mass -> {0},
            Width -> {0}
        }
    },
    {Hp,   {Description -> "Charged Higgs",
            PDG -> {0},
            PDG.IX -> {0},
            Width -> {0},
            Mass -> {0},
            LaTeX -> {"H^+", "H^-"},
            OutputName -> {"Hp", "Hm"},
            ElectricCharge -> 1
        }
    },

    {VP,   {Description -> "Photon"}},
    {VZ,   {Description -> "Z-Boson",
            Goldstone -> Ah
        }
    },
    {VG,   {Description -> "Gluon"}},
    {VWp,  {Description -> "W+ - Boson",
            Goldstone -> Hp
        }
    },
    {gP,   {Description -> "Photon Ghost"}},
    {gWp,  {Description -> "Positive W+ - Boson Ghost"}},
    {gWpC, {Description -> "Negative W+ - Boson Ghost"}},
    {gZ,   {Description -> "Z-Boson Ghost"}},
    {gG,   {Description -> "Gluon Ghost"}},

    {Fd,   {Description -> "Down-Quarks"}},
    {Fu,   {Description -> "Up-Quarks"}},
    {Fe,   {Description -> "Leptons"}}
};

WeylFermionAndIndermediate = {
    (* new fields *)
    {phip, {LaTeX -> "\\phi^+"}},
    {phi0, {LaTeX -> "\\phi^0"}},

```

```
    {CPsi, {LaTeX -> "\\CPsi"}},
    {psip, {LaTeX -> "\\psi'"}},
    {phi, {LaTeX -> "\\phi"}},

    {chi, {LaTeX -> "\\chi"}},
    (* Standard Model *)
    {H, {
            PDG -> {0},
            Width -> 0,
            Mass -> Automatic,
            LaTeX -> "H",
            OutputName -> ""
        }
    },

    {dR,  {LaTeX -> "d_R"}},
    {eR,  {LaTeX -> "e_R"}},
    {l,   {LaTeX -> "l"}},
    {uR,  {LaTeX -> "u_R"}},
    {q,   {LaTeX -> "q"}},
    {eL,  {LaTeX -> "e_L"}},
    {dL,  {LaTeX -> "d_L"}},
    {uL,  {LaTeX -> "u_L"}},
    {vL,  {LaTeX -> "\\nu_L"}},

    {DR,  {LaTeX -> "D_R"}},
    {ER,  {LaTeX -> "E_R"}},
    {UR,  {LaTeX -> "U_R"}},
    {EL,  {LaTeX -> "E_L"}},
    {DL,  {LaTeX -> "D_L"}},
    {UL,  {LaTeX -> "U_L"}}
};
```

## parameters.m

```
ParameterDefinitions = {
    (* new parameters *)
    {mϕ2, {Description -> "BSM parameter mϕ^2",
            LaTeX -> "m_\\phi^2",
            OutputName -> "mph2",
            LesHouches -> "MPHI2",
            Real -> True
        }
    },
    {mΨ, {Description -> "BSM parameter mΨ",
            LaTeX -> "m_\\Psi",
            OutputName -> "mCPsi",
            LesHouches -> {"T13B", 12},
            Real -> True (* not required (Real -> False: CP violation) *)
        }
    },
    {mψψp, {Description -> "BSM parameter mψψ'",
            LaTeX -> "m_{\\psi\\psi'}",
            OutputName -> "mpp",
            LesHouches -> {"T13B", 13},
```

```
            Real -> True (* not required (Real -> False: CP violation) *)
        }
    },
    {λ1, {Description -> "BSM parameter λ1",
            LaTeX -> "\\lambda_1",
            OutputName -> "lam1",
            LesHouches -> "LAM1",
            Real -> True
        }
    },
    {λ2, {Description -> "BSM parameter λ2",
            LaTeX -> "\\lambda_2",
            OutputName -> "lam2",
            LesHouches -> "LAM2",
            Real -> True
        }
    },
    {λ3, {Description -> "BSM parameter λ3",
            LaTeX -> "\\lambda_3",
            OutputName -> "l3",
            LesHouches -> "LAM3",
            Real -> True
        }
    },
    {λ4, {Description -> "BSM parameter λ4",
            LaTeX -> "\\lambda_4",
            OutputName -> "lam4",
            LesHouches -> {"T13B", 24},
            Real -> True (* not required (Real -> False: CP violation) *)
        }
    },
    {λ5, {Description -> "BSM parameter λ5",
            LaTeX -> "\\lambda_5",
            OutputName -> "lam5",
            LesHouches -> {"T13B", 25},
            Real -> True (* not required (Real -> False: CP violation) *)
        }
    },
    {λ6, {Description -> "BSM parameter λ6",
            LaTeX -> "\\lambda_6",
            OutputName -> "lam6",
            LesHouches -> "LAM6",
            Real -> True (* not required (Real -> False: CP violation) *)
        }
    },
    {phasepsi, {Description -> "BSM phase for fermion component ψ-",
            LaTeX -> "p",
            OutputName -> "ppsi",
            LesHouches -> {"T13B", 30}
        }
    },

    (* mixing matrices must have the same Mathematica symbol and OutputName? See
    http://stauby.de/sarah_userforum/viewtopic.php?f=4&t=403 *)
    {Uetp, {Description -> "BSM charged (+1) scalar mixing matrix",
```

```
            LaTeX -> "U_{\\eta^+}",
            OutputName -> "Uetp",
            LesHouches -> "ETAPMIX"
        }
    },
    {Uet0, {Description -> "BSM neutral scalar mixing matrix",
            LaTeX -> "U_{\\eta^0}",
            OutputName -> "Uet0",
            LesHouches -> "ETA0MIX"
        }
    },
    {Uferm, {Description -> "BSM neutral fermion mixing matrix",
            LaTeX -> "U_\\chi",
            OutputName -> "Ufrm",
            LesHouches -> "CHIMIX"
        }
    },

    {Uneu, {Description -> "Neutrino mixing matrix",
            LaTeX -> "U_\\nu",
            OutputName -> "Uneu",
            LesHouches -> "NUMIX"
        }
    },

    (* Standard Model parameters *)
    {g1,        {Description -> "Hypercharge-Coupling"}},
    {g2,        {Description -> "Left-Coupling"}},
    {g3,        {Description -> "Strong-Coupling"}},
    {AlphaS,    {Description -> "Alpha Strong"}},
    {e,         {Description -> "electric charge"}},

    {Gf,        {Description -> "Fermi's constant"}},
    {aEWinv,    {Description -> "inverse weak coupling constant at mZ"}},

    {Yu,        {Description -> "Up-Yukawa-Coupling",
                    DependenceNum -> Sqrt[2]/v * {
                        {Mass[Fu,1], 0, 0},
                        {0, Mass[Fu,2], 0},
                        {0, 0, Mass[Fu,3]}
                    }
                }
    },
    {Yd,        {Description -> "Down-Yukawa-Coupling",
                    DependenceNum -> Sqrt[2]/v * {
                        {Mass[Fd,1], 0, 0},
                        {0, Mass[Fd,2], 0},
                        {0, 0, Mass[Fd,3]}
                    }
                }
    },
    {Ye,        {Description -> "Lepton-Yukawa-Coupling",
                    DependenceNum -> Sqrt[2]/v * {
                        {Mass[Fe,1], 0, 0},
                        {0, Mass[Fe,2], 0},
```

```
                        {0, 0, Mass[Fe,3]}
                    }
                }
    },


    {mu2,         {Description -> "SM Mu Parameter"}},
    {\[Lambda], {Description -> "SM Higgs Selfcouplings",
                    DependenceNum -> Mass[hh]^2/(v^2)
                }
    },
    {v,           {Description -> "EW-VEV",
                    DependenceNum -> Sqrt[4 * Mass[VWp]^2/(g2^2)],
                    DependenceSPheno -> None
                }
    },
    {mH2,         {Description -> "SM Higgs Mass Parameter"}},

    {ThetaW,      {Description -> "Weinberg-Angle",
                    DependenceNum -> ArcSin[Sqrt[1 - Mass[VWp]^2/Mass[VZ]^2]]
                }
    },

    {ZZ,          {Description -> "Photon-Z Mixing Matrix"}},
    {ZW,          {Description -> "W Mixing Matrix",
                    Dependence -> 1/Sqrt[2] {
                        {1, 1},
                        {\[ImaginaryI], -\[ImaginaryI]}
                    }
                }
    },

    {Vu,          {Description -> "Left-Up-Mixing-Matrix"}},
    {Vd,          {Description -> "Left-Down-Mixing-Matrix"}},
    {Uu,          {Description -> "Right-Up-Mixing-Matrix"}},
    {Ud,          {Description -> "Right-Down-Mixing-Matrix"}},
    {Ve,          {Description -> "Left-Lepton-Mixing-Matrix"}},
    {Ue,          {Description -> "Right-Lepton-Mixing-Matrix"}}
};
```

## SPheno.m

```
OnlyLowEnergySPheno = True;
(* add Block TREELEVELUNITARITY to SPheno output spectrum file *)
(* AddTreeLevelUnitarityLimits = True; *)

MINPAR = {
    {1, lambdaInput},
    (*{2, mHInput},*)
    {12, mCPsiInput},
    {13, mpsipsipInput},
    {24, lambda4Input},
    {25, lambda5Input}
};

```

```
BoundaryLowScaleInput = {
    {mϕ2, LHInput[mϕ2]},
    {mΨ, mCPsiInput},
    {mψψp, mpsipsipInput},
    {λ1, LHInput[λ1]},
    (*{λ2, LHInput[λ2]},*)
    (*{λ3, LHInput[λ3]},*)
    {λ4, lambda4Input},
    {λ5, lambda5Input},
    {λ6, LHInput[λ6]},
    (* Standard Model *)
    {λ, lambdaInput}
    (*{λ, mHInput^2 / vSM^2}*)
};

DEFINITION[MatchingConditions] = {
    {Ye, YeSM},
    {Yd, YdSM},
    {Yu, YuSM},
    {g1, g1SM},
    {g2, g2SM},
    {g3, g3SM},
    {v, vSM}
};

DefaultInputValues = {
    mϕ2[1, 1] -> 510^2,
    mϕ2[2, 2] -> 520^2,
    mCPsiInput -> 610,
    mpsipsipInput -> 720,
    lambda4Input -> 0.014,
    lambda5Input -> 0.015,
    λ6[1, 1] -> 0.016,
    λ6[2, 1] -> 0.016,
    λ6[3, 1] -> 0.016,
    λ6[1, 2] -> 0.016,
    λ6[2, 2] -> 0.016,
    λ6[3, 2] -> 0.016,
    (* Standard Model *)
    (*mHInput -> 125.09,*) (* see PDG *)
    lambdaInput -> 0.2612 (* ≈ 125.09² / 244.74² *)
};

ParametersToSolveTadpoles = {mu2};

ListDecayParticles = {Fu, Fe, Fd, hh, Hp, etap, eta0, Fchi, Fpsi};
ListDecayParticles3B = {{Fu, "Fu.f90"}, {Fe, "Fe.f90"}, {Fd, "Fd.f90"}};
```

## Acknowledgements

I would like to thank Sonja Esch for countless interesting and useful discussion as well as a fun time in the office, Karol Kovařík for answering many questions and a lot of enlightening information, and Prof. Michael Klasen for offering me to work on this thesis in his group. Moreover, I am grateful to Valentin Kunz for discussions on some mathematical details as well as Alexandra Everwand for her constant support.